\documentclass[prd,aps,nofootinbib,floatfix,11pt]{revtex4}

\usepackage{amsmath,graphicx,epsfig,amssymb,dsfont,mathtools,hyperref}
\usepackage[usenames]{color}
\usepackage{ulem} 
\usepackage{bigstrut}
\usepackage{slashed}
\usepackage{multirow}
\usepackage{makecell}
\usepackage{verbatim}
\usepackage{amsmath,graphicx,color,epsfig}
\usepackage{longtable}
  \allowdisplaybreaks[1]

\allowdisplaybreaks
\renewcommand{\thesubsection}{\Roman{subsection}}

\begin{document}
\title{Weak Decays of Stable Open-bottom Tetraquark by SU(3) Symmetry Analysis }
\author{
Ye Xing$^1$~\footnote{Email:xingye\_guang@sjtu.edu.cn}, Fu-Sheng Yu$^{2,3}$~\footnote{Email:yufsh@lzu.edu.cn}, Ruilin Zhu$^{4}$~\footnote{Email:rlzhu@njnu.edu.cn}
}
\affiliation{
 $^1$ INPAC,  Shanghai Key Laboratory for Particle Physics and Cosmology, MOE Key Laboratory for Particle Physics, School of Physics and Astronomy, \\ Shanghai Jiao Tong University, Shanghai  200240, China}
\affiliation{
$^2$ School of Nuclear Science and Technology, Lanzhou University, Lanzhou 730000, China\\
$^3$ Research Center for Hadron and CSR Physics, Lanzhou University and Institute of Modern Physics of CAS, Lanzhou 730000, China
 }

 \affiliation{
$^4$Department of Physics and Institute of Theoretical Physics,
Nanjing Normal University, Nanjing, Jiangsu 210023, China
 }

\begin{abstract}
The exotic state $X(5568)$ which was observed by D0 Collaboration is very likely to be a tetraquark state with four different valence quark flavors, though the existence was not confirmed by other collaborations. The possibility of such state still generate lots of interests in theory. In the paper, we will study the properties of the state under the SU(3) flavor symmetry. This four quark state with a heavy bottom quark and three light quarks(anti-quark) can form a $6$ or $\overline {15}$ representation. The weak decays can be dominant and should be discussed carefully while such state is stable against the strong interaction. Therefor we will study the multi-body semileptonic and nonleptonic weak decays systematically. With the help of SU(3) flavor symmetry, we can give the Hamiltonian in the hadronic level, then obtain the parameterized irreducible amplitudes and the relations of different channels. At the end of the article, we collect some Cabibbo allowed two-body and three-body weak decay channels which can be used to reconstruct $X_{b6}$ states at the branching fraction up to be $10^{-5}$.
\end{abstract}

\maketitle

\section{Introduction}
The naive quark model in which the hadron is composed of  a quark-antiquark pair or triquarks has achieved lots of successes over the past fifty years.
Quantum Chromodynamics (QCD) theory also allows the existence of the exotic states such as tetraquarks, pentaquarks, moleculars and baryonium.
Such exotic states attract wide interests in both experiments and theories (see the recent reviews in \cite{Chen:2016qju,Olsen:2014qna,Brambilla:2010cs,Ali:2017jda,Esposito:2016noz,Lebed:2016hpi,Guo:2017jvc}).
Current experimental data strongly indicated the possibility of the existence of hidden heavy flavor tetraquarks and pentaquarks.
The discovery of X(3872) in $B^{\pm}\to K^{\pm}X(X(3872)\to J/\psi \pi^+\pi^-)$ at Belle in 2003~\cite{Choi:2003ue} fired the
first shot in the studies of the hidden charm tetraquarks. The hidden bottom tetraquarks $ Z_b^{\pm}(10610,10650)$ were first observed in $e^+e^-\to \Upsilon(nS)\pi^+\pi^-(n=1,2,3)$ and $e^+e^-\to h_b(nP)\pi^+\pi^-(n=1,2)$ at Belle in 2011~\cite{Belle:2011aa}. Two $P_c$ states discovered in $\Lambda_b^0\to J/\psi K^- p$ at LHCb in 2015 could be treated as the candidates of the hidden charm pentaquarks~\cite{Aaij:2015tga}.

Quite unexpectedly, the D0 Collaboration reported the signal for the tetraquark $X(5568)$ with four different flavors in the decay $X(5568)\to B_s^0\pi^{\pm}$ in 2016~\cite{D0:2016mwd}. However, efforts by LHCb~\cite{Aaij:2016iev}, CMS~\cite{Sirunyan:2017ofq}, CDF~\cite{Aaltonen:2017voc} and ATLAS~\cite{Aaboud:2018hgx} Collaborations to confirm the state provide no supporting evidence for its existence in the identical channel. Recently, the D0 Collaboration have reconfirmed the existence of $X(5568)$ via the semileptonic decays of $B_s$~\cite{Abazov:2017poh}. If the observation of $X(5568)$ is true, it would be the first discovery of tetraquark states with four different valence quark flavors.
The theoretical papers have appeared to study the properties of the open-flavor tetraquark state~\cite{Ali:2016gdg,Agaev:2016mjb,Chen:2016mqt,Wang:2016wkj,
Guo:2016nhb,Wang:2016tsi,Burns:2016gvy,He:2016yhd,Tang:2016pcf,Chen:2017rhl,
Yu:2017pmn,Chen:2018hts,Huang:2019otd}.

The studies on stable open-flavor tetraquarks or pentaquarks are of even higher interests in literatures, especially on the doubly heavy flavor tetraquarks~\cite{Ali:2018xfq,Ali:2018ifm,Yan:2018gik,Ebert:2007rn,Karliner:2017qjm,Eichten:2017ffp,Francis:2016hui,Bicudo:2016ooe,Mehen:2017nrh,Maiani:2017kyi,Xing:2018bqt}, Theoretical calculations on the masses of $bb\bar q_1\bar q_2$ with $q_i=u$, $d$ or $s$ always imply these states are stable against strong decays. However, the productions of such doubly heavy-flavor tetraquark states are too rare in experiments, especially at the current stage \cite{Ali:2018xfq,Ali:2018ifm}.
To search for stable open-flavor tetraquarks,
one of the present authors proposed the possible tetraquark composed of $bs\bar{u}\bar{d}$ in Ref.~\cite{Yu:2017pmn}, which is supported by the quark delocalization color screening model~\cite{Huang:2019otd}. Such stable tetraquark state has a unique advantage in the experimental searches. It has a lifetime as large as ordinary $B$ mesons, so that it decays at a secondary vertex at the proton-proton colliders which rejects most of the backgrounds from the preliminary vertex. Besides, its production is much larger than the $bb$-tetraquarks. Therefore the searches of such stable singly heavy-flavor tetraquarks would be more promising.

To hunting for the possible stable open-bottom tetraquarks, a systematical analysis of decay modes is helpful. A successful example is the discovery of the $\Xi_{cc}^{++}$ via the final state of $\Lambda_c^+ K^- \pi^+\pi^+$ \cite{Aaij:2017ueg}, which was first pointed out as the most favorable mode in \cite{Yu:2017zst}.
Although the theoretical calculation of such multi-body decay modes has to deal with the nonperturbative  strongly-coupled gauge dynamics, some dynamics-screening approaches  always reduce the difficulties.  In this paper, we will adopt the light quark SU(3) flavor symmetry to deal with  the weak decays of the open-bottom tetraquarks. SU(3) flavor symmetry analysis  has been applied into the studies of B meson and heavy baryon decays~\cite{Savage:1989ub,Gronau:1995hm,He:1998rq,He:2000ys,Chiang:2004nm,Li:2007bh,Wang:2009azc,Cheng:2011qh,Hsiao:2015iiu,Lu:2016ogy,He:2016xvd,Wang:2017mqp,Wang:2017azm,Hu:2017dzi,Shi:2017dto,Zhu:2018epc,Wang:2018utj,He:2018php,Wang:2017gxe,Geng:2017mxn,Geng:2018upx} and it provides a general insight for different decay modes.

The open-bottom four quark multiplets $Qq_i\bar q_j\bar q_k$ can form a $\bar{3}$, $6$ or $\overline {15}$ representation in the SU(3) flavor symmetry. It will be seen in the next section that the fifteen-fold states are the excited tetraquark states which can hadronic decay into the the sextet states, and the anti-triplet can usually electromagnetic decay into B meson. Therefore we will study the possible weak decays of sextet states (denoted as $X_{b6}$) on the main body of the paper. If the $X_{b6}$ multiplets can be stable against the strong and electromagnetic decays, the weak decay analysis of such states will be the key tool for searching these exotic states in experiment. By constructing the weak decay Hamiltonian in hadronic level and parameterizing the amplitudes into some irreducible parts, we will give the weak decay amplitude expressions for the $X_{b6}$ states and obtain the relations among different decay channels.

The paper is organized as followed. We give the multiplets of the open-bottom tetraquarks and the related hadrons in the SU(3) flavor symmetry  in Sec.~\ref{sec:particle_multiplet}.  We will discuss the energy thresholds of sextet open-bottom tetraquarks, and list the possible masses from corresponding literatures in Sec.~\ref{sec:Xb6_thresholds}. From Sec.~\ref{sec:semileptonic_decay} to Sec.~\ref{sec:nonleptonic_decay}, we mainly study the semi-leptonic and non-leptonic weak decays of the $X_{b6}$ states. In Sec.~\ref{sec:golden_channels}, we will select some possible golden channels which may be performed in future experiments. We summarize and conclude in the end.

\section{Particle Multiplets}
\label{sec:particle_multiplet}

In the  SU(3) flavor symmetry, the open-bottom tetraquark ($bq_i\bar{q_j}\bar{q_k}$, $q_x=u,d,s$)~\footnote{Throughout the paper, the charge conjugation is assumed and one can get the corresponding properties for $\bar{b}\bar{q_i}q_j q_k$ under the charge conjugation.} can form  $\bar{3}$, $6$ and $\overline {15}$ by the decomposition of $3\bigotimes \bar{3}\bigotimes \bar{3}=\bar{3}\bigoplus \bar{3}\bigoplus 6\bigoplus \overline {15}$.  The sextet tetraquark state can be expressed as $(X_{b6})_{[jk]}^i=\epsilon_{\alpha jk}(X_{b6})^{\{\alpha i\}}=\epsilon^{i\alpha\beta}(X_{b6})_{\{[\alpha\beta],[jk]\}}$ (antisymmetric tensor $\epsilon_{123}=-1\ and\ \epsilon^{123}=1 )$. While the $\overline {15}$ states usually strong decay into the sextets and we will not consider them here. On the other hand, the $\bar{3}$ tetraquarks
have a quark-antiquark pair and can electromagnetic decay into B meson, we will only focus on the sextet tetraquark states. The traceless sextet $(X_{b6})_{j,k}^i$ satisfies the condition that the flavor components are antisymmetric under the exchange of j and k ~\cite{He:2016yhd}, which can be written explicitly
\begin{eqnarray}
&&(X_{b6})_{[2,3]}^1= \frac{1}{\sqrt{2}} X_{u\bar{d}\bar{s}},~(X_{b6})_{[3,1]}^2= \frac{1}{\sqrt{2}} X_{d\bar{s}\bar{u}},~(X_{b6})_{[1,2]}^3= \frac{1}{\sqrt{2}} X_{s\bar{u}\bar{d}},\\\nonumber
&&(X_{b6})_{[1,2]}^1=(X_{b6})_{[2,3]}^3= \frac{1}{2} Y_{(u\bar{u},s\bar{s})\bar{d}},~(X_{b6})_{[3,1]}^1=(X_{b6})_{[2,3]}^2= \frac{1}{2} Y_{(u\bar{u},d\bar{d})\bar{s}},\\\nonumber
&& (X_{b6})_{[1,2]}^2=(X_{b6})_{[3,1]}^3= \frac{1}{2} Y_{(d\bar{d},s\bar{s})\bar{u}}.
\end{eqnarray}

The light pseudo-scalars form an octet plus a singlet, and the octet mesons can be written as
\begin{eqnarray}
 M_{8}=\begin{pmatrix}
 \frac{\pi^0}{\sqrt{2}}+\frac{\eta}{\sqrt{6}}
 &\pi^+ & K^+\\
 \pi^-&-\frac{\pi^0}{\sqrt{2}}+\frac{\eta}{\sqrt{6}}&{K^0}\\
 K^-&\bar K^0 &-2\frac{\eta}{\sqrt{6}}
 \end{pmatrix}.
\end{eqnarray}
For the bottom mesons, they form  two triplets with  $B_i=\left(\begin{array}{ccc} B^-, & \overline B^0, &\overline B^0_s  \end{array} \right)$ and
$\overline B^i=\left(\begin{array}{ccc}  B^+, & B^0, & B^0_s  \end{array} \right).$
For the charm mesons, two similar triplets are given as $D_i=\left(\begin{array}{ccc} D^0, & D^+, & D^+_s  \end{array} \right)$ and $\overline D^i=\left(\begin{array}{ccc}\overline D^0, & D^-, & D^-_s  \end{array} \right)$.

For the baryon states,
the light anti-baryons form an octet and an anti-decuplet representation. The octet can be written as
\begin{eqnarray}
\overline {T_8}= \left(\begin{array}{ccc} \frac{1}{\sqrt{2}}\overline \Sigma^0+\frac{1}{\sqrt{6}}\overline \Lambda^0 & \overline \Sigma^+  &  \overline \Xi^+   \\ \overline \Sigma^-  &  -\frac{1}{\sqrt{2}}\overline \Sigma^0+\frac{1}{\sqrt{6}}\overline \Lambda^0 & \overline \Xi^0 \\ \overline p   & \overline n  & -\sqrt{\frac{2}{3}}\overline \Lambda^0
  \end{array} \right),
\end{eqnarray}
while the light anti-decuplet become
\begin{eqnarray}
(T_{\overline {10}})^{111} &=&  \overline \Delta^{--},\;\;\; (T_{\overline {10}})^{112}= (T_{\overline {10}})^{121}=(T_{\overline {10}})^{211}= \frac{1}{\sqrt3}  \overline \Delta^-,\nonumber\\
(T_{\overline {10}})^{222} &=&   \overline \Delta^{+},\;\;\; (T_{\overline {10}})^{122}= (T_{\overline {10}})^{212}=(T_{\overline {10}})^{221}= \frac{1}{\sqrt3} \overline \Delta^0, \nonumber\\
(T_{\overline {10}})^{113} &=& (T_{\overline {10}})^{131}=(T_{\overline {10}})^{311}= \frac{1}{\sqrt3} \overline \Sigma^{\prime-},\;\;(T_{\overline {10}})^{223} = (T_{\overline {10}})^{232}=(T_{\overline {10}})^{322}= \frac{1}{\sqrt3} \overline \Sigma^{\prime+},\nonumber\\
(T_{\overline {10}})^{123} &=& (T_{\overline {10}})^{132}=(T_{\overline {10}})^{213}=(T_{\overline {10}})^{231}=(T_{\overline {10}})^{312}=(T_{\overline {10}})^{321}= \frac{1}{\sqrt6} \overline \Sigma^{\prime0},\nonumber\\
(T_{\overline {10}})^{133} &=& (T_{\overline {10}})^{313}=(T_{\overline {10}})^{331}= \frac{1}{\sqrt3} \overline \Xi^{\prime0},\;\;(T_{\overline {10}})^{233} = (T_{\overline {10}})^{323}=(T_{\overline {10}})^{332}= \frac{1}{\sqrt3}   \overline \Xi^{\prime+}, \nonumber\\
(T_{\overline {10}})^{333}&=& \overline  \Omega^+.
\end{eqnarray}

For  the anti-charmed anti-baryons, they form a triplet and an anti-sextet
\begin{eqnarray}
 T_{\bf{\bar{c} 3}}= \left(\begin{array}{ccc} 0 & \overline \Lambda_{\bar{c}}^-  &  \overline \Xi_{\bar{c}}^-  \\ -\overline \Lambda_{\bar{c}}^- & 0 & \overline \Xi_{\bar{c}}^0 \\ -\overline \Xi_{\bar{c}}^-   &  -\overline \Xi_{\bar{c}}^0  & 0
  \end{array} \right), \;\;
 T_{\bf{\bar{c}\bar{6}}} = \left(\begin{array}{ccc} \overline \Sigma_{\bar{c}}^{--} &  \frac{1}{\sqrt{2}}\overline \Sigma_{\bar{c}}^-   & \frac{1}{\sqrt{2}} \overline \Xi_{\bar{c}}^{\prime-}\\
  \frac{1}{\sqrt{2}}\overline \Sigma_{\bar{c}}^-& \overline \Sigma_{\bar{c}}^{0} & \frac{1}{\sqrt{2}} \overline \Xi_{\bar{c}}^{\prime0} \\
  \frac{1}{\sqrt{2}} \overline \Xi_{\bar{c}}^{\prime-}   &  \frac{1}{\sqrt{2}} \overline \Xi_{\bar{c}}^{\prime0}  & \overline \Omega_{\bar{c}}^0
  \end{array} \right)\,.
\end{eqnarray}
It is easily to get the components for the singly charmed baryons which form an anti-triplet $T_{\bf{c\bar 3}}$ and a sextet $ T_{\bf{c6}}$. Their explicit expressions can be found in Refs.~\cite{Shi:2017dto,Wang:2017azm,Wang:2018utj,Xing:2018bqt}.

\section{strong decay thresholds for the $X_{b6}$  tetraquarks}
\label{sec:Xb6_thresholds}

In this section, we will look at the strong and electromagnetic decay thresholds for the  $X_{b6}$  tetraquarks. One should note that the states of $X_{s\bar u\bar d}$, $X_{u\bar d\bar s}$ and $X_{d\bar s\bar u}$ are purely open flavor states and can not electromagnetic decay into B meson.
For the other three states of $Y_{(u\bar{u},d\bar{d})\bar{s}}$, $Y_{(u\bar{u},s\bar{s})\bar{d}}$ and $Y_{(d\bar{d},s\bar{s})\bar{u}}$, even though they have a quark-antiquark pair,  they will not
electromagnetic decay into B meson because of  the antisymmetric quark  structure for $X_{b6}$.
 The  $X_{b6}$  tetraquarks may have different spin-parities, and the ground states of the $X_{b6}$ states
 will have the spin-parity with $J^{P}=0^+$. Higher excited  $X_{b6}$ states will strong decay into the ground states. Thus we only focus on the $j^{P}=0^+$ $X_{b6}$ ground states.

The $B K$ mass threshold at $5.77$GeV will influence the decay properties if the mass of $X_{s\bar u\bar d}$ state is higher than $5.77$GeV. The $B_s \pi$ mass threshold at $5.51$GeV will influence the decay properties if the masses of $X_{u\bar d\bar s}$, $X_{d\bar s\bar u}$ and $Y_{(u\bar{u},d\bar{d})\bar{s}}$ states are higher than $5.51$GeV. While for the $Y_{(u\bar{u},s\bar{s})\bar{d}}$ and $Y_{(d\bar{d},s\bar{s})\bar{u}}$ states, their strong decay threshold lies in the $B\pi$ mass at $5.41$GeV. Therefor, it can be assigned a stable tetraquark against strong decays only if the mass is below the strong decay threshold. The stable tetraquark has to decay weakly, but most of weak decay channels can be tested in the future experiment inversely.

In literatures, there are lots of theoretical predictions for these open-bottom tetraquarks.
Using the diquark-antidiquark model, the mass of $X_{s\bar u\bar d}$ state was predicted at 5.637 GeV in Ref.~\cite{Yu:2017pmn} by one of the present authors, while the states of $X_{u\bar d\bar s}$ and $X_{d\bar s\bar u}$ were predicted with the mass at 5.7 GeV in Ref.~\cite{Wang:2016tsi} by another one of the present authors and the collaborator. Using the simple constituent quark model, the the mass of $X_{s\bar u\bar d}$ state was predicted at 6.119 GeV~\cite{Yu:2017pmn}.

The masses of the lowest-lying open-flavor bottom tetraquarks have been explored in recent articles, so we listed them  in Tab.~\ref{tab:mass_tetraquark}.

\begin{table}
\newcommand{\tabincell}[2]{\begin{tabular}{@{}#1@{}}#2\end{tabular}}
\caption{The masses (in GeV) of the lowest-lying open bottom $X_{b6}$ tetraquarks with $J^{P}=0^+$ were predicted by different models such as diquark-antidiquark model~\cite{Wang:2016tsi,Yu:2017pmn}, simple constituent quark model~\cite{Yu:2017pmn}, QCD sum rules~\cite{Chen:2017rhl,Agaev:2016mjb},
quark delocalization color screening model~\cite{Huang:2019otd}. The predictions for the masses of the $Y_{(u\bar{u},d\bar{d})\bar{s}}$, $Y_{(u\bar{u},s\bar{s})\bar{d}}$ and $Y_{(d\bar{d},s\bar{s})\bar{u}}$ states are missing.}\label{tab:mass_tetraquark}
\begin{tabular}{|c|c|c|c|}
  \hline\hline
  $X_{u\bar d\bar s}$        &  $X_{d\bar s\bar u}$  & $X_{s\bar u\bar d}$ \\\hline
  \tabincell{c}{5.708~\cite{Wang:2016tsi}\\$5.59\pm0.18$~\cite{Chen:2017rhl} }&
   \tabincell{c}{5.708~\cite{Wang:2016tsi} \\$5.86\pm0.20$~\cite{Chen:2017rhl}\\$5.58\pm0.14$~~\cite{Agaev:2016mjb}  }& \tabincell{c}{ 5.637(Mod-I)~\cite{Yu:2017pmn}\\ 6.119(Mod-II)~\cite{Yu:2017pmn}\\
   ~5.701\cite{Huang:2019otd}}\\
  \hline\hline
\end{tabular}
\end{table}

From the predictions of the mass spectra of  $X_{b6}$ states, one can see some predictions support
the existence of the $X_{b6}$ states blow the strong decay thresholds.  The processes of weak decays are dominant once the stable tetraquarks are confirmed. It is very useful to study the weak decays by the SU(3) symmetry analysis without any assumptions of factorization or dynamic information. The semi-leptonic and nonleptonic decay amplitudes  have been parameterized in terms of SU(3) irreducible representations. For completeness, the weak two body and three body decays of open bottom tetraquarks will be studied in the following.

\section{Semi-leptonic decays}
\renewcommand\thesubsection{(\Roman{subsection})}
\label{sec:semileptonic_decay}

\begin{figure}
\includegraphics[width=0.8\columnwidth]{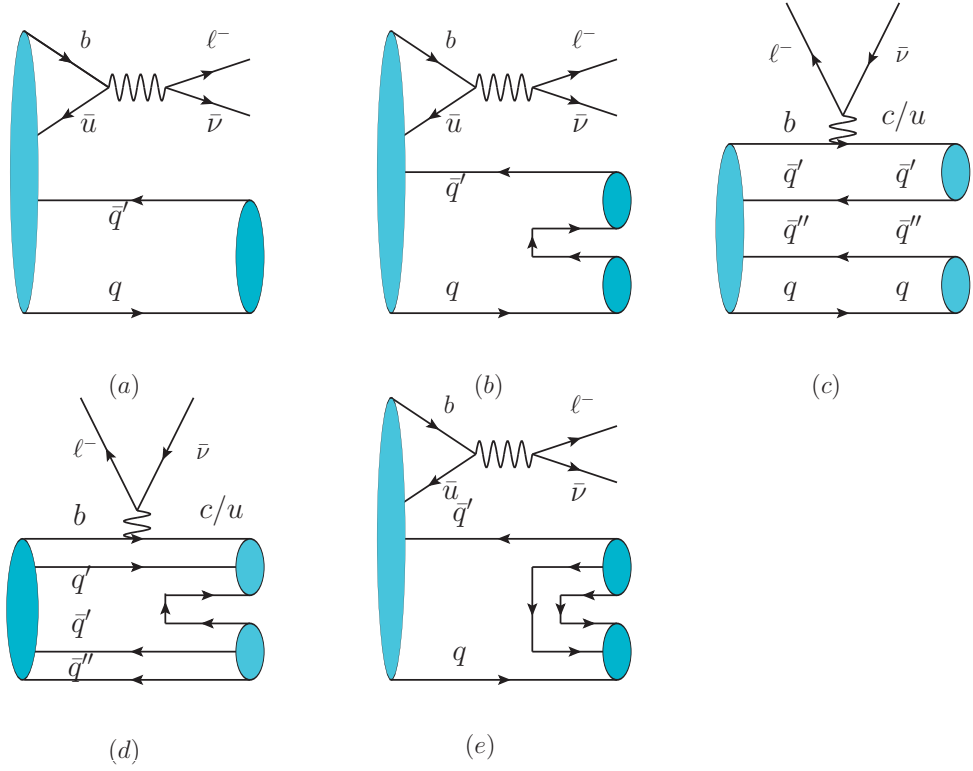}
\caption{Feynman diagrams for semileptonic decays of open-flavor bottom tetraquarks. Panel (a) represents the semileptonic decays into one meson. In panels (b,c), the final states include two mesons. Panels (d,e) denote the processes including baryonic states. The suppressed annihilation topologies are panels (a,c,e). }
\label{fig:topology-semileptonic}
\end{figure}

\subsubsection{$b\to q \ell \overline \nu_{\ell}$:  Semi-leptonic decays into mesons}
In the following, we will focus on the bottom quark decays. For the  $b$ quark decay, the electro-weak  Hamiltonian  can be expressed as
\begin{eqnarray}
 {\cal H}_{eff} &=& \frac{G_F}{\sqrt2} \left[V_{q'b} \bar q' \gamma^\mu(1-\gamma_5)b \bar  \ell\gamma_\mu(1-\gamma_5) \nu_{\ell}\right] +h.c.,
\end{eqnarray}
with $q'=u,c$.  Therein the $b\to c$ transition forms a singlet in SU(3) flavor symmetry, while the $b\to u$ transition becomes a SU(3) triplet $H_{3}'$ with $(H_3')^1=-(H_3')_{23}=(H_3')_{32}=1$ and $(H_3')^{2,3}=\epsilon^{(2,3)ij}(H_3')_{ij}=0,\ (i,j=1,2,3)$.

Firstly, the effective Hamiltonian for $X_{b6}$  decays to a light meson and $\ell \overline \nu_{\ell}$ in hadron level can be easily written as
\begin{eqnarray}
{\cal H}_{{eff}}&=&a_1 (X_{b6})_{[jk]}^i (H_3')^j M^{k}_i ~\bar \ell\nu_{\ell},
\end{eqnarray}
where and in the following the coefficient $a_{i}$ represents the nonperturbative parameters. The related Feynman diagram is plotted  in Fig.~\ref{fig:topology-semileptonic}. (a). Expanding the above hamiltonian, one can obtain the effective amplitudes for different decay channels, which are given in Tab.~\ref{tab:b6_Mlv}. From it one can see that all amplitudes of six different decay channels  are proportional to $a_1$. Therefor, we can derive the decay width relations for different channels when ignoring the small effects of phase space.  For the tetraquark semi-leptonic decays into  a light meson, the relations for  different decay widths  are
\begin{eqnarray*}
\Gamma(X_{d\bar{s}\bar{u}}^{-}\to K^0  l^-\bar\nu)= \Gamma(X_{s\bar{u}\bar{d}}^{-}\to \overline K^0  l^-\bar\nu)=2\Gamma(Y_{(u\bar{u},s\bar{s})\bar{d}}^{0}\to \pi^+ l^-\bar\nu)\\=2\Gamma(Y_{(u\bar{u},d\bar{d})\bar{s}}^{0}\to K^+  l^-\bar\nu)
=4\Gamma(Y_{(d\bar{d},s\bar{s})\bar{u}}^{-}\to \pi^0  l^-\bar\nu )=\frac{4}{3}\Gamma(Y_{(d\bar{d},s\bar{s})\bar{u}}^{-}\to \eta  l^-\bar\nu ).
\end{eqnarray*}

The effective Hamiltonian for  $X_{b6}$ semileptonic decays into two mesons  can also be constructed as
\begin{eqnarray}
{\cal H}_{{eff}}&=&
a_2 (X_{b6})^{\{ij\}} (H_3')_{[il]} M^{k}_j M^l_k ~\bar \ell\nu_{\ell}
+a_3 (X_{b6})_{[jk]}^i  M^j_i (\overline D)^k~\bar \ell\nu_{\ell} .
\end{eqnarray}
The corresponding Feynman diagrams are given in Fig.~\ref{fig:topology-semileptonic}.(b,c). The full term $a_2 (X_{b6})^{\{ij\}} (H_3')_{[il]} M^{k}_j M^l_k ~\bar \ell\nu_{\ell}+a_2^{\prime} (X_{b6})^{\{ij\}} (H_3')_{[ij]} M^{k}_l M^l_k ~\bar \ell\nu_{\ell}$ can be expressed as two new terms $a_2 \big(\; 2(X_{b6})^i_{[kl]}(H_3')^j M^k_iM^l_j-4(X_{b6})^i_{[jl]}(H_3')^j M^k_iM^l_k \;\big) ~\bar \ell\nu_{\ell}$ which are related to Fig.~\ref{fig:topology-semileptonic}.(c) and Fig.~\ref{fig:topology-semileptonic}.(b) respectively. It should be mentioned that the process of $X_{u\bar d\bar s} \to \pi^+K^+ \ell^-\bar \nu$ is trivial because of the exchangeable antisymmetric antiquarks $\bar d$ and $\bar s$ in the initial state. One can obtain the decay amplitudes from the effective Hamiltonian. The results for a light  meson plus a charmed meson in final states are given in Tab.~\ref{tab:b6_D_M_lv}, while the ones for two light mesons in final states are given in  Tab.~\ref{tab:b6_2M_lv}.

For the semi-leptonic decays into a charmed meson and a light meson, one has the relations  as
\begin{eqnarray*}
&&\Gamma(X_{u\bar{d}\bar{s}}^{+}\to \pi^+    D^+_s l^-\bar\nu)= \Gamma(X_{u\bar{d}\bar{s}}^{+}\to K^+    D^+ l^-\bar\nu)=\Gamma(X_{d\bar{s}\bar{u}}^{-}\to \pi^-    D^+_s l^-\bar\nu )\\
&=&\Gamma(X_{s\bar{u}\bar{d}}^{-}\to \overline K^0    D^0 l^-\bar\nu)=\Gamma(X_{s\bar{u}\bar{d}}^{-}\to K^-    D^+ l^-\bar\nu )=2\Gamma(Y_{(u\bar{u},s\bar{s})\bar{d}}^{0}\to \pi^+    D^0 l^-\bar\nu)\\
&=&4\Gamma(Y_{(u\bar{u},s\bar{s})\bar{d}}^{0}\to \pi^0    D^+ l^-\bar\nu)=2\Gamma(Y_{(u\bar{u},s\bar{s})\bar{d}}^{0}\to \overline K^0    D^+_s l^-\bar\nu)= \frac{4}{3}\Gamma(Y_{(u\bar{u},s\bar{s})\bar{d}}^{0}\to \eta    D^+ l^-\bar\nu)\\
&=&\Gamma(Y_{(u\bar{u},d\bar{d})\bar{s}}^{0}\to \pi^0    D^+_s l^-\bar\nu)=2\Gamma(Y_{(u\bar{u},d\bar{d})\bar{s}}^{0}\to K^+    D^0 l^-\bar\nu)=2\Gamma(Y_{(u\bar{u},d\bar{d})\bar{s}}^{0}\to K^0    D^+ l^-\bar\nu )\\
&=&4\Gamma(Y_{(d\bar{d},s\bar{s})\bar{u}}^{-}\to \pi^0    D^0 l^-\bar\nu )=2\Gamma(Y_{(d\bar{d},s\bar{s})\bar{u}}^{-}\to \pi^-    D^+ l^-\bar\nu)=2\Gamma(Y_{(d\bar{d},s\bar{s})\bar{u}}^{-}\to K^-    D^+_s l^-\bar\nu  )\\
&=& \frac{4}{3}\Gamma(Y_{(d\bar{d},s\bar{s})\bar{u}}^{-}\to \eta    D^0 l^-\bar\nu )=\Gamma(X_{d\bar{s}\bar{u}}^{-}\to K^0    D^0 l^-\bar\nu).
\end{eqnarray*}
For the semi-leptonic decays into two light mesons, one has
\begin{eqnarray*}
&&\Gamma(X_{d\bar{s}\bar{u}}^{-}\to K^+   \pi^-  l^-\bar\nu )=2\Gamma(X_{d\bar{s}\bar{u}}^{-}\to K^0   \pi^0  l^-\bar\nu )=6\Gamma(X_{d\bar{s}\bar{u}}^{-}\to \eta   K^0  l^-\bar\nu )\\
&=&\Gamma(X_{s\bar{u}\bar{d}}^{-}\to K^-   \pi^+  l^-\bar\nu )=6\Gamma(X_{s\bar{u}\bar{d}}^{-}\to \eta   \overline K^0  l^-\bar\nu)=2\Gamma(Y_{(u\bar{u},s\bar{s})\bar{d}}^{0}\to \pi^+   \eta  l^-\bar\nu )\\
&=&4\Gamma(Y_{(u\bar{u},d\bar{d})\bar{s}}^{0}\to K^+   \pi^0  l^-\bar\nu)=2\Gamma(Y_{(u\bar{u},d\bar{d})\bar{s}}^{0}\to K^0   \pi^+  l^-\bar\nu )=12\Gamma(Y_{(u\bar{u},d\bar{d})\bar{s}}^{0}\to \eta   K^+  l^-\bar\nu)\\
&=&2\Gamma(Y_{(d\bar{d},s\bar{s})\bar{u}}^{-}\to \pi^+   \pi^-  l^-\bar\nu )
=2\Gamma(Y_{(d\bar{d},s\bar{s})\bar{u}}^{-}\to \pi^0   \pi^0  l^-\bar\nu)
=2\Gamma(Y_{(d\bar{d},s\bar{s})\bar{u}}^{-}\to K^-   K^+  l^-\bar\nu )\\
&=&6\Gamma(Y_{(d\bar{d},s\bar{s})\bar{u}}^{-}\to \eta   \pi^0  l^-\bar\nu)=2\Gamma(Y_{(d\bar{d},s\bar{s})\bar{u}}^{-}\to \eta   \eta  l^-\bar\nu )=2\Gamma(X_{s\bar{u}\bar{d}}^{-}\to \overline K^0   \pi^0  l^-\bar\nu)\\
&=&3\Gamma(Y_{(u\bar{u},s\bar{s})\bar{d}}^{0}\to K^+   \overline K^0  l^-\bar\nu).
\end{eqnarray*}
\begin{table}
\caption{Amplitudes for open bottom tetraquark $X_{b6}$ semi-leptonic decays into a light meson.}\label{tab:b6_Mlv}\begin{tabular}{|cc|cc|}
\hline
channel & amplitude($/V_{ub}$) &channel &amplitude($/V_{ub}$) \\\hline
$X_{d\bar{s}\bar{u}}^{-}\to K^0  l^-\bar\nu $ & $ -\frac{a_1 }{\sqrt{2}}$&
$X_{s\bar{u}\bar{d}}^{-}\to \overline K^0  l^-\bar\nu $ & $ \frac{a_1 }{\sqrt{2}}$\\\hline
$Y_{(u\bar{u},s\bar{s})\bar{d}}^{0}\to \pi^+  l^-\bar\nu $ & $ \frac{a_1 }{2}$&
$Y_{(u\bar{u},d\bar{d})\bar{s}}^{0}\to K^+  l^-\bar\nu $ & $ -\frac{1}{2} a_1 $\\\hline
$Y_{(d\bar{d},s\bar{s})\bar{u}}^{-}\to \pi^0  l^-\bar\nu $ & $ -\frac{a_1 }{2 \sqrt{2}}$&
$Y_{(d\bar{d},s\bar{s})\bar{u}}^{-}\to \eta  l^-\bar\nu $ & $ \frac{1}{2} \sqrt{\frac{3}{2}} a_1 $\\\hline
\hline
\end{tabular}
\end{table}
\begin{table}
\caption{Amplitudes for open bottom tetraquark $X_{b6}$ semi-leptonic decays into a charmed meson and a light meson.}\label{tab:b6_D_M_lv}\begin{tabular}{|cc|cc|}\hline\hline
channel & amplitude($/V_{cb}$) & channel &amplitude($/V_{cb}$) \\\hline
$X_{u\bar{d}\bar{s}}^{+}\to \pi^+    D^+_s l^-\bar\nu $ & $ \frac{a_3}{\sqrt{2}}$&
$X_{u\bar{d}\bar{s}}^{+}\to K^+    D^+ l^-\bar\nu $ & $ -\frac{a_3}{\sqrt{2}}$\\\hline
$X_{d\bar{s}\bar{u}}^{-}\to \pi^-    D^+_s l^-\bar\nu $ & $ -\frac{a_3}{\sqrt{2}}$&
$X_{d\bar{s}\bar{u}}^{-}\to K^0    D^0 l^-\bar\nu $ & $ \frac{a_3}{\sqrt{2}}$\\\hline
$X_{s\bar{u}\bar{d}}^{-}\to \overline K^0    D^0 l^-\bar\nu $ & $ -\frac{a_3}{\sqrt{2}}$&
$X_{s\bar{u}\bar{d}}^{-}\to K^-    D^+ l^-\bar\nu $ & $ \frac{a_3}{\sqrt{2}}$\\\hline
$Y_{(u\bar{u},s\bar{s})\bar{d}}^{0}\to \pi^+    D^0 l^-\bar\nu $ & $ -\frac{a_3}{2}$&
$Y_{(u\bar{u},s\bar{s})\bar{d}}^{0}\to \pi^0    D^+ l^-\bar\nu $ & $ \frac{a_3}{2 \sqrt{2}}$\\\hline
$Y_{(u\bar{u},s\bar{s})\bar{d}}^{0}\to \overline K^0    D^+_s l^-\bar\nu $ & $ \frac{a_3}{2}$&
$Y_{(u\bar{u},s\bar{s})\bar{d}}^{0}\to \eta    D^+ l^-\bar\nu $ & $ \frac{1}{2} \sqrt{\frac{3}{2}} a_3$\\\hline
$Y_{(u\bar{u},d\bar{d})\bar{s}}^{0}\to \pi^0    D^+_s l^-\bar\nu $ & $ -\frac{a_3}{\sqrt{2}}$&
$Y_{(u\bar{u},d\bar{d})\bar{s}}^{0}\to K^+    D^0 l^-\bar\nu $ & $ \frac{a_3}{2}$\\\hline
$Y_{(u\bar{u},d\bar{d})\bar{s}}^{0}\to K^0    D^+ l^-\bar\nu $ & $ -\frac{a_3}{2}$&
$Y_{(d\bar{d},s\bar{s})\bar{u}}^{-}\to \pi^0    D^0 l^-\bar\nu $ & $ \frac{a_3}{2 \sqrt{2}}$\\\hline
$Y_{(d\bar{d},s\bar{s})\bar{u}}^{-}\to \pi^-    D^+ l^-\bar\nu $ & $ \frac{a_3}{2}$&
$Y_{(d\bar{d},s\bar{s})\bar{u}}^{-}\to K^-    D^+_s l^-\bar\nu $ & $ -\frac{a_3}{2}$\\\hline
$Y_{(d\bar{d},s\bar{s})\bar{u}}^{-}\to \eta    D^0 l^-\bar\nu $ & $ -\frac{1}{2} \sqrt{\frac{3}{2}} a_3$& &\\
\hline
\end{tabular}
\end{table}
\begin{table}
\caption{Amplitudes for open bottom tetraquark $X_{b6}$ semi-leptonic decays into  two light mesons.}\label{tab:b6_2M_lv}\begin{tabular}{|cc|cc|}\hline\hline
channel & amplitude(/$V_{\text{ub}}$) &channel &amplitude(/$V_{\text{ub}}$) \\\hline
$X_{d\bar{s}\bar{u}}^{-}\to K^+   \pi^-  l^-\bar\nu $ & $ -\frac{a_2 }{\sqrt{2}}$&
$X_{d\bar{s}\bar{u}}^{-}\to K^0   \pi^0  l^-\bar\nu $ & $ \frac{a_2}{2} $\\\hline
$X_{d\bar{s}\bar{u}}^{-}\to \eta   K^0  l^-\bar\nu $ & $ \frac{a_2 }{2 \sqrt{3}}$&
$X_{s\bar{u}\bar{d}}^{-}\to \overline K^0   \pi^0  l^-\bar\nu $ & $ -\frac{a_2}{2} $\\\hline
$X_{s\bar{u}\bar{d}}^{-}\to K^-   \pi^+  l^-\bar\nu $ & $ \frac{a_2 }{\sqrt{2}}$&
$X_{s\bar{u}\bar{d}}^{-}\to \eta   \overline K^0  l^-\bar\nu $ & $ -\frac{a_2 }{2 \sqrt{3}}$\\\hline
$Y_{(u\bar{u},s\bar{s})\bar{d}}^{0}\to \overline K^0   K^+  l^-\bar\nu $ & $ \frac{a_2}{2} $&
$Y_{(u\bar{u},s\bar{s})\bar{d}}^{0}\to \eta   \pi^+  l^-\bar\nu $ & $ \frac{a_2 }{\sqrt{6}}$\\\hline
$Y_{(u\bar{u},d\bar{d})\bar{s}}^{0}\to K^+   \pi^0  l^-\bar\nu $ & $ -\frac{a_2 }{2 \sqrt{2}}$&
$Y_{(u\bar{u},d\bar{d})\bar{s}}^{0}\to K^0   \pi^+  l^-\bar\nu $ & $ -\frac{a_2}{2} $\\\hline
$Y_{(u\bar{u},d\bar{d})\bar{s}}^{0}\to \eta   K^+  l^-\bar\nu $ & $ \frac{a_2 }{2 \sqrt{6}}$&
$Y_{(d\bar{d},s\bar{s})\bar{u}}^{-}\to \pi^+   \pi^-  l^-\bar\nu $ & $ \frac{a_2}{2} $\\\hline
$Y_{(d\bar{d},s\bar{s})\bar{u}}^{-}\to \pi^0   \pi^0  l^-\bar\nu $ & $ \frac{a_2}{2} $&
$Y_{(d\bar{d},s\bar{s})\bar{u}}^{-}\to K^-   K^+  l^-\bar\nu $ & $ -\frac{a_2}{2} $\\\hline
$Y_{(d\bar{d},s\bar{s})\bar{u}}^{-}\to \eta   \pi^0  l^-\bar\nu $ & $ -\frac{a_2 }{2 \sqrt{3}}$&
$Y_{(d\bar{d},s\bar{s})\bar{u}}^{-}\to \eta   \eta  l^-\bar\nu $ & $ -\frac{a_2}{2} $\\\hline
\hline
\end{tabular}
\end{table}

\subsubsection{$b\to q \ell \overline \nu_{\ell}$: Semileptonic decays into a light baryon plus a light anti-baryon}
The $X_{b6}$ tetraquark can also decay into a light baryon which is from an octet or anti-decuplet plus a light anti-baryon. Thus there are four different combinations for final states.  The irreducible Hamiltonian can be constructed as follows
  \begin{eqnarray}
  \mathcal{H}_{eff}&=&b_1 (X_{b6})_{[jk]}^i (H_3)^j (T_8)^k_l (\overline T_8)^l_i \bar{\ell} \nu_{\ell}+b_2 (X_{b6})_{[jk]}^i (H_3)^j (T_8)^l_i (\overline T_8)^k_l \bar{\ell} \nu_{\ell}\nonumber\\
  &&+b_3 (X_{b6})_{[jk]}^i (H_3)^l (T_8)^j_l (\overline T_8)^k_i \bar{\ell} \nu_{\ell}+b_4 (X_{b6})_{[jk]}^i (H_3)^l (T_8)^j_i (\overline T_8)^k_l \bar{\ell} \nu_{\ell}\nonumber\\
  &&
  +c_1 (X_{b6})_{[ij],[kl]} (H_3)^i (T_8)^k_m (\overline T_{\overline{10}})^{\{jlm\}} \bar{\ell} \nu_{\ell}
  -d_1 (X_{b6})^{\{ij\}} (H_3)^k (T_{10})_{\{ijl\}} (\overline T_{8})^l_k \bar{\ell} \nu_{\ell}\nonumber\\
  &&-d_2 (X_{b6})^{\{ij\}} (H_3)^k (T_{10})_{\{ikl\}} (\overline T_{8})^l_j \bar{\ell} \nu_{\ell}+f_1 (X_{b6})^{i}_{[jk]} (H_3)^j (T_{10})_{\{ilm\}} (\overline T_{\overline {10}})^{\{klm\}} \bar{\ell} \nu_{\ell}.
  \end{eqnarray}
The four kinds of amplitudes are given in Tab.~\ref{tab:b6_T8_Tbar8_lv} and Tab.~\ref{tab:b6_T10_Tbar8_lv}.
We labelled the different final states as class \uppercase\expandafter{\romannumeral1} for an octet anti-baryon plus an octet baryon, class \uppercase\expandafter{\romannumeral2} for an octet anti-baryon plus a decuplet baryon, class \uppercase\expandafter{\romannumeral3} for an anti-decuplet anti-baryon plus an octet baryon, and class \uppercase\expandafter{\romannumeral4} for an anti-decuplet anti-baryon plus a decuplet baryon.  The decay amplitudes of $X_{u\bar d\bar s}^- $ in Class IV disappear. In topology level, the corresponding Feynman diagrams are shown in Fig.~\ref{fig:topology-semileptonic}. (d,e).
For class \uppercase\expandafter{\romannumeral1}, the corresponding results for the decay widths become
\small
\begin{eqnarray*}
&&\Gamma(X_{d\bar{s}\bar{u}}^{-}\to {p} \overline \Sigma^-l^-\bar\nu)=\Gamma(X_{s\bar{u}\bar{d}}^{-}\to \Sigma^+ \overline pl^-\bar\nu)=2\Gamma(X_{s\bar{u}\bar{d}}^{-}\to \Sigma^0 \overline nl^-\bar\nu)=2\Gamma(Y_{(d\bar{d},s\bar{s})\bar{u}}^{-}\to {p} \overline pl^-\bar\nu)\\&&=2\Gamma(Y_{(d\bar{d},s\bar{s})\bar{u}}^{-}\to \Sigma^+ \overline \Sigma^-l^-\bar\nu), \ \ \ \Gamma(X_{u\bar{d}\bar{s}}^{+}\to {p} \overline \Sigma^+l^-\bar\nu)= { }\Gamma(X_{u\bar{d}\bar{s}}^{+}\to \Sigma^+ \overline \Xi^+l^-\bar\nu),\\
&&\Gamma(X_{d\bar{s}\bar{u}}^{-}\to {n} \overline \Sigma^0l^-\bar\nu)= { }\Gamma(Y_{(u\bar{u},s\bar{s})\bar{d}}^{0}\to \Xi^0 \overline \Xi^+l^-\bar\nu)=\Gamma(Y_{(u\bar{u},d\bar{d})\bar{s}}^{0}\to {n} \overline \Sigma^+l^-\bar\nu),\\
&&\Gamma(X_{s\bar{u}\bar{d}}^{-}\to \Xi^- \overline \Sigma^+l^-\bar\nu)= \Gamma(X_{d\bar{s}\bar{u}}^{-}\to \Sigma^- \overline \Xi^+l^-\bar\nu)=2\Gamma(Y_{(d\bar{d},s\bar{s})\bar{u}}^{-}\to \Xi^- \overline \Xi^+l^-\bar\nu)=2\Gamma(X_{s\bar{u}\bar{d}}^{-}\to \Xi^0 \overline \Sigma^0l^-\bar\nu)\\&&=2\Gamma(Y_{(d\bar{d},s\bar{s})\bar{u}}^{-}\to \Sigma^- \overline \Sigma^+l^-\bar\nu),\ \ \
\Gamma(Y_{(u\bar{u},s\bar{s})\bar{d}}^{0}\to \Sigma^+ \overline \Sigma^0l^-\bar\nu)= { }\Gamma(Y_{(u\bar{u},s\bar{s})\bar{d}}^{0}\to \Sigma^0 \overline \Sigma^+l^-\bar\nu),\\
&&\Gamma(Y_{(u\bar{u},s\bar{s})\bar{d}}^{0}\to {p} \overline nl^-\bar\nu)= { }\Gamma(X_{d\bar{s}\bar{u}}^{-}\to \Sigma^0 \overline \Xi^0l^-\bar\nu)=\Gamma(Y_{(u\bar{u},d\bar{d})\bar{s}}^{0}\to \Sigma^+ \overline \Xi^0l^-\bar\nu),\\
&&\Gamma(Y_{(u\bar{u},d\bar{d})\bar{s}}^{0}\to \Lambda^0 \overline \Xi^+l^-\bar\nu)= \frac{1}{2}\Gamma(X_{d\bar{s}\bar{u}}^{-}\to \Lambda^0 \overline \Xi^0l^-\bar\nu),
       \Gamma(Y_{(d\bar{d},s\bar{s})\bar{u}}^{-}\to \Lambda^0 \overline \Lambda^0l^-\bar\nu)= { }\Gamma(Y_{(d\bar{d},s\bar{s})\bar{u}}^{-}\to \Sigma^0 \overline \Sigma^0l^-\bar\nu),\\
&&\Gamma(Y_{(d\bar{d},s\bar{s})\bar{u}}^{-}\to \Lambda^0 \overline \Sigma^0l^-\bar\nu)= \frac{1}{2}\Gamma(Y_{(u\bar{u},s\bar{s})\bar{d}}^{0}\to \Lambda^0 \overline \Sigma^+l^-\bar\nu),
            \Gamma(Y_{(d\bar{d},s\bar{s})\bar{u}}^{-}\to {n} \overline nl^-\bar\nu)= { }\Gamma(Y_{(d\bar{d},s\bar{s})\bar{u}}^{-}\to \Xi^0 \overline \Xi^0l^-\bar\nu),\\
&&\Gamma(Y_{(u\bar{u},s\bar{s})\bar{d}}^{0}\to \Sigma^+ \overline \Lambda^0l^-\bar\nu)= 2\Gamma(Y_{(d\bar{d},s\bar{s})\bar{u}}^{-}\to \Sigma^0 \overline \Lambda^0l^-\bar\nu),
     \Gamma(Y_{(u\bar{u},d\bar{d})\bar{s}}^{0}\to {p} \overline \Lambda^0l^-\bar\nu)= \frac{1}{2}\Gamma(X_{d\bar{s}\bar{u}}^{-}\to {n} \overline \Lambda^0l^-\bar\nu).
\end{eqnarray*}
The relations for class \uppercase\expandafter{\romannumeral2} are
\begin{eqnarray*}
&&2\Gamma(X_{d\bar{s}\bar{u}}^{-}\to \Lambda^0  \overline \Xi^{\prime0} l^-\bar\nu )=6\Gamma(X_{d\bar{s}\bar{u}}^{-}\to \Sigma^0  \overline \Xi^{\prime0} l^-\bar\nu)
=3\Gamma(X_{d\bar{s}\bar{u}}^{-}\to \Sigma^-  \overline \Xi^{\prime+} l^-\bar\nu)\\
&&=6\Gamma(X_{d\bar{s}\bar{u}}^{-}\to {n}  \overline \Sigma^{\prime0} l^-\bar\nu)
=\Gamma(X_{d\bar{s}\bar{u}}^{-}\to \Xi^-  \overline \Omega^+ l^-\bar\nu )
=3\Gamma(X_{s\bar{u}\bar{d}}^{-}\to \Sigma^+  \overline \Delta^{-} l^-\bar\nu )\\
&&=\frac{3}{2}\Gamma(X_{s\bar{u}\bar{d}}^{-}\to \Sigma^0  \overline \Delta^{0} l^-\bar\nu )
=\Gamma(X_{s\bar{u}\bar{d}}^{-}\to \Sigma^-  \overline \Delta^{+} l^-\bar\nu)
=3\Gamma(X_{s\bar{u}\bar{d}}^{-}\to \Xi^-  \overline \Sigma^{\prime+} l^-\bar\nu)\\
&&=6\Gamma(X_{s\bar{u}\bar{d}}^{-}\to \Xi^0  \overline \Sigma^{\prime0} l^-\bar\nu )
=4\Gamma(Y_{(u\bar{u},s\bar{s})\bar{d}}^{0}\to \Lambda^0  \overline \Sigma^{\prime+} l^-\bar\nu)
=12\Gamma(Y_{(u\bar{u},s\bar{s})\bar{d}}^{0}\to \Sigma^+  \overline \Sigma^{\prime0} l^-\bar\nu)\\
&&=12\Gamma(Y_{(u\bar{u},s\bar{s})\bar{d}}^{0}\to \Sigma^0  \overline \Sigma^{\prime+} l^-\bar\nu )
=6\Gamma(Y_{(u\bar{u},s\bar{s})\bar{d}}^{0}\to {p}  \overline \Delta^{0} l^-\bar\nu )
=2\Gamma(Y_{(u\bar{u},s\bar{s})\bar{d}}^{0}\to {n}  \overline \Delta^{+} l^-\bar\nu )\\
&&=6\Gamma(Y_{(u\bar{u},s\bar{s})\bar{d}}^{0}\to \Xi^0  \overline \Xi^{\prime+} l^-\bar\nu)
=4\Gamma(Y_{(u\bar{u},d\bar{d})\bar{s}}^{0}\to \Lambda^0  \overline\Xi^{\prime+} l^-\bar\nu )
=6\Gamma(Y_{(u\bar{u},d\bar{d})\bar{s}}^{0}\to \Sigma^+  \overline \Xi^{\prime0} l^-\bar\nu )\\
&&=12\Gamma(Y_{(u\bar{u},d\bar{d})\bar{s}}^{0}\to \Sigma^0  \overline\Xi^{\prime+} l^-\bar\nu )
=12\Gamma(Y_{(u\bar{u},d\bar{d})\bar{s}}^{0}\to {p}  \overline \Sigma^{\prime0} l^-\bar\nu )
=6\Gamma(Y_{(u\bar{u},d\bar{d})\bar{s}}^{0}\to {n}  \overline \Sigma^{\prime+} l^-\bar\nu)\\
&&=2\Gamma(Y_{(u\bar{u},d\bar{d})\bar{s}}^{0}\to \Xi^0  \overline\Omega^+ l^-\bar\nu )
=8\Gamma(Y_{(d\bar{d},s\bar{s})\bar{u}}^{-}\to \Lambda^0  \overline \Sigma^{\prime0} l^-\bar\nu )
=6\Gamma(Y_{(d\bar{d},s\bar{s})\bar{u}}^{-}\to \Sigma^+  \overline\Sigma^{\prime-} l^-\bar\nu)\\
&&=\frac{8}{3}\Gamma(Y_{(d\bar{d},s\bar{s})\bar{u}}^{-}\to \Sigma^0  \overline \Sigma^{\prime0} l^-\bar\nu)
=\frac{3}{2}\Gamma(Y_{(d\bar{d},s\bar{s})\bar{u}}^{-}\to \Sigma^-  \overline \Sigma^{\prime+} l^-\bar\nu)
=6\Gamma(Y_{(d\bar{d},s\bar{s})\bar{u}}^{-}\to {p}  \overline\Delta^{-} l^-\bar\nu)\\
&&=6\Gamma(Y_{(d\bar{d},s\bar{s})\bar{u}}^{-}\to {n}  \overline \Delta^{0} l^-\bar\nu)
=\frac{3}{2}\Gamma(Y_{(d\bar{d},s\bar{s})\bar{u}}^{-}\to \Xi^-  \overline\Xi^{\prime+} l^-\bar\nu)
=6\Gamma(Y_{(d\bar{d},s\bar{s})\bar{u}}^{-}\to \Xi^0  \overline \Xi^{\prime0} l^-\bar\nu  )\\
&&=3\Gamma(X_{d\bar{s}\bar{u}}^{-}\to {p}  \overline \Sigma^{\prime-} l^-\bar\nu ).
\end{eqnarray*}
For class \uppercase\expandafter{\romannumeral3}, the relations are
\begin{eqnarray*}
&&\Gamma(X_{u\bar{d}\bar{s}}^{+}\to \Delta^{++} \overline \Lambda^0l^-\bar\nu)
=\frac{1}{3}\Gamma(X_{u\bar{d}\bar{s}}^{+}\to \Delta^{++} \overline \Sigma^0l^-\bar\nu)
=\frac{1}{2}\Gamma(X_{u\bar{d}\bar{s}}^{+}\to \Delta^{+} \overline \Sigma^+l^-\bar\nu)\\
&&=\frac{1}{2}\Gamma(X_{u\bar{d}\bar{s}}^{+}\to \Sigma^{\prime+} \overline \Xi^+l^-\bar\nu)
=3\Gamma(X_{d\bar{s}\bar{u}}^{-}\to \Delta^{0} \overline \Lambda^0l^-\bar\nu)
=\frac{3}{2}\Gamma(Y_{(u\bar{u},d\bar{d})\bar{s}}^{0}\to \Delta^{+} \overline \Lambda^0l^-\bar\nu),\\
&&\Gamma(X_{d\bar{s}\bar{u}}^{-}\to \Delta^{+} \overline \Sigma^-l^-\bar\nu)
=2\Gamma(X_{d\bar{s}\bar{u}}^{-}\to \Sigma^{\prime0} \overline \Xi^0l^-\bar\nu)
=\Gamma(X_{s\bar{u}\bar{d}}^{-}\to \Sigma^{\prime+} \overline pl^-\bar\nu)\\
&&=2\Gamma(X_{s\bar{u}\bar{d}}^{-}\to \Sigma^{\prime0} \overline nl^-\bar\nu)
=\frac{2}{3}\Gamma(Y_{(u\bar{u},s\bar{s})\bar{d}}^{0}\to \Delta^{++} \overline pl^-\bar\nu)
=2\Gamma(Y_{(u\bar{u},s\bar{s})\bar{d}}^{0}\to \Delta^{+} \overline nl^-\bar\nu)\\
&&=2\Gamma(Y_{(u\bar{u},d\bar{d})\bar{s}}^{0}\to \Sigma^{\prime+} \overline \Xi^0l^-\bar\nu)
=2\Gamma(Y_{(d\bar{d},s\bar{s})\bar{u}}^{-}\to \Delta^{+} \overline pl^-\bar\nu)
=2\Gamma(Y_{(d\bar{d},s\bar{s})\bar{u}}^{-}\to \Delta^{0} \overline nl^-\bar\nu)\\
&&=2\Gamma(Y_{(d\bar{d},s\bar{s})\bar{u}}^{-}\to \Xi^{\prime0} \overline \Xi^0l^-\bar\nu)
=\frac{2}{3}\Gamma(Y_{(u\bar{u},d\bar{d})\bar{s}}^{0}\to \Delta^{++} \overline \Sigma^-l^-\bar\nu)
=2\Gamma(Y_{(d\bar{d},s\bar{s})\bar{u}}^{-}\to \Sigma^{\prime+} \overline \Sigma^-l^-\bar\nu),\\
&&\Gamma(X_{d\bar{s}\bar{u}}^{-}\to \Delta^{-} \overline \Sigma^+l^-\bar\nu)
=3\Gamma(X_{d\bar{s}\bar{u}}^{-}\to \Sigma^{\prime-} \overline \Xi^+l^-\bar\nu)
=\Gamma(X_{s\bar{u}\bar{d}}^{-}\to \Omega^- \overline \Xi^+l^-\bar\nu)\\
&&=\frac{3}{2}\Gamma(Y_{(d\bar{d},s\bar{s})\bar{u}}^{-}\to \Xi^{\prime-} \overline \Xi^+l^-\bar\nu)
=6\Gamma(X_{s\bar{u}\bar{d}}^{-}\to \Xi^{\prime0} \overline \Sigma^0l^-\bar\nu)
=3\Gamma(X_{s\bar{u}\bar{d}}^{-}\to \Xi^{\prime-} \overline \Sigma^+l^-\bar\nu)\\
&&= \frac{3}{2}\Gamma(Y_{(d\bar{d},s\bar{s})\bar{u}}^{-}\to \Sigma^{\prime-} \overline \Sigma^+l^-\bar\nu)
=3\Gamma(Y_{(u\bar{u},d\bar{d})\bar{s}}^{0}\to \Delta^{+} \overline \Sigma^0l^-\bar\nu),\\
&&\Gamma(Y_{(u\bar{u},s\bar{s})\bar{d}}^{0}\to \Sigma^{\prime+} \overline \Lambda^0l^-\bar\nu)
= \frac{2}{3}\Gamma(Y_{(d\bar{d},s\bar{s})\bar{u}}^{-}\to \Sigma^{\prime0} \overline \Sigma^0l^-\bar\nu)=2\Gamma(Y_{(d\bar{d},s\bar{s})\bar{u}}^{-}\to \Sigma^{\prime0} \overline \Lambda^0l^-\bar\nu),\\
&&\Gamma(Y_{(u\bar{u},s\bar{s})\bar{d}}^{0}\to \Sigma^{\prime+} \overline \Sigma^0l^-\bar\nu)
= { }\Gamma(Y_{(u\bar{u},s\bar{s})\bar{d}}^{0}\to \Sigma^{\prime0} \overline \Sigma^+l^-\bar\nu)
=\frac{1}{2}\Gamma(Y_{(u\bar{u},s\bar{s})\bar{d}}^{0}\to \Xi^{\prime0} \overline \Xi^+l^-\bar\nu)\\
&&=\frac{1}{2}\Gamma(Y_{(u\bar{u},d\bar{d})\bar{s}}^{0}\to \Delta^{0} \overline \Sigma^+l^-\bar\nu)
= \Gamma(Y_{(u\bar{u},d\bar{d})\bar{s}}^{0}\to \Sigma^{\prime0} \overline \Xi^+l^-\bar\nu).
\end{eqnarray*}
For class \uppercase\expandafter{\romannumeral4}, the relations become
\begin{eqnarray*}
&&\frac{9}{2}\Gamma(X_{d\bar{s}\bar{u}}^{-}\to \Delta^{+}  \overline\Sigma^{\prime-} l^-\bar\nu )
=\frac{9}{4}\Gamma(X_{d\bar{s}\bar{u}}^{-}\to \Delta^{0}  \overline \Sigma^{\prime0} l^-\bar\nu)
=\frac{3}{2}\Gamma(X_{d\bar{s}\bar{u}}^{-}\to \Delta^{-}  \overline \Sigma^{\prime+} l^-\bar\nu )\\
&&=\frac{9}{4}\Gamma(X_{d\bar{s}\bar{u}}^{-}\to \Sigma^{\prime0}  \overline \Xi^{\prime0} l^-\bar\nu)
=\frac{9}{8}\Gamma(X_{d\bar{s}\bar{u}}^{-}\to \Sigma^{\prime-}  \overline\Xi^{\prime+} l^-\bar\nu )
=\frac{3}{2}\Gamma(X_{d\bar{s}\bar{u}}^{-}\to \Xi^{\prime-}  \overline\Omega^+ l^-\bar\nu )\\
&&=\frac{9}{2}\Gamma(X_{s\bar{u}\bar{d}}^{-}\to \Sigma^{\prime+}  \overline\Delta^{-} l^-\bar\nu )
=\frac{9}{4}\Gamma(X_{s\bar{u}\bar{d}}^{-}\to \Sigma^{\prime0}  \overline \Delta^{0} l^-\bar\nu)
=\frac{3}{2}\Gamma(X_{s\bar{u}\bar{d}}^{-}\to \Sigma^{\prime-}  \overline\Delta^{+} l^-\bar\nu )\\
&&=\frac{9}{4}\Gamma(X_{s\bar{u}\bar{d}}^{-}\to \Xi^{\prime0}  \overline \Sigma^{\prime0} l^-\bar\nu )
=\frac{9}{8}\Gamma(X_{s\bar{u}\bar{d}}^{-}\to \Xi^{\prime-}  \overline \Sigma^{\prime+} l^-\bar\nu )
=\frac{3}{2}\Gamma(X_{s\bar{u}\bar{d}}^{-}\to \Omega^-  \overline\Xi^{\prime+} l^-\bar\nu )\\
&&=3\Gamma(Y_{(u\bar{u},s\bar{s})\bar{d}}^{0}\to \Delta^{++}  \overline\Delta^{-} l^-\bar\nu )
=\frac{9}{4}\Gamma(Y_{(u\bar{u},s\bar{s})\bar{d}}^{0}\to \Delta^{+}  \overline \Delta^{0} l^-\bar\nu )
=3\Gamma(Y_{(u\bar{u},s\bar{s})\bar{d}}^{0}\to \Delta^{0}  \overline\Delta^{+} l^-\bar\nu )\\
&&=\frac{9}{2}\Gamma(Y_{(u\bar{u},s\bar{s})\bar{d}}^{0}\to \Sigma^{\prime+}  \overline \Sigma^{\prime0} l^-\bar\nu)
=\frac{9}{2}\Gamma(Y_{(u\bar{u},s\bar{s})\bar{d}}^{0}\to \Sigma^{\prime0}  \overline \Sigma^{\prime+} l^-\bar\nu )
=9\Gamma(Y_{(u\bar{u},s\bar{s})\bar{d}}^{0}\to \Xi^{\prime0}  \overline\Xi^{\prime+} l^-\bar\nu )\\
&&=3\Gamma(Y_{(u\bar{u},d\bar{d})\bar{s}}^{0}\to \Delta^{++}  \overline\Sigma^{\prime-} l^-\bar\nu )
=\frac{9}{2}\Gamma(Y_{(u\bar{u},d\bar{d})\bar{s}}^{0}\to \Delta^{+}  \overline \Sigma^{\prime0} l^-\bar\nu)
=9\Gamma(Y_{(u\bar{u},d\bar{d})\bar{s}}^{0}\to \Delta^{0}  \overline \Sigma^{\prime+} l^-\bar\nu )\\
&&=\frac{9}{4}\Gamma(Y_{(u\bar{u},d\bar{d})\bar{s}}^{0}\to \Sigma^{\prime+}  \overline \Xi^{\prime0} l^-\bar\nu)
=\frac{9}{2}\Gamma(Y_{(u\bar{u},d\bar{d})\bar{s}}^{0}\to \Sigma^{\prime0}  \overline\Xi^{\prime+} l^-\bar\nu )
=3\Gamma(Y_{(u\bar{u},d\bar{d})\bar{s}}^{0}\to \Xi^{\prime0}  \overline\Omega^+ l^-\bar\nu
)\\
&&=9\Gamma(Y_{(d\bar{d},s\bar{s})\bar{u}}^{-}\to \Delta^{+}  \overline\Delta^{-} l^-\bar\nu )
=\frac{9}{4}\Gamma(Y_{(d\bar{d},s\bar{s})\bar{u}}^{-}\to \Delta^{0}  \overline \Delta^{0} l^-\bar\nu )
=\Gamma(Y_{(d\bar{d},s\bar{s})\bar{u}}^{-}\to \Delta^{-}  \overline\Delta^{+} l^-\bar\nu )\\
&&=9\Gamma(Y_{(d\bar{d},s\bar{s})\bar{u}}^{-}\to \Sigma^{\prime+}  \overline\Sigma^{\prime-} l^-\bar\nu )
=9\Gamma(Y_{(d\bar{d},s\bar{s})\bar{u}}^{-}\to \Sigma^{\prime-}  \overline \Sigma^{\prime+} l^-\bar\nu )
=\frac{9}{4}\Gamma(Y_{(d\bar{d},s\bar{s})\bar{u}}^{-}\to \Xi^{\prime0}  \overline \Xi^{\prime0} l^-\bar\nu )\\
&&=9\Gamma(Y_{(d\bar{d},s\bar{s})\bar{u}}^{-}\to \Xi^{\prime-}  \overline\Xi^{\prime+} l^-\bar\nu )
=\Gamma(Y_{(d\bar{d},s\bar{s})\bar{u}}^{-}\to \Omega^-  \overline\Omega^+ l^-\bar\nu ).
\end{eqnarray*}
\begin{table}
\caption{Amplitudes for open bottom tetraquark $X_{b6}$ semi-leptonic decays into a light baryon octet plus a light anti-baryon octet for class I, a light baryon octet and an anti-baryon anti-decuplet for class II}\label{tab:b6_T8_Tbar8_lv}\begin{tabular}{|cc|cc|}\hline\hline
\multicolumn{4}{|c|} {Class \uppercase\expandafter{\romannumeral1}} \\\hline
channel & amplitude(/$V_{\text{ub}}$) &channel & amplitude(/$V_{\text{ub}}$)\\\hline
$X_{u\bar{d}\bar{s}}^{+}\to \Sigma^+  \overline \Xi^+ l^-\bar\nu $ & $ \frac{\left(b_3+b_4\right) }{\sqrt{2}}$&
$X_{u\bar{d}\bar{s}}^{+}\to {p}  \overline \Sigma^+ l^-\bar\nu $ & $ -\frac{\left(b_3+b_4\right) }{\sqrt{2}}$\\\hline
$X_{d\bar{s}\bar{u}}^{-}\to \Lambda^0  \overline \Xi^0 l^-\bar\nu $ & $ \frac{\left(2 b_1-b_2-b_3\right) }{2 \sqrt{3}}$&
$X_{d\bar{s}\bar{u}}^{-}\to \Sigma^0  \overline \Xi^0 l^-\bar\nu $ & $ \frac{1}{2} \left(b_2-b_3\right) $\\\hline
$X_{d\bar{s}\bar{u}}^{-}\to \Sigma^-  \overline \Xi^+ l^-\bar\nu $ & $ -\frac{\left(b_2+b_4\right) }{\sqrt{2}}$&
$X_{d\bar{s}\bar{u}}^{-}\to {p}  \overline \Sigma^- l^-\bar\nu $ & $ -\frac{\left(b_1-b_3\right) }{\sqrt{2}}$\\\hline
$X_{d\bar{s}\bar{u}}^{-}\to {n}  \overline \Lambda^0 l^-\bar\nu $ & $ -\frac{\left(b_1-2 b_2-b_4\right) }{2 \sqrt{3}}$&
$X_{d\bar{s}\bar{u}}^{-}\to {n}  \overline \Sigma^0 l^-\bar\nu $ & $ \frac{1}{2} \left(b_1+b_4\right) $\\\hline
$X_{s\bar{u}\bar{d}}^{-}\to \Lambda^0  \overline n l^-\bar\nu $ & $ \frac{\left(b_1-2 b_2+b_3\right) }{2 \sqrt{3}}$&
$X_{s\bar{u}\bar{d}}^{-}\to \Sigma^+  \overline p l^-\bar\nu $ & $ \frac{\left(b_1-b_3\right) }{\sqrt{2}}$\\\hline
$X_{s\bar{u}\bar{d}}^{-}\to \Sigma^0  \overline n l^-\bar\nu $ & $ -\frac{1}{2} \left(b_1-b_3\right) $&
$X_{s\bar{u}\bar{d}}^{-}\to \Xi^-  \overline \Sigma^+ l^-\bar\nu $ & $ \frac{\left(b_2+b_4\right) }{\sqrt{2}}$\\\hline
$X_{s\bar{u}\bar{d}}^{-}\to \Xi^0  \overline \Lambda^0 l^-\bar\nu $ & $ -\frac{\left(2 b_1-b_2+b_4\right) }{2 \sqrt{3}}$&
$X_{s\bar{u}\bar{d}}^{-}\to \Xi^0  \overline \Sigma^0 l^-\bar\nu $ & $ -\frac{1}{2} \left(b_2+b_4\right) $\\\hline
$Y_{(u\bar{u},s\bar{s})\bar{d}}^{0}\to \Lambda^0  \overline \Sigma^+ l^-\bar\nu $ & $ \frac{\sqrt{6}}{12} \left( b_1+ b_2+ b_3+3 b_4\right) $&
$Y_{(u\bar{u},s\bar{s})\bar{d}}^{0}\to \Sigma^+  \overline \Lambda^0 l^-\bar\nu $ & $ \frac{\left(b_1+b_2-3 b_3-b_4\right) }{2 \sqrt{6}}$\\\hline
$Y_{(u\bar{u},s\bar{s})\bar{d}}^{0}\to \Sigma^+  \overline \Sigma^0 l^-\bar\nu $ & $ \frac{\left(b_1-b_2-b_3-b_4\right) }{2 \sqrt{2}}$&
$Y_{(u\bar{u},s\bar{s})\bar{d}}^{0}\to \Sigma^0  \overline \Sigma^+ l^-\bar\nu $ & $ -\frac{\left(b_1-b_2-b_3-b_4\right) }{2 \sqrt{2}}$\\\hline
$Y_{(u\bar{u},s\bar{s})\bar{d}}^{0}\to {p}  \overline n l^-\bar\nu $ & $ \frac{1}{2} \left(b_2-b_3\right) $&
$Y_{(u\bar{u},s\bar{s})\bar{d}}^{0}\to \Xi^0  \overline \Xi^+ l^-\bar\nu $ & $ \frac{1}{2} \left(b_1+b_4\right) $\\\hline
$Y_{(u\bar{u},d\bar{d})\bar{s}}^{0}\to \Lambda^0  \overline \Xi^+ l^-\bar\nu $ & $ \frac{\left(2 b_1-b_2-b_3\right) }{2 \sqrt{6}}$&
$Y_{(u\bar{u},d\bar{d})\bar{s}}^{0}\to \Sigma^+  \overline \Xi^0 l^-\bar\nu $ & $ -\frac{1}{2} \left(b_2-b_3\right) $\\\hline
$Y_{(u\bar{u},d\bar{d})\bar{s}}^{0}\to \Sigma^0  \overline \Xi^+ l^-\bar\nu $ & $ -\frac{\left(b_2+b_3+2 b_4\right) }{2 \sqrt{2}}$&
$Y_{(u\bar{u},d\bar{d})\bar{s}}^{0}\to {p}  \overline \Lambda^0 l^-\bar\nu $ & $ -\frac{\left(b_1-2 b_2-b_4\right) }{2 \sqrt{6}}$\\\hline
$Y_{(u\bar{u},d\bar{d})\bar{s}}^{0}\to {p}  \overline \Sigma^0 l^-\bar\nu $ & $ -\frac{\left(b_1-2 b_3-b_4\right) }{2 \sqrt{2}}$&
$Y_{(u\bar{u},d\bar{d})\bar{s}}^{0}\to {n}  \overline \Sigma^+ l^-\bar\nu $ & $ -\frac{1}{2} \left(b_1+b_4\right) $\\\hline
$Y_{(d\bar{d},s\bar{s})\bar{u}}^{-}\to \Lambda^0  \overline \Lambda^0 l^-\bar\nu $ & $ -\frac{1}{4} \left(b_1+b_2-b_3+b_4\right) $&
$Y_{(d\bar{d},s\bar{s})\bar{u}}^{-}\to \Lambda^0  \overline \Sigma^0 l^-\bar\nu $ & $ -\frac{\left(b_1+b_2+b_3+3 b_4\right) }{4 \sqrt{3}}$\\\hline
$Y_{(d\bar{d},s\bar{s})\bar{u}}^{-}\to \Sigma^+  \overline \Sigma^- l^-\bar\nu $ & $ \frac{1}{2} \left(b_1-b_3\right) $&
$Y_{(d\bar{d},s\bar{s})\bar{u}}^{-}\to \Sigma^0  \overline \Lambda^0 l^-\bar\nu $ & $ -\frac{\left(b_1+b_2-3 b_3-b_4\right) }{4 \sqrt{3}}$\\\hline
$Y_{(d\bar{d},s\bar{s})\bar{u}}^{-}\to \Sigma^0  \overline \Sigma^0 l^-\bar\nu $ & $ \frac{1}{4} \left(b_1+b_2-b_3+b_4\right) $&
$Y_{(d\bar{d},s\bar{s})\bar{u}}^{-}\to \Sigma^-  \overline \Sigma^+ l^-\bar\nu $ & $ \frac{1}{2} \left(b_2+b_4\right) $\\\hline
$Y_{(d\bar{d},s\bar{s})\bar{u}}^{-}\to {p}  \overline p l^-\bar\nu $ & $ -\frac{1}{2} \left(b_1-b_3\right) $&
$Y_{(d\bar{d},s\bar{s})\bar{u}}^{-}\to {n}  \overline n l^-\bar\nu $ & $ -\frac{1}{2} \left(b_1-b_2\right) $\\\hline
$Y_{(d\bar{d},s\bar{s})\bar{u}}^{-}\to \Xi^-  \overline \Xi^+ l^-\bar\nu $ & $ -\frac{1}{2} \left(b_2+b_4\right) $&
$Y_{(d\bar{d},s\bar{s})\bar{u}}^{-}\to \Xi^0  \overline \Xi^0 l^-\bar\nu $ & $ \frac{1}{2} \left(b_1-b_2\right) $\\\hline
\hline
%
\multicolumn{4}{|c|} {Class \uppercase\expandafter{\romannumeral2}} \\\hline
$X_{d\bar{s}\bar{u}}^{-}\to \Lambda^0  \overline \Xi^{\prime0} l^-\bar\nu $ & $ \frac{c_1 }{2}$&
$X_{d\bar{s}\bar{u}}^{-}\to \Sigma^0  \overline \Xi^{\prime0} l^-\bar\nu $ & $ \frac{c_1 }{2 \sqrt{3}}$\\\hline
$X_{d\bar{s}\bar{u}}^{-}\to \Sigma^-  \overline \Xi^{\prime+} l^-\bar\nu $ & $ \frac{c_1 }{\sqrt{6}}$&
$X_{d\bar{s}\bar{u}}^{-}\to {p}  \overline \Sigma^{\prime-} l^-\bar\nu $ & $ -\frac{c_1 }{\sqrt{6}}$\\\hline
$X_{d\bar{s}\bar{u}}^{-}\to {n}  \overline \Sigma^{\prime0} l^-\bar\nu $ & $ -\frac{c_1 }{2 \sqrt{3}}$&
$X_{d\bar{s}\bar{u}}^{-}\to \Xi^-  \overline \Omega^+ l^-\bar\nu $ & $ \frac{c_1 }{\sqrt{2}}$\\\hline
$X_{s\bar{u}\bar{d}}^{-}\to \Sigma^+  \overline \Delta^{-} l^-\bar\nu $ & $ -\frac{c_1 }{\sqrt{6}}$&
$X_{s\bar{u}\bar{d}}^{-}\to \Sigma^0  \overline \Delta^{0} l^-\bar\nu $ & $ \frac{c_1 }{\sqrt{3}}$\\\hline
$X_{s\bar{u}\bar{d}}^{-}\to \Sigma^-  \overline \Delta^{+} l^-\bar\nu $ & $ \frac{c_1 }{\sqrt{2}}$&
$X_{s\bar{u}\bar{d}}^{-}\to \Xi^-  \overline \Sigma^{\prime+} l^-\bar\nu $ & $ \frac{c_1 }{\sqrt{6}}$\\\hline
$X_{s\bar{u}\bar{d}}^{-}\to \Xi^0  \overline \Sigma^{\prime0} l^-\bar\nu $ & $ -\frac{c_1 }{2 \sqrt{3}}$&
$Y_{(u\bar{u},s\bar{s})\bar{d}}^{0}\to \Lambda^0  \overline \Sigma^{\prime+} l^-\bar\nu $ & $ \frac{c_1 }{2 \sqrt{2}}$\\\hline
$Y_{(u\bar{u},s\bar{s})\bar{d}}^{0}\to \Sigma^+  \overline \Sigma^{\prime0} l^-\bar\nu $ & $ \frac{c_1 }{2 \sqrt{6}}$&
$Y_{(u\bar{u},s\bar{s})\bar{d}}^{0}\to \Sigma^0  \overline \Sigma^{\prime+} l^-\bar\nu $ & $ -\frac{c_1 }{2 \sqrt{6}}$\\\hline
$Y_{(u\bar{u},s\bar{s})\bar{d}}^{0}\to {p}  \overline \Delta^{0} l^-\bar\nu $ & $ -\frac{c_1 }{2 \sqrt{3}}$&
$Y_{(u\bar{u},s\bar{s})\bar{d}}^{0}\to {n}  \overline \Delta^{+} l^-\bar\nu $ & $ -\frac{1}{2} c_1 $\\\hline
$Y_{(u\bar{u},s\bar{s})\bar{d}}^{0}\to \Xi^0  \overline \Xi^{\prime+} l^-\bar\nu $ & $ \frac{c_1 }{2 \sqrt{3}}$&
$Y_{(u\bar{u},d\bar{d})\bar{s}}^{0}\to \Lambda^0  \overline\Xi^{\prime+} l^-\bar\nu $ & $ -\frac{c_1 }{2 \sqrt{2}}$\\\hline
$Y_{(u\bar{u},d\bar{d})\bar{s}}^{0}\to \Sigma^+  \overline \Xi^{\prime0} l^-\bar\nu $ & $ -\frac{c_1 }{2 \sqrt{3}}$&
$Y_{(u\bar{u},d\bar{d})\bar{s}}^{0}\to \Sigma^0  \overline\Xi^{\prime+} l^-\bar\nu $ & $ \frac{c_1 }{2 \sqrt{6}}$\\\hline
$Y_{(u\bar{u},d\bar{d})\bar{s}}^{0}\to {p}  \overline \Sigma^{\prime0} l^-\bar\nu $ & $ \frac{c_1 }{2 \sqrt{6}}$&
$Y_{(u\bar{u},d\bar{d})\bar{s}}^{0}\to {n}  \overline \Sigma^{\prime+} l^-\bar\nu $ & $ \frac{c_1 }{2 \sqrt{3}}$\\\hline
$Y_{(u\bar{u},d\bar{d})\bar{s}}^{0}\to \Xi^0  \overline\Omega^+ l^-\bar\nu $ & $ -\frac{1}{2} c_1 $&
$Y_{(d\bar{d},s\bar{s})\bar{u}}^{-}\to \Lambda^0  \overline \Sigma^{\prime0} l^-\bar\nu $ & $ -\frac{1}{4} c_1 $\\\hline
$Y_{(d\bar{d},s\bar{s})\bar{u}}^{-}\to \Sigma^+  \overline\Sigma^{\prime-} l^-\bar\nu $ & $ \frac{c_1 }{2 \sqrt{3}}$&
$Y_{(d\bar{d},s\bar{s})\bar{u}}^{-}\to \Sigma^0  \overline \Sigma^{\prime0} l^-\bar\nu $ & $ -\frac{1}{4} \sqrt{3} c_1 $\\\hline
$Y_{(d\bar{d},s\bar{s})\bar{u}}^{-}\to \Sigma^-  \overline \Sigma^{\prime+} l^-\bar\nu $ & $ -\frac{c_1 }{\sqrt{3}}$&
$Y_{(d\bar{d},s\bar{s})\bar{u}}^{-}\to {p}  \overline\Delta^{-} l^-\bar\nu $ & $ \frac{c_1 }{2 \sqrt{3}}$\\\hline
$Y_{(d\bar{d},s\bar{s})\bar{u}}^{-}\to {n}  \overline \Delta^{0} l^-\bar\nu $ & $ \frac{c_1 }{2 \sqrt{3}}$&
$Y_{(d\bar{d},s\bar{s})\bar{u}}^{-}\to \Xi^-  \overline\Xi^{\prime+} l^-\bar\nu $ & $ -\frac{c_1 }{\sqrt{3}}$\\\hline
$Y_{(d\bar{d},s\bar{s})\bar{u}}^{-}\to \Xi^0  \overline \Xi^{\prime0} l^-\bar\nu $ & $ \frac{c_1 }{2 \sqrt{3}}$& &\\\hline
\hline
\end{tabular}
\end{table}

\begin{table}
\caption{Amplitudes for open bottom tetraquark $X_{b6}$ semi-leptonic decays into a light baryon decuplet and an anti-baryon octet for class III, a light baryon decuplet and an anti-baryon anti-decuplet for class IV.}\label{tab:b6_T10_Tbar8_lv}\begin{tabular}{|cc|cc|}\hline\hline
\multicolumn{4}{|c|} {Class \uppercase\expandafter{\romannumeral3}} \\\hline
channel & amplitude($/V_{\text{ub}}$) &channel &amplitude($/V_{\text{ub}}$)\\\hline
$X_{u\bar{d}\bar{s}}^{+}\to \Delta^{++}  \overline \Lambda^0 l^-\bar\nu $ & $ \frac{\left(d_1+d_2\right) }{2 \sqrt{3}}$&
$X_{u\bar{d}\bar{s}}^{+}\to \Delta^{++}  \overline \Sigma^0 l^-\bar\nu $ & $ \frac{1}{2} \left(d_1+d_2\right) $\\\hline
$X_{u\bar{d}\bar{s}}^{+}\to \Delta^{+}  \overline \Sigma^+ l^-\bar\nu $ & $ \frac{\left(d_1+d_2\right) }{\sqrt{6}}$&
$X_{u\bar{d}\bar{s}}^{+}\to \Sigma^{\prime+}  \overline \Xi^+ l^-\bar\nu $ & $ \frac{\left(d_1+d_2\right) }{\sqrt{6}}$\\\hline
$X_{d\bar{s}\bar{u}}^{-}\to \Delta^{+}  \overline \Sigma^- l^-\bar\nu $ & $ \frac{d_2 }{\sqrt{6}}$&
$X_{d\bar{s}\bar{u}}^{-}\to \Delta^{0}  \overline \Lambda^0 l^-\bar\nu $ & $ \frac{1}{6} \left(d_1+d_2\right) $\\\hline
$X_{d\bar{s}\bar{u}}^{-}\to \Delta^{0}  \overline \Sigma^0 l^-\bar\nu $ & $ \frac{\left(d_1-d_2\right) }{2 \sqrt{3}}$&
$X_{d\bar{s}\bar{u}}^{-}\to \Delta^{-}  \overline \Sigma^+ l^-\bar\nu $ & $ \frac{d_1 }{\sqrt{2}}$\\\hline
$X_{d\bar{s}\bar{u}}^{-}\to \Sigma^{\prime0}  \overline \Xi^0 l^-\bar\nu $ & $ \frac{d_2 }{2 \sqrt{3}}$&
$X_{d\bar{s}\bar{u}}^{-}\to \Sigma^{\prime-}  \overline \Xi^+ l^-\bar\nu $ & $ \frac{d_1 }{\sqrt{6}}$\\\hline
$X_{s\bar{u}\bar{d}}^{-}\to \Sigma^{\prime+}  \overline p l^-\bar\nu $ & $ \frac{d_2 }{\sqrt{6}}$&
$X_{s\bar{u}\bar{d}}^{-}\to \Sigma^{\prime0}  \overline n l^-\bar\nu $ & $ \frac{d_2 }{2 \sqrt{3}}$\\\hline
$X_{s\bar{u}\bar{d}}^{-}\to \Xi^{\prime0}  \overline \Lambda^0 l^-\bar\nu $ & $ \frac{1}{6} \left(d_1-2 d_2\right) $&
$X_{s\bar{u}\bar{d}}^{-}\to \Xi^{\prime0}  \overline \Sigma^0 l^-\bar\nu $ & $ \frac{d_1 }{2 \sqrt{3}}$\\\hline
$X_{s\bar{u}\bar{d}}^{-}\to \Xi^{\prime-}  \overline \Sigma^+ l^-\bar\nu $ & $ \frac{d_1 }{\sqrt{6}}$&
$X_{s\bar{u}\bar{d}}^{-}\to \Omega^-  \overline \Xi^+ l^-\bar\nu $ & $ \frac{d_1 }{\sqrt{2}}$\\\hline
$Y_{(u\bar{u},s\bar{s})\bar{d}}^{0}\to \Delta^{++}  \overline p l^-\bar\nu $ & $ \frac{d_2 }{2}$&
$Y_{(u\bar{u},s\bar{s})\bar{d}}^{0}\to \Delta^{+}  \overline n l^-\bar\nu $ & $ \frac{d_2 }{2 \sqrt{3}}$\\\hline
$Y_{(u\bar{u},s\bar{s})\bar{d}}^{0}\to \Sigma^{\prime+}  \overline \Lambda^0 l^-\bar\nu $ & $ \frac{\left(2 d_1-d_2\right) }{6 \sqrt{2}}$&
$Y_{(u\bar{u},s\bar{s})\bar{d}}^{0}\to \Sigma^{\prime+}  \overline \Sigma^0 l^-\bar\nu $ & $ \frac{\left(2 d_1+d_2\right) }{2 \sqrt{6}}$\\\hline
$Y_{(u\bar{u},s\bar{s})\bar{d}}^{0}\to \Sigma^{\prime0}  \overline \Sigma^+ l^-\bar\nu $ & $ \frac{\left(2 d_1+d_2\right) }{2 \sqrt{6}}$&
$Y_{(u\bar{u},s\bar{s})\bar{d}}^{0}\to \Xi^{\prime0}  \overline \Xi^+ l^-\bar\nu $ & $ \frac{\left(2 d_1+d_2\right) }{2 \sqrt{3}}$\\\hline
$Y_{(u\bar{u},d\bar{d})\bar{s}}^{0}\to \Delta^{++}  \overline \Sigma^- l^-\bar\nu $ & $ \frac{d_2 }{2}$&
$Y_{(u\bar{u},d\bar{d})\bar{s}}^{0}\to \Delta^{+}  \overline \Lambda^0 l^-\bar\nu $ & $ \frac{\left(d_1+d_2\right) }{3 \sqrt{2}}$\\\hline
$Y_{(u\bar{u},d\bar{d})\bar{s}}^{0}\to \Delta^{+}  \overline \Sigma^0 l^-\bar\nu $ & $ \frac{d_1 }{\sqrt{6}}$&
$Y_{(u\bar{u},d\bar{d})\bar{s}}^{0}\to \Delta^{0}  \overline \Sigma^+ l^-\bar\nu $ & $ \frac{\left(2 d_1+d_2\right) }{2 \sqrt{3}}$\\\hline
$Y_{(u\bar{u},d\bar{d})\bar{s}}^{0}\to \Sigma^{\prime+}  \overline \Xi^0 l^-\bar\nu $ & $ \frac{d_2 }{2 \sqrt{3}}$&
$Y_{(u\bar{u},d\bar{d})\bar{s}}^{0}\to \Sigma^{\prime0}  \overline \Xi^+ l^-\bar\nu $ & $ \frac{\left(2 d_1+d_2\right) }{2 \sqrt{6}}$\\\hline
$Y_{(d\bar{d},s\bar{s})\bar{u}}^{-}\to \Delta^{+}  \overline p l^-\bar\nu $ & $ \frac{d_2 }{2 \sqrt{3}}$&
$Y_{(d\bar{d},s\bar{s})\bar{u}}^{-}\to \Delta^{0}  \overline n l^-\bar\nu $ & $ \frac{d_2 }{2 \sqrt{3}}$\\\hline
$Y_{(d\bar{d},s\bar{s})\bar{u}}^{-}\to \Sigma^{\prime+}  \overline \Sigma^- l^-\bar\nu $ & $ \frac{d_2 }{2 \sqrt{3}}$&
$Y_{(d\bar{d},s\bar{s})\bar{u}}^{-}\to \Sigma^{\prime0}  \overline \Lambda^0 l^-\bar\nu $ & $ \frac{1}{12} \left(2 d_1-d_2\right) $\\\hline
$Y_{(d\bar{d},s\bar{s})\bar{u}}^{-}\to \Sigma^{\prime0}  \overline \Sigma^0 l^-\bar\nu $ & $ \frac{\left(2 d_1-d_2\right) }{4 \sqrt{3}}$&
$Y_{(d\bar{d},s\bar{s})\bar{u}}^{-}\to \Sigma^{\prime-}  \overline \Sigma^+ l^-\bar\nu $ & $ \frac{d_1 }{\sqrt{3}}$\\\hline
$Y_{(d\bar{d},s\bar{s})\bar{u}}^{-}\to \Xi^{\prime0}  \overline \Xi^0 l^-\bar\nu $ & $ \frac{d_2 }{2 \sqrt{3}}$&
$Y_{(d\bar{d},s\bar{s})\bar{u}}^{-}\to \Xi^{\prime-}  \overline \Xi^+ l^-\bar\nu $ & $ \frac{d_1 }{\sqrt{3}}$\\\hline
\hline
\multicolumn{4}{|c|} {Class \uppercase\expandafter{\romannumeral4}} \\\hline
$X_{d\bar{s}\bar{u}}^{-}\to \Delta^{+}  \overline\Sigma^{\prime-} l^-\bar\nu $ & $ -\frac{f_1 }{3 \sqrt{2}}$&
$X_{d\bar{s}\bar{u}}^{-}\to \Delta^{0}  \overline \Sigma^{\prime0} l^-\bar\nu $ & $ -\frac{1}{3} f_1 $\\\hline
$X_{d\bar{s}\bar{u}}^{-}\to \Delta^{-}  \overline \Sigma^{\prime+} l^-\bar\nu $ & $ -\frac{f_1 }{\sqrt{6}}$&
$X_{d\bar{s}\bar{u}}^{-}\to \Sigma^{\prime0}  \overline \Xi^{\prime0} l^-\bar\nu $ & $ -\frac{1}{3} f_1 $\\\hline
$X_{d\bar{s}\bar{u}}^{-}\to \Sigma^{\prime-}  \overline\Xi^{\prime+} l^-\bar\nu $ & $ -\frac{1}{3} \sqrt{2} f_1 $&
$X_{d\bar{s}\bar{u}}^{-}\to \Xi^{\prime-}  \overline\Omega^+ l^-\bar\nu $ & $ -\frac{f_1 }{\sqrt{6}}$\\\hline
$X_{s\bar{u}\bar{d}}^{-}\to \Sigma^{\prime+}  \overline\Delta^{-} l^-\bar\nu $ & $ \frac{f_1 }{3 \sqrt{2}}$&
$X_{s\bar{u}\bar{d}}^{-}\to \Sigma^{\prime0}  \overline \Delta^{0} l^-\bar\nu $ & $ \frac{f_1 }{3}$\\\hline
$X_{s\bar{u}\bar{d}}^{-}\to \Sigma^{\prime-}  \overline\Delta^{+} l^-\bar\nu $ & $ \frac{f_1 }{\sqrt{6}}$&
$X_{s\bar{u}\bar{d}}^{-}\to \Xi^{\prime0}  \overline \Sigma^{\prime0} l^-\bar\nu $ & $ \frac{f_1 }{3}$\\\hline
$X_{s\bar{u}\bar{d}}^{-}\to \Xi^{\prime-}  \overline \Sigma^{\prime+} l^-\bar\nu $ & $ \frac{1}{3} \sqrt{2} f_1 $&
$X_{s\bar{u}\bar{d}}^{-}\to \Omega^-  \overline\Xi^{\prime+} l^-\bar\nu $ & $ \frac{f_1 }{\sqrt{6}}$\\\hline
$Y_{(u\bar{u},s\bar{s})\bar{d}}^{0}\to \Delta^{++}  \overline\Delta^{-} l^-\bar\nu $ & $ \frac{f_1 }{2 \sqrt{3}}$&
$Y_{(u\bar{u},s\bar{s})\bar{d}}^{0}\to \Delta^{+}  \overline \Delta^{0} l^-\bar\nu $ & $ \frac{f_1 }{3}$\\\hline
$Y_{(u\bar{u},s\bar{s})\bar{d}}^{0}\to \Delta^{0}  \overline\Delta^{+} l^-\bar\nu $ & $ \frac{f_1 }{2 \sqrt{3}}$&
$Y_{(u\bar{u},s\bar{s})\bar{d}}^{0}\to \Sigma^{\prime+}  \overline \Sigma^{\prime0} l^-\bar\nu $ & $ \frac{f_1 }{3 \sqrt{2}}$\\\hline
$Y_{(u\bar{u},s\bar{s})\bar{d}}^{0}\to \Sigma^{\prime0}  \overline \Sigma^{\prime+} l^-\bar\nu $ & $ \frac{f_1 }{3 \sqrt{2}}$&
$Y_{(u\bar{u},s\bar{s})\bar{d}}^{0}\to \Xi^{\prime0}  \overline\Xi^{\prime+} l^-\bar\nu $ & $ \frac{f_1 }{6}$\\\hline
$Y_{(u\bar{u},d\bar{d})\bar{s}}^{0}\to \Delta^{++}  \overline\Sigma^{\prime-} l^-\bar\nu $ & $ -\frac{f_1 }{2 \sqrt{3}}$&
$Y_{(u\bar{u},d\bar{d})\bar{s}}^{0}\to \Delta^{+}  \overline \Sigma^{\prime0} l^-\bar\nu $ & $ -\frac{f_1 }{3 \sqrt{2}}$\\\hline
$Y_{(u\bar{u},d\bar{d})\bar{s}}^{0}\to \Delta^{0}  \overline \Sigma^{\prime+} l^-\bar\nu $ & $ -\frac{1}{6} f_1 $&
$Y_{(u\bar{u},d\bar{d})\bar{s}}^{0}\to \Sigma^{\prime+}  \overline \Xi^{\prime0} l^-\bar\nu $ & $ -\frac{1}{3} f_1 $\\\hline
$Y_{(u\bar{u},d\bar{d})\bar{s}}^{0}\to \Sigma^{\prime0}  \overline\Xi^{\prime+} l^-\bar\nu $ & $ -\frac{f_1 }{3 \sqrt{2}}$&
$Y_{(u\bar{u},d\bar{d})\bar{s}}^{0}\to \Xi^{\prime0}  \overline\Omega^+ l^-\bar\nu $ & $ -\frac{f_1 }{2 \sqrt{3}}$\\\hline
$Y_{(d\bar{d},s\bar{s})\bar{u}}^{-}\to \Delta^{+}  \overline\Delta^{-} l^-\bar\nu $ & $ \frac{f_1 }{6}$&
$Y_{(d\bar{d},s\bar{s})\bar{u}}^{-}\to \Delta^{0}  \overline \Delta^{0} l^-\bar\nu $ & $ \frac{f_1 }{3}$\\\hline
$Y_{(d\bar{d},s\bar{s})\bar{u}}^{-}\to \Delta^{-}  \overline\Delta^{+} l^-\bar\nu $ & $ \frac{f_1 }{2}$&
$Y_{(d\bar{d},s\bar{s})\bar{u}}^{-}\to \Sigma^{\prime+}  \overline\Sigma^{\prime-} l^-\bar\nu $ & $ -\frac{1}{6} f_1 $\\\hline
$Y_{(d\bar{d},s\bar{s})\bar{u}}^{-}\to \Sigma^{\prime-}  \overline \Sigma^{\prime+} l^-\bar\nu $ & $ \frac{f_1 }{6}$&
$Y_{(d\bar{d},s\bar{s})\bar{u}}^{-}\to \Xi^{\prime0}  \overline \Xi^{\prime0} l^-\bar\nu $ & $ -\frac{1}{3} f_1 $\\\hline
$Y_{(d\bar{d},s\bar{s})\bar{u}}^{-}\to \Xi^{\prime-}  \overline\Xi^{\prime+} l^-\bar\nu $ & $ -\frac{1}{6} f_1 $&
$Y_{(d\bar{d},s\bar{s})\bar{u}}^{-}\to \Omega^-  \overline\Omega^+ l^-\bar\nu $ & $ -\frac{1}{2} f_1 $\\\hline
\hline
\end{tabular}
\end{table}

\subsubsection{$b\to c \ell \overline \nu_{\ell}$: semi-leptonic decays into a charmed  baryon plus a light anti-baryon}

$b\to c \ell \overline \nu_{\ell}$ transition belongs to the SU(3) flavor singlet. $X_{b6}$ can decay into a light anti-baryon  and a charmed  anti-triplet or sextet baryon. According to the symmetry of indexes, the processes with the final states of light anti-decuplet anti-baryon are not allowed. Therefor the effective Hamiltonian is written as
  \begin{eqnarray}
  \mathcal{H}_{eff}&=&-b_{1} (X_{b6})^{\{ij\}} (T_{c\bar3})_{[il]}(\overline T_8)^l_j \bar{\ell} \nu_{\ell}-c_1 (X_{b6})^{\{ij\}} (T_{c6})_{\{il\}}(\overline T_8)^l_j \bar{\ell} \nu_{\ell}.
  \end{eqnarray}
The amplitudes for different channels are given in Tab.~\ref{tab:b6_Tc3bar_Tbar8_lv} for a charmed anti-triplet baryon plus an octet anti-baryon and a charmed  sextet baryon plus an octet anti-baryon. We plotted the Feynman diagram in  Fig.~\ref{fig:topology-semileptonic}. (d).

The relations of decay widths for class  \uppercase\expandafter{\romannumeral1} are
\begin{eqnarray*}
&&\Gamma(X_{u\bar{d}\bar{s}}^{+}\to \Lambda_c^+  \overline \Sigma^+ l^-\bar\nu )
=\Gamma(X_{u\bar{d}\bar{s}}^{+}\to \Xi_c^+  \overline \Xi^+ l^-\bar\nu )
=\Gamma(X_{d\bar{s}\bar{u}}^{-}\to \Lambda_c^+  \overline \Sigma^- l^-\bar\nu)\\
&&=\Gamma(X_{d\bar{s}\bar{u}}^{-}\to \Xi_c^0  \overline \Xi^0 l^-\bar\nu )
=\Gamma(X_{s\bar{u}\bar{d}}^{-}\to \Xi_c^+  \overline p l^-\bar\nu )
=\Gamma(X_{s\bar{u}\bar{d}}^{-}\to \Xi_c^0  \overline n l^-\bar\nu )\\
&&=2\Gamma(Y_{(u\bar{u},s\bar{s})\bar{d}}^{0}\to \Lambda_c^+  \overline n l^-\bar\nu )
=\frac{4}{3}\Gamma(Y_{(u\bar{u},s\bar{s})\bar{d}}^{0}\to \Xi_c^+  \overline \Lambda^0 l^-\bar\nu)
=4\Gamma(Y_{(u\bar{u},s\bar{s})\bar{d}}^{0}\to \Xi_c^+  \overline \Sigma^0 l^-\bar\nu)\\
&&=2\Gamma(Y_{(u\bar{u},s\bar{s})\bar{d}}^{0}\to \Xi_c^0  \overline \Sigma^+ l^-\bar\nu)
=\Gamma(Y_{(u\bar{u},d\bar{d})\bar{s}}^{0}\to \Lambda_c^+  \overline \Sigma^0 l^-\bar\nu)
=2\Gamma(Y_{(u\bar{u},d\bar{d})\bar{s}}^{0}\to \Xi_c^+  \overline \Xi^0 l^-\bar\nu )\\
&&=2\Gamma(Y_{(u\bar{u},d\bar{d})\bar{s}}^{0}\to \Xi_c^0  \overline \Xi^+ l^-\bar\nu)
=2\Gamma(Y_{(d\bar{d},s\bar{s})\bar{u}}^{-}\to \Lambda_c^+  \overline p l^-\bar\nu )
=2\Gamma(Y_{(d\bar{d},s\bar{s})\bar{u}}^{-}\to \Xi_c^+  \overline \Sigma^- l^-\bar\nu )\\
&&=\frac{4}{3}\Gamma(Y_{(d\bar{d},s\bar{s})\bar{u}}^{-}\to \Xi_c^0  \overline \Lambda^0 l^-\bar\nu)
=4\Gamma(Y_{(d\bar{d},s\bar{s})\bar{u}}^{-}\to \Xi_c^0  \overline \Sigma^0 l^-\bar\nu).
\end{eqnarray*}
The relations of decay widths for class  \uppercase\expandafter{\romannumeral2} are
\begin{eqnarray*}
&&3\Gamma(X_{u\bar{d}\bar{s}}^{+}\to \Sigma_{c}^{++}  \overline \Lambda^0 l^-\bar\nu)
=\Gamma(X_{u\bar{d}\bar{s}}^{+}\to \Sigma_{c}^{++}  \overline \Sigma^0 l^-\bar\nu )
=\Gamma(X_{u\bar{d}\bar{s}}^{+}\to \Sigma_{c}^{+}  \overline \Sigma^+ l^-\bar\nu )\\
&&=\Gamma(X_{u\bar{d}\bar{s}}^{+}\to \Xi_{c}^{\prime+}  \overline \Xi^+ l^-\bar\nu)
=\Gamma(X_{d\bar{s}\bar{u}}^{-}\to \Sigma_{c}^{+}  \overline \Sigma^- l^-\bar\nu )
=3\Gamma(X_{d\bar{s}\bar{u}}^{-}\to \Sigma_{c}^{0}  \overline \Lambda^0 l^-\bar\nu )\\
&&=\Gamma(X_{d\bar{s}\bar{u}}^{-}\to \Sigma_{c}^{0}  \overline \Sigma^0 l^-\bar\nu )
=\Gamma(X_{d\bar{s}\bar{u}}^{-}\to \Xi_{c}^{\prime0}  \overline \Xi^0 l^-\bar\nu )
=\Gamma(X_{s\bar{u}\bar{d}}^{-}\to \Xi_{c}^{\prime+}  \overline p l^-\bar\nu )\\
&&=\Gamma(X_{s\bar{u}\bar{d}}^{-}\to \Xi_{c}^{\prime0}  \overline n l^-\bar\nu )
=\frac{3}{4}\Gamma(X_{s\bar{u}\bar{d}}^{-}\to \Omega_{c}^{0}  \overline \Lambda^0 l^-\bar\nu)
=\Gamma(Y_{(u\bar{u},s\bar{s})\bar{d}}^{0}\to \Sigma_{c}^{++}  \overline p l^-\bar\nu )\\
&&=2\Gamma(Y_{(u\bar{u},s\bar{s})\bar{d}}^{0}\to \Sigma_{c}^{+}  \overline n l^-\bar\nu )
=12\Gamma(Y_{(u\bar{u},s\bar{s})\bar{d}}^{0}\to \Xi_{c}^{\prime+}  \overline \Lambda^0 l^-\bar\nu )
=4\Gamma(Y_{(u\bar{u},s\bar{s})\bar{d}}^{0}\to \Xi_{c}^{\prime+}  \overline \Sigma^0 l^-\bar\nu )\\
&&=2\Gamma(Y_{(u\bar{u},s\bar{s})\bar{d}}^{0}\to \Xi_{c}^{\prime0}  \overline \Sigma^+ l^-\bar\nu)
=\Gamma(Y_{(u\bar{u},s\bar{s})\bar{d}}^{0}\to \Omega_{c}^{0}  \overline \Xi^+ l^-\bar\nu )
=\Gamma(Y_{(u\bar{u},d\bar{d})\bar{s}}^{0}\to \Sigma_{c}^{++}  \overline \Sigma^- l^-\bar\nu )\\
&&=3\Gamma(Y_{(u\bar{u},d\bar{d})\bar{s}}^{0}\to \Sigma_{c}^{+}  \overline \Lambda^0 l^-\bar\nu )
=\Gamma(Y_{(u\bar{u},d\bar{d})\bar{s}}^{0}\to \Sigma_{c}^{0}  \overline \Sigma^+ l^-\bar\nu)
=2\Gamma(Y_{(u\bar{u},d\bar{d})\bar{s}}^{0}\to \Xi_{c}^{\prime+}  \overline \Xi^0 l^-\bar\nu)\\
&&=2\Gamma(Y_{(u\bar{u},d\bar{d})\bar{s}}^{0}\to \Xi_{c}^{\prime0}  \overline \Xi^+ l^-\bar\nu )
=2\Gamma(Y_{(d\bar{d},s\bar{s})\bar{u}}^{-}\to \Sigma_{c}^{+}  \overline p l^-\bar\nu )
=\Gamma(Y_{(d\bar{d},s\bar{s})\bar{u}}^{-}\to \Sigma_{c}^{0}  \overline n l^-\bar\nu )\\
&&=2\Gamma(Y_{(d\bar{d},s\bar{s})\bar{u}}^{-}\to \Xi_{c}^{\prime+}  \overline \Sigma^- l^-\bar\nu )
=12\Gamma(Y_{(d\bar{d},s\bar{s})\bar{u}}^{-}\to \Xi_{c}^{\prime0}  \overline \Lambda^0 l^-\bar\nu )
=4\Gamma(Y_{(d\bar{d},s\bar{s})\bar{u}}^{-}\to \Xi_{c}^{\prime0}  \overline \Sigma^0 l^-\bar\nu )\\
&&=\Gamma(Y_{(d\bar{d},s\bar{s})\bar{u}}^{-}\to \Omega_{c}^{0}  \overline \Xi^0 l^-\bar\nu).
\end{eqnarray*}
\begin{table}
\caption{Amplitudes for open bottom tetraquark $X_{b6}$ semi-leptonic decays into  a charmed  baryon anti-triplet and an anti-baryon octet for class I, a charmed  baryon sextet and an anti-baryon octet for class II.}\label{tab:b6_Tc3bar_Tbar8_lv}\begin{tabular}{|cc|cc|}\hline\hline
\multicolumn{4}{|c|} {Class \uppercase\expandafter{\romannumeral1}} \\\hline
channel & amplitude($/V_{cb}$) &channel & amplitude($/V_{cb}$)\\\hline
$X_{u\bar{d}\bar{s}}^{+}\to \Lambda_c^+  \overline \Sigma^+ l^-\bar\nu $ & $ \frac{b_1}{\sqrt{2}}$&
$X_{u\bar{d}\bar{s}}^{+}\to \Xi_c^+  \overline \Xi^+ l^-\bar\nu $ & $ \frac{b_1}{\sqrt{2}}$\\\hline
$X_{d\bar{s}\bar{u}}^{-}\to \Lambda_c^+  \overline \Sigma^- l^-\bar\nu $ & $ -\frac{b_1}{\sqrt{2}}$&
$X_{d\bar{s}\bar{u}}^{-}\to \Xi_c^0  \overline \Xi^0 l^-\bar\nu $ & $ \frac{b_1}{\sqrt{2}}$\\\hline
$X_{s\bar{u}\bar{d}}^{-}\to \Xi_c^+  \overline p l^-\bar\nu $ & $ -\frac{b_1}{\sqrt{2}}$&
$X_{s\bar{u}\bar{d}}^{-}\to \Xi_c^0  \overline n l^-\bar\nu $ & $ -\frac{b_1}{\sqrt{2}}$\\\hline
$Y_{(u\bar{u},s\bar{s})\bar{d}}^{0}\to \Lambda_c^+  \overline n l^-\bar\nu $ & $ \frac{b_1}{2}$&
$Y_{(u\bar{u},s\bar{s})\bar{d}}^{0}\to \Xi_c^+  \overline \Lambda^0 l^-\bar\nu $ & $ -\frac{1}{2} \sqrt{\frac{3}{2}} b_1$\\\hline
$Y_{(u\bar{u},s\bar{s})\bar{d}}^{0}\to \Xi_c^+  \overline \Sigma^0 l^-\bar\nu $ & $ -\frac{b_1}{2 \sqrt{2}}$&
$Y_{(u\bar{u},s\bar{s})\bar{d}}^{0}\to \Xi_c^0  \overline \Sigma^+ l^-\bar\nu $ & $ -\frac{b_1}{2}$\\\hline
$Y_{(u\bar{u},d\bar{d})\bar{s}}^{0}\to \Lambda_c^+  \overline \Sigma^0 l^-\bar\nu $ & $ -\frac{b_1}{\sqrt{2}}$&
$Y_{(u\bar{u},d\bar{d})\bar{s}}^{0}\to \Xi_c^+  \overline \Xi^0 l^-\bar\nu $ & $ \frac{b_1}{2}$\\\hline
$Y_{(u\bar{u},d\bar{d})\bar{s}}^{0}\to \Xi_c^0  \overline \Xi^+ l^-\bar\nu $ & $ \frac{b_1}{2}$&
$Y_{(d\bar{d},s\bar{s})\bar{u}}^{-}\to \Lambda_c^+  \overline p l^-\bar\nu $ & $ -\frac{b_1}{2}$\\\hline
$Y_{(d\bar{d},s\bar{s})\bar{u}}^{-}\to \Xi_c^+  \overline \Sigma^- l^-\bar\nu $ & $ -\frac{b_1}{2}$&
$Y_{(d\bar{d},s\bar{s})\bar{u}}^{-}\to \Xi_c^0  \overline \Lambda^0 l^-\bar\nu $ & $ -\frac{1}{2} \sqrt{\frac{3}{2}} b_1$\\\hline
$Y_{(d\bar{d},s\bar{s})\bar{u}}^{-}\to \Xi_c^0  \overline \Sigma^0 l^-\bar\nu $ & $ \frac{b_1}{2 \sqrt{2}}$& &\\
\hline
\multicolumn{4}{|c|} {Class \uppercase\expandafter{\romannumeral2}} \\\hline
$X_{u\bar{d}\bar{s}}^{+}\to \Sigma_{c}^{++}  \overline \Lambda^0 l^-\bar\nu $ & $ \frac{c_1}{2 \sqrt{3}}$&
$X_{u\bar{d}\bar{s}}^{+}\to \Sigma_{c}^{++}  \overline \Sigma^0 l^-\bar\nu $ & $ \frac{c_1}{2}$\\\hline
$X_{u\bar{d}\bar{s}}^{+}\to \Sigma_{c}^{+}  \overline \Sigma^+ l^-\bar\nu $ & $ \frac{c_1}{2}$&
$X_{u\bar{d}\bar{s}}^{+}\to \Xi_{c}^{\prime+}  \overline \Xi^+ l^-\bar\nu $ & $ \frac{c_1}{2}$\\\hline
$X_{d\bar{s}\bar{u}}^{-}\to \Sigma_{c}^{+}  \overline \Sigma^- l^-\bar\nu $ & $ \frac{c_1}{2}$&
$X_{d\bar{s}\bar{u}}^{-}\to \Sigma_{c}^{0}  \overline \Lambda^0 l^-\bar\nu $ & $ \frac{c_1}{2 \sqrt{3}}$\\\hline
$X_{d\bar{s}\bar{u}}^{-}\to \Sigma_{c}^{0}  \overline \Sigma^0 l^-\bar\nu $ & $ -\frac{c_1}{2}$&
$X_{d\bar{s}\bar{u}}^{-}\to \Xi_{c}^{\prime0}  \overline \Xi^0 l^-\bar\nu $ & $ \frac{c_1}{2}$\\\hline
$X_{s\bar{u}\bar{d}}^{-}\to \Xi_{c}^{\prime+}  \overline p l^-\bar\nu $ & $ \frac{c_1}{2}$&
$X_{s\bar{u}\bar{d}}^{-}\to \Xi_{c}^{\prime0}  \overline n l^-\bar\nu $ & $ \frac{c_1}{2}$\\\hline
$X_{s\bar{u}\bar{d}}^{-}\to \Omega_{c}^{0}  \overline \Lambda^0 l^-\bar\nu $ & $ -\frac{c_1}{\sqrt{3}}$&
$Y_{(u\bar{u},s\bar{s})\bar{d}}^{0}\to \Sigma_{c}^{++}  \overline p l^-\bar\nu $ & $ \frac{c_1}{2}$\\\hline
$Y_{(u\bar{u},s\bar{s})\bar{d}}^{0}\to \Sigma_{c}^{+}  \overline n l^-\bar\nu $ & $ \frac{c_1}{2 \sqrt{2}}$&
$Y_{(u\bar{u},s\bar{s})\bar{d}}^{0}\to \Xi_{c}^{\prime+}  \overline \Lambda^0 l^-\bar\nu $ & $ -\frac{c_1}{4 \sqrt{3}}$\\\hline
$Y_{(u\bar{u},s\bar{s})\bar{d}}^{0}\to \Xi_{c}^{\prime+}  \overline \Sigma^0 l^-\bar\nu $ & $ \frac{c_1}{4}$&
$Y_{(u\bar{u},s\bar{s})\bar{d}}^{0}\to \Xi_{c}^{\prime0}  \overline \Sigma^+ l^-\bar\nu $ & $ \frac{c_1}{2 \sqrt{2}}$\\\hline
$Y_{(u\bar{u},s\bar{s})\bar{d}}^{0}\to \Omega_{c}^{0}  \overline \Xi^+ l^-\bar\nu $ & $ \frac{c_1}{2}$&
$Y_{(u\bar{u},d\bar{d})\bar{s}}^{0}\to \Sigma_{c}^{++}  \overline \Sigma^- l^-\bar\nu $ & $ \frac{c_1}{2}$\\\hline
$Y_{(u\bar{u},d\bar{d})\bar{s}}^{0}\to \Sigma_{c}^{+}  \overline \Lambda^0 l^-\bar\nu $ & $ \frac{c_1}{2 \sqrt{3}}$&
$Y_{(u\bar{u},d\bar{d})\bar{s}}^{0}\to \Sigma_{c}^{0}  \overline \Sigma^+ l^-\bar\nu $ & $ \frac{c_1}{2}$\\\hline
$Y_{(u\bar{u},d\bar{d})\bar{s}}^{0}\to \Xi_{c}^{\prime+}  \overline \Xi^0 l^-\bar\nu $ & $ \frac{c_1}{2 \sqrt{2}}$&
$Y_{(u\bar{u},d\bar{d})\bar{s}}^{0}\to \Xi_{c}^{\prime0}  \overline \Xi^+ l^-\bar\nu $ & $ \frac{c_1}{2 \sqrt{2}}$\\\hline
$Y_{(d\bar{d},s\bar{s})\bar{u}}^{-}\to \Sigma_{c}^{+}  \overline p l^-\bar\nu $ & $ \frac{c_1}{2 \sqrt{2}}$&
$Y_{(d\bar{d},s\bar{s})\bar{u}}^{-}\to \Sigma_{c}^{0}  \overline n l^-\bar\nu $ & $ \frac{c_1}{2}$\\\hline
$Y_{(d\bar{d},s\bar{s})\bar{u}}^{-}\to \Xi_{c}^{\prime+}  \overline \Sigma^- l^-\bar\nu $ & $ \frac{c_1}{2 \sqrt{2}}$&
$Y_{(d\bar{d},s\bar{s})\bar{u}}^{-}\to \Xi_{c}^{\prime0}  \overline \Lambda^0 l^-\bar\nu $ & $ -\frac{c_1}{4 \sqrt{3}}$\\\hline
$Y_{(d\bar{d},s\bar{s})\bar{u}}^{-}\to \Xi_{c}^{\prime0}  \overline \Sigma^0 l^-\bar\nu $ & $ -\frac{c_1}{4}$&
$Y_{(d\bar{d},s\bar{s})\bar{u}}^{-}\to \Omega_{c}^{0}  \overline \Xi^0 l^-\bar\nu $ & $ \frac{c_1}{2}$\\\hline
\hline
\end{tabular}
\end{table}

\begin{figure}
\includegraphics[width=0.9\columnwidth]{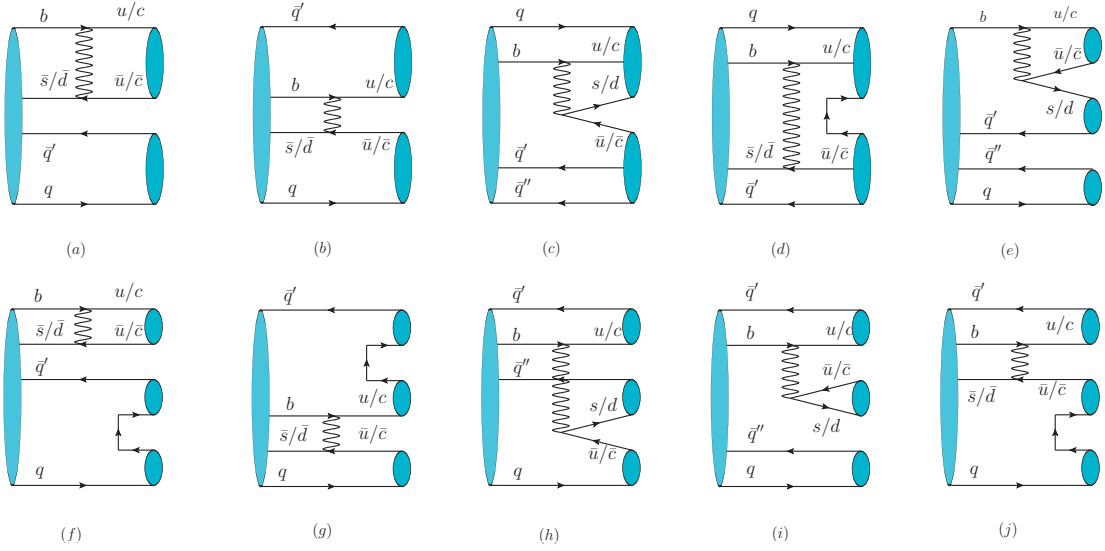}
\caption{Feynman diagrams for nonleptonic decays of open-flavor bottom tetraquark. Panels (a,b) represent the decays into two mesons in the final states. In panels (e-j), the final states include three mesons. Panels (c,d) denote the process of baryonic states. }
\label{fig:topology-nonleptonic}
\end{figure}
\section{Non-leptonic $X_{b6}$ decays}
\label{sec:nonleptonic_decay}


In quark-level transitions, the bottom quark non-leptonic weak decays can be divided into four different kinds
\begin{eqnarray}
b\to c\bar c d/s, \;
b\to c \bar u d/s, \;
b\to u \bar c d/s, \;
b\to q_1 \bar q_2 q_3,
\end{eqnarray}
where $q_{1,2,3}$ represent the light quark.
For these four kinds of transitions, the weak effective Hamiltonian ${\cal H}_{eff}$ are given separately
\begin{eqnarray}
 {\cal H}_{eff}&=&\frac{G_{F}}{\sqrt{2}}
     V_{cb} V_{cq}^{*} \big[
     C_{1}  O^{\bar cc}_{1}
  +  C_{2}  O^{\bar cc}_{2}\Big] +{\rm h.c.} ,\nonumber \\
  {\cal H}_{eff}&=&\frac{G_{F}}{\sqrt{2}}
     V_{cb} V_{uq}^{*} \big[
     C_{1}  O^{\bar cu}_{1}
  +  C_{2}  O^{\bar cu}_{2}\Big] +{\rm h.c.} ,\nonumber \\
 {\cal H}_{eff}&=&\frac{G_{F}}{\sqrt{2}}
     V_{ub} V_{cq}^{*} \big[
     C_{1}  O^{\bar uc}_{1}
  +  C_{2}  O^{\bar uc}_{2}\Big] +{\rm h.c.} ,\nonumber \\
 {\cal H}_{eff}&=&\frac{G_{F}}{\sqrt{2}}
     \bigg\{ V_{ub} V_{uq}^{*} \big[
     C_{1}  O^{\bar uu}_{1}
  +  C_{2}  O^{\bar uu}_{2}\Big]- V_{tb} V_{tq}^{*} \big[{\sum\limits_{i=3}^{10}} C_{i}  O_{i} \Big]\bigg\}+ \mbox{h.c.}
\end{eqnarray}
where $O_i$ is the four-quark effective operators and $C_i$ is the Wilson short-distance coefficient. $O_1$, $O_2$ are tree level operators while $O_3 - O_{10}$ are called as penguin operators.
In hadron level, it is easily to discuss the $X_{b6}$ non-leptonic decay modes when writing down the effective Hamiltonian using the SU(3) flavor symmetry.

\subsection{$b\to c\bar c d/s$ transition}
\subsubsection{Two-body decays into mesons}

The  $b\to c\bar c d/s$ transition leads to form a SU(3) triplet operator. For $X_{b6}$ decays to two mesons, the effective Hamiltonian  is constructed as
  \begin{eqnarray}
  \mathcal{H}_{eff}&=&a_1 (X_{b6})_{[jk]}^{i}(H_3)^j M^k_i  J/\psi +a_2 (X_{b6})_{[jk]}^i (H_3)^j  (\overline D)^k  D_i,
  \end{eqnarray}
where $(H_{  3})_{2}=-(H_{  3})^{13}=(H_{  3})^{31}=V_{cd}^*$ and $(H_{  3})_{3}=(H_{  3})^{12}=-(H_{  3})^{21}=V_{cs}^*$.
The amplitudes for the decays to the $J/\psi$ plus a light meson  are given in Tab.~\ref{tab:b6_Jpsi_M}. The amplitudes for the decays to the charmed meson plus an anti-charmed meson are given in Tab.~\ref{tab:b6_D_Dbar}. The corresponding Feynman diagrams are given in Fig.~\ref{fig:topology-nonleptonic}. (a,b), which are denoted as the W-exchange processes.

The relations for decay widths are  given as respectively
\begin{eqnarray*}
    \Gamma(X_{u\bar{d}\bar{s}}^{+}\to \pi^+ J/\psi)= { }\Gamma(X_{d\bar{s}\bar{u}}^{-}\to \pi^- J/\psi)=
2\Gamma(Y_{(u\bar{u},s\bar{s})\bar{d}}^{0}\to \overline K^0 J/\psi)=
\Gamma(Y_{(u\bar{u},d\bar{d})\bar{s}}^{0}\to \pi^0 J/\psi)\\=
2\Gamma(Y_{(d\bar{d},s\bar{s})\bar{u}}^{-}\to K^- J/\psi),\\
    \Gamma(X_{u\bar{d}\bar{s}}^{+}\to K^+ J/\psi)= { }\Gamma(X_{s\bar{u}\bar{d}}^{-}\to K^- J/\psi)=
\frac{4}{3}\Gamma(Y_{(u\bar{u},s\bar{s})\bar{d}}^{0}\to \eta J/\psi)=
 2\Gamma(Y_{(u\bar{u},d\bar{d})\bar{s}}^{0}\to K^0 J/\psi)\\=
4\Gamma(Y_{(u\bar{u},s\bar{s})\bar{d}}^{0}\to \pi^0 J/\psi)=
2\Gamma(Y_{(d\bar{d},s\bar{s})\bar{u}}^{-}\to \pi^- J/\psi),\\
\frac{\Gamma(X_{u\bar{d}\bar{s}}^{+}\to \pi^+ J/\psi)}{\Gamma(X_{u\bar{d}\bar{s}}^{+}\to K^+ J/\psi)}=\frac{|V_{cs}|^2}{|V_{cd}|^2},
\end{eqnarray*}
and
\begin{eqnarray*}
\frac{\Gamma(X_{u\bar{d}\bar{s}}^{+}\to  D^+\overline D^0)}{ \Gamma(X_{u\bar{d}\bar{s}}^{+}\to  D^+_s\overline D^0)}=\frac{\Gamma(X_{d\bar{s}\bar{u}}^{-}\to  D^0D^-)}{\Gamma(X_{s\bar{u}\bar{d}}^{-}\to  D^0 D^-_s)}=\frac{2\Gamma(Y_{(u\bar{u},s\bar{s})\bar{d}}^{0}\to  D^+ D^-_s)}{2\Gamma(Y_{(u\bar{u},s\bar{s})\bar{d}}^{0}\to  D^+_s D^-_s)}=\frac{2\Gamma(Y_{(u\bar{u},d\bar{d})\bar{s}}^{0}\to  D^+D^-)}{2\Gamma(Y_{(u\bar{u},d\bar{d})\bar{s}}^{0}\to  D^+_sD^-)}\\
=\frac{2\Gamma(Y_{(d\bar{d},s\bar{s})\bar{u}}^{-}\to  D^0 D^-_s)}{2\Gamma(Y_{(d\bar{d},s\bar{s})\bar{u}}^{-}\to  D^0D^-)}=\frac{2\Gamma(Y_{(u\bar{u},d\bar{d})\bar{s}}^{0}\to  D^0\overline D^0)}{2\Gamma(Y_{(u\bar{u},s\bar{s})\bar{d}}^{0}\to  D^0\overline D^0)}=\frac{|V_{cs}|^2}{|V_{cd}|^2}.
\end{eqnarray*}
\begin{table}
\caption{Open bottom tetraquark $X_{b6}$ decays into  $J/\psi$ and a light meson.}\label{tab:b6_Jpsi_M}\begin{tabular}{|cc|cc|}\hline\hline
channel & amplitude($/V_{\text{cb}}$) &channel & amplitude($/V_{\text{cb}}$)\\\hline
$X_{u\bar{d}\bar{s}}^{+}\to   \pi^+   J/\psi $ & $ -\frac{a_1 V_{\text{cs}}^*}{\sqrt{2}}$&
$X_{u\bar{d}\bar{s}}^{+}\to   K^+   J/\psi $ & $ \frac{a_1 V^*_{cd}}{\sqrt{2}}$\\\hline
$X_{d\bar{s}\bar{u}}^{-}\to   \pi^-   J/\psi $ & $ \frac{a_1 V_{\text{cs}}^*}{\sqrt{2}}$&
$X_{s\bar{u}\bar{d}}^{-}\to   K^-   J/\psi $ & $ -\frac{a_1 V^*_{cd}}{\sqrt{2}}$\\\hline
$Y_{(u\bar{u},s\bar{s})\bar{d}}^{0}\to   \pi^0   J/\psi $ & $ -\frac{a_1 V^*_{cd}}{2 \sqrt{2}}$&
$Y_{(u\bar{u},s\bar{s})\bar{d}}^{0}\to   \overline K^0   J/\psi $ & $ -\frac{1}{2} a_1 V_{\text{cs}}^*$\\\hline
$Y_{(u\bar{u},s\bar{s})\bar{d}}^{0}\to   \eta   J/\psi $ & $ -\frac{1}{2} \sqrt{\frac{3}{2}} a_1 V^*_{cd}$&
$Y_{(u\bar{u},d\bar{d})\bar{s}}^{0}\to   \pi^0   J/\psi $ & $ \frac{a_1 V_{\text{cs}}^*}{\sqrt{2}}$\\\hline
$Y_{(u\bar{u},d\bar{d})\bar{s}}^{0}\to   K^0   J/\psi $ & $ \frac{1}{2} a_1 V^*_{cd}$&
$Y_{(d\bar{d},s\bar{s})\bar{u}}^{-}\to   \pi^-   J/\psi $ & $ -\frac{1}{2} a_1 V^*_{cd}$\\\hline
$Y_{(d\bar{d},s\bar{s})\bar{u}}^{-}\to   K^-   J/\psi $ & $ \frac{1}{2} a_1 V_{\text{cs}}^*$& &\\
\hline
\end{tabular}
\end{table}
\begin{table}
\caption{Open bottom tetraquark $X_{b6}$ decays into a charmed meson and an anti-charmed meson.}\label{tab:b6_D_Dbar}\begin{tabular}{|cc|cc|}\hline\hline
channel & amplitude($/V_{\text{cb}}$) &channel &amplitude($/V_{\text{cb}}$)\\\hline
$X_{u\bar{d}\bar{s}}^{+}\to    D^+  \overline D^0 $ & $ -\frac{a_2 V^*_{cs}}{\sqrt{2}}$&
$X_{u\bar{d}\bar{s}}^{+}\to    D^+_s  \overline D^0 $ & $ \frac{a_2 V^*_{cd}}{\sqrt{2}}$\\\hline
$X_{d\bar{s}\bar{u}}^{-}\to    D^0  D^- $ & $ \frac{a_2 V^*_{cs}}{\sqrt{2}}$&
$X_{s\bar{u}\bar{d}}^{-}\to    D^0   D^-_s $ & $ -\frac{a_2 V^*_{cd}}{\sqrt{2}}$\\\hline
$Y_{(u\bar{u},s\bar{s})\bar{d}}^{0}\to    D^0  \overline D^0 $ & $ -\frac{1}{2} a_2 V^*_{cd}$&
$Y_{(u\bar{u},s\bar{s})\bar{d}}^{0}\to    D^+   D^-_s $ & $ -\frac{1}{2} a_2 V^*_{cs}$\\\hline
$Y_{(u\bar{u},s\bar{s})\bar{d}}^{0}\to    D^+_s   D^-_s $ & $ \frac{1}{2} a_2 V^*_{cd}$&
$Y_{(u\bar{u},d\bar{d})\bar{s}}^{0}\to    D^0  \overline D^0 $ & $ \frac{1}{2} a_2 V^*_{cs}$\\\hline
$Y_{(u\bar{u},d\bar{d})\bar{s}}^{0}\to    D^+  D^- $ & $ -\frac{1}{2} a_2 V^*_{cs}$&
$Y_{(u\bar{u},d\bar{d})\bar{s}}^{0}\to    D^+_s  D^- $ & $ \frac{1}{2} a_2 V^*_{cd}$\\\hline
$Y_{(d\bar{d},s\bar{s})\bar{u}}^{-}\to    D^0  D^- $ & $ -\frac{1}{2} a_2 V^*_{cd}$&
$Y_{(d\bar{d},s\bar{s})\bar{u}}^{-}\to    D^0   D^-_s $ & $ \frac{1}{2} a_2 V^*_{cs}$\\\hline
\hline
\end{tabular}
\end{table}
\subsubsection{Two-body decays into a charmed baryon and an anti-charmed anti-baryon}

The transition $b\to c\bar c d/s$ can also produce a charmed baryon and an anti-charmed anti-baryon.  Charmed baryons can be an anti-triplet or a sextet, while the anti-charmed anti-baryons can form  the triplet or anti-sextet baryons. The effective Hamiltonian in hadron level is constructed as
  \begin{eqnarray}
  \mathcal{H}_{eff}&=&b_1 (X_{b6})^{\{ij\}}(H_3)_{[ik]} (T_{c\bar3})_{[jl]} (T_{\bar{c}3})^{[kl]} +c_1 (X_{b6})_{[jk]}^i (H_3)^j (T_{c6})_{\{il\}} (T_{\bar{c}3})^{[kl]}\nonumber\\
  &&+c_2 (X_{b6})_{[jk]}^i (H_3)^l (T_{c6})_{\{il\}} (T_{\bar{c}3})^{[jk]}+d_1 (X_{b6})_{[jk]}^i (H_3)^j (T_{c\bar3})_{[il]} (T_{\bar{c}\bar6})^{\{kl\}} \nonumber\\
  &&+f_1 (X_{b6})_{[jk]}^i (H_3)^j (T_{c6})_{\{il\}} (T_{\bar{c}\bar6})^{\{kl\}}.
  \end{eqnarray}
We listed the  decay amplitudes in Tab.\ref{tab:b6_Tc3bar_Tbarc3} and Tab.\ref{tab:b6_Tc3bar_Tbarc6bar}, which  represent anti-triplet charmed baryon plus the triplet anti-charmed anti-baryon, sextet charmed baryon plus the triplet anti-charmed anti-baryon, anti-triplet charmed baryon plus the anti-sextet anti-charmed anti-baryon, sextet charmed baryon plus the anti-sextet anti-charmed anti-baryon, respectively. The corresponding topologies are given in Fig.~\ref{fig:topology-nonleptonic}. (c,d).

The relations of decay widths for class \uppercase\expandafter{\romannumeral1} are
\begin{eqnarray*}
&&\frac{\Gamma(X_{u\bar{d}\bar{s}}^{+}\to   \Lambda_c^+  \overline \Xi_{\bar{c}}^0 )}{\Gamma(X_{u\bar{d}\bar{s}}^{+}\to   \Xi_c^+  \overline \Xi_{\bar{c}}^0 )}=\frac{\Gamma(X_{s\bar{u}\bar{d}}^{-}\to   \Xi_c^0 \overline \Lambda_{\bar{c}}^-)}{\Gamma(X_{d\bar{s}\bar{u}}^{-}\to   \Xi_c^0  \overline \Xi_{\bar{c}}^- )}=\frac{2\Gamma(Y_{(u\bar{u},s\bar{s})\bar{d}}^{0}\to   \Lambda_c^+  \overline \Lambda_{\bar{c}}^-)}{2\Gamma(Y_{(u\bar{u},s\bar{s})\bar{d}}^{0}\to   \Xi_c^+  \overline \Lambda_{\bar{c}}^- )}=\frac{2\Gamma(Y_{(u\bar{u},s\bar{s})\bar{d}}^{0}\to   \Xi_c^0  \overline \Xi_{\bar{c}}^0 )}{2\Gamma(Y_{(u\bar{u},d\bar{d})\bar{s}}^{0}\to   \Xi_c^+  \overline \Xi_{\bar{c}}^- )}\\
&=&\frac{2\Gamma(Y_{(u\bar{u},d\bar{d})\bar{s}}^{0}\to   \Lambda_c^+  \overline \Xi_{\bar{c}}^-)}{2\Gamma(Y_{(u\bar{u},d\bar{d})\bar{s}}^{0}\to   \Xi_c^0  \overline \Xi_{\bar{c}}^0)}=\frac{2\Gamma(Y_{(d\bar{d},s\bar{s})\bar{u}}^{-}\to   \Xi_c^0  \overline \Xi_{\bar{c}}^-)}{2\Gamma(Y_{(d\bar{d},s\bar{s})\bar{u}}^{-}\to   \Xi_c^0  \overline \Lambda_{\bar{c}}^- )}=\frac{|V_{cd}|^2}{|V_{cs}|^2}.
\end{eqnarray*}
The relations of decay widths for class \uppercase\expandafter{\romannumeral2} are
\begin{eqnarray*}
&&\frac{\Gamma(X_{u\bar{d}\bar{s}}^{+}\to   \Sigma_{c}^{+}  \overline \Xi_{\bar{c}}^0 )}{\Gamma(X_{u\bar{d}\bar{s}}^{+}\to   \Xi_{c}^{\prime+}  \overline \Xi_{\bar{c}}^0 )}=\frac{\Gamma(X_{s\bar{u}\bar{d}}^{-}\to   \Xi_{c}^{\prime0}  \overline \Lambda_{\bar{c}}^-)}{\Gamma(X_{d\bar{s}\bar{u}}^{-}\to   \Xi_{c}^{\prime0}  \overline \Xi_{\bar{c}}^- )}=\frac{2\Gamma(Y_{(u\bar{u},s\bar{s})\bar{d}}^{0}\to   \Sigma_{c}^{+}  \overline \Lambda_{\bar{c}}^-)}{2\Gamma(Y_{(u\bar{u},d\bar{d})\bar{s}}^{0}\to   \Xi_{c}^{\prime0}  \overline \Xi_{\bar{c}}^0)}=\frac{2\Gamma(Y_{(u\bar{u},s\bar{s})\bar{d}}^{0}\to   \Xi_{c}^{\prime0}  \overline \Xi_{\bar{c}}^0 )}{2\Gamma(Y_{(u\bar{u},d\bar{d})\bar{s}}^{0}\to   \Xi_{c}^{\prime+}  \overline \Xi_{\bar{c}}^- )}\\
&=&\frac{\Gamma(Y_{(d\bar{d},s\bar{s})\bar{u}}^{-}\to   \Sigma_{c}^{0}  \overline \Lambda_{\bar{c}}^-)}{\Gamma(Y_{(u\bar{u},s\bar{s})\bar{d}}^{0}\to   \Omega_{c}^{0}  \overline \Xi_{\bar{c}}^0 )}=\frac{\Gamma(Y_{(u\bar{u},d\bar{d})\bar{s}}^{0}\to   \Sigma_{c}^{0}  \overline \Xi_{\bar{c}}^0 )}{\Gamma(Y_{(d\bar{d},s\bar{s})\bar{u}}^{-}\to   \Omega_{c}^{0}  \overline \Xi_{\bar{c}}^- )}=\frac{|V_{cd}|^2}{|V_{cs}|^2},\\
&&\frac{\Gamma(X_{u\bar{d}\bar{s}}^{+}\to   \Sigma_{c}^{++}  \overline \Xi_{\bar{c}}^-)}{\Gamma(X_{u\bar{d}\bar{s}}^{+}\to   \Sigma_{c}^{++}  \overline \Lambda_{\bar{c}}^-)}=\frac{\Gamma(X_{s\bar{u}\bar{d}}^{-}\to   \Omega_{c}^{0}  \overline \Xi_{\bar{c}}^-  )}{\Gamma(X_{d\bar{s}\bar{u}}^{-}\to   \Sigma_{c}^{0}  \overline \Lambda_{\bar{c}}^- )}=\frac{\Gamma(Y_{(u\bar{u},s\bar{s})\bar{d}}^{0}\to   \Xi_{c}^{\prime+}  \overline \Xi_{\bar{c}}^-)}{\Gamma(Y_{(u\bar{u},d\bar{d})\bar{s}}^{0}\to   \Sigma_{c}^{+}  \overline \Lambda_{\bar{c}}^- )}=\frac{|V_{cd}|^2}{|V_{cs}|^2},\\
&&\frac{\Gamma(Y_{(u\bar{u},d\bar{d})\bar{s}}^{0}\to   \Sigma_{c}^{+}  \overline \Xi_{\bar{c}}^- )}{\Gamma(Y_{(u\bar{u},s\bar{s})\bar{d}}^{0}\to   \Xi_{c}^{\prime+}  \overline \Lambda_{\bar{c}}^-)}=\frac{\Gamma(Y_{(d\bar{d},s\bar{s})\bar{u}}^{-}\to   \Xi_{c}^{\prime0}  \overline \Xi_{\bar{c}}^- )}{\Gamma(Y_{(d\bar{d},s\bar{s})\bar{u}}^{-}\to   \Xi_{c}^{\prime0}  \overline \Lambda_{\bar{c}}^- )}=\frac{|V_{cd}|^2}{|V_{cs}|^2},\\
&&\frac{\Gamma(X_{d\bar{s}\bar{u}}^{-}\to   \Sigma_{c}^{0}  \overline \Xi_{\bar{c}}^- )}{\Gamma(X_{s\bar{u}\bar{d}}^{-}\to   \Omega_{c}^{0}  \overline \Lambda_{\bar{c}}^-)}=\frac{|V_{cd}|^2}{|V_{cs}|^2}.
\end{eqnarray*}
 For class \uppercase\expandafter{\romannumeral3}, the relations of decay widths are
\begin{eqnarray*}
&&\frac{\Gamma(X_{u\bar{d}\bar{s}}^{+}\to   \Xi_c^+  \overline \Omega_{\bar{c}}^{0})}{\Gamma(X_{u\bar{d}\bar{s}}^{+}\to   \Lambda_c^+  \overline \Sigma_{\bar{c}}^{0} )}=\frac{\Gamma(X_{s\bar{u}\bar{d}}^{-}\to   \Xi_c^+  \overline \Sigma_{\bar{c}}^{--})}{\Gamma(X_{d\bar{s}\bar{u}}^{-}\to   \Lambda_c^+  \overline \Sigma_{\bar{c}}^{--})}=\frac{\Gamma(Y_{(u\bar{u},s\bar{s})\bar{d}}^{0}\to   \Xi_c^+  \overline \Xi_{\bar{c}}^{\prime-} )}{\Gamma(Y_{(u\bar{u},d\bar{d})\bar{s}}^{0}\to   \Lambda_c^+  \overline \Sigma_{\bar{c}}^- )}=\frac{2\Gamma(X_{u\bar{d}\bar{s}}^{+}\to   \Lambda_c^+  \overline \Xi_{\bar{c}}^{\prime0})}{2\Gamma(Y_{(d\bar{d},s\bar{s})\bar{u}}^{-}\to   \Xi_c^+  \overline \Sigma_{\bar{c}}^{--} )}\\
&=&\frac{2\Gamma(X_{s\bar{u}\bar{d}}^{-}\to   \Xi_c^0  \overline \Sigma_{\bar{c}}^- )}{2\Gamma(X_{u\bar{d}\bar{s}}^{+}\to   \Xi_c^+  \overline \Xi_{\bar{c}}^{\prime0} )}=\frac{2\Gamma(Y_{(u\bar{u},d\bar{d})\bar{s}}^{0}\to   \Xi_c^0  \overline \Omega_{\bar{c}}^{0} )}{2\Gamma(X_{d\bar{s}\bar{u}}^{-}\to   \Xi_c^0  \overline \Xi_{\bar{c}}^{\prime-})}=\frac{2\Gamma(Y_{(d\bar{d},s\bar{s})\bar{u}}^{-}\to   \Lambda_c^+  \overline \Sigma_{\bar{c}}^{--} )}{2\Gamma(Y_{(u\bar{u},s\bar{s})\bar{d}}^{0}\to   \Xi_c^0  \overline \Sigma_{\bar{c}}^{0} )}=\frac{4\Gamma(Y_{(u\bar{u},s\bar{s})\bar{d}}^{0}\to   \Lambda_c^+  \overline \Sigma_{\bar{c}}^- )}{4\Gamma(Y_{(u\bar{u},s\bar{s})\bar{d}}^{0}\to   \Xi_c^+  \overline \Sigma_{\bar{c}}^-)}\\
&=&\frac{4\Gamma(Y_{(u\bar{u},s\bar{s})\bar{d}}^{0}\to   \Xi_c^0  \overline \Xi_{\bar{c}}^{\prime0})}{4\Gamma(Y_{(u\bar{u},d\bar{d})\bar{s}}^{0}\to   \Xi_c^+  \overline \Xi_{\bar{c}}^{\prime-} )}=\frac{4\Gamma(Y_{(u\bar{u},d\bar{d})\bar{s}}^{0}\to   \Lambda_c^+  \overline \Xi_{\bar{c}}^{\prime-})}{4\Gamma(Y_{(u\bar{u},d\bar{d})\bar{s}}^{0}\to   \Xi_c^0  \overline \Xi_{\bar{c}}^{\prime0} )}=\frac{4\Gamma(Y_{(d\bar{d},s\bar{s})\bar{u}}^{-}\to   \Xi_c^0  \overline \Xi_{\bar{c}}^{\prime-})}{4\Gamma(Y_{(d\bar{d},s\bar{s})\bar{u}}^{-}\to   \Xi_c^0  \overline \Sigma_{\bar{c}}^-)}=\frac{|V_{cd}|^2}{|V_{cs}|^2}.
\end{eqnarray*}
For class \uppercase\expandafter{\romannumeral4}, the relations of decay widths are
\begin{eqnarray*}
&&\frac{\Gamma(X_{u\bar{d}\bar{s}}^{+}\to   \Sigma_{c}^{++}  \overline \Sigma_{\bar{c}}^-)}{\Gamma(X_{u\bar{d}\bar{s}}^{+}\to   \Sigma_{c}^{++}  \overline \Xi_{\bar{c}}^{\prime-} )}=\frac{\Gamma(X_{u\bar{d}\bar{s}}^{+}\to   \Sigma_{c}^{+}  \overline \Sigma_{\bar{c}}^{0} )}{\Gamma(X_{u\bar{d}\bar{s}}^{+}\to   \Xi_{c}^{\prime+}  \overline \Omega_{\bar{c}}^{0})}=\frac{\Gamma(X_{d\bar{s}\bar{u}}^{-}\to   \Sigma_{c}^{+}  \overline \Sigma_{\bar{c}}^{--} )}{\Gamma(X_{s\bar{u}\bar{d}}^{-}\to   \Xi_{c}^{\prime+}  \overline \Sigma_{\bar{c}}^{--})}=\frac{\Gamma(X_{d\bar{s}\bar{u}}^{-}\to   \Sigma_{c}^{0}  \overline \Sigma_{\bar{c}}^-)}{\Gamma(X_{s\bar{u}\bar{d}}^{-}\to   \Omega_{c}^{0}  \overline \Xi_{\bar{c}}^{\prime-})}\\
&=&\frac{\Gamma(Y_{(u\bar{u},d\bar{d})\bar{s}}^{0}\to   \Sigma_{c}^{++}  \overline \Sigma_{\bar{c}}^{--})}{\Gamma(Y_{(u\bar{u},s\bar{s})\bar{d}}^{0}\to   \Sigma_{c}^{++}  \overline \Sigma_{\bar{c}}^{--})}=\frac{\Gamma(Y_{(u\bar{u},d\bar{d})\bar{s}}^{0}\to   \Sigma_{c}^{0}  \overline \Sigma_{\bar{c}}^{0} )}{\Gamma(Y_{(u\bar{u},s\bar{s})\bar{d}}^{0}\to   \Omega_{c}^{0}  \overline \Omega_{\bar{c}}^{0})}=\frac{2\Gamma(X_{u\bar{d}\bar{s}}^{+}\to   \Xi_{c}^{\prime+}  \overline \Xi_{\bar{c}}^{\prime0})}{2\Gamma(X_{s\bar{u}\bar{d}}^{-}\to   \Xi_{c}^{\prime0}  \overline \Sigma_{\bar{c}}^- )}=\frac{2\Gamma(X_{d\bar{s}\bar{u}}^{-}\to   \Xi_{c}^{\prime0}  \overline \Xi_{\bar{c}}^{\prime-} )}{2\Gamma(X_{u\bar{d}\bar{s}}^{+}\to   \Sigma_{c}^{+}  \overline \Xi_{\bar{c}}^{\prime0} )}\\
&=&\frac{2\Gamma(Y_{(u\bar{u},s\bar{s})\bar{d}}^{0}\to   \Xi_{c}^{\prime0}  \overline \Sigma_{\bar{c}}^{0})}{2\Gamma(Y_{(u\bar{u},d\bar{d})\bar{s}}^{0}\to   \Sigma_{c}^{0}  \overline \Xi_{\bar{c}}^{\prime0} )}=\frac{2\Gamma(Y_{(u\bar{u},s\bar{s})\bar{d}}^{0}\to   \Omega_{c}^{0}  \overline \Xi_{\bar{c}}^{\prime0})}{2\Gamma(Y_{(u\bar{u},d\bar{d})\bar{s}}^{0}\to   \Xi_{c}^{\prime0}  \overline \Omega_{\bar{c}}^{0})}=\frac{2\Gamma(Y_{(d\bar{d},s\bar{s})\bar{u}}^{-}\to   \Xi_{c}^{\prime+}  \overline \Sigma_{\bar{c}}^{--})}{2\Gamma(Y_{(d\bar{d},s\bar{s})\bar{u}}^{-}\to   \Sigma_{c}^{+}  \overline \Sigma_{\bar{c}}^{--} )}=\frac{2\Gamma(Y_{(d\bar{d},s\bar{s})\bar{u}}^{-}\to   \Omega_{c}^{0}  \overline \Xi_{\bar{c}}^{\prime-} )}{2\Gamma(Y_{(d\bar{d},s\bar{s})\bar{u}}^{-}\to   \Sigma_{c}^{0}  \overline \Sigma_{\bar{c}}^-)}\\
&=&\frac{4\Gamma(Y_{(d\bar{d},s\bar{s})\bar{u}}^{-}\to   \Xi_{c}^{\prime0}  \overline \Sigma_{\bar{c}}^-)}{4\Gamma(Y_{(u\bar{u},d\bar{d})\bar{s}}^{0}\to   \Sigma_{c}^{+}  \overline \Xi_{\bar{c}}^{\prime-} )}=\frac{4\Gamma(Y_{(u\bar{u},d\bar{d})\bar{s}}^{0}\to   \Xi_{c}^{\prime+}  \overline \Xi_{\bar{c}}^{\prime-})}{4\Gamma(Y_{(d\bar{d},s\bar{s})\bar{u}}^{-}\to   \Xi_{c}^{\prime0}  \overline \Xi_{\bar{c}}^{\prime-})}=\frac{4\Gamma(Y_{(u\bar{u},d\bar{d})\bar{s}}^{0}\to   \Xi_{c}^{\prime0}  \overline \Xi_{\bar{c}}^{\prime0})}{4\Gamma(Y_{(u\bar{u},s\bar{s})\bar{d}}^{0}\to   \Sigma_{c}^{+}  \overline \Sigma_{\bar{c}}^-)}=\frac{4\Gamma(Y_{(u\bar{u},s\bar{s})\bar{d}}^{0}\to   \Xi_{c}^{\prime+}  \overline \Sigma_{\bar{c}}^-)}{4\Gamma(Y_{(u\bar{u},s\bar{s})\bar{d}}^{0}\to   \Xi_{c}^{\prime0}  \overline \Xi_{\bar{c}}^{\prime0} )}\\
&=&\frac{|V_{cs}|^2}{|V_{cd}|^2}\\.
\end{eqnarray*}
\begin{table}
\caption{Open bottom tetraquark $X_{b6}$ decays into a charmed baryon anti-triplet and an anti-charmed anti-baryon triplet for class I, a charmed baryon sextet and an anti-charmed anti-baryon triplet for class II.}\label{tab:b6_Tc3bar_Tbarc3}\begin{tabular}{|cc|cc|}\hline\hline
\multicolumn{4}{|c|} {Class \uppercase\expandafter{\romannumeral1}} \\\hline
channel & amplitude($/V_{\text{cb}}$)&channel & amplitude($/V_{\text{cb}}$) \\\hline
$X_{u\bar{d}\bar{s}}^{+}\to   \Lambda_c^+  \overline \Xi_{\bar{c}}^0 $ & $ -\frac{b_1 V^*_{cd}}{\sqrt{2}}$&
$X_{u\bar{d}\bar{s}}^{+}\to   \Xi_c^+  \overline \Xi_{\bar{c}}^0 $ & $ -\frac{b_1 V^*_{cs}}{\sqrt{2}}$\\\hline
$X_{d\bar{s}\bar{u}}^{-}\to   \Xi_c^0  \overline \Xi_{\bar{c}}^- $ & $ \frac{b_1 V^*_{cs}}{\sqrt{2}}$&
$X_{s\bar{u}\bar{d}}^{-}\to   \Xi_c^0  \overline \Lambda_{\bar{c}}^- $ & $ \frac{b_1 V^*_{cd}}{\sqrt{2}}$\\\hline
$Y_{(u\bar{u},s\bar{s})\bar{d}}^{0}\to   \Lambda_c^+  \overline \Lambda_{\bar{c}}^- $ & $ -\frac{1}{2} b_1 V^*_{cd}$&
$Y_{(u\bar{u},s\bar{s})\bar{d}}^{0}\to   \Xi_c^+  \overline \Lambda_{\bar{c}}^- $ & $ -\frac{1}{2} b_1 V^*_{cs}$\\\hline
$Y_{(u\bar{u},s\bar{s})\bar{d}}^{0}\to   \Xi_c^0  \overline \Xi_{\bar{c}}^0 $ & $ \frac{1}{2} b_1 V^*_{cd}$&
$Y_{(u\bar{u},d\bar{d})\bar{s}}^{0}\to   \Lambda_c^+  \overline \Xi_{\bar{c}}^- $ & $ \frac{1}{2} b_1 V^*_{cd}$\\\hline
$Y_{(u\bar{u},d\bar{d})\bar{s}}^{0}\to   \Xi_c^+  \overline \Xi_{\bar{c}}^- $ & $ \frac{1}{2} b_1 V^*_{cs}$&
$Y_{(u\bar{u},d\bar{d})\bar{s}}^{0}\to   \Xi_c^0  \overline \Xi_{\bar{c}}^0 $ & $ -\frac{1}{2} b_1 V^*_{cs}$\\\hline
$Y_{(d\bar{d},s\bar{s})\bar{u}}^{-}\to   \Xi_c^0  \overline \Lambda_{\bar{c}}^- $ & $ -\frac{1}{2} b_1 V^*_{cs}$&
$Y_{(d\bar{d},s\bar{s})\bar{u}}^{-}\to   \Xi_c^0  \overline \Xi_{\bar{c}}^- $ & $ -\frac{1}{2} b_1 V^*_{cd}$\\\hline
\hline
\multicolumn{4}{|c|} {Class \uppercase\expandafter{\romannumeral2}} \\\hline
$X_{u\bar{d}\bar{s}}^{+}\to   \Sigma_{c}^{++}  \overline \Lambda_{\bar{c}}^- $ & $ \frac{c_1 V^*_{cs}}{\sqrt{2}}$&
$X_{u\bar{d}\bar{s}}^{+}\to   \Sigma_{c}^{++}  \overline \Xi_{\bar{c}}^- $ & $ -\frac{c_1 V^*_{cd}}{\sqrt{2}}$\\\hline
$X_{u\bar{d}\bar{s}}^{+}\to   \Sigma_{c}^{+}  \overline \Xi_{\bar{c}}^0 $ & $ -\frac{1}{2} \left(c_1-2 c_2\right) V^*_{cd}$&
$X_{u\bar{d}\bar{s}}^{+}\to   \Xi_{c}^{\prime+}  \overline \Xi_{\bar{c}}^0 $ & $ -\frac{1}{2} \left(c_1-2 c_2\right) V^*_{cs}$\\\hline
$X_{d\bar{s}\bar{u}}^{-}\to   \Sigma_{c}^{0}  \overline \Lambda_{\bar{c}}^- $ & $ \frac{c_1 V^*_{cs}}{\sqrt{2}}$&
$X_{d\bar{s}\bar{u}}^{-}\to   \Sigma_{c}^{0}  \overline \Xi_{\bar{c}}^- $ & $ -\sqrt{2} c_2 V^*_{cd}$\\\hline
$X_{d\bar{s}\bar{u}}^{-}\to   \Xi_{c}^{\prime0}  \overline \Xi_{\bar{c}}^- $ & $ \frac{1}{2} \left(c_1-2 c_2\right) V^*_{cs}$&
$X_{s\bar{u}\bar{d}}^{-}\to   \Xi_{c}^{\prime0}  \overline \Lambda_{\bar{c}}^- $ & $ -\frac{1}{2} \left(c_1-2 c_2\right) V^*_{cd}$\\\hline
$X_{s\bar{u}\bar{d}}^{-}\to   \Omega_{c}^{0}  \overline \Lambda_{\bar{c}}^- $ & $ \sqrt{2} c_2 V^*_{cs}$&
$X_{s\bar{u}\bar{d}}^{-}\to   \Omega_{c}^{0}  \overline \Xi_{\bar{c}}^- $ & $ -\frac{c_1 V^*_{cd}}{\sqrt{2}}$\\\hline
$Y_{(u\bar{u},s\bar{s})\bar{d}}^{0}\to   \Sigma_{c}^{+}  \overline \Lambda_{\bar{c}}^- $ & $ -\frac{\left(c_1-2 c_2\right) V^*_{cd}}{2 \sqrt{2}}$&
$Y_{(u\bar{u},s\bar{s})\bar{d}}^{0}\to   \Xi_{c}^{\prime+}  \overline \Lambda_{\bar{c}}^- $ & $ \frac{\left(c_1+2 c_2\right) V^*_{cs}}{2 \sqrt{2}}$\\\hline
$Y_{(u\bar{u},s\bar{s})\bar{d}}^{0}\to   \Xi_{c}^{\prime+}  \overline \Xi_{\bar{c}}^- $ & $ -\frac{c_1 V^*_{cd}}{\sqrt{2}}$&
$Y_{(u\bar{u},s\bar{s})\bar{d}}^{0}\to   \Xi_{c}^{\prime0}  \overline \Xi_{\bar{c}}^0 $ & $ -\frac{\left(c_1-2 c_2\right) V^*_{cd}}{2 \sqrt{2}}$\\\hline
$Y_{(u\bar{u},s\bar{s})\bar{d}}^{0}\to   \Omega_{c}^{0}  \overline \Xi_{\bar{c}}^0 $ & $ -\frac{1}{2} \left(c_1-2 c_2\right) V^*_{cs}$&
$Y_{(u\bar{u},d\bar{d})\bar{s}}^{0}\to   \Sigma_{c}^{+}  \overline \Lambda_{\bar{c}}^- $ & $ \frac{c_1 V^*_{cs}}{\sqrt{2}}$\\\hline
$Y_{(u\bar{u},d\bar{d})\bar{s}}^{0}\to   \Sigma_{c}^{+}  \overline \Xi_{\bar{c}}^- $ & $ -\frac{\left(c_1+2 c_2\right) V^*_{cd}}{2 \sqrt{2}}$&
$Y_{(u\bar{u},d\bar{d})\bar{s}}^{0}\to   \Sigma_{c}^{0}  \overline \Xi_{\bar{c}}^0 $ & $ -\frac{1}{2} \left(c_1-2 c_2\right) V^*_{cd}$\\\hline
$Y_{(u\bar{u},d\bar{d})\bar{s}}^{0}\to   \Xi_{c}^{\prime+}  \overline \Xi_{\bar{c}}^- $ & $ \frac{\left(c_1-2 c_2\right) V^*_{cs}}{2 \sqrt{2}}$&
$Y_{(u\bar{u},d\bar{d})\bar{s}}^{0}\to   \Xi_{c}^{\prime0}  \overline \Xi_{\bar{c}}^0 $ & $ -\frac{\left(c_1-2 c_2\right) V^*_{cs}}{2 \sqrt{2}}$\\\hline
$Y_{(d\bar{d},s\bar{s})\bar{u}}^{-}\to   \Sigma_{c}^{0}  \overline \Lambda_{\bar{c}}^- $ & $ -\frac{1}{2} \left(c_1-2 c_2\right) V^*_{cd}$&
$Y_{(d\bar{d},s\bar{s})\bar{u}}^{-}\to   \Xi_{c}^{\prime0}  \overline \Lambda_{\bar{c}}^- $ & $ \frac{\left(c_1+2 c_2\right) V^*_{cs}}{2 \sqrt{2}}$\\\hline
$Y_{(d\bar{d},s\bar{s})\bar{u}}^{-}\to   \Xi_{c}^{\prime0}  \overline \Xi_{\bar{c}}^- $ & $ -\frac{\left(c_1+2 c_2\right) V^*_{cd}}{2 \sqrt{2}}$&
$Y_{(d\bar{d},s\bar{s})\bar{u}}^{-}\to   \Omega_{c}^{0}  \overline \Xi_{\bar{c}}^- $ & $ \frac{1}{2} \left(c_1-2 c_2\right) V^*_{cs}$\\\hline
\hline
\end{tabular}
\end{table}
\begin{table}
\caption{Open bottom tetraquark $X_{b}$ decays into a charmed baryon anti-triplet and an anti-charmed anti-baryon anti-sextet for class III, a charmed baryon sextet and an anti-charmed anti-baryon anti-sextet for class IV.}\label{tab:b6_Tc3bar_Tbarc6bar}\begin{tabular}{|cc|cc|}\hline\hline
\multicolumn{4}{|c|} {Class \uppercase\expandafter{\romannumeral3}} \\\hline
channel & amplitude($/V_{\text{cb}}$)&channel & amplitude($/V_{\text{cb}}$) \\\hline
$X_{u\bar{d}\bar{s}}^{+}\to   \Lambda_c^+  \overline \Sigma_{\bar{c}}^{0} $ & $ -\frac{d_1 V^*_{cs}}{\sqrt{2}}$&
$X_{u\bar{d}\bar{s}}^{+}\to   \Lambda_c^+  \overline \Xi_{\bar{c}}^{\prime0} $ & $ \frac{1}{2} d_1 V^*_{cd}$\\\hline
$X_{u\bar{d}\bar{s}}^{+}\to   \Xi_c^+  \overline \Xi_{\bar{c}}^{\prime0} $ & $ -\frac{1}{2} d_1 V^*_{cs}$&
$X_{u\bar{d}\bar{s}}^{+}\to   \Xi_c^+  \overline \Omega_{\bar{c}}^{0} $ & $ \frac{d_1 V^*_{cd}}{\sqrt{2}}$\\\hline
$X_{d\bar{s}\bar{u}}^{-}\to   \Lambda_c^+  \overline \Sigma_{\bar{c}}^{--} $ & $ -\frac{d_1 V^*_{cs}}{\sqrt{2}}$&
$X_{d\bar{s}\bar{u}}^{-}\to   \Xi_c^0  \overline \Xi_{\bar{c}}^{\prime-} $ & $ \frac{1}{2} d_1 V^*_{cs}$\\\hline
$X_{s\bar{u}\bar{d}}^{-}\to   \Xi_c^+  \overline \Sigma_{\bar{c}}^{--} $ & $ \frac{d_1 V^*_{cd}}{\sqrt{2}}$&
$X_{s\bar{u}\bar{d}}^{-}\to   \Xi_c^0  \overline \Sigma_{\bar{c}}^- $ & $ \frac{1}{2} d_1 V^*_{cd}$\\\hline
$Y_{(u\bar{u},s\bar{s})\bar{d}}^{0}\to   \Lambda_c^+  \overline \Sigma_{\bar{c}}^- $ & $ -\frac{d_1 V^*_{cd}}{2 \sqrt{2}}$&
$Y_{(u\bar{u},s\bar{s})\bar{d}}^{0}\to   \Xi_c^+  \overline \Sigma_{\bar{c}}^- $ & $ \frac{d_1 V^*_{cs}}{2 \sqrt{2}}$\\\hline
$Y_{(u\bar{u},s\bar{s})\bar{d}}^{0}\to   \Xi_c^+  \overline \Xi_{\bar{c}}^{\prime-} $ & $ -\frac{d_1 V^*_{cd}}{\sqrt{2}}$&
$Y_{(u\bar{u},s\bar{s})\bar{d}}^{0}\to   \Xi_c^0  \overline \Sigma_{\bar{c}}^{0} $ & $ \frac{1}{2} d_1 V^*_{cs}$\\\hline
$Y_{(u\bar{u},s\bar{s})\bar{d}}^{0}\to   \Xi_c^0  \overline \Xi_{\bar{c}}^{\prime0} $ & $ -\frac{d_1 V^*_{cd}}{2 \sqrt{2}}$&
$Y_{(u\bar{u},d\bar{d})\bar{s}}^{0}\to   \Lambda_c^+  \overline \Sigma_{\bar{c}}^- $ & $ \frac{d_1 V^*_{cs}}{\sqrt{2}}$\\\hline
$Y_{(u\bar{u},d\bar{d})\bar{s}}^{0}\to   \Lambda_c^+  \overline \Xi_{\bar{c}}^{\prime-} $ & $ -\frac{d_1 V^*_{cd}}{2 \sqrt{2}}$&
$Y_{(u\bar{u},d\bar{d})\bar{s}}^{0}\to   \Xi_c^+  \overline \Xi_{\bar{c}}^{\prime-} $ & $ \frac{d_1 V^*_{cs}}{2 \sqrt{2}}$\\\hline
$Y_{(u\bar{u},d\bar{d})\bar{s}}^{0}\to   \Xi_c^0  \overline \Xi_{\bar{c}}^{\prime0} $ & $ -\frac{d_1 V^*_{cs}}{2 \sqrt{2}}$&
$Y_{(u\bar{u},d\bar{d})\bar{s}}^{0}\to   \Xi_c^0  \overline \Omega_{\bar{c}}^{0} $ & $ \frac{1}{2} d_1 V^*_{cd}$\\\hline
$Y_{(d\bar{d},s\bar{s})\bar{u}}^{-}\to   \Lambda_c^+  \overline \Sigma_{\bar{c}}^{--} $ & $ \frac{1}{2} d_1 V^*_{cd}$&
$Y_{(d\bar{d},s\bar{s})\bar{u}}^{-}\to   \Xi_c^+  \overline \Sigma_{\bar{c}}^{--} $ & $ -\frac{1}{2} d_1 V^*_{cs}$\\\hline
$Y_{(d\bar{d},s\bar{s})\bar{u}}^{-}\to   \Xi_c^0  \overline \Sigma_{\bar{c}}^- $ & $ -\frac{d_1 V^*_{cs}}{2 \sqrt{2}}$&
$Y_{(d\bar{d},s\bar{s})\bar{u}}^{-}\to   \Xi_c^0  \overline \Xi_{\bar{c}}^{\prime-} $ & $ -\frac{d_1 V^*_{cd}}{2 \sqrt{2}}$\\\hline
\hline
\multicolumn{4}{|c|} {Class \uppercase\expandafter{\romannumeral4}} \\\hline
$X_{u\bar{d}\bar{s}}^{+}\to   \Sigma_{c}^{++}  \overline \Sigma_{\bar{c}}^- $ & $ -\frac{1}{2} f_1 V^*_{cs}$&
$X_{u\bar{d}\bar{s}}^{+}\to   \Sigma_{c}^{++}  \overline \Xi_{\bar{c}}^{\prime-} $ & $ \frac{1}{2} f_1 V^*_{cd}$\\\hline
$X_{u\bar{d}\bar{s}}^{+}\to   \Sigma_{c}^{+}  \overline \Sigma_{\bar{c}}^{0} $ & $ -\frac{1}{2} f_1 V^*_{cs}$&
$X_{u\bar{d}\bar{s}}^{+}\to   \Sigma_{c}^{+}  \overline \Xi_{\bar{c}}^{\prime0} $ & $ \frac{f_1 V^*_{cd}}{2 \sqrt{2}}$\\\hline
$X_{u\bar{d}\bar{s}}^{+}\to   \Xi_{c}^{\prime+}  \overline \Xi_{\bar{c}}^{\prime0} $ & $ -\frac{f_1 V^*_{cs}}{2 \sqrt{2}}$&
$X_{u\bar{d}\bar{s}}^{+}\to   \Xi_{c}^{\prime+}  \overline \Omega_{\bar{c}}^{0} $ & $ \frac{1}{2} f_1 V^*_{cd}$\\\hline
$X_{d\bar{s}\bar{u}}^{-}\to   \Sigma_{c}^{+}  \overline \Sigma_{\bar{c}}^{--} $ & $ \frac{1}{2} f_1 V^*_{cs}$&
$X_{d\bar{s}\bar{u}}^{-}\to   \Sigma_{c}^{0}  \overline \Sigma_{\bar{c}}^- $ & $ \frac{1}{2} f_1 V^*_{cs}$\\\hline
$X_{d\bar{s}\bar{u}}^{-}\to   \Xi_{c}^{\prime0}  \overline \Xi_{\bar{c}}^{\prime-} $ & $ \frac{f_1 V^*_{cs}}{2 \sqrt{2}}$&
$X_{s\bar{u}\bar{d}}^{-}\to   \Xi_{c}^{\prime+}  \overline \Sigma_{\bar{c}}^{--} $ & $ -\frac{1}{2} f_1 V^*_{cd}$\\\hline
$X_{s\bar{u}\bar{d}}^{-}\to   \Xi_{c}^{\prime0}  \overline \Sigma_{\bar{c}}^- $ & $ -\frac{f_1 V^*_{cd}}{2 \sqrt{2}}$&
$X_{s\bar{u}\bar{d}}^{-}\to   \Omega_{c}^{0}  \overline \Xi_{\bar{c}}^{\prime-} $ & $ -\frac{1}{2} f_1 V^*_{cd}$\\\hline
$Y_{(u\bar{u},s\bar{s})\bar{d}}^{0}\to   \Sigma_{c}^{++}  \overline \Sigma_{\bar{c}}^{--} $ & $ -\frac{1}{2} f_1 V^*_{cd}$&
$Y_{(u\bar{u},s\bar{s})\bar{d}}^{0}\to   \Sigma_{c}^{+}  \overline \Sigma_{\bar{c}}^- $ & $ -\frac{1}{4} f_1 V^*_{cd}$\\\hline
$Y_{(u\bar{u},s\bar{s})\bar{d}}^{0}\to   \Xi_{c}^{\prime+}  \overline \Sigma_{\bar{c}}^- $ & $ -\frac{1}{4} f_1 V^*_{cs}$&
$Y_{(u\bar{u},s\bar{s})\bar{d}}^{0}\to   \Xi_{c}^{\prime0}  \overline \Sigma_{\bar{c}}^{0} $ & $ -\frac{f_1 V^*_{cs}}{2 \sqrt{2}}$\\\hline
$Y_{(u\bar{u},s\bar{s})\bar{d}}^{0}\to   \Xi_{c}^{\prime0}  \overline \Xi_{\bar{c}}^{\prime0} $ & $ \frac{1}{4} f_1 V^*_{cd}$&
$Y_{(u\bar{u},s\bar{s})\bar{d}}^{0}\to   \Omega_{c}^{0}  \overline \Xi_{\bar{c}}^{\prime0} $ & $ -\frac{f_1 V^*_{cs}}{2 \sqrt{2}}$\\\hline
$Y_{(u\bar{u},s\bar{s})\bar{d}}^{0}\to   \Omega_{c}^{0}  \overline \Omega_{\bar{c}}^{0} $ & $ \frac{1}{2} f_1 V^*_{cd}$&
$Y_{(u\bar{u},d\bar{d})\bar{s}}^{0}\to   \Sigma_{c}^{++}  \overline \Sigma_{\bar{c}}^{--} $ & $ \frac{1}{2} f_1 V^*_{cs}$\\\hline
$Y_{(u\bar{u},d\bar{d})\bar{s}}^{0}\to   \Sigma_{c}^{+}  \overline \Xi_{\bar{c}}^{\prime-} $ & $ \frac{1}{4} f_1 V^*_{cd}$&
$Y_{(u\bar{u},d\bar{d})\bar{s}}^{0}\to   \Sigma_{c}^{0}  \overline \Sigma_{\bar{c}}^{0} $ & $ -\frac{1}{2} f_1 V^*_{cs}$\\\hline
$Y_{(u\bar{u},d\bar{d})\bar{s}}^{0}\to   \Sigma_{c}^{0}  \overline \Xi_{\bar{c}}^{\prime0} $ & $ \frac{f_1 V^*_{cd}}{2 \sqrt{2}}$&
$Y_{(u\bar{u},d\bar{d})\bar{s}}^{0}\to   \Xi_{c}^{\prime+}  \overline \Xi_{\bar{c}}^{\prime-} $ & $ \frac{1}{4} f_1 V^*_{cs}$\\\hline
$Y_{(u\bar{u},d\bar{d})\bar{s}}^{0}\to   \Xi_{c}^{\prime0}  \overline \Xi_{\bar{c}}^{\prime0} $ & $ -\frac{1}{4} f_1 V^*_{cs}$&
$Y_{(u\bar{u},d\bar{d})\bar{s}}^{0}\to   \Xi_{c}^{\prime0}  \overline \Omega_{\bar{c}}^{0} $ & $ \frac{f_1 V^*_{cd}}{2 \sqrt{2}}$\\\hline
$Y_{(d\bar{d},s\bar{s})\bar{u}}^{-}\to   \Sigma_{c}^{+}  \overline \Sigma_{\bar{c}}^{--} $ & $ -\frac{f_1 V^*_{cd}}{2 \sqrt{2}}$&
$Y_{(d\bar{d},s\bar{s})\bar{u}}^{-}\to   \Sigma_{c}^{0}  \overline \Sigma_{\bar{c}}^- $ & $ -\frac{f_1 V^*_{cd}}{2 \sqrt{2}}$\\\hline
$Y_{(d\bar{d},s\bar{s})\bar{u}}^{-}\to   \Xi_{c}^{\prime+}  \overline \Sigma_{\bar{c}}^{--} $ & $ \frac{f_1 V^*_{cs}}{2 \sqrt{2}}$&
$Y_{(d\bar{d},s\bar{s})\bar{u}}^{-}\to   \Xi_{c}^{\prime0}  \overline \Sigma_{\bar{c}}^- $ & $ \frac{1}{4} f_1 V^*_{cs}$\\\hline
$Y_{(d\bar{d},s\bar{s})\bar{u}}^{-}\to   \Xi_{c}^{\prime0}  \overline \Xi_{\bar{c}}^{\prime-} $ & $ -\frac{1}{4} f_1 V^*_{cd}$&
$Y_{(d\bar{d},s\bar{s})\bar{u}}^{-}\to   \Omega_{c}^{0}  \overline \Xi_{\bar{c}}^{\prime-} $ & $ \frac{f_1 V^*_{cs}}{2 \sqrt{2}}$\\\hline
\hline
\end{tabular}
\end{table}

\subsubsection{Three-body decays into mesons}

The $b\to c\bar c d/s$ transition can also lead to three-body meson decays in which the effective Hamiltonian in hadron level is
\begin{eqnarray}
  {\cal H}_{{eff}}&= &  a_1 (X_{b6})^{\{ij\}} (H_{  3})_{[il]} M^{k}_{j} M^{l}_k~ J/\psi+ a_2 (X_{b6})_{[jk]}^i (H_{  3})^j M^{k}_{i} D_l (\overline D)^{l}+ a_3 (X_{b6})_{[jk]}^i (H_{  3})^j M^{l}_{i} D_l (\overline D)^{k}\nonumber\\
  &&+a_4 (X_{b6})_{[jk]}^i (H_{  3})^l M^{j}_{i} D_l (\overline D)^{k}+a_5 (X_{b6})_{[jk]}^i (H_{  3})^j M^{k}_{l} D_i (\overline D)^{l}+a_6 (X_{b6})_{[jk]}^i (H_{  3})^l M^{j}_{l} D_i (\overline D)^{k}.
\end{eqnarray}
The Feynman diagrams are shown in Fig.~\ref{fig:topology-nonleptonic}. (e-j). It should be noted that the diagram in Fig.~\ref{fig:topology-nonleptonic}.(e)  can lead to the process $X_{s\bar u\bar d}\to \overline K^0 K^- J/\psi$. However,  
the total amplitude of the process vanishes for the antisymmetric lower indexes in $X_{b6}$. The decay amplitudes of  $X_{b6}$ decays to $J/\psi$ and two light mesons are given in Tab.~\ref{tab:b6_Jpsi_2M}, while the amplitudes of  $X_{b6}$ decays to an anti-charmed meson plus a charmed  meson and a light meson are given in Tab.~\ref{tab:b6_D_antiD_M}. From them,  we obtain the results for these decay widths
\small
\begin{eqnarray*}
&&\Gamma(X_{u\bar{d}\bar{s}}^{+}\to   \pi^+   K^0   J/\psi )=2\Gamma(X_{u\bar{d}\bar{s}}^{+}\to   \pi^0   K^+   J/\psi )=6\Gamma(X_{u\bar{d}\bar{s}}^{+}\to   K^+   \eta   J/\psi )\\
&=&2\Gamma(X_{s\bar{u}\bar{d}}^{-}\to   \pi^0   K^-   J/\psi)=\Gamma(X_{s\bar{u}\bar{d}}^{-}\to   \pi^-   \overline K^0   J/\psi )=6\Gamma(X_{s\bar{u}\bar{d}}^{-}\to   K^-   \eta   J/\psi)\\
&=&2\Gamma(Y_{(u\bar{u},s\bar{s})\bar{d}}^{0}\to   \pi^+   \pi^-   J/\psi)=\Gamma(Y_{(u\bar{u},s\bar{s})\bar{d}}^{0}\to   \pi^0   \pi^0   J/\psi)=6\Gamma(Y_{(u\bar{u},s\bar{s})\bar{d}}^{0}\to   \pi^0   \eta   J/\psi )\\
&=&2\Gamma(Y_{(u\bar{u},s\bar{s})\bar{d}}^{0}\to   K^0   \overline K^0   J/\psi)=\Gamma(Y_{(u\bar{u},s\bar{s})\bar{d}}^{0}\to   \eta   \eta   J/\psi)=4\Gamma(Y_{(u\bar{u},d\bar{d})\bar{s}}^{0}\to   \pi^0   K^0   J/\psi)\\
&=&2\Gamma(Y_{(u\bar{u},d\bar{d})\bar{s}}^{0}\to   \pi^-   K^+   J/\psi)=12\Gamma(Y_{(u\bar{u},d\bar{d})\bar{s}}^{0}\to   K^0   \eta   J/\psi)=3\Gamma(Y_{(d\bar{d},s\bar{s})\bar{u}}^{-}\to   \pi^-   \eta   J/\psi)\\
&=&2\Gamma(Y_{(d\bar{d},s\bar{s})\bar{u}}^{-}\to   K^0   K^-   J/\psi ),\\
&&\frac{3}{2}\Gamma(X_{u\bar{d}\bar{s}}^{+}\to   \pi^+   \eta   J/\psi)=\Gamma(X_{u\bar{d}\bar{s}}^{+}\to   K^+   \overline K^0   J/\psi )=\frac{3}{2}\Gamma(X_{d\bar{s}\bar{u}}^{-}\to   \pi^-   \eta   J/\psi )\\
&=&\Gamma(X_{d\bar{s}\bar{u}}^{-}\to   K^0   K^-   J/\psi)=2\Gamma(Y_{(u\bar{u},s\bar{s})\bar{d}}^{0}\to   \pi^+   K^-   J/\psi)=4\Gamma(Y_{(u\bar{u},s\bar{s})\bar{d}}^{0}\to   \pi^0   \overline K^0   J/\psi)\\
&=&12\Gamma(Y_{(u\bar{u},s\bar{s})\bar{d}}^{0}\to   \overline K^0   \eta   J/\psi )=\frac{3}{2}\Gamma(Y_{(u\bar{u},d\bar{d})\bar{s}}^{0}\to   \pi^0   \eta   J/\psi)=2\Gamma(Y_{(u\bar{u},d\bar{d})\bar{s}}^{0}\to   K^+   K^-   J/\psi)\\
&=&2\Gamma(Y_{(u\bar{u},d\bar{d})\bar{s}}^{0}\to   K^0   \overline K^0   J/\psi )=4\Gamma(Y_{(d\bar{d},s\bar{s})\bar{u}}^{-}\to   \pi^0   K^-   J/\psi )=2\Gamma(Y_{(d\bar{d},s\bar{s})\bar{u}}^{-}\to   \pi^-   \overline K^0   J/\psi )\\
&=&12\Gamma(Y_{(d\bar{d},s\bar{s})\bar{u}}^{-}\to   K^-   \eta   J/\psi ),\\
\end{eqnarray*}
and
\small
\begin{eqnarray*}
&&\Gamma(X_{u\bar{d}\bar{s}}^{+}\to  D^+ \pi^0 \overline D^0)= { }\Gamma(X_{d\bar{s}\bar{u}}^{-}\to  D^0 \pi^0 D^-)=\Gamma(Y_{(u\bar{u},d\bar{d})\bar{s}}^{0}\to  D^0 \pi^+ D^-)=\Gamma(Y_{(u\bar{u},d\bar{d})\bar{s}}^{0}\to  D^+ \pi^- \overline D^0),\\
&&\Gamma(X_{u\bar{d}\bar{s}}^{+}\to  D^+_s \overline K^0 \overline D^0)= { }\Gamma(X_{d\bar{s}\bar{u}}^{-}\to  D^+_s K^- D^-)=2\Gamma(Y_{(u\bar{u},d\bar{d})\bar{s}}^{0}\to  D^+_s \overline K^0 D^-)= 2\Gamma(Y_{(u\bar{u},d\bar{d})\bar{s}}^{0}\to  D^+_s K^- \overline D^0),\\
&&\Gamma(X_{d\bar{s}\bar{u}}^{-}\to  D^0 \pi^- \overline D^0)={ }\Gamma(X_{u\bar{d}\bar{s}}^{+}\to  D^+ \pi^+ D^-)=
2\Gamma(Y_{(u\bar{u},s\bar{s})\bar{d}}^{0}\to  D^+ \overline K^0 D^-)=2\Gamma(Y_{(d\bar{d},s\bar{s})\bar{u}}^{-}\to  D^0 K^- \overline D^0),\\
&&\Gamma(X_{s\bar{u}\bar{d}}^{-}\to  D^0 K^- \overline D^0)={ }\Gamma(X_{u\bar{d}\bar{s}}^{+}\to  D^+_s K^+  D^-_s)=
2\Gamma(Y_{(u\bar{u},d\bar{d})\bar{s}}^{0}\to  D^+_s K^0  D^-_s)=2\Gamma(Y_{(d\bar{d},s\bar{s})\bar{u}}^{-}\to  D^0 \pi^- \overline D^0),\\
&&\Gamma(X_{u\bar{d}\bar{s}}^{+}\to  D^+ K^0 \overline D^0)= 2\Gamma(X_{s\bar{u}\bar{d}}^{-}\to  D^0 \pi^0  D^-_s)=
\Gamma(X_{s\bar{u}\bar{d}}^{-}\to  D^+ \pi^-  D^-_s)=2\Gamma(Y_{(u\bar{u},s\bar{s})\bar{d}}^{0}\to  D^+ K^0  D^-_s)\\&=&2\Gamma(Y_{(u\bar{u},s\bar{s})\bar{d}}^{0}\to  D^+ \pi^- \overline D^0),\\
&&\Gamma(Y_{(u\bar{u},d\bar{d})\bar{s}}^{0}\to  D^0 K^0 \overline D^0)= 2\Gamma(Y_{(u\bar{u},s\bar{s})\bar{d}}^{0}\to  D^+_s \pi^0  D^-_s)=
{ }\Gamma(Y_{(d\bar{d},s\bar{s})\bar{u}}^{-}\to  D^+_s \pi^-  D^-_s),\\
&&\Gamma(X_{u\bar{d}\bar{s}}^{+}\to  D^+ \eta \overline D^0)= { }\Gamma(X_{d\bar{s}\bar{u}}^{-}\to  D^0 \eta D^-)=
 2\Gamma(Y_{(u\bar{u},d\bar{d})\bar{s}}^{0}\to  D^+ \eta D^-)=2\Gamma(Y_{(u\bar{u},d\bar{d})\bar{s}}^{0}\to  D^0 \eta \overline D^0),\\
&&\Gamma(X_{u\bar{d}\bar{s}}^{+}\to  D^0 \pi^+ \overline D^0)= { }\Gamma(X_{d\bar{s}\bar{u}}^{-}\to  D^+ \pi^- D^-),
       \Gamma(X_{u\bar{d}\bar{s}}^{+}\to  D^0 K^+ \overline D^0)= { }\Gamma(X_{s\bar{u}\bar{d}}^{-}\to  D^+_s K^-  D^-_s),\\
&&\Gamma(X_{u\bar{d}\bar{s}}^{+}\to  D^+_s \pi^0 \overline D^0)= { }\Gamma(Y_{(d\bar{d},s\bar{s})\bar{u}}^{-}\to  D^0 K^0  D^-_s),
       \Gamma(X_{u\bar{d}\bar{s}}^{+}\to  D^+_s \eta \overline D^0)= 2\Gamma(Y_{(u\bar{u},d\bar{d})\bar{s}}^{0}\to  D^+_s \eta D^-),\\
&&\Gamma(Y_{(u\bar{u},s\bar{s})\bar{d}}^{0}\to  D^0 \overline K^0 \overline D^0)= { }\Gamma(Y_{(d\bar{d},s\bar{s})\bar{u}}^{-}\to  D^+ K^- D^-),
       \Gamma(Y_{(u\bar{u},s\bar{s})\bar{d}}^{0}\to  D^+ K^- \overline D^0)= { }\Gamma(Y_{(d\bar{d},s\bar{s})\bar{u}}^{-}\to  D^0 \overline K^0 D^-),\\
&&\Gamma(Y_{(u\bar{u},s\bar{s})\bar{d}}^{0}\to  D^+_s K^- \overline D^0)= { }\Gamma(Y_{(u\bar{u},s\bar{s})\bar{d}}^{0}\to  D^0 K^+  D^-_s),
       \Gamma(Y_{(u\bar{u},d\bar{d})\bar{s}}^{0}\to  D^0 \pi^0 \overline D^0)= { }\Gamma(Y_{(u\bar{u},d\bar{d})\bar{s}}^{0}\to  D^+ \pi^0 D^-),\\
&&\Gamma(Y_{(u\bar{u},d\bar{d})\bar{s}}^{0}\to  D^+_s \pi^- \overline D^0)= { }\Gamma(Y_{(d\bar{d},s\bar{s})\bar{u}}^{-}\to  D^0 K^0  D^-_s),\\
             \end{eqnarray*}
\begin{table}
\caption{Open bottom tetraquark $X_{b6}$ decays into a $J/\psi$ and two light mesons.}\label{tab:b6_Jpsi_2M}\begin{tabular}{|cc|cc|}\hline\hline
channel & amplitude($/V_{\text{cb}}$) &channel & amplitude($/V_{\text{cb}}$)\\\hline
$X_{u\bar{d}\bar{s}}^{+}\to   \pi^+   K^0   J/\psi $ & $ \frac{a_1 V^*_{cd}}{\sqrt{2}}$&
$X_{u\bar{d}\bar{s}}^{+}\to   \pi^+   \eta   J/\psi $ & $ -\frac{a_1 V^*_{cs}}{\sqrt{3}}$\\\hline
$X_{u\bar{d}\bar{s}}^{+}\to   \pi^0   K^+   J/\psi $ & $ \frac{1}{2} a_1 V^*_{cd}$&
$X_{u\bar{d}\bar{s}}^{+}\to   K^+   \overline K^0   J/\psi $ & $ -\frac{a_1 V^*_{cs}}{\sqrt{2}}$\\\hline
$X_{u\bar{d}\bar{s}}^{+}\to   K^+   \eta   J/\psi $ & $ -\frac{a_1 V^*_{cd}}{2 \sqrt{3}}$&
$X_{d\bar{s}\bar{u}}^{-}\to   \pi^-   \eta   J/\psi $ & $ \frac{a_1 V^*_{cs}}{\sqrt{3}}$\\\hline
$X_{d\bar{s}\bar{u}}^{-}\to   K^0   K^-   J/\psi $ & $ \frac{a_1 V^*_{cs}}{\sqrt{2}}$&
$X_{s\bar{u}\bar{d}}^{-}\to   \pi^0   K^-   J/\psi $ & $ -\frac{1}{2} a_1 V^*_{cd}$\\\hline
$X_{s\bar{u}\bar{d}}^{-}\to   \pi^-   \overline K^0   J/\psi $ & $ -\frac{a_1 V^*_{cd}}{\sqrt{2}}$&
$X_{s\bar{u}\bar{d}}^{-}\to   K^-   \eta   J/\psi $ & $ \frac{a_1 V^*_{cd}}{2 \sqrt{3}}$\\\hline
$Y_{(u\bar{u},s\bar{s})\bar{d}}^{0}\to   \pi^+   \pi^-   J/\psi $ & $ -\frac{1}{2} a_1 V^*_{cd}$&
$Y_{(u\bar{u},s\bar{s})\bar{d}}^{0}\to   \pi^+   K^-   J/\psi $ & $ -\frac{1}{2} a_1 V^*_{cs}$\\\hline
$Y_{(u\bar{u},s\bar{s})\bar{d}}^{0}\to   \pi^0   \pi^0   J/\psi $ & $ -\frac{1}{2} a_1 V^*_{cd}$&
$Y_{(u\bar{u},s\bar{s})\bar{d}}^{0}\to   \pi^0   \overline K^0   J/\psi $ & $ \frac{a_1 V^*_{cs}}{2 \sqrt{2}}$\\\hline
$Y_{(u\bar{u},s\bar{s})\bar{d}}^{0}\to   \pi^0   \eta   J/\psi $ & $ -\frac{a_1 V^*_{cd}}{2 \sqrt{3}}$&
$Y_{(u\bar{u},s\bar{s})\bar{d}}^{0}\to   K^0   \overline K^0   J/\psi $ & $ \frac{1}{2} a_1 V^*_{cd}$\\\hline
$Y_{(u\bar{u},s\bar{s})\bar{d}}^{0}\to   \overline K^0   \eta   J/\psi $ & $ \frac{a_1 V^*_{cs}}{2 \sqrt{6}}$&
$Y_{(u\bar{u},s\bar{s})\bar{d}}^{0}\to   \eta   \eta   J/\psi $ & $ \frac{1}{2} a_1 V^*_{cd}$\\\hline
$Y_{(u\bar{u},d\bar{d})\bar{s}}^{0}\to   \pi^0   K^0   J/\psi $ & $ -\frac{a_1 V^*_{cd}}{2 \sqrt{2}}$&
$Y_{(u\bar{u},d\bar{d})\bar{s}}^{0}\to   \pi^0   \eta   J/\psi $ & $ \frac{a_1 V^*_{cs}}{\sqrt{3}}$\\\hline
$Y_{(u\bar{u},d\bar{d})\bar{s}}^{0}\to   \pi^-   K^+   J/\psi $ & $ \frac{1}{2} a_1 V^*_{cd}$&
$Y_{(u\bar{u},d\bar{d})\bar{s}}^{0}\to   K^+   K^-   J/\psi $ & $ \frac{1}{2} a_1 V^*_{cs}$\\\hline
$Y_{(u\bar{u},d\bar{d})\bar{s}}^{0}\to   K^0   \overline K^0   J/\psi $ & $ -\frac{1}{2} a_1 V^*_{cs}$&
$Y_{(u\bar{u},d\bar{d})\bar{s}}^{0}\to   K^0   \eta   J/\psi $ & $ -\frac{a_1 V^*_{cd}}{2 \sqrt{6}}$\\\hline
$Y_{(d\bar{d},s\bar{s})\bar{u}}^{-}\to   \pi^0   K^-   J/\psi $ & $ \frac{a_1 V^*_{cs}}{2 \sqrt{2}}$&
$Y_{(d\bar{d},s\bar{s})\bar{u}}^{-}\to   \pi^-   \overline K^0   J/\psi $ & $ \frac{1}{2} a_1 V^*_{cs}$\\\hline
$Y_{(d\bar{d},s\bar{s})\bar{u}}^{-}\to   \pi^-   \eta   J/\psi $ & $ -\frac{a_1 V^*_{cd}}{\sqrt{6}}$&
$Y_{(d\bar{d},s\bar{s})\bar{u}}^{-}\to   K^0   K^-   J/\psi $ & $ -\frac{1}{2} a_1 V^*_{cd}$\\\hline
$Y_{(d\bar{d},s\bar{s})\bar{u}}^{-}\to   K^-   \eta   J/\psi $ & $ -\frac{a_1 V^*_{cs}}{2 \sqrt{6}}$& &\\
\hline
\end{tabular}
\end{table}
\begin{table}
\footnotesize
\caption{Open bottom tetraquark $X_{b6}$ decays into a charmed meson, an anti-charmed meson and a light meson.}\label{tab:b6_D_antiD_M}\begin{tabular}{|cc|cc|}\hline\hline
channel & amplitude($/V_{\text{cb}}$) &channel & amplitude($/V_{\text{cb}}$)\\\hline
$X_{u\bar{d}\bar{s}}^{+}\to    D^0  \pi^+   \overline D^0 $ & $ -\frac{\left(a_2+a_5\right) V^*_{cs}}{\sqrt{2}}$&
$X_{u\bar{d}\bar{s}}^{+}\to    D^0  K^+   \overline D^0 $ & $ \frac{\left(a_2+a_5\right) V^*_{cd}}{\sqrt{2}}$\\\hline
$X_{u\bar{d}\bar{s}}^{+}\to    D^+  \pi^+   D^- $ & $ -\frac{\left(a_2+a_3\right) V^*_{cs}}{\sqrt{2}}$&
$X_{u\bar{d}\bar{s}}^{+}\to    D^+  \pi^0   \overline D^0 $ & $ \frac{1}{2} \left(a_5-a_3\right) V^*_{cs}$\\\hline
$X_{u\bar{d}\bar{s}}^{+}\to    D^+  K^+   D^- $ & $ \frac{\left(a_2-a_4\right) V^*_{cd}}{\sqrt{2}}$&
$X_{u\bar{d}\bar{s}}^{+}\to    D^+  K^+    D^-_s $ & $ -\frac{\left(a_3+a_4\right) V^*_{cs}}{\sqrt{2}}$\\\hline
$X_{u\bar{d}\bar{s}}^{+}\to    D^+  K^0   \overline D^0 $ & $ \frac{\left(a_5-a_6\right) V^*_{cd}}{\sqrt{2}}$&
$X_{u\bar{d}\bar{s}}^{+}\to    D^+  \eta   \overline D^0 $ & $ -\frac{\left(a_3+a_5-2 a_6\right) V^*_{cs}}{2 \sqrt{3}}$\\\hline
$X_{u\bar{d}\bar{s}}^{+}\to    D^+_s  \pi^+   D^- $ & $ \frac{\left(a_3+a_4\right) V^*_{cd}}{\sqrt{2}}$&
$X_{u\bar{d}\bar{s}}^{+}\to    D^+_s  \pi^+    D^-_s $ & $ \frac{\left(a_4-a_2\right) V^*_{cs}}{\sqrt{2}}$\\\hline
$X_{u\bar{d}\bar{s}}^{+}\to    D^+_s  \pi^0   \overline D^0 $ & $ \frac{1}{2} \left(a_3-a_6\right) V^*_{cd}$&
$X_{u\bar{d}\bar{s}}^{+}\to    D^+_s  K^+    D^-_s $ & $ \frac{a_2 V^*_{cd}}{\sqrt{2}}+\frac{a_3 V^*_{cd}}{\sqrt{2}}$\\\hline
$X_{u\bar{d}\bar{s}}^{+}\to    D^+_s  \overline K^0   \overline D^0 $ & $ \frac{\left(a_6-a_5\right) V^*_{cs}}{\sqrt{2}}$&
$X_{u\bar{d}\bar{s}}^{+}\to    D^+_s  \eta   \overline D^0 $ & $ \frac{\left(a_3-2 a_5+a_6\right) V^*_{cd}}{2 \sqrt{3}}$\\\hline
$X_{d\bar{s}\bar{u}}^{-}\to    D^0  \pi^0   D^- $ & $ \frac{1}{2} \left(a_5-a_3\right) V^*_{cs}$&
$X_{d\bar{s}\bar{u}}^{-}\to    D^0  \pi^-   \overline D^0 $ & $ \frac{\left(a_2+a_3\right) V^*_{cs}}{\sqrt{2}}$\\\hline
$X_{d\bar{s}\bar{u}}^{-}\to    D^0  K^0   D^- $ & $ \frac{\left(a_4+a_6\right) V^*_{cd}}{\sqrt{2}}$&
$X_{d\bar{s}\bar{u}}^{-}\to    D^0  K^0    D^-_s $ & $ \frac{\left(a_3+a_4\right) V^*_{cs}}{\sqrt{2}}$\\\hline
$X_{d\bar{s}\bar{u}}^{-}\to    D^0  \eta   D^- $ & $ \frac{\left(a_3+a_5-2 a_6\right) V^*_{cs}}{2 \sqrt{3}}$&
$X_{d\bar{s}\bar{u}}^{-}\to    D^+  \pi^-   D^- $ & $ \frac{\left(a_2+a_5\right) V^*_{cs}}{\sqrt{2}}$\\\hline
$X_{d\bar{s}\bar{u}}^{-}\to    D^+_s  \pi^-   D^- $ & $ -\frac{\left(a_4+a_6\right) V^*_{cd}}{\sqrt{2}}$&
$X_{d\bar{s}\bar{u}}^{-}\to    D^+_s  \pi^-    D^-_s $ & $ \frac{\left(a_2-a_4\right) V^*_{cs}}{\sqrt{2}}$\\\hline
$X_{d\bar{s}\bar{u}}^{-}\to    D^+_s  K^-   D^- $ & $ \frac{\left(a_5-a_6\right) V^*_{cs}}{\sqrt{2}}$&
$X_{s\bar{u}\bar{d}}^{-}\to    D^0  \pi^0    D^-_s $ & $ \frac{1}{2} \left(a_6-a_5\right) V^*_{cd}$\\\hline
$X_{s\bar{u}\bar{d}}^{-}\to    D^0  \overline K^0   D^- $ & $ -\frac{\left(a_3+a_4\right) V^*_{cd}}{\sqrt{2}}$&
$X_{s\bar{u}\bar{d}}^{-}\to    D^0  \overline K^0    D^-_s $ & $ -\frac{\left(a_4+a_6\right) V^*_{cs}}{\sqrt{2}}$\\\hline
$X_{s\bar{u}\bar{d}}^{-}\to    D^0  K^-   \overline D^0 $ & $ -\frac{a_2 V^*_{cd}}{\sqrt{2}}-\frac{a_3 V^*_{cd}}{\sqrt{2}}$&
$X_{s\bar{u}\bar{d}}^{-}\to    D^0  \eta    D^-_s $ & $ \frac{\left(2 a_3-a_5-a_6\right) V^*_{cd}}{2 \sqrt{3}}$\\\hline
$X_{s\bar{u}\bar{d}}^{-}\to    D^+  \pi^-    D^-_s $ & $ \frac{\left(a_6-a_5\right) V^*_{cd}}{\sqrt{2}}$&
$X_{s\bar{u}\bar{d}}^{-}\to    D^+  K^-   D^- $ & $ \frac{\left(a_4-a_2\right) V^*_{cd}}{\sqrt{2}}$\\\hline
$X_{s\bar{u}\bar{d}}^{-}\to    D^+  K^-    D^-_s $ & $ \frac{\left(a_4+a_6\right) V^*_{cs}}{\sqrt{2}}$&
$X_{s\bar{u}\bar{d}}^{-}\to    D^+_s  K^-    D^-_s $ & $ -\frac{\left(a_2+a_5\right) V^*_{cd}}{\sqrt{2}}$\\\hline
$Y_{(u\bar{u},s\bar{s})\bar{d}}^{0}\to    D^0  \pi^+   D^- $ & $ -\frac{1}{2} \left(a_3+a_4\right) V^*_{cd}$&
$Y_{(u\bar{u},s\bar{s})\bar{d}}^{0}\to    D^0  \pi^+    D^-_s $ & $ -\frac{1}{2} \left(a_4+a_5\right) V^*_{cs}$\\\hline
$Y_{(u\bar{u},s\bar{s})\bar{d}}^{0}\to    D^0  \pi^0   \overline D^0 $ & $ -\frac{\left(a_2+a_3+a_5-a_6\right) V^*_{cd}}{2 \sqrt{2}}$&
$Y_{(u\bar{u},s\bar{s})\bar{d}}^{0}\to    D^0  K^+    D^-_s $ & $ \frac{1}{2} \left(a_5-a_3\right) V^*_{cd}$\\\hline
$Y_{(u\bar{u},s\bar{s})\bar{d}}^{0}\to    D^0  \overline K^0   \overline D^0 $ & $ -\frac{1}{2} \left(a_2+a_6\right) V^*_{cs}$&
$Y_{(u\bar{u},s\bar{s})\bar{d}}^{0}\to    D^0  \eta   \overline D^0 $ & $ -\frac{\left(3 a_2+a_3+a_5+a_6\right) V^*_{cd}}{2 \sqrt{6}}$\\\hline
$Y_{(u\bar{u},s\bar{s})\bar{d}}^{0}\to    D^+  \pi^0   D^- $ & $ \frac{\left(a_4-a_2\right) V^*_{cd}}{2 \sqrt{2}}$&
$Y_{(u\bar{u},s\bar{s})\bar{d}}^{0}\to    D^+  \pi^0    D^-_s $ & $ \frac{\left(a_4+a_5\right) V^*_{cs}}{2 \sqrt{2}}$\\\hline
$Y_{(u\bar{u},s\bar{s})\bar{d}}^{0}\to    D^+  \pi^-   \overline D^0 $ & $ \frac{1}{2} \left(a_6-a_5\right) V^*_{cd}$&
$Y_{(u\bar{u},s\bar{s})\bar{d}}^{0}\to    D^+  K^0    D^-_s $ & $ \frac{1}{2} \left(a_5-a_6\right) V^*_{cd}$\\\hline
$Y_{(u\bar{u},s\bar{s})\bar{d}}^{0}\to    D^+  \overline K^0   D^- $ & $ -\frac{1}{2} \left(a_2+a_3\right) V^*_{cs}$&
$Y_{(u\bar{u},s\bar{s})\bar{d}}^{0}\to    D^+  K^-   \overline D^0 $ & $ \frac{1}{2} \left(a_6-a_3\right) V^*_{cs}$\\\hline
$Y_{(u\bar{u},s\bar{s})\bar{d}}^{0}\to    D^+  \eta   D^- $ & $ \frac{1}{2} \sqrt{\frac{3}{2}} \left(a_4-a_2\right) V^*_{cd}$&
$Y_{(u\bar{u},s\bar{s})\bar{d}}^{0}\to    D^+  \eta    D^-_s $ & $ \frac{\left(2 a_3+3 a_4-a_5+2 a_6\right) V^*_{cs}}{2 \sqrt{6}}$\\\hline
$Y_{(u\bar{u},s\bar{s})\bar{d}}^{0}\to    D^+_s  \pi^0    D^-_s $ & $ -\frac{\left(a_2+a_6\right) V^*_{cd}}{2 \sqrt{2}}$&
$Y_{(u\bar{u},s\bar{s})\bar{d}}^{0}\to    D^+_s  \overline K^0   D^- $ & $ \frac{1}{2} \left(a_3+a_4\right) V^*_{cd}$\\\hline
$Y_{(u\bar{u},s\bar{s})\bar{d}}^{0}\to    D^+_s  \overline K^0    D^-_s $ & $ \frac{1}{2} \left(-a_2+a_4-a_5+a_6\right) V^*_{cs}$&
$Y_{(u\bar{u},s\bar{s})\bar{d}}^{0}\to    D^+_s  K^-   \overline D^0 $ & $ \frac{1}{2} \left(a_3-a_5\right) V^*_{cd}$\\\hline
$Y_{(u\bar{u},s\bar{s})\bar{d}}^{0}\to    D^+_s  \eta    D^-_s $ & $ \frac{\left(-3 a_2-2 a_3-2 a_5+a_6\right) V^*_{cd}}{2 \sqrt{6}}$&
$Y_{(u\bar{u},d\bar{d})\bar{s}}^{0}\to    D^0  \pi^+   D^- $ & $ \frac{1}{2} \left(a_3-a_5\right) V^*_{cs}$\\\hline
$Y_{(u\bar{u},d\bar{d})\bar{s}}^{0}\to    D^0  \pi^0   \overline D^0 $ & $ \frac{\left(2 a_2+a_3+a_5\right) V^*_{cs}}{2 \sqrt{2}}$&
$Y_{(u\bar{u},d\bar{d})\bar{s}}^{0}\to    D^0  K^+   D^- $ & $ \frac{1}{2} \left(a_4+a_5\right) V^*_{cd}$\\\hline
$Y_{(u\bar{u},d\bar{d})\bar{s}}^{0}\to    D^0  K^+    D^-_s $ & $ \frac{1}{2} \left(a_3+a_4\right) V^*_{cs}$&
$Y_{(u\bar{u},d\bar{d})\bar{s}}^{0}\to    D^0  K^0   \overline D^0 $ & $ \frac{1}{2} \left(a_2+a_6\right) V^*_{cd}$\\\hline
$Y_{(u\bar{u},d\bar{d})\bar{s}}^{0}\to    D^0  \eta   \overline D^0 $ & $ \frac{\left(a_3+a_5-2 a_6\right) V^*_{cs}}{2 \sqrt{6}}$&
$Y_{(u\bar{u},d\bar{d})\bar{s}}^{0}\to    D^+  \pi^0   D^- $ & $ \frac{\left(2 a_2+a_3+a_5\right) V^*_{cs}}{2 \sqrt{2}}$\\\hline
$Y_{(u\bar{u},d\bar{d})\bar{s}}^{0}\to    D^+  \pi^-   \overline D^0 $ & $ \frac{1}{2} \left(a_5-a_3\right) V^*_{cs}$&
$Y_{(u\bar{u},d\bar{d})\bar{s}}^{0}\to    D^+  K^0   D^- $ & $ \frac{1}{2} \left(a_2-a_4+a_5-a_6\right) V^*_{cd}$\\\hline
$Y_{(u\bar{u},d\bar{d})\bar{s}}^{0}\to    D^+  K^0    D^-_s $ & $ -\frac{1}{2} \left(a_3+a_4\right) V^*_{cs}$&
$Y_{(u\bar{u},d\bar{d})\bar{s}}^{0}\to    D^+  \eta   D^- $ & $ -\frac{\left(a_3+a_5-2 a_6\right) V^*_{cs}}{2 \sqrt{6}}$\\\hline
$Y_{(u\bar{u},d\bar{d})\bar{s}}^{0}\to    D^+_s  \pi^0   D^- $ & $ -\frac{\left(a_3+2 a_4+a_6\right) V^*_{cd}}{2 \sqrt{2}}$&
$Y_{(u\bar{u},d\bar{d})\bar{s}}^{0}\to    D^+_s  \pi^0    D^-_s $ & $ \frac{\left(a_2-a_4\right) V^*_{cs}}{\sqrt{2}}$\\\hline
$Y_{(u\bar{u},d\bar{d})\bar{s}}^{0}\to    D^+_s  \pi^-   \overline D^0 $ & $ \frac{1}{2} \left(a_3-a_6\right) V^*_{cd}$&
$Y_{(u\bar{u},d\bar{d})\bar{s}}^{0}\to    D^+_s  K^0    D^-_s $ & $ \frac{1}{2} \left(a_2+a_3\right) V^*_{cd}$\\\hline
$Y_{(u\bar{u},d\bar{d})\bar{s}}^{0}\to    D^+_s  \overline K^0   D^- $ & $ \frac{1}{2} \left(a_6-a_5\right) V^*_{cs}$&
$Y_{(u\bar{u},d\bar{d})\bar{s}}^{0}\to    D^+_s  K^-   \overline D^0 $ & $ \frac{1}{2} \left(a_5-a_6\right) V^*_{cs}$\\\hline
$Y_{(u\bar{u},d\bar{d})\bar{s}}^{0}\to    D^+_s  \eta   D^- $ & $ \frac{\left(a_3-2 a_5+a_6\right) V^*_{cd}}{2 \sqrt{6}}$&
$Y_{(d\bar{d},s\bar{s})\bar{u}}^{-}\to    D^0  \pi^0   D^- $ & $ \frac{\left(a_3+a_4-a_5+a_6\right) V^*_{cd}}{2 \sqrt{2}}$\\\hline
$Y_{(d\bar{d},s\bar{s})\bar{u}}^{-}\to    D^0  \pi^0    D^-_s $ & $ \frac{\left(a_4+a_5\right) V^*_{cs}}{2 \sqrt{2}}$&
$Y_{(d\bar{d},s\bar{s})\bar{u}}^{-}\to    D^0  \pi^-   \overline D^0 $ & $ -\frac{1}{2} \left(a_2+a_3\right) V^*_{cd}$\\\hline
$Y_{(d\bar{d},s\bar{s})\bar{u}}^{-}\to    D^0  K^0    D^-_s $ & $ \frac{1}{2} \left(a_6-a_3\right) V^*_{cd}$&
$Y_{(d\bar{d},s\bar{s})\bar{u}}^{-}\to    D^0  \overline K^0   D^- $ & $ \frac{1}{2} \left(a_3-a_6\right) V^*_{cs}$\\\hline
$Y_{(d\bar{d},s\bar{s})\bar{u}}^{-}\to    D^0  K^-   \overline D^0 $ & $ \frac{1}{2} \left(a_2+a_3\right) V^*_{cs}$&
$Y_{(d\bar{d},s\bar{s})\bar{u}}^{-}\to    D^0  \eta   D^- $ & $ -\frac{\left(a_3+3 a_4+a_5+a_6\right) V^*_{cd}}{2 \sqrt{6}}$\\\hline
$Y_{(d\bar{d},s\bar{s})\bar{u}}^{-}\to    D^0  \eta    D^-_s $ & $ \frac{\left(-2 a_3-3 a_4+a_5-2 a_6\right) V^*_{cs}}{2 \sqrt{6}}$&
$Y_{(d\bar{d},s\bar{s})\bar{u}}^{-}\to    D^+  \pi^-   D^- $ & $ \frac{1}{2} \left(-a_2+a_4-a_5+a_6\right) V^*_{cd}$\\\hline
$Y_{(d\bar{d},s\bar{s})\bar{u}}^{-}\to    D^+  \pi^-    D^-_s $ & $ \frac{1}{2} \left(a_4+a_5\right) V^*_{cs}$&
$Y_{(d\bar{d},s\bar{s})\bar{u}}^{-}\to    D^+  K^-   D^- $ & $ \frac{1}{2} \left(a_2+a_6\right) V^*_{cs}$\\\hline
$Y_{(d\bar{d},s\bar{s})\bar{u}}^{-}\to    D^+_s  \pi^-    D^-_s $ & $ -\frac{1}{2} \left(a_2+a_6\right) V^*_{cd}$&
$Y_{(d\bar{d},s\bar{s})\bar{u}}^{-}\to    D^+_s  K^-   D^- $ & $ -\frac{1}{2} \left(a_4+a_5\right) V^*_{cd}$\\\hline
$Y_{(d\bar{d},s\bar{s})\bar{u}}^{-}\to    D^+_s  K^-    D^-_s $ & $ \frac{1}{2} \left(a_2-a_4+a_5-a_6\right) V^*_{cs}$& &\\
\hline
\end{tabular}
\end{table}

\subsection{$b\to c \bar u d/s$ transition}

\subsubsection{Decays into a charmed meson and a light meson}
The $\bar c b \bar q u$  transition can form an octet operator by SU(3) symmetry, in which nonzero entry is
$
(H_{{\bf8}})^2_1 =V_{ud}^*
$
for   the $b\to c\bar ud$ or $b\bar d\to c\bar u$   transition, while  $(H_{{\bf8}})^3_1 =V_{us}^*$ for  the $b\to c\bar
us$ or $b\bar s \to c\bar u$ transition. The  effective Hamiltonian is then written as
  \begin{eqnarray}
  \mathcal{H}_{eff}&=&a_3 (X_{b6})_{[jk]}^i (H_8)^j_i M^k_l (\overline D)^l+a_4 (X_{b6})_{[jk]}^i (H_8)^l_i M^j_l (\overline D)^k+a_5 (X_{b6})_{[jk]}^i (H_8)^j_l M^k_i (\overline D)^l\nonumber\\
  &&+a_6 (X_{b6})_{[jk]}^i (H_8)^j_l M^l_i (\overline D)^k
  \end{eqnarray}
The decays amplitudes are given in Tab.~\ref{tab:b6_D_M}, and the relations among different decay widths become
\begin{eqnarray*}
    \Gamma(X_{u\bar{d}\bar{s}}^{+}\to  D^+\pi^0 )= { }\Gamma(Y_{(u\bar{u},d\bar{d})\bar{s}}^{0}\to  D^+\pi^- ), \Gamma(X_{u\bar{d}\bar{s}}^{+}\to  D^+_s\pi^0 )= { }\Gamma(Y_{(u\bar{u},d\bar{d})\bar{s}}^{0}\to  D^+_s\pi^- ),\\ \Gamma(X_{u\bar{d}\bar{s}}^{+}\to  D^+_s\overline K^0 )= 2\Gamma(Y_{(u\bar{u},d\bar{d})\bar{s}}^{0}\to  D^+_sK^- ), \Gamma(X_{d\bar{s}\bar{u}}^{-}\to  D^0\pi^- )= 2\Gamma(Y_{(d\bar{d},s\bar{s})\bar{u}}^{-}\to  D^0K^- ),\\ \Gamma(Y_{(u\bar{u},s\bar{s})\bar{d}}^{0}\to  D^+\pi^- )= \frac{1}{2}\Gamma(X_{u\bar{d}\bar{s}}^{+}\to  D^+K^0 ), \Gamma(Y_{(d\bar{d},s\bar{s})\bar{u}}^{-}\to  D^0\pi^- )= \frac{1}{2}\Gamma(X_{s\bar{u}\bar{d}}^{-}\to  D^0K^- ),\\
    \Gamma(Y_{(u\bar{u},d\bar{d})\bar{s}}^{0}\to  D^0\eta )= \frac{1}{2}\Gamma(X_{u\bar{d}\bar{s}}^{+}\to  D^+\eta ).
\end{eqnarray*}

\begin{table}
\caption{Open bottom tetraquark $X_{b6}$ decays into a charmed meson and a light meson.}\label{tab:b6_D_M}\begin{tabular}{|cc|cc|}\hline\hline
channel & amplitude($/V_{\text{cb}}$) &channel&amplitude($/V_{\text{cb}}$)\\\hline
$X_{u\bar{d}\bar{s}}^{+}\to    D^0  \pi^+  $ & $ -\frac{\left(a_3+a_5\right) V_{\text{us}}^*}{\sqrt{2}}$&
$X_{u\bar{d}\bar{s}}^{+}\to    D^0  K^+  $ & $ \frac{\left(a_3+a_5\right) V_{\text{ud}}^*}{\sqrt{2}}$\\\hline
$X_{u\bar{d}\bar{s}}^{+}\to    D^+  \pi^0  $ & $ \frac{1}{2} \left(a_3-a_6\right) V_{\text{us}}^*$&
$X_{u\bar{d}\bar{s}}^{+}\to    D^+  K^0  $ & $ \frac{\left(a_3-a_4\right) V_{\text{ud}}^*}{\sqrt{2}}$\\\hline
$X_{u\bar{d}\bar{s}}^{+}\to    D^+  \eta  $ & $ -\frac{\left(a_3-2 a_4+a_6\right) V^*_{us}}{2 \sqrt{3}}$&
$X_{u\bar{d}\bar{s}}^{+}\to    D^+_s  \pi^0  $ & $ \frac{1}{2} \left(a_6-a_4\right) V_{\text{ud}}^*$\\\hline
$X_{u\bar{d}\bar{s}}^{+}\to    D^+_s  \overline K^0  $ & $ \frac{\left(a_4-a_3\right) V^*_{us}}{\sqrt{2}}$&
$X_{u\bar{d}\bar{s}}^{+}\to    D^+_s  \eta  $ & $ \frac{\left(-2 a_3+a_4+a_6\right) V_{\text{ud}}^*}{2 \sqrt{3}}$\\\hline
$X_{d\bar{s}\bar{u}}^{-}\to    D^0  \pi^-  $ & $ \frac{\left(a_5+a_6\right) V^*_{us}}{\sqrt{2}}$&
$X_{s\bar{u}\bar{d}}^{-}\to    D^0  K^-  $ & $ -\frac{\left(a_5+a_6\right) V_{\text{ud}}^*}{\sqrt{2}}$\\\hline
$Y_{(u\bar{u},s\bar{s})\bar{d}}^{0}\to    D^0  \pi^0  $ & $ -\frac{\left(a_3-a_4+a_5+a_6\right) V_{\text{ud}}^*}{2 \sqrt{2}}$&
$Y_{(u\bar{u},s\bar{s})\bar{d}}^{0}\to    D^0  \overline K^0  $ & $ -\frac{1}{2} \left(a_4+a_5\right) V^*_{us}$\\\hline
$Y_{(u\bar{u},s\bar{s})\bar{d}}^{0}\to    D^0  \eta  $ & $ -\frac{\left(a_3+a_4+3 a_5+a_6\right) V_{\text{ud}}^*}{2 \sqrt{6}}$&
$Y_{(u\bar{u},s\bar{s})\bar{d}}^{0}\to    D^+  \pi^-  $ & $ \frac{1}{2} \left(a_4-a_3\right) V_{\text{ud}}^*$\\\hline
$Y_{(u\bar{u},s\bar{s})\bar{d}}^{0}\to    D^+  K^-  $ & $ \frac{1}{2} \left(a_4-a_6\right) V^*_{us}$&
$Y_{(u\bar{u},s\bar{s})\bar{d}}^{0}\to    D^+_s  K^-  $ & $ \frac{1}{2} \left(a_6-a_3\right) V_{\text{ud}}^*$\\\hline
$Y_{(u\bar{u},d\bar{d})\bar{s}}^{0}\to    D^0  \pi^0  $ & $ \frac{\left(a_3+2 a_5+a_6\right) V^*_{us}}{2 \sqrt{2}}$&
$Y_{(u\bar{u},d\bar{d})\bar{s}}^{0}\to    D^0  K^0  $ & $ \frac{1}{2} \left(a_4+a_5\right) V_{\text{ud}}^*$\\\hline
$Y_{(u\bar{u},d\bar{d})\bar{s}}^{0}\to    D^0  \eta  $ & $ \frac{\left(a_3-2 a_4+a_6\right) V^*_{us}}{2 \sqrt{6}}$&
$Y_{(u\bar{u},d\bar{d})\bar{s}}^{0}\to    D^+  \pi^-  $ & $ \frac{1}{2} \left(a_3-a_6\right) V^*_{us}$\\\hline
$Y_{(u\bar{u},d\bar{d})\bar{s}}^{0}\to    D^+_s  \pi^-  $ & $ \frac{1}{2} \left(a_6-a_4\right) V^*_{ud}$&
$Y_{(u\bar{u},d\bar{d})\bar{s}}^{0}\to    D^+_s  K^-  $ & $ \frac{1}{2} \left(a_3-a_4\right) V^*_{us}$\\\hline
$Y_{(d\bar{d},s\bar{s})\bar{u}}^{-}\to    D^0  \pi^-  $ & $ -\frac{1}{2} \left(a_5+a_6\right) V^*_{ud}$&
$Y_{(d\bar{d},s\bar{s})\bar{u}}^{-}\to    D^0  K^-  $ & $ \frac{1}{2} \left(a_5+a_6\right) V^*_{us}$\\\hline
\hline
\end{tabular}
\end{table}

\subsubsection{Two-body decays into a charmed baryon and an anti-baryon}
Concerning the two-body decays into a charmed baryon and an anti-baryon, the efficient Hamiltonian which includes four different kinds of combinations in final states can be written as
  \begin{eqnarray}
  \mathcal{H}_{eff}&=&-b_1 (X_{b6})^{\{ij\}} (H_8)^k_i (T_{c\bar3})_{[jl]}(\overline T_{8})^{l}_k -b_2 (X_{b6})^{\{ij\}} (H_8)^k_i (T_{c\bar3})_{[kl]}(\overline T_{8})^{l}_j\nonumber\\
  &&-b_3 (X_{b6})^{\{ij\}} (H_8)^k_l (T_{c\bar3})_{[ik]}(\overline T_{8})^{l}_j-c_1 (X_{b6})^{\{ij\}} (H_8)^k_i (T_{c6})_{\{jl\}}(\overline T_{8})^{l}_k \nonumber\\
  &&-c_2 (X_{b6})^{\{ij\}} (H_8)^k_i (T_{c6})_{\{kl\}}(\overline T_{8})^{l}_j-c_3 (X_{b6})^{\{ij\}} (H_8)^k_l (T_{c6})_{\{ij\}}(\overline T_{8})^{l}_k\nonumber\\
  &&-c_4 (X_{b6})^{\{ij\}} (H_8)^k_l (T_{c6})_{\{ik\}}(\overline T_{8})^{l}_j+d_1 (X_{b6})^i_{[jk]} (H_8)^j_l (T_{c\bar3})_{[im]}(\overline T_{\overline{10}})^{\{klm\}}\nonumber\\
  &&+f_1 (X_{b6})^i_{[jk]} (H_8)^j_i (T_{c6})_{\{lm\}}(\overline T_{\overline{10}})^{\{klm\}}+f_2 (X_{b6})^i_{[jk]} (H_8)^j_l (T_{c6})_{\{im\}}(\overline T_{\overline{10}})^{\{klm\}}.
  \end{eqnarray}
 The related decay amplitudes are given in Tab.~\ref{tab:b6_Tc3bar_Tbar8} and Tab.~\ref{tab:b6_Tc3bar_Tbar10bar}, which are  for anti-triplet charmed baryon plus octet anti-baryon (class \uppercase\expandafter{\romannumeral1}), sextet charmed baryon plus octet anti-baryon class \uppercase\expandafter{\romannumeral2}, anti-triplet charmed baryon plus anti-decuplet anti-baryon class \uppercase\expandafter{\romannumeral3}, sextet charmed baryon plus anti-decuplet anti-baryon class \uppercase\expandafter{\romannumeral4}, respectively.   The relations of decay widths for class \uppercase\expandafter{\romannumeral1} are  given as
\begin{eqnarray*}
    \Gamma(X_{d\bar{s}\bar{u}}^{-}\to \Xi_c^0\overline \Sigma^-)
    = 2\Gamma(Y_{(d\bar{d},s\bar{s})\bar{u}}^{-}\to \Xi_c^0\overline p), \Gamma(Y_{(u\bar{u},s\bar{s})\bar{d}}^{0}\to \Xi_c^+\overline \Sigma^-)
    = \frac{1}{2}\Gamma(X_{u\bar{d}\bar{s}}^{+}\to \Xi_c^+\overline \Xi^0),\\
    \Gamma(Y_{(u\bar{u},d\bar{d})\bar{s}}^{0}\to \Lambda_c^+\overline \Sigma^-)
    = { }\Gamma(X_{u\bar{d}\bar{s}}^{+}\to \Lambda_c^+\overline \Sigma^0), \Gamma(Y_{(u\bar{u},d\bar{d})\bar{s}}^{0}\to \Lambda_c^+\overline p)
    = \frac{1}{2}\Gamma(X_{u\bar{d}\bar{s}}^{+}\to \Lambda_c^+\overline n),\\
    \Gamma(Y_{(u\bar{u},d\bar{d})\bar{s}}^{0}\to \Xi_c^+\overline \Sigma^-)
    = { }\Gamma(X_{u\bar{d}\bar{s}}^{+}\to \Xi_c^+\overline \Sigma^0), \Gamma(Y_{(d\bar{d},s\bar{s})\bar{u}}^{-}\to \Xi_c^0\overline \Sigma^-)
    = \frac{1}{2}\Gamma(X_{s\bar{u}\bar{d}}^{-}\to \Xi_c^0\overline p),\\
    \Gamma(X_{u\bar{d}\bar{s}}^{+}\to \Xi_c^+\overline \Lambda^0)
    = 2\Gamma(Y_{(u\bar{u},d\bar{d})\bar{s}}^{0}\to \Xi_c^0\overline \Lambda^0).
\end{eqnarray*}
The relations of decay widths for class \uppercase\expandafter{\romannumeral2} are
\begin{eqnarray*}
    \Gamma(X_{u\bar{d}\bar{s}}^{+}\to \Sigma_{c}^{0}\overline \Sigma^+)= 2\Gamma(X_{u\bar{d}\bar{s}}^{+}\to \Xi_{c}^{\prime0}\overline \Xi^+)=2\Gamma(Y_{(u\bar{u},s\bar{s})\bar{d}}^{0}\to \Sigma_{c}^{0}\overline n),\\
     \Gamma(X_{u\bar{d}\bar{s}}^{+}\to \Xi_{c}^{\prime0}\overline \Sigma^+)=\frac{1}{2}\Gamma(X_{u\bar{d}\bar{s}}^{+}\to \Omega_{c}^{0}\overline \Xi^+)=\Gamma(Y_{(u\bar{u},d\bar{d})\bar{s}}^{0}\to \Omega_{c}^{0}\overline \Xi^0),\\
      \Gamma(X_{d\bar{s}\bar{u}}^{-}\to \Xi_{c}^{\prime0}\overline \Sigma^-)= 2\Gamma(Y_{(u\bar{u},s\bar{s})\bar{d}}^{0}\to \Omega_{c}^{0}\overline \Sigma^0)=\Gamma(Y_{(d\bar{d},s\bar{s})\bar{u}}^{-}\to \Omega_{c}^{0}\overline \Sigma^-),\\
      \Gamma(Y_{(u\bar{u},d\bar{d})\bar{s}}^{0}\to \Xi_{c}^{\prime+}\overline \Sigma^-)=\Gamma(X_{u\bar{d}\bar{s}}^{+}\to \Xi_{c}^{\prime+}\overline \Sigma^0),
    \Gamma(X_{u\bar{d}\bar{s}}^{+}\to \Sigma_{c}^{+}\overline \Lambda^0)=\Gamma(Y_{(u\bar{u},d\bar{d})\bar{s}}^{0}\to \Sigma_{c}^{0}\overline \Lambda^0),\\
    \Gamma(X_{u\bar{d}\bar{s}}^{+}\to \Sigma_{c}^{+}\overline n)= { }\Gamma(Y_{(u\bar{u},d\bar{d})\bar{s}}^{0}\to \Sigma_{c}^{0}\overline n),
    \Gamma(X_{u\bar{d}\bar{s}}^{+}\to \Xi_{c}^{\prime+}\overline \Lambda^0)= 2\Gamma(Y_{(u\bar{u},d\bar{d})\bar{s}}^{0}\to \Xi_{c}^{\prime0}\overline \Lambda^0),\\
    \Gamma(X_{u\bar{d}\bar{s}}^{+}\to \Xi_{c}^{\prime+}\overline \Xi^0)= { }\Gamma(Y_{(u\bar{u},s\bar{s})\bar{d}}^{0}\to \Omega_{c}^{0}\overline \Xi^0),
    \Gamma(Y_{(d\bar{d},s\bar{s})\bar{u}}^{-}\to \Sigma_{c}^{0}\overline p)= { }\Gamma(X_{s\bar{u}\bar{d}}^{-}\to \Xi_{c}^{\prime0}\overline p).
\end{eqnarray*}
The relations of decay widths for class \uppercase\expandafter{\romannumeral3} become
\begin{eqnarray*}
&&3\Gamma(X_{u\bar{d}\bar{s}}^{+}\to   \Lambda_c^+  \overline \Delta^{0})
=6\Gamma(X_{u\bar{d}\bar{s}}^{+}\to   \Xi_c^+  \overline \Sigma^{\prime0} )
=\Gamma(X_{d\bar{s}\bar{u}}^{-}\to   \Lambda_c^+  \overline \Delta^{--} )
=3\Gamma(X_{d\bar{s}\bar{u}}^{-}\to   \Xi_c^0  \overline \Sigma^{\prime-})\\&&
=6\Gamma(Y_{(u\bar{u},s\bar{s})\bar{d}}^{0}\to   \Xi_c^+  \overline \Delta^{-})
=6\Gamma(Y_{(u\bar{u},s\bar{s})\bar{d}}^{0}\to   \Xi_c^0  \overline \Delta^{0} )
=\frac{3}{2}\Gamma(Y_{(u\bar{u},d\bar{d})\bar{s}}^{0}\to   \Lambda_c^+  \overline \Delta^{-})
=6\Gamma(Y_{(u\bar{u},d\bar{d})\bar{s}}^{0}\to   \Xi_c^+  \overline \Sigma^{\prime-} )\\&&
=12\Gamma(Y_{(u\bar{u},d\bar{d})\bar{s}}^{0}\to   \Xi_c^0  \overline \Sigma^{\prime0})
=2\Gamma(Y_{(d\bar{d},s\bar{s})\bar{u}}^{-}\to   \Xi_c^+  \overline \Delta^{--} )
=6\Gamma(Y_{(d\bar{d},s\bar{s})\bar{u}}^{-}\to   \Xi_c^0  \overline \Delta^{-}),\\
&&6\Gamma(X_{u\bar{d}\bar{s}}^{+}\to   \Lambda_c^+  \overline \Sigma^{\prime0})
=3\Gamma(X_{u\bar{d}\bar{s}}^{+}\to   \Xi_c^+  \overline \Xi^{\prime0})
=\Gamma(X_{s\bar{u}\bar{d}}^{-}\to   \Xi_c^+  \overline \Delta^{--})
=3\Gamma(X_{s\bar{u}\bar{d}}^{-}\to   \Xi_c^0  \overline \Delta^{-})\\&&
=6\Gamma(Y_{(u\bar{u},s\bar{s})\bar{d}}^{0}\to   \Lambda_c^+  \overline \Delta^{-})
=\frac{3}{2}\Gamma(Y_{(u\bar{u},s\bar{s})\bar{d}}^{0}\to   \Xi_c^+  \overline \Sigma^{\prime-})
=12\Gamma(Y_{(u\bar{u},s\bar{s})\bar{d}}^{0}\to   \Xi_c^0  \overline \Sigma^{\prime0})
=6\Gamma(Y_{(u\bar{u},d\bar{d})\bar{s}}^{0}\to   \Lambda_c^+  \overline \Sigma^{\prime-})\\&&
=6\Gamma(Y_{(u\bar{u},d\bar{d})\bar{s}}^{0}\to   \Xi_c^0  \overline \Xi^{\prime0} )
=2\Gamma(Y_{(d\bar{d},s\bar{s})\bar{u}}^{-}\to   \Lambda_c^+  \overline \Delta^{--})
=6\Gamma(Y_{(d\bar{d},s\bar{s})\bar{u}}^{-}\to   \Xi_c^0  \overline \Sigma^{\prime-}).
\end{eqnarray*}
The relations of decay widths for class \uppercase\expandafter{\romannumeral4} become
\begin{eqnarray*}
&&\Gamma(X_{u\bar{d}\bar{s}}^{+}\to   \Sigma_{c}^{0}  \overline \Delta^{+})
=\frac{3}{2}\Gamma(X_{u\bar{d}\bar{s}}^{+}\to   \Xi_{c}^{\prime0}  \overline \Sigma^{\prime+})
=3\Gamma(X_{u\bar{d}\bar{s}}^{+}\to   \Omega_{c}^{0}  \overline \Xi^{\prime+})
=3\Gamma(Y_{(u\bar{u},d\bar{d})\bar{s}}^{0}\to   \Sigma_{c}^{+}  \overline \Delta^{-})\\&&
=6\Gamma(Y_{(u\bar{u},d\bar{d})\bar{s}}^{0}\to   \Omega_{c}^{0}  \overline \Xi^{\prime0}),
\Gamma(X_{d\bar{s}\bar{u}}^{-}\to   \Sigma_{c}^{+}  \overline \Delta^{--})
=\frac{3}{2}\Gamma(X_{d\bar{s}\bar{u}}^{-}\to   \Sigma_{c}^{0}  \overline \Delta^{-})
=3\Gamma(X_{d\bar{s}\bar{u}}^{-}\to   \Xi_{c}^{\prime0}  \overline \Sigma^{\prime-})\\&&
=6\Gamma(Y_{(u\bar{u},s\bar{s})\bar{d}}^{0}\to   \Xi_{c}^{\prime+}  \overline \Delta^{-})
=6\Gamma(Y_{(u\bar{u},s\bar{s})\bar{d}}^{0}\to   \Xi_{c}^{\prime0}  \overline \Delta^{0})
=6\Gamma(Y_{(u\bar{u},s\bar{s})\bar{d}}^{0}\to   \Omega_{c}^{0}  \overline \Sigma^{\prime0})
=2\Gamma(Y_{(d\bar{d},s\bar{s})\bar{u}}^{-}\to   \Xi_{c}^{\prime+}  \overline \Delta^{--})\\&&
=6\Gamma(Y_{(d\bar{d},s\bar{s})\bar{u}}^{-}\to   \Xi_{c}^{\prime0}  \overline \Delta^{-})
=3\Gamma(Y_{(d\bar{d},s\bar{s})\bar{u}}^{-}\to   \Omega_{c}^{0}  \overline \Sigma^{\prime-}),
2\Gamma(X_{u\bar{d}\bar{s}}^{+}\to   \Sigma_{c}^{0}  \overline \Sigma^{\prime+})
=\Gamma(X_{u\bar{d}\bar{s}}^{+}\to   \Xi_{c}^{\prime0}  \overline \Xi^{\prime+})\\&&
=4\Gamma(Y_{(u\bar{u},s\bar{s})\bar{d}}^{0}\to   \Sigma_{c}^{0}  \overline \Delta^{0})
=2\Gamma(Y_{(u\bar{u},s\bar{s})\bar{d}}^{0}\to   \Xi_{c}^{\prime+}  \overline \Sigma^{\prime-})
=\frac{2}{3}\Gamma(X_{u\bar{d}\bar{s}}^{+}\to   \Omega_{c}^{0}  \overline \Omega^+),
\Gamma(X_{s\bar{u}\bar{d}}^{-}\to   \Xi_{c}^{\prime+}  \overline \Delta^{--})\\&&
=3\Gamma(X_{s\bar{u}\bar{d}}^{-}\to   \Xi_{c}^{\prime0}  \overline \Delta^{-})
=\frac{3}{2}\Gamma(X_{s\bar{u}\bar{d}}^{-}\to   \Omega_{c}^{0}  \overline \Sigma^{\prime-})
=6\Gamma(Y_{(u\bar{u},d\bar{d})\bar{s}}^{0}\to   \Sigma_{c}^{+}  \overline \Sigma^{\prime-})
=6\Gamma(Y_{(u\bar{u},d\bar{d})\bar{s}}^{0}\to   \Sigma_{c}^{0}  \overline \Sigma^{\prime0})\\&&
=6\Gamma(Y_{(u\bar{u},d\bar{d})\bar{s}}^{0}\to   \Xi_{c}^{\prime0}  \overline \Sigma^{\prime0})
=2\Gamma(Y_{(d\bar{d},s\bar{s})\bar{u}}^{-}\to   \Sigma_{c}^{+}  \overline \Delta^{--})
=3\Gamma(Y_{(d\bar{d},s\bar{s})\bar{u}}^{-}\to   \Sigma_{c}^{0}  \overline \Delta^{-})
=6\Gamma(Y_{(d\bar{d},s\bar{s})\bar{u}}^{-}\to   \Xi_{c}^{\prime0}  \overline \Sigma^{\prime-}),\\
&&\Gamma(X_{u\bar{d}\bar{s}}^{+}\to   \Sigma_{c}^{+}  \overline \Delta^{0})
=2\Gamma(X_{u\bar{d}\bar{s}}^{+}\to   \Xi_{c}^{\prime+}  \overline \Sigma^{\prime0})
=2\Gamma(Y_{(u\bar{u},d\bar{d})\bar{s}}^{0}\to   \Xi_{c}^{\prime+}  \overline \Sigma^{\prime-}),
2\Gamma(X_{u\bar{d}\bar{s}}^{+}\to   \Sigma_{c}^{+}  \overline \Sigma^{\prime0})\\&&
=\Gamma(X_{u\bar{d}\bar{s}}^{+}\to   \Xi_{c}^{\prime+}  \overline \Xi^{\prime0})
=2\Gamma(Y_{(u\bar{u},s\bar{s})\bar{d}}^{0}\to   \Sigma_{c}^{+}  \overline \Delta^{-}),
\frac{3}{2}\Gamma(X_{u\bar{d}\bar{s}}^{+}\to   \Sigma_{c}^{++}  \overline \Delta^{-})
=\Gamma(Y_{(u\bar{u},d\bar{d})\bar{s}}^{0}\to   \Sigma_{c}^{++}  \overline \Delta^{--}),\\&&
\frac{3}{2}\Gamma(X_{u\bar{d}\bar{s}}^{+}\to   \Sigma_{c}^{++}  \overline \Sigma^{\prime-})
=\Gamma(Y_{(u\bar{u},s\bar{s})\bar{d}}^{0}\to   \Sigma_{c}^{++}  \overline \Delta^{--}).
\end{eqnarray*}
\begin{table}
\caption{Open bottom tetraquark $X_{b6}$ decays into a light anti-baryon octet and a charmed baryon anti-triplet for class I, a light anti-baryon octet and a charmed baryon sextet for class II.}\label{tab:b6_Tc3bar_Tbar8}\begin{tabular}{|cc|cc|}\hline\hline
\multicolumn{4}{|c|} {Class \uppercase\expandafter{\romannumeral1}} \\\hline
channel & amplitude($/V_{cb}$) &channel & amplitude($/V_{cb}$)\\\hline
$X_{u\bar{d}\bar{s}}^{+}\to   \Lambda_c^+  \overline \Lambda^0 $ & $ \frac{\left(b_1-b_2+b_3\right) V^*_{ud}}{2 \sqrt{3}}$&
$X_{u\bar{d}\bar{s}}^{+}\to   \Lambda_c^+  \overline \Sigma^0 $ & $ -\frac{1}{2} \left(b_1+b_2-b_3\right) V^*_{ud}$\\\hline
$X_{u\bar{d}\bar{s}}^{+}\to   \Lambda_c^+  \overline n $ & $ \frac{b_1 V^*_{us}}{\sqrt{2}}$&
$X_{u\bar{d}\bar{s}}^{+}\to   \Xi_c^+  \overline \Lambda^0 $ & $ -\frac{\left(2 b_1+b_2-b_3\right) V^*_{us}}{2 \sqrt{3}}$\\\hline
$X_{u\bar{d}\bar{s}}^{+}\to   \Xi_c^+  \overline \Sigma^0 $ & $ \frac{1}{2} \left(b_3-b_2\right) V^*_{us}$&
$X_{u\bar{d}\bar{s}}^{+}\to   \Xi_c^+  \overline \Xi^0 $ & $ \frac{b_1 V^*_{ud}}{\sqrt{2}}$\\\hline
$X_{u\bar{d}\bar{s}}^{+}\to   \Xi_c^0  \overline \Sigma^+ $ & $ -\frac{b_2 V^*_{us}}{\sqrt{2}}$&
$X_{u\bar{d}\bar{s}}^{+}\to   \Xi_c^0  \overline \Xi^+ $ & $ \frac{b_2 V^*_{ud}}{\sqrt{2}}$\\\hline
$X_{d\bar{s}\bar{u}}^{-}\to   \Xi_c^0  \overline \Sigma^- $ & $ \frac{b_3 V^*_{us}}{\sqrt{2}}$&
$X_{s\bar{u}\bar{d}}^{-}\to   \Xi_c^0  \overline p $ & $ -\frac{b_3 V^*_{ud}}{\sqrt{2}}$\\\hline
$Y_{(u\bar{u},s\bar{s})\bar{d}}^{0}\to   \Lambda_c^+  \overline p $ & $ \frac{1}{2} \left(b_3-b_2\right) V^*_{ud}$&
$Y_{(u\bar{u},s\bar{s})\bar{d}}^{0}\to   \Xi_c^+  \overline \Sigma^- $ & $ -\frac{1}{2} b_1 V^*_{ud}$\\\hline
$Y_{(u\bar{u},s\bar{s})\bar{d}}^{0}\to   \Xi_c^+  \overline p $ & $ -\frac{1}{2} \left(b_1+b_2-b_3\right) V^*_{us}$&
$Y_{(u\bar{u},s\bar{s})\bar{d}}^{0}\to   \Xi_c^0  \overline \Lambda^0 $ & $ -\frac{\left(b_1+2 b_2+b_3\right) V^*_{ud}}{2 \sqrt{6}}$\\\hline
$Y_{(u\bar{u},s\bar{s})\bar{d}}^{0}\to   \Xi_c^0  \overline \Sigma^0 $ & $ \frac{\left(b_1-b_3\right) V^*_{ud}}{2 \sqrt{2}}$&
$Y_{(u\bar{u},s\bar{s})\bar{d}}^{0}\to   \Xi_c^0  \overline n $ & $ -\frac{1}{2} \left(b_1+b_2\right) V^*_{us}$\\\hline
$Y_{(u\bar{u},d\bar{d})\bar{s}}^{0}\to   \Lambda_c^+  \overline \Sigma^- $ & $ -\frac{1}{2} \left(b_1+b_2-b_3\right) V^*_{ud}$&
$Y_{(u\bar{u},d\bar{d})\bar{s}}^{0}\to   \Lambda_c^+  \overline p $ & $ -\frac{1}{2} b_1 V^*_{us}$\\\hline
$Y_{(u\bar{u},d\bar{d})\bar{s}}^{0}\to   \Xi_c^+  \overline \Sigma^- $ & $ \frac{1}{2} \left(b_3-b_2\right) V^*_{us}$&
$Y_{(u\bar{u},d\bar{d})\bar{s}}^{0}\to   \Xi_c^0  \overline \Lambda^0 $ & $ -\frac{\left(2 b_1+b_2-b_3\right) V^*_{us}}{2 \sqrt{6}}$\\\hline
$Y_{(u\bar{u},d\bar{d})\bar{s}}^{0}\to   \Xi_c^0  \overline \Sigma^0 $ & $ \frac{\left(b_2+b_3\right) V^*_{us}}{2 \sqrt{2}}$&
$Y_{(u\bar{u},d\bar{d})\bar{s}}^{0}\to   \Xi_c^0  \overline \Xi^0 $ & $ \frac{1}{2} \left(b_1+b_2\right) V^*_{ud}$\\\hline
$Y_{(d\bar{d},s\bar{s})\bar{u}}^{-}\to   \Xi_c^0  \overline \Sigma^- $ & $ -\frac{1}{2} b_3 V^*_{ud}$&
$Y_{(d\bar{d},s\bar{s})\bar{u}}^{-}\to   \Xi_c^0  \overline p $ & $ \frac{1}{2} b_3 V^*_{us}$\\\hline
\hline
%
\multicolumn{4}{|c|} {Class \uppercase\expandafter{\romannumeral2}} \\\hline
$X_{u\bar{d}\bar{s}}^{+}\to   \Sigma_{c}^{++}  \overline \Sigma^- $ & $ \frac{\left(c_1+c_3\right) V^*_{ud}}{\sqrt{2}}$&
$X_{u\bar{d}\bar{s}}^{+}\to   \Sigma_{c}^{++}  \overline p $ & $ \frac{\left(c_1+c_3\right) V^*_{us}}{\sqrt{2}}$\\\hline
$X_{u\bar{d}\bar{s}}^{+}\to   \Sigma_{c}^{+}  \overline \Lambda^0 $ & $ \frac{\left(c_1+c_2+c_4\right) V^*_{ud}}{2 \sqrt{6}}$&
$X_{u\bar{d}\bar{s}}^{+}\to   \Sigma_{c}^{+}  \overline \Sigma^0 $ & $ \frac{\left(-c_1+c_2+c_4\right) V^*_{ud}}{2 \sqrt{2}}$\\\hline
$X_{u\bar{d}\bar{s}}^{+}\to   \Sigma_{c}^{+}  \overline n $ & $ \frac{1}{2} c_1 V^*_{us}$&
$X_{u\bar{d}\bar{s}}^{+}\to   \Sigma_{c}^{0}  \overline \Sigma^+ $ & $ \frac{c_2 V^*_{ud}}{\sqrt{2}}$\\\hline
$X_{u\bar{d}\bar{s}}^{+}\to   \Xi_{c}^{\prime+}  \overline \Lambda^0 $ & $ \frac{\left(-2 c_1+c_2+c_4\right) V^*_{us}}{2 \sqrt{6}}$&
$X_{u\bar{d}\bar{s}}^{+}\to   \Xi_{c}^{\prime+}  \overline \Sigma^0 $ & $ \frac{\left(c_2+c_4\right) V^*_{us}}{2 \sqrt{2}}$\\\hline
$X_{u\bar{d}\bar{s}}^{+}\to   \Xi_{c}^{\prime+}  \overline \Xi^0 $ & $ \frac{1}{2} c_1 V^*_{ud}$&
$X_{u\bar{d}\bar{s}}^{+}\to   \Xi_{c}^{\prime0}  \overline \Sigma^+ $ & $ \frac{1}{2} c_2 V^*_{us}$\\\hline
$X_{u\bar{d}\bar{s}}^{+}\to   \Xi_{c}^{\prime0}  \overline \Xi^+ $ & $ \frac{1}{2} c_2 V^*_{ud}$&
$X_{u\bar{d}\bar{s}}^{+}\to   \Omega_{c}^{0}  \overline \Xi^+ $ & $ \frac{c_2 V^*_{us}}{\sqrt{2}}$\\\hline
$X_{d\bar{s}\bar{u}}^{-}\to   \Sigma_{c}^{0}  \overline \Sigma^- $ & $ \frac{\left(c_3+c_4\right) V^*_{ud}}{\sqrt{2}}$&
$X_{d\bar{s}\bar{u}}^{-}\to   \Sigma_{c}^{0}  \overline p $ & $ \frac{c_3 V^*_{us}}{\sqrt{2}}$\\\hline
$X_{d\bar{s}\bar{u}}^{-}\to   \Xi_{c}^{\prime0}  \overline \Sigma^- $ & $ \frac{1}{2} c_4 V^*_{us}$&
$X_{s\bar{u}\bar{d}}^{-}\to   \Xi_{c}^{\prime0}  \overline p $ & $ \frac{1}{2} c_4 V^*_{ud}$\\\hline
$X_{s\bar{u}\bar{d}}^{-}\to   \Omega_{c}^{0}  \overline \Sigma^- $ & $ \frac{c_3 V^*_{ud}}{\sqrt{2}}$&
$X_{s\bar{u}\bar{d}}^{-}\to   \Omega_{c}^{0}  \overline p $ & $ \frac{\left(c_3+c_4\right) V^*_{us}}{\sqrt{2}}$\\\hline
$Y_{(u\bar{u},s\bar{s})\bar{d}}^{0}\to   \Sigma_{c}^{+}  \overline p $ & $ \frac{\left(c_2+c_4\right) V^*_{ud}}{2 \sqrt{2}}$&
$Y_{(u\bar{u},s\bar{s})\bar{d}}^{0}\to   \Sigma_{c}^{0}  \overline n $ & $ \frac{1}{2} c_2 V^*_{ud}$\\\hline
$Y_{(u\bar{u},s\bar{s})\bar{d}}^{0}\to   \Xi_{c}^{\prime+}  \overline \Sigma^- $ & $ \frac{\left(c_1+2 c_3\right) V^*_{ud}}{2 \sqrt{2}}$&
$Y_{(u\bar{u},s\bar{s})\bar{d}}^{0}\to   \Xi_{c}^{\prime+}  \overline p $ & $ \frac{\left(c_1+c_2+2 c_3+c_4\right) V^*_{us}}{2 \sqrt{2}}$\\\hline
$Y_{(u\bar{u},s\bar{s})\bar{d}}^{0}\to   \Xi_{c}^{\prime0}  \overline \Lambda^0 $ & $ \frac{\left(c_1-2 c_2+c_4\right) V^*_{ud}}{4 \sqrt{3}}$&
$Y_{(u\bar{u},s\bar{s})\bar{d}}^{0}\to   \Xi_{c}^{\prime0}  \overline \Sigma^0 $ & $ \frac{1}{4} \left(c_4-c_1\right) V^*_{ud}$\\\hline
$Y_{(u\bar{u},s\bar{s})\bar{d}}^{0}\to   \Xi_{c}^{\prime0}  \overline n $ & $ \frac{\left(c_1+c_2\right) V^*_{us}}{2 \sqrt{2}}$&
$Y_{(u\bar{u},s\bar{s})\bar{d}}^{0}\to   \Omega_{c}^{0}  \overline \Lambda^0 $ & $ \frac{\left(-2 c_1-2 c_2+c_4\right) V^*_{us}}{2 \sqrt{6}}$\\\hline
$Y_{(u\bar{u},s\bar{s})\bar{d}}^{0}\to   \Omega_{c}^{0}  \overline \Sigma^0 $ & $ \frac{c_4 V^*_{us}}{2 \sqrt{2}}$&
$Y_{(u\bar{u},s\bar{s})\bar{d}}^{0}\to   \Omega_{c}^{0}  \overline \Xi^0 $ & $ \frac{1}{2} c_1 V^*_{ud}$\\\hline
$Y_{(u\bar{u},d\bar{d})\bar{s}}^{0}\to   \Sigma_{c}^{+}  \overline \Sigma^- $ & $ \frac{\left(c_1+c_2+2 c_3+c_4\right) V^*_{ud}}{2 \sqrt{2}}$&
$Y_{(u\bar{u},d\bar{d})\bar{s}}^{0}\to   \Sigma_{c}^{+}  \overline p $ & $ \frac{\left(c_1+2 c_3\right) V^*_{us}}{2 \sqrt{2}}$\\\hline
$Y_{(u\bar{u},d\bar{d})\bar{s}}^{0}\to   \Sigma_{c}^{0}  \overline \Lambda^0 $ & $ \frac{\left(c_1+c_2+c_4\right) V^*_{ud}}{2 \sqrt{6}}$&
$Y_{(u\bar{u},d\bar{d})\bar{s}}^{0}\to   \Sigma_{c}^{0}  \overline \Sigma^0 $ & $ -\frac{\left(c_1+c_2-c_4\right) V^*_{ud}}{2 \sqrt{2}}$\\\hline
$Y_{(u\bar{u},d\bar{d})\bar{s}}^{0}\to   \Sigma_{c}^{0}  \overline n $ & $ \frac{1}{2} c_1 V^*_{us}$&
$Y_{(u\bar{u},d\bar{d})\bar{s}}^{0}\to   \Xi_{c}^{\prime+}  \overline \Sigma^- $ & $ \frac{\left(c_2+c_4\right) V^*_{us}}{2 \sqrt{2}}$\\\hline
$Y_{(u\bar{u},d\bar{d})\bar{s}}^{0}\to   \Xi_{c}^{\prime0}  \overline \Lambda^0 $ & $ \frac{\left(-2 c_1+c_2+c_4\right) V^*_{us}}{4 \sqrt{3}}$&
$Y_{(u\bar{u},d\bar{d})\bar{s}}^{0}\to   \Xi_{c}^{\prime0}  \overline \Sigma^0 $ & $ \frac{1}{4} \left(c_4-c_2\right) V^*_{us}$\\\hline
$Y_{(u\bar{u},d\bar{d})\bar{s}}^{0}\to   \Xi_{c}^{\prime0}  \overline \Xi^0 $ & $ \frac{\left(c_1+c_2\right) V^*_{ud}}{2 \sqrt{2}}$&
$Y_{(u\bar{u},d\bar{d})\bar{s}}^{0}\to   \Omega_{c}^{0}  \overline \Xi^0 $ & $ \frac{1}{2} c_2 V^*_{us}$\\\hline
$Y_{(d\bar{d},s\bar{s})\bar{u}}^{-}\to   \Sigma_{c}^{0}  \overline p $ & $ \frac{1}{2} c_4 V^*_{ud}$&
$Y_{(d\bar{d},s\bar{s})\bar{u}}^{-}\to   \Xi_{c}^{\prime0}  \overline \Sigma^- $ & $ \frac{\left(2 c_3+c_4\right) V^*_{ud}}{2 \sqrt{2}}$\\\hline
$Y_{(d\bar{d},s\bar{s})\bar{u}}^{-}\to   \Xi_{c}^{\prime0}  \overline p $ & $ \frac{\left(2 c_3+c_4\right) V^*_{us}}{2 \sqrt{2}}$&
$Y_{(d\bar{d},s\bar{s})\bar{u}}^{-}\to   \Omega_{c}^{0}  \overline \Sigma^- $ & $ \frac{1}{2} c_4 V^*_{us}$\\\hline
\hline
\end{tabular}
\end{table}

\begin{table}
\caption{Open bottom tetraquark $X_{b6}$ decays into a light anti-baryon anti-decuplet and a charmed baryon anti-triplet for class III, a light anti-baryon anti-decuplet and a charmed baryon sextet for class IV.}\label{tab:b6_Tc3bar_Tbar10bar}\begin{tabular}{|cc|cc|}\hline\hline
\multicolumn{4}{|c|} {Class \uppercase\expandafter{\romannumeral3}} \\\hline
channel & amplitude($/V_{cb}$) &channel & amplitude($/V_{cb}$)\\\hline
$X_{u\bar{d}\bar{s}}^{+}\to   \Lambda_c^+  \overline \Delta^{0} $ & $ -\frac{d_1 V^*_{us}}{\sqrt{6}}$&
$X_{u\bar{d}\bar{s}}^{+}\to   \Lambda_c^+  \overline \Sigma^{\prime0} $ & $ \frac{d_1 V^*_{ud}}{2 \sqrt{3}}$\\\hline
$X_{u\bar{d}\bar{s}}^{+}\to   \Xi_c^+  \overline \Sigma^{\prime0} $ & $ -\frac{d_1 V^*_{us}}{2 \sqrt{3}}$&
$X_{u\bar{d}\bar{s}}^{+}\to   \Xi_c^+  \overline \Xi^{\prime0} $ & $ \frac{d_1 V^*_{ud}}{\sqrt{6}}$\\\hline
$X_{d\bar{s}\bar{u}}^{-}\to   \Lambda_c^+  \overline \Delta^{--} $ & $ -\frac{d_1 V^*_{us}}{\sqrt{2}}$&
$X_{d\bar{s}\bar{u}}^{-}\to   \Xi_c^0  \overline \Sigma^{\prime-} $ & $ \frac{d_1 V^*_{us}}{\sqrt{6}}$\\\hline
$X_{s\bar{u}\bar{d}}^{-}\to   \Xi_c^+  \overline \Delta^{--} $ & $ \frac{d_1 V^*_{ud}}{\sqrt{2}}$&
$X_{s\bar{u}\bar{d}}^{-}\to   \Xi_c^0  \overline \Delta^{-} $ & $ \frac{d_1 V^*_{ud}}{\sqrt{6}}$\\\hline
$Y_{(u\bar{u},s\bar{s})\bar{d}}^{0}\to   \Lambda_c^+  \overline \Delta^{-} $ & $ -\frac{d_1 V^*_{ud}}{2 \sqrt{3}}$&
$Y_{(u\bar{u},s\bar{s})\bar{d}}^{0}\to   \Xi_c^+  \overline \Delta^{-} $ & $ \frac{d_1 V^*_{us}}{2 \sqrt{3}}$\\\hline
$Y_{(u\bar{u},s\bar{s})\bar{d}}^{0}\to   \Xi_c^+  \overline \Sigma^{\prime-} $ & $ -\frac{d_1 V^*_{ud}}{\sqrt{3}}$&
$Y_{(u\bar{u},s\bar{s})\bar{d}}^{0}\to   \Xi_c^0  \overline \Delta^{0} $ & $ \frac{d_1 V^*_{us}}{2 \sqrt{3}}$\\\hline
$Y_{(u\bar{u},s\bar{s})\bar{d}}^{0}\to   \Xi_c^0  \overline \Sigma^{\prime0} $ & $ -\frac{d_1 V^*_{ud}}{2 \sqrt{6}}$&
$Y_{(u\bar{u},d\bar{d})\bar{s}}^{0}\to   \Lambda_c^+  \overline \Delta^{-} $ & $ \frac{d_1 V^*_{us}}{\sqrt{3}}$\\\hline
$Y_{(u\bar{u},d\bar{d})\bar{s}}^{0}\to   \Lambda_c^+  \overline \Sigma^{\prime-} $ & $ -\frac{d_1 V^*_{ud}}{2 \sqrt{3}}$&
$Y_{(u\bar{u},d\bar{d})\bar{s}}^{0}\to   \Xi_c^+  \overline \Sigma^{\prime-} $ & $ \frac{d_1 V^*_{us}}{2 \sqrt{3}}$\\\hline
$Y_{(u\bar{u},d\bar{d})\bar{s}}^{0}\to   \Xi_c^0  \overline \Sigma^{\prime0} $ & $ -\frac{d_1 V^*_{us}}{2 \sqrt{6}}$&
$Y_{(u\bar{u},d\bar{d})\bar{s}}^{0}\to   \Xi_c^0  \overline \Xi^{\prime0} $ & $ \frac{d_1 V^*_{ud}}{2 \sqrt{3}}$\\\hline
$Y_{(d\bar{d},s\bar{s})\bar{u}}^{-}\to   \Lambda_c^+  \overline \Delta^{--} $ & $ \frac{1}{2} d_1 V^*_{ud}$&
$Y_{(d\bar{d},s\bar{s})\bar{u}}^{-}\to   \Xi_c^+  \overline \Delta^{--} $ & $ -\frac{1}{2} d_1 V^*_{us}$\\\hline
$Y_{(d\bar{d},s\bar{s})\bar{u}}^{-}\to   \Xi_c^0  \overline \Delta^{-} $ & $ -\frac{d_1 V^*_{us}}{2 \sqrt{3}}$&
$Y_{(d\bar{d},s\bar{s})\bar{u}}^{-}\to   \Xi_c^0  \overline \Sigma^{\prime-} $ & $ -\frac{d_1 V^*_{ud}}{2 \sqrt{3}}$\\\hline
\hline
%
\multicolumn{4}{|c|} {Class \uppercase\expandafter{\romannumeral4}} \\\hline
$X_{u\bar{d}\bar{s}}^{+}\to   \Sigma_{c}^{++}  \overline \Delta^{-} $ & $ -\frac{\left(f_1+f_2\right) V^*_{us}}{\sqrt{6}}$&
$X_{u\bar{d}\bar{s}}^{+}\to   \Sigma_{c}^{++}  \overline \Sigma^{\prime-} $ & $ \frac{\left(f_1+f_2\right) V^*_{ud}}{\sqrt{6}}$\\\hline
$X_{u\bar{d}\bar{s}}^{+}\to   \Sigma_{c}^{+}  \overline \Delta^{0} $ & $ -\frac{\left(2 f_1+f_2\right) V^*_{us}}{2 \sqrt{3}}$&
$X_{u\bar{d}\bar{s}}^{+}\to   \Sigma_{c}^{+}  \overline \Sigma^{\prime0} $ & $ \frac{\left(2 f_1+f_2\right) V^*_{ud}}{2 \sqrt{6}}$\\\hline
$X_{u\bar{d}\bar{s}}^{+}\to   \Sigma_{c}^{0}  \overline \Delta^{+} $ & $ -\frac{f_1 V^*_{us}}{\sqrt{2}}$&
$X_{u\bar{d}\bar{s}}^{+}\to   \Sigma_{c}^{0}  \overline \Sigma^{\prime+} $ & $ \frac{f_1 V^*_{ud}}{\sqrt{6}}$\\\hline
$X_{u\bar{d}\bar{s}}^{+}\to   \Xi_{c}^{\prime+}  \overline \Sigma^{\prime0} $ & $ -\frac{\left(2 f_1+f_2\right) V^*_{us}}{2 \sqrt{6}}$&
$X_{u\bar{d}\bar{s}}^{+}\to   \Xi_{c}^{\prime+}  \overline \Xi^{\prime0} $ & $ \frac{\left(2 f_1+f_2\right) V^*_{ud}}{2 \sqrt{3}}$\\\hline
$X_{u\bar{d}\bar{s}}^{+}\to   \Xi_{c}^{\prime0}  \overline \Sigma^{\prime+} $ & $ -\frac{f_1 V^*_{us}}{\sqrt{3}}$&
$X_{u\bar{d}\bar{s}}^{+}\to   \Xi_{c}^{\prime0}  \overline \Xi^{\prime+} $ & $ \frac{f_1 V^*_{ud}}{\sqrt{3}}$\\\hline
$X_{u\bar{d}\bar{s}}^{+}\to   \Omega_{c}^{0}  \overline \Xi^{\prime+} $ & $ -\frac{f_1 V^*_{us}}{\sqrt{6}}$&
$X_{u\bar{d}\bar{s}}^{+}\to   \Omega_{c}^{0}  \overline \Omega^+ $ & $ \frac{f_1 V^*_{ud}}{\sqrt{2}}$\\\hline
$X_{d\bar{s}\bar{u}}^{-}\to   \Sigma_{c}^{+}  \overline \Delta^{--} $ & $ \frac{1}{2} f_2 V^*_{us}$&
$X_{d\bar{s}\bar{u}}^{-}\to   \Sigma_{c}^{0}  \overline \Delta^{-} $ & $ \frac{f_2 V^*_{us}}{\sqrt{6}}$\\\hline
$X_{d\bar{s}\bar{u}}^{-}\to   \Xi_{c}^{\prime0}  \overline \Sigma^{\prime-} $ & $ \frac{f_2 V^*_{us}}{2 \sqrt{3}}$&
$X_{s\bar{u}\bar{d}}^{-}\to   \Xi_{c}^{\prime+}  \overline \Delta^{--} $ & $ -\frac{1}{2} f_2 V^*_{ud}$\\\hline
$X_{s\bar{u}\bar{d}}^{-}\to   \Xi_{c}^{\prime0}  \overline \Delta^{-} $ & $ -\frac{f_2 V^*_{ud}}{2 \sqrt{3}}$&
$X_{s\bar{u}\bar{d}}^{-}\to   \Omega_{c}^{0}  \overline \Sigma^{\prime-} $ & $ -\frac{f_2 V^*_{ud}}{\sqrt{6}}$\\\hline
$Y_{(u\bar{u},s\bar{s})\bar{d}}^{0}\to   \Sigma_{c}^{++}  \overline \Delta^{--} $ & $ -\frac{1}{2} \left(f_1+f_2\right) V^*_{ud}$&
$Y_{(u\bar{u},s\bar{s})\bar{d}}^{0}\to   \Sigma_{c}^{+}  \overline \Delta^{-} $ & $ -\frac{\left(2 f_1+f_2\right) V^*_{ud}}{2 \sqrt{6}}$\\\hline
$Y_{(u\bar{u},s\bar{s})\bar{d}}^{0}\to   \Sigma_{c}^{0}  \overline \Delta^{0} $ & $ -\frac{f_1 V^*_{ud}}{2 \sqrt{3}}$&
$Y_{(u\bar{u},s\bar{s})\bar{d}}^{0}\to   \Xi_{c}^{\prime+}  \overline \Delta^{-} $ & $ -\frac{f_2 V^*_{us}}{2 \sqrt{6}}$\\\hline
$Y_{(u\bar{u},s\bar{s})\bar{d}}^{0}\to   \Xi_{c}^{\prime+}  \overline \Sigma^{\prime-} $ & $ -\frac{f_1 V^*_{ud}}{\sqrt{6}}$&
$Y_{(u\bar{u},s\bar{s})\bar{d}}^{0}\to   \Xi_{c}^{\prime0}  \overline \Delta^{0} $ & $ -\frac{f_2 V^*_{us}}{2 \sqrt{6}}$\\\hline
$Y_{(u\bar{u},s\bar{s})\bar{d}}^{0}\to   \Xi_{c}^{\prime0}  \overline \Sigma^{\prime0} $ & $ \frac{\left(f_2-2 f_1\right) V^*_{ud}}{4 \sqrt{3}}$&
$Y_{(u\bar{u},s\bar{s})\bar{d}}^{0}\to   \Omega_{c}^{0}  \overline \Sigma^{\prime0} $ & $ -\frac{f_2 V^*_{us}}{2 \sqrt{6}}$\\\hline
$Y_{(u\bar{u},s\bar{s})\bar{d}}^{0}\to   \Omega_{c}^{0}  \overline \Xi^{\prime0} $ & $ \frac{\left(f_2-f_1\right) V^*_{ud}}{2 \sqrt{3}}$&
$Y_{(u\bar{u},d\bar{d})\bar{s}}^{0}\to   \Sigma_{c}^{++}  \overline \Delta^{--} $ & $ \frac{1}{2} \left(f_1+f_2\right) V^*_{us}$\\\hline
$Y_{(u\bar{u},d\bar{d})\bar{s}}^{0}\to   \Sigma_{c}^{+}  \overline \Delta^{-} $ & $ \frac{f_1 V^*_{us}}{\sqrt{6}}$&
$Y_{(u\bar{u},d\bar{d})\bar{s}}^{0}\to   \Sigma_{c}^{+}  \overline \Sigma^{\prime-} $ & $ \frac{f_2 V^*_{ud}}{2 \sqrt{6}}$\\\hline
$Y_{(u\bar{u},d\bar{d})\bar{s}}^{0}\to   \Sigma_{c}^{0}  \overline \Delta^{0} $ & $ \frac{\left(f_1-f_2\right) V^*_{us}}{2 \sqrt{3}}$&
$Y_{(u\bar{u},d\bar{d})\bar{s}}^{0}\to   \Sigma_{c}^{0}  \overline \Sigma^{\prime0} $ & $ \frac{f_2 V^*_{ud}}{2 \sqrt{6}}$\\\hline
$Y_{(u\bar{u},d\bar{d})\bar{s}}^{0}\to   \Xi_{c}^{\prime+}  \overline \Sigma^{\prime-} $ & $ \frac{\left(2 f_1+f_2\right) V^*_{us}}{2 \sqrt{6}}$&
$Y_{(u\bar{u},d\bar{d})\bar{s}}^{0}\to   \Xi_{c}^{\prime0}  \overline \Sigma^{\prime0} $ & $ \frac{\left(2 f_1-f_2\right) V^*_{us}}{4 \sqrt{3}}$\\\hline
$Y_{(u\bar{u},d\bar{d})\bar{s}}^{0}\to   \Xi_{c}^{\prime0}  \overline \Xi^{\prime0} $ & $ \frac{f_2 V^*_{ud}}{2 \sqrt{6}}$&
$Y_{(u\bar{u},d\bar{d})\bar{s}}^{0}\to   \Omega_{c}^{0}  \overline \Xi^{\prime0} $ & $ \frac{f_1 V^*_{us}}{2 \sqrt{3}}$\\\hline
$Y_{(d\bar{d},s\bar{s})\bar{u}}^{-}\to   \Sigma_{c}^{+}  \overline \Delta^{--} $ & $ -\frac{f_2 V^*_{ud}}{2 \sqrt{2}}$&
$Y_{(d\bar{d},s\bar{s})\bar{u}}^{-}\to   \Sigma_{c}^{0}  \overline \Delta^{-} $ & $ -\frac{f_2 V^*_{ud}}{2 \sqrt{3}}$\\\hline
$Y_{(d\bar{d},s\bar{s})\bar{u}}^{-}\to   \Xi_{c}^{\prime+}  \overline \Delta^{--} $ & $ \frac{f_2 V^*_{us}}{2 \sqrt{2}}$&
$Y_{(d\bar{d},s\bar{s})\bar{u}}^{-}\to   \Xi_{c}^{\prime0}  \overline \Delta^{-} $ & $ \frac{f_2 V^*_{us}}{2 \sqrt{6}}$\\\hline
$Y_{(d\bar{d},s\bar{s})\bar{u}}^{-}\to   \Xi_{c}^{\prime0}  \overline \Sigma^{\prime-} $ & $ -\frac{f_2 V^*_{ud}}{2 \sqrt{6}}$&
$Y_{(d\bar{d},s\bar{s})\bar{u}}^{-}\to   \Omega_{c}^{0}  \overline \Sigma^{\prime-} $ & $ \frac{f_2 V^*_{us}}{2 \sqrt{3}}$\\\hline
\hline
\end{tabular}
\end{table}

\subsubsection{Three-body decays into mesons }

For the $X_{b6}$ decays into three mesons, one constructs the  effective Hamiltonian as follows
\begin{eqnarray}
  {\cal H}_{{eff}}&=& a_8 (X_{b6})_{[jk]}^i (H_{{\bf8}})^j_i (\overline D)^k M^l_m M^m_l + a_9 (X_{b6})_{[jk]}^i (H_{{\bf8}})^j_i (\overline D)^l M^k_m M^m_l \nonumber \\
&&+a_{10} (X_{b6})_{[jk]}^i (H_{{\bf8}})^l_i (\overline D)^j M^k_m M^m_l +a_{11} (X_{b6})_{[jk]}^i (H_{{\bf8}})^l_i (\overline D)^m M^j_l M^k_m \nonumber \\
&&+a_{12} (X_{b6})_{[jk]}^i (H_{{\bf8}})^j_m (\overline D)^k M^l_i M^m_l+a_{13} (X_{b6})_{[jk]}^i (H_{{\bf8}})^l_m (\overline D)^j M^k_i M^m_l  \nonumber\\
&&+a_{14} (X_{b6})_{[jk]}^i (H_{{\bf8}})^l_m (\overline D)^m M^j_i M^k_l+a_{15} (X_{b6})_{[jk]}^i (H_{{\bf8}})^l_m (\overline D)^j M^m_i M^k_l.
\end{eqnarray}
The decay amplitudes are given in Tab.~\ref{tab:b6_D_2M}. The relations of different decay widths  become
\begin{eqnarray*}
    &&\Gamma(X_{u\bar{d}\bar{s}}^{+}\to  D^0 \pi^+ K^0 )
    = 6\Gamma(X_{u\bar{d}\bar{s}}^{+}\to  D^0 K^+ \eta )
    =2\Gamma(X_{u\bar{d}\bar{s}}^{+}\to  D^0 \pi^0 K^+ ),
     \Gamma(X_{u\bar{d}\bar{s}}^{+}\to  D^+ \pi^0 K^0 )\\&&
     = 3\Gamma(X_{u\bar{d}\bar{s}}^{+}\to  D^+ K^0 \eta )
     =\Gamma(Y_{(u\bar{u},s\bar{s})\bar{d}}^{0}\to  D^+ K^0 K^- ),
      \Gamma(X_{u\bar{d}\bar{s}}^{+}\to  D^+_s \pi^0 \overline K^0 )
      = 3\Gamma(X_{u\bar{d}\bar{s}}^{+}\to  D^+_s \overline K^0 \eta )\\&&
      =6\Gamma(Y_{(u\bar{u},d\bar{d})\bar{s}}^{0}\to  D^+_s K^- \eta )
      =\Gamma(Y_{(u\bar{u},d\bar{d})\bar{s}}^{0}\to  D^+_s \pi^- \overline K^0 ),
      \Gamma(X_{u\bar{d}\bar{s}}^{+}\to  D^0 K^+ \overline K^0 )
     = \frac{3}{2}\Gamma(X_{u\bar{d}\bar{s}}^{+}\to  D^0 \pi^+ \eta ),\\&&
     \Gamma(X_{d\bar{s}\bar{u}}^{-}\to  D^0 \pi^- K^0 )
     = { }\Gamma(X_{s\bar{u}\bar{d}}^{-}\to  D^+ \pi^- K^- )
     =2\Gamma(Y_{(d\bar{d},s\bar{s})\bar{u}}^{-}\to  D^+_s \pi^- K^- )
     =\frac{1}{2}\Gamma(X_{d\bar{s}\bar{u}}^{-}\to  D^+_s \pi^- \pi^- )\\&&
     =2\Gamma(Y_{(u\bar{u},d\bar{d})\bar{s}}^{0}\to  D^+ \pi^- K^0 )
     =4\Gamma(Y_{(u\bar{u},s\bar{s})\bar{d}}^{0}\to  D^+ \pi^0 \pi^- )
     =\Gamma(Y_{(u\bar{u},d\bar{d})\bar{s}}^{0}\to  D^+_s \pi^0 \pi^- )
     =4\Gamma(Y_{(d\bar{d},s\bar{s})\bar{u}}^{-}\to  D^0 \pi^0 \pi^- )\\&&
     =\Gamma(Y_{(d\bar{d},s\bar{s})\bar{u}}^{-}\to  D^+ \pi^- \pi^- ),
         \Gamma(Y_{(u\bar{u},d\bar{d})\bar{s}}^{0}\to  D^+_s K^0 K^- )
         =\frac{3}{2}\Gamma(X_{u\bar{d}\bar{s}}^{+}\to  D^+_s \pi^0 \eta )=\frac{3}{2}\Gamma(Y_{(u\bar{u},d\bar{d})\bar{s}}^{0}\to  D^+_s \pi^- \eta ),\\&&
      \Gamma(X_{d\bar{s}\bar{u}}^{-}\to  D^+_s \pi^- K^- )
      = { }\Gamma(X_{s\bar{u}\bar{d}}^{-}\to  D^0 \overline K^0 K^- )
      =\frac{1}{2}\Gamma(X_{s\bar{u}\bar{d}}^{-}\to  D^+ K^- K^- )
      =2\Gamma(Y_{(u\bar{u},s\bar{s})\bar{d}}^{0}\to  D^+_s \overline K^0 K^- )\\&&
=\Gamma(Y_{(d\bar{d},s\bar{s})\bar{u}}^{-}\to  D^+_s K^- K^- )
=2\Gamma(Y_{(d\bar{d},s\bar{s})\bar{u}}^{-}\to  D^+ \pi^- K^- ),
\Gamma(X_{s\bar{u}\bar{d}}^{-}\to  D^0 \pi^0 K^- )
= 3\Gamma(X_{s\bar{u}\bar{d}}^{-}\to  D^0 K^- \eta )\\&&
=\Gamma(Y_{(d\bar{d},s\bar{s})\bar{u}}^{-}\to  D^0 K^0 K^- ),
           \Gamma(Y_{(d\bar{d},s\bar{s})\bar{u}}^{-}\to  D^0 \pi^0 K^- )
           = \frac{1}{4}\Gamma(X_{d\bar{s}\bar{u}}^{-}\to  D^0 K^0 K^- ),
    \Gamma(Y_{(u\bar{u},s\bar{s})\bar{d}}^{0}\to  D^0 K^+ K^- )\\&&
    = \frac{1}{2}\Gamma(X_{u\bar{d}\bar{s}}^{+}\to  D^+_s K^+ K^- ),
    \Gamma(X_{u\bar{d}\bar{s}}^{+}\to  D^+ \pi^0 \pi^0 )
     = \frac{1}{2}\Gamma(X_{u\bar{d}\bar{s}}^{+}\to  D^+ \pi^+ \pi^- ),
\Gamma(Y_{(u\bar{u},d\bar{d})\bar{s}}^{0}\to  D^0 \pi^+ \pi^- )\\&&
=2\Gamma(Y_{(u\bar{u},d\bar{d})\bar{s}}^{0}\to  D^0 \pi^0 \pi^0 ),
      \Gamma(Y_{(u\bar{u},s\bar{s})\bar{d}}^{0}\to  D^0 \pi^0 \overline K^0 )
      =3\Gamma(Y_{(u\bar{u},s\bar{s})\bar{d}}^{0}\to  D^0 \overline K^0 \eta ),
       \Gamma(Y_{(u\bar{u},s\bar{s})\bar{d}}^{0}\to  D^+_s \pi^0 K^- )\\&&
       = 3\Gamma(Y_{(u\bar{u},s\bar{s})\bar{d}}^{0}\to  D^+_s K^- \eta ),
       \Gamma(Y_{(d\bar{d},s\bar{s})\bar{u}}^{-}\to  D^0 \pi^- \overline K^0 )
       = \frac{3}{4}\Gamma(X_{d\bar{s}\bar{u}}^{-}\to  D^0 \pi^- \eta ),
    \Gamma(X_{u\bar{d}\bar{s}}^{+}\to  D^+ \pi^0 \eta )\\&&
    = { }\Gamma(Y_{(u\bar{u},d\bar{d})\bar{s}}^{0}\to  D^+ \pi^- \eta ),
    \Gamma(Y_{(u\bar{u},d\bar{d})\bar{s}}^{0}\to  D^0 \pi^0 K^0 )
    = 3\Gamma(Y_{(u\bar{u},d\bar{d})\bar{s}}^{0}\to  D^0 K^0 \eta ).
\end{eqnarray*}
\begin{table}
\footnotesize
\caption{Open bottom tetraquark $X_{b6}$ decays into a charmed meson and two light meson.}\label{tab:b6_D_2M}\begin{tabular}{|cc|cc|}\hline\hline
channel & amplitude($/V_{cb}$) &channel &amplitude($/V_{cb}$)\\\hline
$X_{u\bar{d}\bar{s}}^{+}\to    D^0  \pi^+   K^0  $ & $ \frac{\left(a_9-a_{11}+a_{14}\right) V^*_{ud}}{\sqrt{2}}$&
$X_{u\bar{d}\bar{s}}^{+}\to    D^0  \pi^+   \eta  $ & $ -\frac{\left(a_9-a_{11}+a_{14}\right) V^*_{us}}{\sqrt{3}}$\\\hline
$X_{u\bar{d}\bar{s}}^{+}\to    D^0  \pi^0   K^+  $ & $ \frac{\left(a_9-a_{11}+a_{14}\right)V^*_{ud}}{2} $&
$X_{u\bar{d}\bar{s}}^{+}\to    D^0  K^+   \overline K^0  $ & $ -\frac{\left(a_9-a_{11}+a_{14}\right) V^*_{us}}{\sqrt{2}}$\\\hline
$X_{u\bar{d}\bar{s}}^{+}\to    D^0  K^+   \eta  $ & $ -\frac{\left(a_9-a_{11}+a_{14}\right) V^*_{ud}}{2 \sqrt{3}}$&
$X_{u\bar{d}\bar{s}}^{+}\to    D^+  \pi^+   \pi^-  $ & $ -\frac{\left(2 a_8+a_9+a_{12}\right) V^*_{us}}{\sqrt{2}}$\\\hline
$X_{u\bar{d}\bar{s}}^{+}\to    D^+  \pi^0   \pi^0  $ & $ -\frac{\left(2 a_8+a_9+a_{12}\right) V^*_{us}}{\sqrt{2}}$&
$X_{u\bar{d}\bar{s}}^{+}\to    D^+  \pi^0   K^0  $ & $ -\frac{ \left(a_9+a_{10}-a_{15}\right) V^*_{ud}}{2}$\\\hline
$X_{u\bar{d}\bar{s}}^{+}\to    D^+  \pi^0   \eta  $ & $ \frac{\left(a_9-a_{11}-a_{12}-a_{15}\right) V^*_{us}}{\sqrt{6}}$&
$X_{u\bar{d}\bar{s}}^{+}\to    D^+  \pi^-   K^+  $ & $ \frac{\left(a_9+a_{10}+a_{13}\right) V^*_{ud}}{\sqrt{2}}$\\\hline
$X_{u\bar{d}\bar{s}}^{+}\to    D^+  K^+   K^-  $ & $ \frac{\left(-2 a_8+a_{10}-a_{12}+a_{13}\right) V^*_{us}}{\sqrt{2}}$&
$X_{u\bar{d}\bar{s}}^{+}\to    D^+  K^0   \overline K^0  $ & $ \frac{\left(-2 a_8-a_9+a_{10}+a_{11}\right) V^*_{us}}{\sqrt{2}}$\\\hline
$X_{u\bar{d}\bar{s}}^{+}\to    D^+  K^0   \eta  $ & $ -\frac{\left(a_9+a_{10}-a_{15}\right) V^*_{ud}}{2 \sqrt{3}}$&
$X_{u\bar{d}\bar{s}}^{+}\to    D^+  \eta   \eta  $ & $ -\frac{\left(6 a_8+a_9-4 a_{10}-2 a_{11}+a_{12}+2 a_{15}\right) V^*_{us}}{3 \sqrt{2}}$\\\hline
$X_{u\bar{d}\bar{s}}^{+}\to    D^+_s  \pi^+   \pi^-  $ & $ \frac{\left(2 a_8-a_{10}+a_{12}-a_{13}\right) V^*_{ud}}{\sqrt{2}}$&
$X_{u\bar{d}\bar{s}}^{+}\to    D^+_s  \pi^+   K^-  $ & $ -\frac{\left(a_9+a_{10}+a_{13}\right) V^*_{us}}{\sqrt{2}}$\\\hline
$X_{u\bar{d}\bar{s}}^{+}\to    D^+_s  \pi^0   \pi^0  $ & $ \frac{\left(2 a_8-a_{10}+a_{12}+a_{15}\right) V^*_{ud}}{\sqrt{2}}$&
$X_{u\bar{d}\bar{s}}^{+}\to    D^+_s  \pi^0   \overline K^0  $ & $ \frac{ \left(a_9+a_{10}-a_{15}\right) V^*_{us}}{2}$\\\hline
$X_{u\bar{d}\bar{s}}^{+}\to    D^+_s  \pi^0   \eta  $ & $ \frac{\left(a_{10}+a_{11}+a_{12}\right) V^*_{ud}}{\sqrt{6}}$&
$X_{u\bar{d}\bar{s}}^{+}\to    D^+_s  K^+   K^-  $ & $ \frac{\left(2 a_8+a_9+a_{12}\right) V^*_{ud}}{\sqrt{2}}$\\\hline
$X_{u\bar{d}\bar{s}}^{+}\to    D^+_s  K^0   \overline K^0  $ & $ \frac{\left(2 a_8+a_9-a_{10}-a_{11}\right) V^*_{ud}}{\sqrt{2}}$&
$X_{u\bar{d}\bar{s}}^{+}\to    D^+_s  \overline K^0   \eta  $ & $ \frac{\left(a_9+a_{10}-a_{15}\right) V^*_{us}}{2 \sqrt{3}}$\\\hline
$X_{u\bar{d}\bar{s}}^{+}\to    D^+_s  \eta   \eta  $ & $ \frac{\left(6 a_8+4 a_9-a_{10}-2 a_{11}+a_{12}-a_{15}\right) V^*_{ud}}{3 \sqrt{2}}$&
$X_{d\bar{s}\bar{u}}^{-}\to    D^0  \pi^-   K^0  $ & $ -\frac{\left(a_{13}+a_{15}\right) V^*_{ud}}{\sqrt{2}}$\\\hline
$X_{d\bar{s}\bar{u}}^{-}\to    D^0  \pi^-   \eta  $ & $ \frac{\left(a_{12}+a_{14}+a_{15}\right) V^*_{us}}{\sqrt{3}}$&
$X_{d\bar{s}\bar{u}}^{-}\to    D^0  K^0   K^-  $ & $ \frac{\left(a_{12}-a_{13}+a_{14}\right) V^*_{us}}{\sqrt{2}}$\\\hline
$X_{d\bar{s}\bar{u}}^{-}\to    D^+_s  \pi^-   \pi^-  $ & $ \sqrt{2} \left(a_{13}+a_{15}\right) V^*_{ud}$&
$X_{d\bar{s}\bar{u}}^{-}\to    D^+_s  \pi^-   K^-  $ & $ \frac{\left(a_{13}+a_{15}\right) V^*_{us}}{\sqrt{2}}$\\\hline
$X_{s\bar{u}\bar{d}}^{-}\to    D^0  \pi^0   K^-  $ & $ -\frac{\left(a_{12}+a_{14}+a_{15}\right) V^*_{ud}}{2}$&
$X_{s\bar{u}\bar{d}}^{-}\to    D^0  \pi^-   \overline K^0  $ & $ -\frac{\left(a_{12}-a_{13}+a_{14}\right) V^*_{ud}}{\sqrt{2}}$\\\hline
$X_{s\bar{u}\bar{d}}^{-}\to    D^0  \overline K^0   K^-  $ & $ \frac{\left(a_{13}+a_{15}\right) V^*_{us}}{\sqrt{2}}$&
$X_{s\bar{u}\bar{d}}^{-}\to    D^0  K^-   \eta  $ & $ \frac{\left(a_{12}+a_{14}+a_{15}\right) V^*_{ud}}{2 \sqrt{3}}$\\\hline
$X_{s\bar{u}\bar{d}}^{-}\to    D^+  \pi^-   K^-  $ & $ -\frac{\left(a_{13}+a_{15}\right) V^*_{ud}}{\sqrt{2}}$&
$X_{s\bar{u}\bar{d}}^{-}\to    D^+  K^-   K^-  $ & $ -\sqrt{2} \left(a_{13}+a_{15}\right) V^*_{us}$\\\hline
$Y_{(u\bar{u},s\bar{s})\bar{d}}^{0}\to    D^0  \pi^+   \pi^-  $ & $ -\frac{ \left(2 a_8+a_9-a_{10}-a_{11}+a_{12}-a_{13}+a_{14}\right) V^*_{ud}}{2}$&
$Y_{(u\bar{u},s\bar{s})\bar{d}}^{0}\to    D^0  \pi^+   K^-  $ & $ \frac{\left(a_{10}+a_{11}+a_{13}-a_{14}\right) V^*_{us}}{2}$\\\hline
$Y_{(u\bar{u},s\bar{s})\bar{d}}^{0}\to    D^0  \pi^0   \pi^0  $ & $ -\frac{\left(2 a_8+a_9-a_{10}-a_{11}+a_{12}+a_{14}+a_{15}\right) V^*_{ud}}{2}$&
$Y_{(u\bar{u},s\bar{s})\bar{d}}^{0}\to    D^0  \pi^0   \overline K^0  $ & $ \frac{\left(-a_{10}-a_{11}+a_{14}+a_{15}\right) V^*_{us}}{2 \sqrt{2}}$\\\hline
$Y_{(u\bar{u},s\bar{s})\bar{d}}^{0}\to    D^0  \pi^0   \eta  $ & $ -\frac{\left(a_9+a_{10}+a_{12}+a_{14}\right) V^*_{ud}}{2 \sqrt{3}}$&
$Y_{(u\bar{u},s\bar{s})\bar{d}}^{0}\to    D^0  K^+   K^-  $ & $ -\frac{ \left(2 a_8+a_9+a_{12}\right) V^*_{ud}}{2}$\\\hline
$Y_{(u\bar{u},s\bar{s})\bar{d}}^{0}\to    D^0  K^0   \overline K^0  $ & $ \frac{ \left(-2 a_8+a_{10}+a_{14}\right) V^*_{ud}}{2}$&
$Y_{(u\bar{u},s\bar{s})\bar{d}}^{0}\to    D^0  \overline K^0   \eta  $ & $ \frac{\left(-a_{10}-a_{11}+a_{14}+a_{15}\right) V^*_{us}}{2 \sqrt{6}}$\\\hline
$Y_{(u\bar{u},s\bar{s})\bar{d}}^{0}\to    D^0  \eta   \eta  $ & $ -\frac{\left(6 a_8+a_9-a_{10}+a_{11}+a_{12}-3 a_{14}-a_{15}\right) V^*_{ud}}{6}$&
$Y_{(u\bar{u},s\bar{s})\bar{d}}^{0}\to    D^+  \pi^0   \pi^-  $ & $ -\frac{\left(a_{13}+a_{15}\right) V^*_{ud}}{2 \sqrt{2}}$\\\hline
$Y_{(u\bar{u},s\bar{s})\bar{d}}^{0}\to    D^+  \pi^0   K^-  $ & $ -\frac{\left(a_{10}+a_{11}+a_{12}+a_{13}+a_{15}\right) V^*_{us}}{2 \sqrt{2}}$&
$Y_{(u\bar{u},s\bar{s})\bar{d}}^{0}\to    D^+  \pi^-   \overline K^0  $ & $ -\frac{ \left(a_{10}+a_{11}+a_{12}\right) V^*_{us}}{2}$\\\hline
$Y_{(u\bar{u},s\bar{s})\bar{d}}^{0}\to    D^+  \pi^-   \eta  $ & $ -\frac{\left(2 a_9+2 a_{10}+3 a_{13}+a_{15}\right) V^*_{ud}}{2 \sqrt{6}}$&
$Y_{(u\bar{u},s\bar{s})\bar{d}}^{0}\to    D^+  K^0   K^-  $ & $ -\frac{ \left(a_9+a_{10}-a_{15}\right) V^*_{ud}}{2}$\\\hline
$Y_{(u\bar{u},s\bar{s})\bar{d}}^{0}\to    D^+  K^-   \eta  $ & $ \frac{\left(a_{10}+a_{11}+a_{12}-3 a_{13}-3 a_{15}\right) V^*_{us}}{2 \sqrt{6}}$&
$Y_{(u\bar{u},s\bar{s})\bar{d}}^{0}\to    D^+_s  \pi^0   K^-  $ & $ \frac{\left(-a_9+a_{11}+a_{12}+a_{15}\right) V^*_{ud}}{2 \sqrt{2}}$\\\hline
$Y_{(u\bar{u},s\bar{s})\bar{d}}^{0}\to    D^+_s  \pi^-   \overline K^0  $ & $ \frac{1}{2} \left(-a_9+a_{11}+a_{12}-a_{13}\right) V^*_{ud}$&
$Y_{(u\bar{u},s\bar{s})\bar{d}}^{0}\to    D^+_s  \overline K^0   K^-  $ & $ -\frac{ \left(a_{13}+a_{15}\right) V^*_{us}}{2}$\\\hline
$Y_{(u\bar{u},s\bar{s})\bar{d}}^{0}\to    D^+_s  K^-   \eta  $ & $ \frac{\left(a_9-a_{11}-a_{12}-a_{15}\right) V^*_{ud}}{2 \sqrt{6}}$&
$Y_{(u\bar{u},d\bar{d})\bar{s}}^{0}\to    D^0  \pi^+   \pi^-  $ & $ \frac{ \left(2 a_8+a_9+a_{12}\right) V^*_{us}}{2}$\\\hline
$Y_{(u\bar{u},d\bar{d})\bar{s}}^{0}\to    D^0  \pi^0   \pi^0  $ & $ \frac{ \left(2 a_8+a_9+a_{12}\right) V^*_{us}}{2}$&
$Y_{(u\bar{u},d\bar{d})\bar{s}}^{0}\to    D^0  \pi^0   K^0  $ & $ \frac{\left(a_{10}+a_{11}-a_{14}-a_{15}\right) V^*_{ud}}{2 \sqrt{2}}$\\\hline
$Y_{(u\bar{u},d\bar{d})\bar{s}}^{0}\to    D^0  \pi^0   \eta  $ & $ \frac{\left(a_9-a_{11}+a_{12}+2 a_{14}+a_{15}\right) V^*_{us}}{2 \sqrt{3}}$&
$Y_{(u\bar{u},d\bar{d})\bar{s}}^{0}\to    D^0  \pi^-   K^+  $ & $ -\frac{ \left(a_{10}+a_{11}+a_{13}-a_{14}\right) V^*_{ud}}{2}$\\\hline
$Y_{(u\bar{u},d\bar{d})\bar{s}}^{0}\to    D^0  K^+   K^-  $ & $ \frac{ \left(2 a_8+a_9-a_{10}-a_{11}+a_{12}-a_{13}+a_{14}\right) V^*_{us}}{2}$&
$Y_{(u\bar{u},d\bar{d})\bar{s}}^{0}\to    D^0  K^0   \overline K^0  $ & $ \frac{\left(2 a_8-a_{10}-a_{14}\right) V^*_{us}}{2} $\\\hline
$Y_{(u\bar{u},d\bar{d})\bar{s}}^{0}\to    D^0  K^0   \eta  $ & $ \frac{\left(a_{10}+a_{11}-a_{14}-a_{15}\right) V^*_{ud}}{2 \sqrt{6}}$&
$Y_{(u\bar{u},d\bar{d})\bar{s}}^{0}\to    D^0  \eta   \eta  $ & $ \frac{ \left(6 a_8+a_9-4 a_{10}-2 a_{11}+a_{12}+2 a_{15}\right) V^*_{us}}{6}$\\\hline
$Y_{(u\bar{u},d\bar{d})\bar{s}}^{0}\to    D^+  \pi^-   K^0  $ & $ \frac{\left(a_{13}+a_{15}\right) V^*_{ud}}{2} $&
$Y_{(u\bar{u},d\bar{d})\bar{s}}^{0}\to    D^+  \pi^-   \eta  $ & $ \frac{\left(a_9-a_{11}-a_{12}-a_{15}\right) V^*_{us}}{\sqrt{6}}$\\\hline
$Y_{(u\bar{u},d\bar{d})\bar{s}}^{0}\to    D^+  K^0   K^-  $ & $ \frac{ \left(a_9-a_{11}-a_{12}+a_{13}\right) V^*_{us}}{2}$&
$Y_{(u\bar{u},d\bar{d})\bar{s}}^{0}\to    D^+_s  \pi^0   \pi^-  $ & $ \frac{\left(a_{13}+a_{15}\right) V^*_{ud}}{\sqrt{2}}$\\\hline
$Y_{(u\bar{u},d\bar{d})\bar{s}}^{0}\to    D^+_s  \pi^0   K^-  $ & $ \frac{\left(a_9+a_{10}+2 a_{13}+a_{15}\right) V^*_{us}}{2 \sqrt{2}}$&
$Y_{(u\bar{u},d\bar{d})\bar{s}}^{0}\to    D^+_s  \pi^-   \overline K^0  $ & $ \frac{1}{2} \left(a_9+a_{10}-a_{15}\right) V^*_{us}$\\\hline
$Y_{(u\bar{u},d\bar{d})\bar{s}}^{0}\to    D^+_s  \pi^-   \eta  $ & $ \frac{\left(a_{10}+a_{11}+a_{12}\right) V^*_{ud}}{\sqrt{6}}$&
$Y_{(u\bar{u},d\bar{d})\bar{s}}^{0}\to    D^+_s  K^0   K^-  $ & $ \frac{ \left(a_{10}+a_{11}+a_{12}\right) V^*_{ud}}{2}$\\\hline
$Y_{(u\bar{u},d\bar{d})\bar{s}}^{0}\to    D^+_s  K^-   \eta  $ & $ -\frac{\left(a_9+a_{10}-a_{15}\right) V^*_{us}}{2 \sqrt{6}}$&
$Y_{(d\bar{d},s\bar{s})\bar{u}}^{-}\to    D^0  \pi^0   \pi^-  $ & $ -\frac{\left(a_{13}+a_{15}\right) V^*_{ud}}{2 \sqrt{2}}$\\\hline
$Y_{(d\bar{d},s\bar{s})\bar{u}}^{-}\to    D^0  \pi^0   K^-  $ & $ \frac{\left(a_{12}-a_{13}+a_{14}\right) V^*_{us}}{2 \sqrt{2}}$&
$Y_{(d\bar{d},s\bar{s})\bar{u}}^{-}\to    D^0  \pi^-   \overline K^0  $ & $ \frac{1}{2} \left(a_{12}+a_{14}+a_{15}\right) V^*_{us}$\\\hline
$Y_{(d\bar{d},s\bar{s})\bar{u}}^{-}\to    D^0  \pi^-   \eta  $ & $ \frac{\left(-2 a_{12}+3 a_{13}-2 a_{14}+a_{15}\right) V^*_{ud}}{2 \sqrt{6}}$&
$Y_{(d\bar{d},s\bar{s})\bar{u}}^{-}\to    D^0  K^0   K^-  $ & $ -\frac{ \left(a_{12}+a_{14}+a_{15}\right) V^*_{ud}}{2}$\\\hline
$Y_{(d\bar{d},s\bar{s})\bar{u}}^{-}\to    D^0  K^-   \eta  $ & $ -\frac{\left(a_{12}-3 a_{13}+a_{14}-2 a_{15}\right) V^*_{us}}{2 \sqrt{6}}$&
$Y_{(d\bar{d},s\bar{s})\bar{u}}^{-}\to    D^+  \pi^-   \pi^-  $ & $ -\left(a_{13}+a_{15}\right) V^*_{ud}$\\\hline
$Y_{(d\bar{d},s\bar{s})\bar{u}}^{-}\to    D^+  \pi^-   K^-  $ & $ -\frac{ \left(a_{13}+a_{15}\right) V^*_{us}}{2}$&
$Y_{(d\bar{d},s\bar{s})\bar{u}}^{-}\to    D^+_s  \pi^-   K^-  $ & $ \frac{ \left(a_{13}+a_{15}\right) V^*_{ud}}{2}$\\\hline
$Y_{(d\bar{d},s\bar{s})\bar{u}}^{-}\to    D^+_s  K^-   K^-  $ & $ \left(a_{13}+a_{15}\right) V^*_{us}$& &\\
\hline
\end{tabular}
\end{table}

\subsection{$b\to u \bar c d/s$ transition}

\subsubsection{Two-body decays into mesons}
The  $(\bar ub)(\bar qc)$  transition can lead to an anti-symmetric ${\bf  \bar 3}$ and a symmetric ${\bf  6}$ tensor representations.
 The anti-symmetric tensor $H_{\bar 3}''$ and the symmetric tensor
$H_{ 6}$ have the nonzero components
$  (H_{\bar 3}'')^{13} =- (H_{\bar 3}'')^{31} =V_{cs}^*$, $ (H_{\bar 6})^{13}=(H_{\bar 6})^{31} =V_{cs}^*,
$
for the $b\to u\bar cs$ or $b\bar s\to u\bar c$ transition.
When interchanging  $2\leftrightarrow 3$ in the
subscripts and replacing $V_{cs}$ by $V_{cd}$ at the same time, one gets the nonzero components  for the $b\to
u\bar cd$ transition.
The effective Hamiltonian for $X_{b6}$ produce two mesons by $b\to u \bar c d/s$ transition is
  \begin{eqnarray}
  \mathcal{H}_{eff}&=&a_7 (X_{b6})_{[jk]}^i (H_{\bar3}^{\prime\prime})^{[jk]}  M^l_i D_l +a_8 (X_{b6})_{[jk]}^i (H_{\bar3}^{\prime\prime})^{[jl]}  M^k_i D_l+a_9 (X_{b6})_{[jk]}^i (H_{\bar3}^{\prime\prime})^{[jl]}  M^k_l D_i\nonumber\\
  &&+a_{10} (X_{b6})_{[jk]}^i (H_6)^{\{jl\}}  M^k_i D_l+a_{11} (X_{b6})_{[jk]}^i (H_6)^{\{jl\}}  M^k_l D_i.
  \end{eqnarray}
Decay amplitudes for different channels are shown in Tab.~\ref{tab:b6_antiD_M}, which leads to the following relations
\begin{eqnarray*}
    \Gamma(X_{d\bar{s}\bar{u}}^{-}\to \overline D^0\pi^- )= 2\Gamma(Y_{(d\bar{d},s\bar{s})\bar{u}}^{-}\to \overline D^0K^- ),& \Gamma(Y_{(u\bar{u},s\bar{s})\bar{d}}^{0}\to D^-\pi^+ )= \frac{1}{2}\Gamma(X_{s\bar{u}\bar{d}}^{-}\to D^-\overline K^0 ),\\ \Gamma(Y_{(u\bar{u},s\bar{s})\bar{d}}^{0}\to  D^-_s\pi^+ )= 2\Gamma(Y_{(d\bar{d},s\bar{s})\bar{u}}^{-}\to  D^-_s\pi^0 ),& \Gamma(Y_{(u\bar{u},d\bar{d})\bar{s}}^{0}\to D^-\pi^+ )= { }\Gamma(X_{d\bar{s}\bar{u}}^{-}\to D^-\pi^0 ),\\ \Gamma(Y_{(u\bar{u},d\bar{d})\bar{s}}^{0}\to  D^-_sK^+ )= \frac{1}{2}\Gamma(X_{d\bar{s}\bar{u}}^{-}\to  D^-_sK^0 ),& \Gamma(Y_{(d\bar{d},s\bar{s})\bar{u}}^{-}\to \overline D^0\pi^- )= \frac{1}{2}\Gamma(X_{s\bar{u}\bar{d}}^{-}\to \overline D^0K^- ),\\
    \Gamma(Y_{(u\bar{u},d\bar{d})\bar{s}}^{0}\to \overline D^0\eta )= \frac{1}{2}\Gamma(X_{d\bar{s}\bar{u}}^{-}\to D^-\eta ).
\end{eqnarray*}
\begin{table}
\caption{Open bottom tetraquark $X_{b6}$ decays into a anti-charmed meson and a light meson.}\label{tab:b6_antiD_M}\begin{tabular}{|cc|cc|}\hline\hline
channel & amplitude($/V_{ub}$) &channel&amplitude($/V_{ub}$) \\\hline
$X_{u\bar{d}\bar{s}}^{+}\to   \overline D^0  \pi^+  $ & $ \frac{\left(a_8+a_9-a_{10}-a_{11}\right) V^*_{cs}}{\sqrt{2}}$&
$X_{u\bar{d}\bar{s}}^{+}\to   \overline D^0  K^+  $ & $ \frac{\left(-a_8-a_9+a_{10}+a_{11}\right) V^*_{cd}}{\sqrt{2}}$\\\hline
$X_{d\bar{s}\bar{u}}^{-}\to   \overline D^0  \pi^-  $ & $ \frac{\left(-2 a_7-a_8+a_{10}\right) V^*_{cs}}{\sqrt{2}}$&
$X_{d\bar{s}\bar{u}}^{-}\to   D^-  \pi^0  $ & $ \frac{1}{2} \left(2 a_7-a_9+a_{11}\right) V^*_{cs}$\\\hline
$X_{d\bar{s}\bar{u}}^{-}\to   D^-  K^0  $ & $ -\frac{\left(a_8+a_9+a_{10}+a_{11}\right) V^*_{cd}}{\sqrt{2}}$&
$X_{d\bar{s}\bar{u}}^{-}\to   D^-  \eta  $ & $ \frac{\left(-2 a_7+a_9+3 a_{11}\right) V^*_{cs}}{2 \sqrt{3}}$\\\hline
$X_{d\bar{s}\bar{u}}^{-}\to    D^-_s  K^0  $ & $ -\frac{\left(2 a_7+a_8+a_{10}\right) V^*_{cs}}{\sqrt{2}}$&
$X_{s\bar{u}\bar{d}}^{-}\to   \overline D^0  K^-  $ & $ \frac{\left(2 a_7+a_8-a_{10}\right) V^*_{cd}}{\sqrt{2}}$\\\hline
$X_{s\bar{u}\bar{d}}^{-}\to   D^-  \overline K^0  $ & $ \frac{\left(2 a_7+a_8+a_{10}\right) V^*_{cd}}{\sqrt{2}}$&
$X_{s\bar{u}\bar{d}}^{-}\to    D^-_s  \pi^0  $ & $ -a_{11} V^*_{cd}$\\\hline
$X_{s\bar{u}\bar{d}}^{-}\to    D^-_s  \overline K^0  $ & $ \frac{\left(a_8+a_9+a_{10}+a_{11}\right) V^*_{cs}}{\sqrt{2}}$&
$X_{s\bar{u}\bar{d}}^{-}\to    D^-_s  \eta  $ & $ \frac{\left(a_9-2 a_7\right) V^*_{cd}}{\sqrt{3}}$\\\hline
$Y_{(u\bar{u},s\bar{s})\bar{d}}^{0}\to   \overline D^0  \pi^0  $ & $ \frac{\left(2 a_7+a_8-a_{10}-2 a_{11}\right) V^*_{cd}}{2 \sqrt{2}}$&
$Y_{(u\bar{u},s\bar{s})\bar{d}}^{0}\to   \overline D^0  \overline K^0  $ & $ \frac{1}{2} \left(a_8+a_9-a_{10}+a_{11}\right) V^*_{cs}$\\\hline
$Y_{(u\bar{u},s\bar{s})\bar{d}}^{0}\to   \overline D^0  \eta  $ & $ \frac{\left(2 a_7+3 a_8+2 a_9-3 a_{10}\right) V^*_{cd}}{2 \sqrt{6}}$&
$Y_{(u\bar{u},s\bar{s})\bar{d}}^{0}\to   D^-  \pi^+  $ & $ \frac{1}{2} \left(2 a_7+a_8+a_{10}\right) V^*_{cd}$\\\hline
$Y_{(u\bar{u},s\bar{s})\bar{d}}^{0}\to    D^-_s  \pi^+  $ & $ \frac{1}{2} \left(a_8+a_9+a_{10}-a_{11}\right) V^*_{cs}$&
$Y_{(u\bar{u},s\bar{s})\bar{d}}^{0}\to    D^-_s  K^+  $ & $ \frac{1}{2} \left(2 a_7-a_9+a_{11}\right) V^*_{cd}$\\\hline
$Y_{(u\bar{u},d\bar{d})\bar{s}}^{0}\to   \overline D^0  \pi^0  $ & $ -\frac{\left(2 a_7+2 a_8+a_9-2 a_{10}-a_{11}\right) V^*_{cs}}{2 \sqrt{2}}$&
$Y_{(u\bar{u},d\bar{d})\bar{s}}^{0}\to   \overline D^0  K^0  $ & $ -\frac{1}{2} \left(a_8+a_9-a_{10}+a_{11}\right) V^*_{cd}$\\\hline
$Y_{(u\bar{u},d\bar{d})\bar{s}}^{0}\to   \overline D^0  \eta  $ & $ \frac{\left(-2 a_7+a_9+3 a_{11}\right) V^*_{cs}}{2 \sqrt{6}}$&
$Y_{(u\bar{u},d\bar{d})\bar{s}}^{0}\to   D^-  \pi^+  $ & $ -\frac{1}{2} \left(2 a_7-a_9+a_{11}\right) V^*_{cs}$\\\hline
$Y_{(u\bar{u},d\bar{d})\bar{s}}^{0}\to   D^-  K^+  $ & $ -\frac{1}{2} \left(a_8+a_9+a_{10}-a_{11}\right) V^*_{cd}$&
$Y_{(u\bar{u},d\bar{d})\bar{s}}^{0}\to    D^-_s  K^+  $ & $ -\frac{1}{2} \left(2 a_7+a_8+a_{10}\right) V^*_{cs}$\\\hline
$Y_{(d\bar{d},s\bar{s})\bar{u}}^{-}\to   \overline D^0  \pi^-  $ & $ \frac{1}{2} \left(2 a_7+a_8-a_{10}\right) V^*_{cd}$&
$Y_{(d\bar{d},s\bar{s})\bar{u}}^{-}\to   \overline D^0  K^-  $ & $ -\frac{1}{2} \left(2 a_7+a_8-a_{10}\right) V^*_{cs}$\\\hline
$Y_{(d\bar{d},s\bar{s})\bar{u}}^{-}\to   D^-  \pi^0  $ & $ -\frac{\left(2 a_7+a_8+a_{10}+2 a_{11}\right) V^*_{cd}}{2 \sqrt{2}}$&
$Y_{(d\bar{d},s\bar{s})\bar{u}}^{-}\to   D^-  \overline K^0  $ & $ \frac{1}{2} \left(-2 a_7+a_9+a_{11}\right) V^*_{cs}$\\\hline
$Y_{(d\bar{d},s\bar{s})\bar{u}}^{-}\to   D^-  \eta  $ & $ \frac{\left(2 a_7+3 a_8+2 a_9+3 a_{10}\right) V^*_{cd}}{2 \sqrt{6}}$&
$Y_{(d\bar{d},s\bar{s})\bar{u}}^{-}\to    D^-_s  \pi^0  $ & $ -\frac{\left(a_8+a_9+a_{10}-a_{11}\right) V^*_{cs}}{2 \sqrt{2}}$\\\hline
$Y_{(d\bar{d},s\bar{s})\bar{u}}^{-}\to    D^-_s  K^0  $ & $ \frac{1}{2} \left(2 a_7-a_9-a_{11}\right) V^*_{cd}$&
$Y_{(d\bar{d},s\bar{s})\bar{u}}^{-}\to    D^-_s  \eta  $ & $ \frac{\left(4 a_7+3 a_8+a_9+3 a_{10}+3 a_{11}\right) V^*_{cs}}{2 \sqrt{6}}$\\\hline
\hline
\end{tabular}
\end{table}

\subsubsection{Two-body decays into a baryon and an anti-baryon}
The  $b\to u \bar c d/s$ transition can lead to the decays into an anti-charmed anti-baryon plus a light baryon. The effective Hamiltonian in hadron level is
then written as
\begin{eqnarray}
  \mathcal{H}_{eff}&=&b_1 (T_{b6})_{[ij],[kl]} (H_{\bar3}^{\prime\prime})^{[ij]} (T_8)^k_m (T_{\bar{c}3})^{[lm]}
  +b_2 (T_{b6})_{[ij],[kl]} (H_{\bar3}^{\prime\prime})^{[im]} (T_8)^j_m (T_{\bar{c}3})^{[kl]}\nonumber\\
  &&+b_3 (T_{b6})_{[ij],[kl]} (H_6^{\prime\prime})^{\{ik\}} (T_8)^j_m (T_{\bar{c}3})^{[lm]}
  +b_4 (T_{b6})_{[ij],[kl]} (H_6^{\prime\prime})^{\{im\}} (T_8)^j_m (T_{\bar{c}3})^{[kl]}\nonumber\\
  &&+c_1 (T_{b6})_{[ij],[kl]} (H_{\bar3}^{\prime\prime})^{[ij]} (T_8)^k_m (T_{\bar{c}\bar6})^{\{lm\}}
  +c_2 (T_{b6})_{[ij],[kl]} (H_{\bar3}^{\prime\prime})^{[im]} (T_8)^k_m (T_{\bar{c}\bar6})^{\{jl\}}\nonumber\\
  &&+c_3 (T_{b6})_{[ij],[kl]} (H_6^{\prime\prime})^{\{ik\}} (T_8)^j_m (T_{\bar{c}\bar6})^{\{lm\}}
  +c_4 (T_{b6})_{[ij],[kl]} (H_6^{\prime\prime})^{\{im\}} (T_8)^k_m (T_{\bar{c}\bar6})^{\{jl\}}\nonumber\\
  &&+d_1 (T_{b6})_{[jk]}^i (H_{\bar3}^{\prime\prime})^{[jl]} (T_{10})_{\{ilm\}} (T_{\bar{c}3})^{[km]}+d_2 (T_{b6})_{[jk]}^i (H_6^{\prime\prime})^{\{jl\}} (T_{10})_{\{ilm\}} (T_{\bar{c}3})^{[km]}\nonumber\\
  &&+d_3 (T_{b6})_{[jk]}^i (H_6^{\prime\prime})^{\{lm\}} (T_{10})_{\{ilm\}} (T_{\bar{c}3})^{[jk]}+f_1 (T_{b6})_{[jk]}^i (H_{\bar3}^{\prime\prime})^{[jk]} (T_{10})_{\{ilm\}} (T_{\bar{c}\bar6})^{\{lm\}}\nonumber\\
  &&+f_2 (T_{b6})_{[jk]}^i (H_{\bar3}^{\prime\prime})^{[jl]} (T_{10})_{\{ilm\}} (T_{\bar{c}\bar6})^{\{km\}}+f_3 (T_{b6})_{[jk]}^i (H_6^{\prime\prime})^{\{jl\}} (T_{10})_{\{ilm\}} (T_{\bar{c}\bar6})^{\{km\}}.
\end{eqnarray}
The related decay amplitudes for different channels are given in Tab.~\ref{tab:b6_T8_Tbarc3} and Tab.~\ref{tab:b6_T10_Tbarc3}. Therein class \uppercase\expandafter{\romannumeral1} corresponds with the processes with octet baryon plus triplet anti-baryon, class \uppercase\expandafter{\romannumeral2} corresponds with  the processes with octet baryon plus anti-sextet anti-baryon, class \uppercase\expandafter{\romannumeral3} corresponds with  the processes with decuplet baryon plus triplet anti-baryon, class \uppercase\expandafter{\romannumeral4} corresponds with the processes with decuplet  baryon plus anti-sextet anti-baryon. The relation of decay widths for class \uppercase\expandafter{\romannumeral1} are given as:
\begin{eqnarray*}
    \Gamma(X_{d\bar{s}\bar{u}}^{-}\to \Lambda^0\overline \Xi_{\bar{c}}^-)= 2\Gamma(Y_{(u\bar{u},d\bar{d})\bar{s}}^{0}\to \Lambda^0\overline \Xi_{\bar{c}}^0),
    \Gamma(X_{d\bar{s}\bar{u}}^{-}\to \Sigma^-\overline \Xi_{\bar{c}}^0)= 2\Gamma(Y_{(d\bar{d},s\bar{s})\bar{u}}^{-}\to \Xi^-\overline \Xi_{\bar{c}}^0),\\
    \Gamma(Y_{(u\bar{u},s\bar{s})\bar{d}}^{0}\to \Sigma^+\overline \Lambda_{\bar{c}}^-)= 2\Gamma(Y_{(d\bar{d},s\bar{s})\bar{u}}^{-}\to \Sigma^0\overline \Lambda_{\bar{c}}^-),
    \Gamma(Y_{(u\bar{u},s\bar{s})\bar{d}}^{0}\to \Sigma^+\overline \Xi_{\bar{c}}^-)= \frac{1}{2}\Gamma(X_{s\bar{u}\bar{d}}^{-}\to \Xi^0\overline \Xi_{\bar{c}}^-),\\
    \Gamma(Y_{(u\bar{u},d\bar{d})\bar{s}}^{0}\to \Sigma^+\overline \Xi_{\bar{c}}^-)= { }\Gamma(X_{d\bar{s}\bar{u}}^{-}\to \Sigma^0\overline \Xi_{\bar{c}}^-),
    \Gamma(Y_{(u\bar{u},d\bar{d})\bar{s}}^{0}\to {p}\overline \Lambda_{\bar{c}}^-)= \frac{1}{2}\Gamma(X_{d\bar{s}\bar{u}}^{-}\to {n}\overline \Lambda_{\bar{c}}^-),\\
    \Gamma(Y_{(d\bar{d},s\bar{s})\bar{u}}^{-}\to \Sigma^-\overline \Xi_{\bar{c}}^0)= \frac{1}{2}\Gamma(X_{s\bar{u}\bar{d}}^{-}\to \Xi^-\overline \Xi_{\bar{c}}^0).
\end{eqnarray*}
The relations of decay widths for class \uppercase\expandafter{\romannumeral2} are given as
\begin{eqnarray*}
    \Gamma(X_{u\bar{d}\bar{s}}^{+}\to {p}\overline \Sigma_{\bar{c}}^{0})= 2\Gamma(X_{u\bar{d}\bar{s}}^{+}\to \Sigma^+\overline \Xi_{\bar{c}}^{\prime0}),
    \Gamma(X_{u\bar{d}\bar{s}}^{+}\to {p}\overline \Xi_{\bar{c}}^{\prime0})= \frac{1}{2}\Gamma(X_{u\bar{d}\bar{s}}^{+}\to \Sigma^+\overline \Omega_{\bar{c}}^{0}),\\
    \Gamma(X_{d\bar{s}\bar{u}}^{-}\to \Lambda^0\overline \Xi_{\bar{c}}^{\prime-})= 2\Gamma(Y_{(u\bar{u},d\bar{d})\bar{s}}^{0}\to \Lambda^0\overline \Xi_{\bar{c}}^{\prime0}),
    \Gamma(X_{d\bar{s}\bar{u}}^{-}\to {n}\overline \Sigma_{\bar{c}}^-)= { }\Gamma(Y_{(u\bar{u},d\bar{d})\bar{s}}^{0}\to {n}\overline \Sigma_{\bar{c}}^{0}),\\
    \Gamma(X_{s\bar{u}\bar{d}}^{-}\to \Xi^-\overline \Xi_{\bar{c}}^{\prime0})= { }\Gamma(Y_{(d\bar{d},s\bar{s})\bar{u}}^{-}\to \Xi^-\overline \Omega_{\bar{c}}^{0}),
     \Gamma(X_{s\bar{u}\bar{d}}^{-}\to \Xi^0\overline \Sigma_{\bar{c}}^-)= \frac{1}{2}\Gamma(X_{s\bar{u}\bar{d}}^{-}\to \Xi^-\overline \Sigma_{\bar{c}}^{0}),\\
      \Gamma(X_{s\bar{u}\bar{d}}^{-}\to \Xi^0\overline \Xi_{\bar{c}}^{\prime-})= { }\Gamma(Y_{(u\bar{u},s\bar{s})\bar{d}}^{0}\to \Xi^0\overline \Omega_{\bar{c}}^{0}),
      \Gamma(Y_{(u\bar{u},s\bar{s})\bar{d}}^{0}\to \Sigma^+\overline \Sigma_{\bar{c}}^-)= { }\Gamma(Y_{(u\bar{u},s\bar{s})\bar{d}}^{0}\to \Sigma^0\overline \Sigma_{\bar{c}}^{0}),\\
      \Gamma(Y_{(d\bar{d},s\bar{s})\bar{u}}^{-}\to \Lambda^0\overline \Sigma_{\bar{c}}^-)= \frac{1}{2}\Gamma(Y_{(u\bar{u},s\bar{s})\bar{d}}^{0}\to \Lambda^0\overline \Sigma_{\bar{c}}^{0}), \Gamma(Y_{(d\bar{d},s\bar{s})\bar{u}}^{-}\to \Sigma^-\overline \Sigma_{\bar{c}}^{0})= { }\Gamma(X_{d\bar{s}\bar{u}}^{-}\to \Sigma^-\overline \Xi_{\bar{c}}^{\prime0}),\\
    \Gamma(X_{d\bar{s}\bar{u}}^{-}\to \Xi^-\overline \Omega_{\bar{c}}^{0})= 2\Gamma(Y_{(u\bar{u},d\bar{d})\bar{s}}^{0}\to \Xi^0\overline \Omega_{\bar{c}}^{0}),
    \Gamma(X_{s\bar{u}\bar{d}}^{-}\to \Sigma^+\overline \Sigma_{\bar{c}}^{--})= 2\Gamma(Y_{(d\bar{d},s\bar{s})\bar{u}}^{-}\to {p}\overline \Sigma_{\bar{c}}^{--}),\\
    \Gamma(X_{s\bar{u}\bar{d}}^{-}\to \Sigma^-\overline \Sigma_{\bar{c}}^{0})= 2\Gamma(Y_{(u\bar{u},s\bar{s})\bar{d}}^{0}\to {n}\overline \Sigma_{\bar{c}}^{0}),
     \Gamma(Y_{(u\bar{u},d\bar{d})\bar{s}}^{0}\to \Sigma^+\overline \Xi_{\bar{c}}^{\prime-})= { }\Gamma(X_{d\bar{s}\bar{u}}^{-}\to \Sigma^0\overline \Xi_{\bar{c}}^{\prime-}),\\
      \Gamma(Y_{(d\bar{d},s\bar{s})\bar{u}}^{-}\to \Sigma^+\overline \Sigma_{\bar{c}}^{--})= \frac{1}{2}\Gamma(X_{d\bar{s}\bar{u}}^{-}\to {p}\overline \Sigma_{\bar{c}}^{--}).
\end{eqnarray*}
The relations of decay widths for class \uppercase\expandafter{\romannumeral3} are given as
\begin{eqnarray*}
     \Gamma(Y_{(u\bar{u},s\bar{s})\bar{d}}^{0}\to \Sigma^{\prime+}\overline \Lambda_{\bar{c}}^-)= 2\Gamma(Y_{(d\bar{d},s\bar{s})\bar{u}}^{-}\to \Sigma^{\prime0}\overline \Lambda_{\bar{c}}^-),
      \Gamma(Y_{(u\bar{u},d\bar{d})\bar{s}}^{0}\to \Delta^{+}\overline \Xi_{\bar{c}}^-)=
      2\Gamma(Y_{(d\bar{d},s\bar{s})\bar{u}}^{-}\to \Sigma^{\prime0}\overline \Xi_{\bar{c}}^-),\\
       \Gamma(Y_{(u\bar{u},d\bar{d})\bar{s}}^{0}\to \Delta^{0}\overline \Xi_{\bar{c}}^0)= 2\Gamma(Y_{(u\bar{u},s\bar{s})\bar{d}}^{0}\to \Sigma^{\prime0}\overline \Xi_{\bar{c}}^0),
      \Gamma(Y_{(u\bar{u},d\bar{d})\bar{s}}^{0}\to \Sigma^{\prime0}\overline \Xi_{\bar{c}}^0)= \frac{1}{2}\Gamma(Y_{(u\bar{u},s\bar{s})\bar{d}}^{0}\to \Xi^{\prime0}\overline \Xi_{\bar{c}}^0),\\
    \Gamma(X_{u\bar{d}\bar{s}}^{+}\to \Delta^{++}\overline \Lambda_{\bar{c}}^-)= 3\Gamma(X_{d\bar{s}\bar{u}}^{-}\to \Delta^{0}\overline \Lambda_{\bar{c}}^-)=\frac{3}{2}\Gamma(Y_{(u\bar{u},d\bar{d})\bar{s}}^{0}\to \Delta^{+}\overline \Lambda_{\bar{c}}^-),\\
     \Gamma(X_{u\bar{d}\bar{s}}^{+}\to \Delta^{++}\overline \Xi_{\bar{c}}^-)= 3\Gamma(X_{s\bar{u}\bar{d}}^{-}\to \Xi^{\prime0}\overline \Xi_{\bar{c}}^-)=
 \frac{3}{2}\Gamma(Y_{(u\bar{u},s\bar{s})\bar{d}}^{0}\to \Sigma^{\prime+}\overline \Xi_{\bar{c}}^-),\\
     \Gamma(X_{d\bar{s}\bar{u}}^{-}\to \Delta^{-}\overline \Xi_{\bar{c}}^0)= 3\Gamma(X_{s\bar{u}\bar{d}}^{-}\to \Xi^{\prime-}\overline \Xi_{\bar{c}}^0)=\frac{3}{2}\Gamma(Y_{(d\bar{d},s\bar{s})\bar{u}}^{-}\to \Sigma^{\prime-}\overline \Xi_{\bar{c}}^0),\\
     \Gamma(X_{d\bar{s}\bar{u}}^{-}\to \Sigma^{\prime0}\overline \Xi_{\bar{c}}^-)= \Gamma(Y_{(d\bar{d},s\bar{s})\bar{u}}^{-}\to \Xi^{\prime0}\overline \Xi_{\bar{c}}^-)=\Gamma(Y_{(u\bar{u},d\bar{d})\bar{s}}^{0}\to \Sigma^{\prime+}\overline \Xi_{\bar{c}}^-),\\
     \Gamma(X_{d\bar{s}\bar{u}}^{-}\to \Sigma^{\prime-}\overline \Xi_{\bar{c}}^0)= \frac{1}{3}\Gamma(X_{s\bar{u}\bar{d}}^{-}\to \Omega^-\overline \Xi_{\bar{c}}^0)=\frac{1}{2}\Gamma(Y_{(d\bar{d},s\bar{s})\bar{u}}^{-}\to \Xi^{\prime-}\overline \Xi_{\bar{c}}^0),\\
     \Gamma(Y_{(u\bar{u},s\bar{s})\bar{d}}^{0}\to \Delta^{+}\overline \Lambda_{\bar{c}}^-)=\Gamma(X_{s\bar{u}\bar{d}}^{-}\to \Sigma^{\prime0}\overline \Lambda_{\bar{c}}^-)=\Gamma(Y_{(d\bar{d},s\bar{s})\bar{u}}^{-}\to \Delta^{0}\overline \Lambda_{\bar{c}}^-).
\end{eqnarray*}
The relations of decay widths for class \uppercase\expandafter{\romannumeral4} are given as:
\begin{eqnarray*}
    \Gamma(X_{u\bar{d}\bar{s}}^{+}\to \Delta^{++}\overline \Sigma_{\bar{c}}^-)= \frac{3}{2}\Gamma(X_{u\bar{d}\bar{s}}^{+}\to \Delta^{+}\overline \Sigma_{\bar{c}}^{0})=3\Gamma(X_{u\bar{d}\bar{s}}^{+}\to \Sigma^{\prime+}\overline \Xi_{\bar{c}}^{\prime0}),\\
    \Gamma(X_{u\bar{d}\bar{s}}^{+}\to \Delta^{++}\overline \Xi_{\bar{c}}^{\prime-})= 3\Gamma(X_{u\bar{d}\bar{s}}^{+}\to \Delta^{+}\overline \Xi_{\bar{c}}^{\prime0})=\frac{3}{2}\Gamma(X_{u\bar{d}\bar{s}}^{+}\to \Sigma^{\prime+}\overline \Omega_{\bar{c}}^{0}),\\
    \Gamma(X_{d\bar{s}\bar{u}}^{-}\to \Delta^{0}\overline \Xi_{\bar{c}}^{\prime-})= \frac{1}{3}\Gamma(X_{d\bar{s}\bar{u}}^{-}\to \Delta^{-}\overline \Xi_{\bar{c}}^{\prime0})=\frac{1}{2}\Gamma(X_{d\bar{s}\bar{u}}^{-}\to \Sigma^{\prime-}\overline \Omega_{\bar{c}}^{0}),\\
      \Gamma(X_{d\bar{s}\bar{u}}^{-}\to \Delta^{-}\overline \Sigma_{\bar{c}}^{0})=3\Gamma(Y_{(d\bar{d},s\bar{s})\bar{u}}^{-}\to \Xi^{\prime-}\overline \Xi_{\bar{c}}^{\prime0})=3\Gamma(Y_{(u\bar{u},d\bar{d})\bar{s}}^{0}\to \Delta^{+}\overline \Sigma_{\bar{c}}^-),\\
        \Gamma(X_{s\bar{u}\bar{d}}^{-}\to \Xi^{\prime0}\overline \Sigma_{\bar{c}}^-)= \frac{1}{2}\Gamma(X_{s\bar{u}\bar{d}}^{-}\to \Xi^{\prime-}\overline \Sigma_{\bar{c}}^{0})= \frac{1}{3}\Gamma(X_{s\bar{u}\bar{d}}^{-}\to \Omega^-\overline \Xi_{\bar{c}}^{\prime0}),\\
        \Gamma(Y_{(u\bar{u},s\bar{s})\bar{d}}^{0}\to \Sigma^{\prime+}\overline \Sigma_{\bar{c}}^-)= { }\Gamma(Y_{(u\bar{u},s\bar{s})\bar{d}}^{0}\to \Sigma^{\prime0}\overline \Sigma_{\bar{c}}^{0})= \Gamma(Y_{(u\bar{u},s\bar{s})\bar{d}}^{0}\to \Xi^{\prime0}\overline \Xi_{\bar{c}}^{\prime0}),\\
         \Gamma(Y_{(u\bar{u},s\bar{s})\bar{d}}^{0}\to \Sigma^{\prime+}\overline \Xi_{\bar{c}}^{\prime-})= \frac{1}{3}\Gamma(X_{s\bar{u}\bar{d}}^{-}\to \Omega^-\overline \Omega_{\bar{c}}^{0})=
\Gamma(Y_{(d\bar{d},s\bar{s})\bar{u}}^{-}\to \Sigma^{\prime-}\overline \Xi_{\bar{c}}^{\prime0}),\\
            \Gamma(Y_{(u\bar{u},d\bar{d})\bar{s}}^{0}\to \Delta^{+}\overline \Xi_{\bar{c}}^{\prime-})= { }\Gamma(Y_{(u\bar{u},d\bar{d})\bar{s}}^{0}\to \Delta^{0}\overline \Xi_{\bar{c}}^{\prime0})=
\Gamma(Y_{(u\bar{u},d\bar{d})\bar{s}}^{0}\to \Sigma^{\prime0}\overline \Omega_{\bar{c}}^{0}),\\
    \Gamma(X_{d\bar{s}\bar{u}}^{-}\to \Delta^{+}\overline \Sigma_{\bar{c}}^{--})=2\Gamma(Y_{(d\bar{d},s\bar{s})\bar{u}}^{-}\to \Sigma^{\prime+}\overline \Sigma_{\bar{c}}^{--})=\frac{2}{3}\Gamma(Y_{(u\bar{u},d\bar{d})\bar{s}}^{0}\to \Delta^{++}\overline \Sigma_{\bar{c}}^{--}),\\
    \Gamma(X_{d\bar{s}\bar{u}}^{-}\to \Sigma^{\prime0}\overline \Xi_{\bar{c}}^{\prime-})= { }\Gamma(Y_{(d\bar{d},s\bar{s})\bar{u}}^{-}\to \Xi^{\prime0}\overline \Xi_{\bar{c}}^{\prime-})=\Gamma(Y_{(u\bar{u},d\bar{d})\bar{s}}^{0}\to \Sigma^{\prime+}\overline \Xi_{\bar{c}}^{\prime-}),\\
    \Gamma(X_{d\bar{s}\bar{u}}^{-}\to \Xi^{\prime-}\overline \Omega_{\bar{c}}^{0})= \frac{2}{3}\Gamma(Y_{(d\bar{d},s\bar{s})\bar{u}}^{-}\to \Omega^-\overline \Omega_{\bar{c}}^{0})=2\Gamma(Y_{(u\bar{u},d\bar{d})\bar{s}}^{0}\to \Xi^{\prime0}\overline \Omega_{\bar{c}}^{0}),\\
     \Gamma(Y_{(u\bar{u},s\bar{s})\bar{d}}^{0}\to \Delta^{++}\overline \Sigma_{\bar{c}}^{--})= \frac{3}{2}\Gamma(X_{s\bar{u}\bar{d}}^{-}\to \Sigma^{\prime+}\overline \Sigma_{\bar{c}}^{--})=
3\Gamma(Y_{(d\bar{d},s\bar{s})\bar{u}}^{-}\to \Delta^{+}\overline \Sigma_{\bar{c}}^{--}),\\
    \Gamma(Y_{(u\bar{u},s\bar{s})\bar{d}}^{0}\to \Delta^{+}\overline \Sigma_{\bar{c}}^-)= { }\Gamma(X_{s\bar{u}\bar{d}}^{-}\to \Sigma^{\prime0}\overline \Sigma_{\bar{c}}^-)=
\Gamma(Y_{(d\bar{d},s\bar{s})\bar{u}}^{-}\to \Delta^{0}\overline \Sigma_{\bar{c}}^-),\\
      \Gamma(Y_{(u\bar{u},s\bar{s})\bar{d}}^{0}\to \Delta^{0}\overline \Sigma_{\bar{c}}^{0})= \frac{1}{2}\Gamma(X_{s\bar{u}\bar{d}}^{-}\to \Sigma^{\prime-}\overline \Sigma_{\bar{c}}^{0})=
\frac{1}{3}\Gamma(Y_{(d\bar{d},s\bar{s})\bar{u}}^{-}\to \Delta^{-}\overline \Sigma_{\bar{c}}^{0}).
\end{eqnarray*}
\begin{table}
\scriptsize
\caption{Open bottom tetraquark $X_{b6}$ decays into a anti-charmed anti-baryon triplet and a light baryon octet for class I, a anti-charmed anti-baryon anti-sextet and a light baryon octet for class II.}
\label{tab:b6_T8_Tbarc3}\begin{tabular}{|cc|cc|}\hline\hline
\multicolumn{4}{|c|} {Class \uppercase\expandafter{\romannumeral1}} \\\hline
channel & amplitude($/V_{ub}$) &channel & amplitude($/V_{ub}$)\\\hline
$X_{u\bar{d}\bar{s}}^{+}\to   \Sigma^+  \overline \Xi_{\bar{c}}^0 $ & $ \sqrt{2} \left(b_2-b_4\right) V^*_{cs}$&
$X_{u\bar{d}\bar{s}}^{+}\to   {p}  \overline \Xi_{\bar{c}}^0 $ & $ \sqrt{2} \left(b_4-b_2\right) V^*_{cd}$\\\hline
$X_{d\bar{s}\bar{u}}^{-}\to   \Lambda^0  \overline \Xi_{\bar{c}}^- $ & $ \frac{\left(2 b_1-2 b_2+3 b_3-6 b_4\right) V^*_{cs}}{2 \sqrt{3}}$&
$X_{d\bar{s}\bar{u}}^{-}\to   \Sigma^0  \overline \Xi_{\bar{c}}^- $ & $ \frac{1}{2} \left(-2 b_1+2 b_2+b_3-2 b_4\right) V^*_{cs}$\\\hline
$X_{d\bar{s}\bar{u}}^{-}\to   \Sigma^-  \overline \Xi_{\bar{c}}^0 $ & $ \frac{\left(b_3-2 b_1\right) V^*_{cs}}{\sqrt{2}}$&
$X_{d\bar{s}\bar{u}}^{-}\to   {n}  \overline \Lambda_{\bar{c}}^- $ & $ -\frac{\left(2 b_1+b_3\right) V^*_{cs}}{\sqrt{2}}$\\\hline
$X_{d\bar{s}\bar{u}}^{-}\to   {n}  \overline \Xi_{\bar{c}}^- $ & $ \sqrt{2} \left(b_2+b_4\right) V^*_{cd}$&
$X_{s\bar{u}\bar{d}}^{-}\to   \Lambda^0  \overline \Lambda_{\bar{c}}^- $ & $ \frac{2 \left(b_2-b_1\right) V^*_{cd}}{\sqrt{3}}$\\\hline
$X_{s\bar{u}\bar{d}}^{-}\to   \Sigma^0  \overline \Lambda_{\bar{c}}^- $ & $ \left(b_3-2 b_4\right) V^*_{cd}$&
$X_{s\bar{u}\bar{d}}^{-}\to   \Xi^-  \overline \Xi_{\bar{c}}^0 $ & $ \frac{\left(2 b_1-b_3\right) V^*_{cd}}{\sqrt{2}}$\\\hline
$X_{s\bar{u}\bar{d}}^{-}\to   \Xi^0  \overline \Lambda_{\bar{c}}^- $ & $ \sqrt{2} \left(b_2+b_4\right) V^*_{cs}$&
$X_{s\bar{u}\bar{d}}^{-}\to   \Xi^0  \overline \Xi_{\bar{c}}^- $ & $ -\frac{\left(2 b_1+b_3\right) V^*_{cd}}{\sqrt{2}}$\\\hline
$Y_{(u\bar{u},s\bar{s})\bar{d}}^{0}\to   \Lambda^0  \overline \Xi_{\bar{c}}^0 $ & $ \frac{\left(2 b_1+4 b_2-3 b_3\right) V^*_{cd}}{2 \sqrt{6}}$&
$Y_{(u\bar{u},s\bar{s})\bar{d}}^{0}\to   \Sigma^+  \overline \Lambda_{\bar{c}}^- $ & $ \left(b_2+b_3-b_4\right) V^*_{cs}$\\\hline
$Y_{(u\bar{u},s\bar{s})\bar{d}}^{0}\to   \Sigma^+  \overline \Xi_{\bar{c}}^- $ & $ -\frac{1}{2} \left(2 b_1+b_3\right) V^*_{cd}$&
$Y_{(u\bar{u},s\bar{s})\bar{d}}^{0}\to   \Sigma^0  \overline \Xi_{\bar{c}}^0 $ & $ \frac{\left(2 b_1+b_3-4 b_4\right) V^*_{cd}}{2 \sqrt{2}}$\\\hline
$Y_{(u\bar{u},s\bar{s})\bar{d}}^{0}\to   {p}  \overline \Lambda_{\bar{c}}^- $ & $ \frac{1}{2} \left(2 b_1-2 b_2-b_3+2 b_4\right) V^*_{cd}$&
$Y_{(u\bar{u},s\bar{s})\bar{d}}^{0}\to   \Xi^0  \overline \Xi_{\bar{c}}^0 $ & $ \left(b_2-b_3+b_4\right) V^*_{cs}$\\\hline
$Y_{(u\bar{u},d\bar{d})\bar{s}}^{0}\to   \Lambda^0  \overline \Xi_{\bar{c}}^0 $ & $ \frac{\left(-2 b_1+2 b_2-3 b_3+6 b_4\right) V^*_{cs}}{2 \sqrt{6}}$&
$Y_{(u\bar{u},d\bar{d})\bar{s}}^{0}\to   \Sigma^+  \overline \Xi_{\bar{c}}^- $ & $ \frac{1}{2} \left(2 b_1-2 b_2-b_3+2 b_4\right) V^*_{cs}$\\\hline
$Y_{(u\bar{u},d\bar{d})\bar{s}}^{0}\to   \Sigma^0  \overline \Xi_{\bar{c}}^0 $ & $ \frac{\left(-2 b_1-2 b_2+b_3+2 b_4\right) V^*_{cs}}{2 \sqrt{2}}$&
$Y_{(u\bar{u},d\bar{d})\bar{s}}^{0}\to   {p}  \overline \Lambda_{\bar{c}}^- $ & $ -\frac{1}{2} \left(2 b_1+b_3\right) V^*_{cs}$\\\hline
$Y_{(u\bar{u},d\bar{d})\bar{s}}^{0}\to   {p}  \overline \Xi_{\bar{c}}^- $ & $ \left(b_2+b_3-b_4\right) V^*_{cd}$&
$Y_{(u\bar{u},d\bar{d})\bar{s}}^{0}\to   {n}  \overline \Xi_{\bar{c}}^0 $ & $ -\left(b_2-b_3+b_4\right) V^*_{cd}$\\\hline
$Y_{(d\bar{d},s\bar{s})\bar{u}}^{-}\to   \Lambda^0  \overline \Lambda_{\bar{c}}^- $ & $ \frac{\left(2 b_1+b_2+3 b_4\right) V^*_{cs}}{\sqrt{6}}$&
$Y_{(d\bar{d},s\bar{s})\bar{u}}^{-}\to   \Lambda^0  \overline \Xi_{\bar{c}}^- $ & $ -\frac{\left(2 b_1+4 b_2+3 b_3\right) V^*_{cd}}{2 \sqrt{6}}$\\\hline
$Y_{(d\bar{d},s\bar{s})\bar{u}}^{-}\to   \Sigma^0  \overline \Lambda_{\bar{c}}^- $ & $ -\frac{\left(b_2+b_3-b_4\right) V^*_{cs}}{\sqrt{2}}$&
$Y_{(d\bar{d},s\bar{s})\bar{u}}^{-}\to   \Sigma^0  \overline \Xi_{\bar{c}}^- $ & $ \frac{\left(2 b_1-b_3+4 b_4\right) V^*_{cd}}{2 \sqrt{2}}$\\\hline
$Y_{(d\bar{d},s\bar{s})\bar{u}}^{-}\to   \Sigma^-  \overline \Xi_{\bar{c}}^0 $ & $ \frac{1}{2} \left(2 b_1-b_3\right) V^*_{cd}$&
$Y_{(d\bar{d},s\bar{s})\bar{u}}^{-}\to   {n}  \overline \Lambda_{\bar{c}}^- $ & $ \frac{1}{2} \left(2 b_1-2 b_2+b_3-2 b_4\right) V^*_{cd}$\\\hline
$Y_{(d\bar{d},s\bar{s})\bar{u}}^{-}\to   \Xi^-  \overline \Xi_{\bar{c}}^0 $ & $ \frac{1}{2} \left(b_3-2 b_1\right) V^*_{cs}$&
$Y_{(d\bar{d},s\bar{s})\bar{u}}^{-}\to   \Xi^0  \overline \Xi_{\bar{c}}^- $ & $ \frac{1}{2} \left(2 b_1-2 b_2+b_3-2 b_4\right) V^*_{cs}$\\\hline
\hline
\multicolumn{4}{|c|} {Class \uppercase\expandafter{\romannumeral2}} \\\hline
$X_{u\bar{d}\bar{s}}^{+}\to   \Sigma^+  \overline \Xi_{\bar{c}}^{\prime0} $ & $ \frac{1}{2} \left(c_2-c_4\right) V^*_{cs}$&
$X_{u\bar{d}\bar{s}}^{+}\to   \Sigma^+  \overline \Omega_{\bar{c}}^{0} $ & $ \frac{\left(c_4-c_2\right) V^*_{cd}}{\sqrt{2}}$\\\hline
$X_{u\bar{d}\bar{s}}^{+}\to   {p}  \overline \Sigma_{\bar{c}}^{0} $ & $ \frac{\left(c_4-c_2\right) V^*_{cs}}{\sqrt{2}}$&
$X_{u\bar{d}\bar{s}}^{+}\to   {p}  \overline \Xi_{\bar{c}}^{\prime0} $ & $ \frac{1}{2} \left(c_2-c_4\right) V^*_{cd}$\\\hline
$X_{d\bar{s}\bar{u}}^{-}\to   \Lambda^0  \overline \Xi_{\bar{c}}^{\prime-} $ & $ \frac{\left(6 c_1+3 c_2+c_3+c_4\right) V^*_{cs}}{2 \sqrt{6}}$&
$X_{d\bar{s}\bar{u}}^{-}\to   \Sigma^0  \overline \Xi_{\bar{c}}^{\prime-} $ & $ \frac{\left(2 c_1+c_2-c_3-c_4\right) V^*_{cs}}{2 \sqrt{2}}$\\\hline
$X_{d\bar{s}\bar{u}}^{-}\to   \Sigma^-  \overline \Xi_{\bar{c}}^{\prime0} $ & $ \frac{1}{2} \left(2 c_1-c_3\right) V^*_{cs}$&
$X_{d\bar{s}\bar{u}}^{-}\to   \Sigma^-  \overline \Omega_{\bar{c}}^{0} $ & $ \frac{\left(c_2+c_4\right) V^*_{cd}}{\sqrt{2}}$\\\hline
$X_{d\bar{s}\bar{u}}^{-}\to   {p}  \overline \Sigma_{\bar{c}}^{--} $ & $ -\frac{\left(2 c_1+c_2+c_3-c_4\right) V^*_{cs}}{\sqrt{2}}$&
$X_{d\bar{s}\bar{u}}^{-}\to   {n}  \overline \Sigma_{\bar{c}}^- $ & $ -\frac{1}{2} \left(2 c_1+c_3\right) V^*_{cs}$\\\hline
$X_{d\bar{s}\bar{u}}^{-}\to   {n}  \overline \Xi_{\bar{c}}^{\prime-} $ & $ -\frac{1}{2} \left(c_2+c_4\right) V^*_{cd}$&
$X_{d\bar{s}\bar{u}}^{-}\to   \Xi^-  \overline \Omega_{\bar{c}}^{0} $ & $ \frac{\left(2 c_1+c_2-c_3+c_4\right) V^*_{cs}}{\sqrt{2}}$\\\hline
$X_{s\bar{u}\bar{d}}^{-}\to   \Lambda^0  \overline \Sigma_{\bar{c}}^- $ & $ -\frac{\left(c_3+c_4\right) V^*_{cd}}{\sqrt{6}}$&
$X_{s\bar{u}\bar{d}}^{-}\to   \Sigma^+  \overline \Sigma_{\bar{c}}^{--} $ & $ -\frac{\left(2 c_1+c_2+c_3-c_4\right) V^*_{cd}}{\sqrt{2}}$\\\hline
$X_{s\bar{u}\bar{d}}^{-}\to   \Sigma^0  \overline \Sigma_{\bar{c}}^- $ & $ \frac{\left(2 c_1+c_2\right) V^*_{cd}}{\sqrt{2}}$&
$X_{s\bar{u}\bar{d}}^{-}\to   \Sigma^-  \overline \Sigma_{\bar{c}}^{0} $ & $ \frac{\left(2 c_1+c_2-c_3+c_4\right) V^*_{cd}}{\sqrt{2}}$\\\hline
$X_{s\bar{u}\bar{d}}^{-}\to   \Xi^-  \overline \Sigma_{\bar{c}}^{0} $ & $ \frac{\left(c_2+c_4\right) V^*_{cs}}{\sqrt{2}}$&
$X_{s\bar{u}\bar{d}}^{-}\to   \Xi^-  \overline \Xi_{\bar{c}}^{\prime0} $ & $ \frac{1}{2} \left(2 c_1-c_3\right) V^*_{cd}$\\\hline
$X_{s\bar{u}\bar{d}}^{-}\to   \Xi^0  \overline \Sigma_{\bar{c}}^- $ & $ -\frac{1}{2} \left(c_2+c_4\right) V^*_{cs}$&
$X_{s\bar{u}\bar{d}}^{-}\to   \Xi^0  \overline \Xi_{\bar{c}}^{\prime-} $ & $ -\frac{1}{2} \left(2 c_1+c_3\right) V^*_{cd}$\\\hline
$Y_{(u\bar{u},s\bar{s})\bar{d}}^{0}\to   \Lambda^0  \overline \Sigma_{\bar{c}}^{0} $ & $ \frac{\left(3 c_2-2 c_3+c_4\right) V^*_{cs}}{2 \sqrt{6}}$&
$Y_{(u\bar{u},s\bar{s})\bar{d}}^{0}\to   \Lambda^0  \overline \Xi_{\bar{c}}^{\prime0} $ & $ \frac{\left(6 c_1-c_3+2 c_4\right) V^*_{cd}}{4 \sqrt{3}}$\\\hline
$Y_{(u\bar{u},s\bar{s})\bar{d}}^{0}\to   \Sigma^+  \overline \Sigma_{\bar{c}}^- $ & $ -\frac{\left(c_2+2 c_3-c_4\right) V^*_{cs}}{2 \sqrt{2}}$&
$Y_{(u\bar{u},s\bar{s})\bar{d}}^{0}\to   \Sigma^+  \overline \Xi_{\bar{c}}^{\prime-} $ & $ \frac{\left(2 c_1+2 c_2+c_3-2 c_4\right) V^*_{cd}}{2 \sqrt{2}}$\\\hline
$Y_{(u\bar{u},s\bar{s})\bar{d}}^{0}\to   \Sigma^0  \overline \Sigma_{\bar{c}}^{0} $ & $ \frac{\left(c_2+2 c_3-c_4\right) V^*_{cs}}{2 \sqrt{2}}$&
$Y_{(u\bar{u},s\bar{s})\bar{d}}^{0}\to   \Sigma^0  \overline \Xi_{\bar{c}}^{\prime0} $ & $ -\frac{1}{4} \left(2 c_1+2 c_2+c_3\right) V^*_{cd}$\\\hline
$Y_{(u\bar{u},s\bar{s})\bar{d}}^{0}\to   {p}  \overline \Sigma_{\bar{c}}^- $ & $ \frac{\left(-2 c_1-c_2+c_3+c_4\right) V^*_{cd}}{2 \sqrt{2}}$&
$Y_{(u\bar{u},s\bar{s})\bar{d}}^{0}\to   {n}  \overline \Sigma_{\bar{c}}^{0} $ & $ -\frac{1}{2} \left(2 c_1+c_2-c_3+c_4\right) V^*_{cd}$\\\hline
$Y_{(u\bar{u},s\bar{s})\bar{d}}^{0}\to   \Xi^0  \overline \Xi_{\bar{c}}^{\prime0} $ & $ \frac{\left(c_2-2 c_3+c_4\right) V^*_{cs}}{2 \sqrt{2}}$&
$Y_{(u\bar{u},s\bar{s})\bar{d}}^{0}\to   \Xi^0  \overline \Omega_{\bar{c}}^{0} $ & $ \frac{1}{2} \left(2 c_1+c_3\right) V^*_{cd}$\\\hline
$Y_{(u\bar{u},d\bar{d})\bar{s}}^{0}\to   \Lambda^0  \overline \Xi_{\bar{c}}^{\prime0} $ & $ -\frac{\left(6 c_1+3 c_2+c_3+c_4\right) V^*_{cs}}{4 \sqrt{3}}$&
$Y_{(u\bar{u},d\bar{d})\bar{s}}^{0}\to   \Lambda^0  \overline \Omega_{\bar{c}}^{0} $ & $ \frac{\left(2 c_3-c_4\right) V^*_{cd}}{\sqrt{6}}$\\\hline
$Y_{(u\bar{u},d\bar{d})\bar{s}}^{0}\to   \Sigma^+  \overline \Xi_{\bar{c}}^{\prime-} $ & $ \frac{\left(-2 c_1-c_2+c_3+c_4\right) V^*_{cs}}{2 \sqrt{2}}$&
$Y_{(u\bar{u},d\bar{d})\bar{s}}^{0}\to   \Sigma^0  \overline \Xi_{\bar{c}}^{\prime0} $ & $ \frac{1}{4} \left(2 c_1-c_2-c_3+c_4\right) V^*_{cs}$\\\hline
$Y_{(u\bar{u},d\bar{d})\bar{s}}^{0}\to   \Sigma^0  \overline \Omega_{\bar{c}}^{0} $ & $ \frac{c_2 V^*_{cd}}{\sqrt{2}}$&
$Y_{(u\bar{u},d\bar{d})\bar{s}}^{0}\to   {p}  \overline \Sigma_{\bar{c}}^- $ & $ \frac{\left(2 c_1+2 c_2+c_3-2 c_4\right) V^*_{cs}}{2 \sqrt{2}}$\\\hline
$Y_{(u\bar{u},d\bar{d})\bar{s}}^{0}\to   {p}  \overline \Xi_{\bar{c}}^{\prime-} $ & $ -\frac{\left(c_2+2 c_3-c_4\right) V^*_{cd}}{2 \sqrt{2}}$&
$Y_{(u\bar{u},d\bar{d})\bar{s}}^{0}\to   {n}  \overline \Sigma_{\bar{c}}^{0} $ & $ \frac{1}{2} \left(2 c_1+c_3\right) V^*_{cs}$\\\hline
$Y_{(u\bar{u},d\bar{d})\bar{s}}^{0}\to   {n}  \overline \Xi_{\bar{c}}^{\prime0} $ & $ \frac{\left(c_2-2 c_3+c_4\right) V^*_{cd}}{2 \sqrt{2}}$&
$Y_{(u\bar{u},d\bar{d})\bar{s}}^{0}\to   \Xi^0  \overline \Omega_{\bar{c}}^{0} $ & $ -\frac{1}{2} \left(2 c_1+c_2-c_3+c_4\right) V^*_{cs}$\\\hline
$Y_{(d\bar{d},s\bar{s})\bar{u}}^{-}\to   \Lambda^0  \overline \Sigma_{\bar{c}}^- $ & $ -\frac{\left(3 c_2-2 c_3+c_4\right) V^*_{cs}}{4 \sqrt{3}}$&
$Y_{(d\bar{d},s\bar{s})\bar{u}}^{-}\to   \Lambda^0  \overline \Xi_{\bar{c}}^{\prime-} $ & $ -\frac{\left(6 c_1+c_3-2 c_4\right) V^*_{cd}}{4 \sqrt{3}}$\\\hline
$Y_{(d\bar{d},s\bar{s})\bar{u}}^{-}\to   \Sigma^+  \overline \Sigma_{\bar{c}}^{--} $ & $ \frac{1}{2} \left(2 c_1+c_2+c_3-c_4\right) V^*_{cs}$&
$Y_{(d\bar{d},s\bar{s})\bar{u}}^{-}\to   \Sigma^0  \overline \Sigma_{\bar{c}}^- $ & $ -\frac{1}{4} \left(4 c_1+c_2-c_4\right) V^*_{cs}$\\\hline
$Y_{(d\bar{d},s\bar{s})\bar{u}}^{-}\to   \Sigma^0  \overline \Xi_{\bar{c}}^{\prime-} $ & $ \frac{1}{4} \left(-2 c_1-2 c_2+c_3\right) V^*_{cd}$&
$Y_{(d\bar{d},s\bar{s})\bar{u}}^{-}\to   \Sigma^-  \overline \Sigma_{\bar{c}}^{0} $ & $ \frac{1}{2} \left(c_3-2 c_1\right) V^*_{cs}$\\\hline
$Y_{(d\bar{d},s\bar{s})\bar{u}}^{-}\to   \Sigma^-  \overline \Xi_{\bar{c}}^{\prime0} $ & $ \frac{\left(-2 c_1-2 c_2+c_3-2 c_4\right) V^*_{cd}}{2 \sqrt{2}}$&
$Y_{(d\bar{d},s\bar{s})\bar{u}}^{-}\to   {p}  \overline \Sigma_{\bar{c}}^{--} $ & $ \frac{1}{2} \left(2 c_1+c_2+c_3-c_4\right) V^*_{cd}$\\\hline
$Y_{(d\bar{d},s\bar{s})\bar{u}}^{-}\to   {n}  \overline \Sigma_{\bar{c}}^- $ & $ \frac{\left(2 c_1+c_2+c_3+c_4\right) V^*_{cd}}{2 \sqrt{2}}$&
$Y_{(d\bar{d},s\bar{s})\bar{u}}^{-}\to   \Xi^-  \overline \Xi_{\bar{c}}^{\prime0} $ & $ \frac{\left(-2 c_1-2 c_2+c_3-2 c_4\right) V^*_{cs}}{2 \sqrt{2}}$\\\hline
$Y_{(d\bar{d},s\bar{s})\bar{u}}^{-}\to   \Xi^-  \overline \Omega_{\bar{c}}^{0} $ & $ \frac{1}{2} \left(c_3-2 c_1\right) V^*_{cd}$&
$Y_{(d\bar{d},s\bar{s})\bar{u}}^{-}\to   \Xi^0  \overline \Xi_{\bar{c}}^{\prime-} $ & $ \frac{\left(2 c_1+c_2+c_3+c_4\right) V^*_{cs}}{2 \sqrt{2}}$\\\hline
\hline
\end{tabular}
\end{table}
\begin{table}
\scriptsize
\caption{Open bottom tetraquark $X_{b6}$ decays into an anti-charmed anti-baryon triplet and a light baryon decuplet for class III, an anti-charmed anti-baryon anti-sextet and a light baryon decuplet for class IV.}\label{tab:b6_T10_Tbarc3}\begin{tabular}{|cc|cc|}\hline\hline
\multicolumn{4}{|c|} {Class \uppercase\expandafter{\romannumeral3}} \\\hline
channel & amplitude($/V_{ub}$) &channel & amplitude($/V_{ub}$)\\\hline
$X_{u\bar{d}\bar{s}}^{+}\to   \Delta^{++}  \overline \Lambda_{\bar{c}}^- $ & $ \frac{\left(d_2-d_1\right) V^*_{cs}}{\sqrt{2}}$&
$X_{u\bar{d}\bar{s}}^{+}\to   \Delta^{++}  \overline \Xi_{\bar{c}}^- $ & $ \frac{\left(d_1-d_2\right) V^*_{cd}}{\sqrt{2}}$\\\hline
$X_{u\bar{d}\bar{s}}^{+}\to   \Delta^{+}  \overline \Xi_{\bar{c}}^0 $ & $ \frac{\left(d_1-d_2+4 d_3\right) V^*_{cd}}{\sqrt{6}}$&
$X_{u\bar{d}\bar{s}}^{+}\to   \Sigma^{\prime+}  \overline \Xi_{\bar{c}}^0 $ & $ \frac{\left(d_1-d_2+4 d_3\right) V^*_{cs}}{\sqrt{6}}$\\\hline
$X_{d\bar{s}\bar{u}}^{-}\to   \Delta^{0}  \overline \Lambda_{\bar{c}}^- $ & $ \frac{\left(d_2-d_1\right) V^*_{cs}}{\sqrt{6}}$&
$X_{d\bar{s}\bar{u}}^{-}\to   \Delta^{0}  \overline \Xi_{\bar{c}}^- $ & $ \frac{\left(d_1+d_2-4 d_3\right) V^*_{cd}}{\sqrt{6}}$\\\hline
$X_{d\bar{s}\bar{u}}^{-}\to   \Delta^{-}  \overline \Xi_{\bar{c}}^0 $ & $ \frac{\left(d_1+d_2\right) V^*_{cd}}{\sqrt{2}}$&
$X_{d\bar{s}\bar{u}}^{-}\to   \Sigma^{\prime0}  \overline \Xi_{\bar{c}}^- $ & $ \frac{\left(d_2-2 d_3\right) V^*_{cs}}{\sqrt{3}}$\\\hline
$X_{d\bar{s}\bar{u}}^{-}\to   \Sigma^{\prime-}  \overline \Xi_{\bar{c}}^0 $ & $ \frac{\left(d_1+d_2\right) V^*_{cs}}{\sqrt{6}}$&
$X_{s\bar{u}\bar{d}}^{-}\to   \Sigma^{\prime0}  \overline \Lambda_{\bar{c}}^- $ & $ -\frac{\left(d_2-2 d_3\right) V^*_{cd}}{\sqrt{3}}$\\\hline
$X_{s\bar{u}\bar{d}}^{-}\to   \Xi^{\prime0}  \overline \Lambda_{\bar{c}}^- $ & $ -\frac{\left(d_1+d_2-4 d_3\right) V^*_{cs}}{\sqrt{6}}$&
$X_{s\bar{u}\bar{d}}^{-}\to   \Xi^{\prime0}  \overline \Xi_{\bar{c}}^- $ & $ \frac{\left(d_1-d_2\right) V^*_{cd}}{\sqrt{6}}$\\\hline
$X_{s\bar{u}\bar{d}}^{-}\to   \Xi^{\prime-}  \overline \Xi_{\bar{c}}^0 $ & $ \frac{\left(d_1+d_2\right) V^*_{cd}}{\sqrt{6}}$&
$X_{s\bar{u}\bar{d}}^{-}\to   \Omega^-  \overline \Xi_{\bar{c}}^0 $ & $ \frac{\left(d_1+d_2\right) V^*_{cs}}{\sqrt{2}}$\\\hline
$Y_{(u\bar{u},s\bar{s})\bar{d}}^{0}\to   \Delta^{+}  \overline \Lambda_{\bar{c}}^- $ & $ -\frac{\left(d_2-2 d_3\right) V^*_{cd}}{\sqrt{3}}$&
$Y_{(u\bar{u},s\bar{s})\bar{d}}^{0}\to   \Sigma^{\prime+}  \overline \Lambda_{\bar{c}}^- $ & $ -\frac{\left(d_1-2 d_3\right) V^*_{cs}}{\sqrt{3}}$\\\hline
$Y_{(u\bar{u},s\bar{s})\bar{d}}^{0}\to   \Sigma^{\prime+}  \overline \Xi_{\bar{c}}^- $ & $ \frac{\left(d_1-d_2\right) V^*_{cd}}{\sqrt{3}}$&
$Y_{(u\bar{u},s\bar{s})\bar{d}}^{0}\to   \Sigma^{\prime0}  \overline \Xi_{\bar{c}}^0 $ & $ \frac{\left(d_1+2 d_3\right) V^*_{cd}}{\sqrt{6}}$\\\hline
$Y_{(u\bar{u},s\bar{s})\bar{d}}^{0}\to   \Xi^{\prime0}  \overline \Xi_{\bar{c}}^0 $ & $ \frac{\left(d_1+2 d_3\right) V^*_{cs}}{\sqrt{3}}$&
$Y_{(u\bar{u},d\bar{d})\bar{s}}^{0}\to   \Delta^{+}  \overline \Lambda_{\bar{c}}^- $ & $ \frac{\left(d_2-d_1\right) V^*_{cs}}{\sqrt{3}}$\\\hline
$Y_{(u\bar{u},d\bar{d})\bar{s}}^{0}\to   \Delta^{+}  \overline \Xi_{\bar{c}}^- $ & $ \frac{\left(d_1-2 d_3\right) V^*_{cd}}{\sqrt{3}}$&
$Y_{(u\bar{u},d\bar{d})\bar{s}}^{0}\to   \Delta^{0}  \overline \Xi_{\bar{c}}^0 $ & $ \frac{\left(d_1+2 d_3\right) V^*_{cd}}{\sqrt{3}}$\\\hline
$Y_{(u\bar{u},d\bar{d})\bar{s}}^{0}\to   \Sigma^{\prime+}  \overline \Xi_{\bar{c}}^- $ & $ \frac{\left(d_2-2 d_3\right) V^*_{cs}}{\sqrt{3}}$&
$Y_{(u\bar{u},d\bar{d})\bar{s}}^{0}\to   \Sigma^{\prime0}  \overline \Xi_{\bar{c}}^0 $ & $ \frac{\left(d_1+2 d_3\right) V^*_{cs}}{\sqrt{6}}$\\\hline
$Y_{(d\bar{d},s\bar{s})\bar{u}}^{-}\to   \Delta^{0}  \overline \Lambda_{\bar{c}}^- $ & $ -\frac{\left(d_2-2 d_3\right) V^*_{cd}}{\sqrt{3}}$&
$Y_{(d\bar{d},s\bar{s})\bar{u}}^{-}\to   \Sigma^{\prime0}  \overline \Lambda_{\bar{c}}^- $ & $ -\frac{\left(d_1-2 d_3\right) V^*_{cs}}{\sqrt{6}}$\\\hline
$Y_{(d\bar{d},s\bar{s})\bar{u}}^{-}\to   \Sigma^{\prime0}  \overline \Xi_{\bar{c}}^- $ & $ \frac{\left(d_1-2 d_3\right) V^*_{cd}}{\sqrt{6}}$&
$Y_{(d\bar{d},s\bar{s})\bar{u}}^{-}\to   \Sigma^{\prime-}  \overline \Xi_{\bar{c}}^0 $ & $ \frac{\left(d_1+d_2\right) V^*_{cd}}{\sqrt{3}}$\\\hline
$Y_{(d\bar{d},s\bar{s})\bar{u}}^{-}\to   \Xi^{\prime0}  \overline \Xi_{\bar{c}}^- $ & $ \frac{\left(d_2-2 d_3\right) V^*_{cs}}{\sqrt{3}}$&
$Y_{(d\bar{d},s\bar{s})\bar{u}}^{-}\to   \Xi^{\prime-}  \overline \Xi_{\bar{c}}^0 $ & $ \frac{\left(d_1+d_2\right) V^*_{cs}}{\sqrt{3}}$\\\hline
\hline
\multicolumn{4}{|c|} {Class \uppercase\expandafter{\romannumeral4}} \\\hline
$X_{u\bar{d}\bar{s}}^{+}\to   \Delta^{++}  \overline \Sigma_{\bar{c}}^- $ & $ \frac{1}{2} \left(f_2-f_3\right) V^*_{cs}$&
$X_{u\bar{d}\bar{s}}^{+}\to   \Delta^{++}  \overline \Xi_{\bar{c}}^{\prime-} $ & $ \frac{1}{2} \left(f_3-f_2\right) V^*_{cd}$\\\hline
$X_{u\bar{d}\bar{s}}^{+}\to   \Delta^{+}  \overline \Sigma_{\bar{c}}^{0} $ & $ \frac{\left(f_2-f_3\right) V^*_{cs}}{\sqrt{6}}$&
$X_{u\bar{d}\bar{s}}^{+}\to   \Delta^{+}  \overline \Xi_{\bar{c}}^{\prime0} $ & $ \frac{\left(f_3-f_2\right) V^*_{cd}}{2 \sqrt{3}}$\\\hline
$X_{u\bar{d}\bar{s}}^{+}\to   \Sigma^{\prime+}  \overline \Xi_{\bar{c}}^{\prime0} $ & $ \frac{\left(f_2-f_3\right) V^*_{cs}}{2 \sqrt{3}}$&
$X_{u\bar{d}\bar{s}}^{+}\to   \Sigma^{\prime+}  \overline \Omega_{\bar{c}}^{0} $ & $ \frac{\left(f_3-f_2\right) V^*_{cd}}{\sqrt{6}}$\\\hline
$X_{d\bar{s}\bar{u}}^{-}\to   \Delta^{+}  \overline \Sigma_{\bar{c}}^{--} $ & $ \frac{\left(-2 f_1-f_2+f_3\right) V^*_{cs}}{\sqrt{6}}$&
$X_{d\bar{s}\bar{u}}^{-}\to   \Delta^{0}  \overline \Sigma_{\bar{c}}^- $ & $ -\frac{\left(4 f_1+f_2-f_3\right) V^*_{cs}}{2 \sqrt{3}}$\\\hline
$X_{d\bar{s}\bar{u}}^{-}\to   \Delta^{0}  \overline \Xi_{\bar{c}}^{\prime-} $ & $ -\frac{\left(f_2+f_3\right) V^*_{cd}}{2 \sqrt{3}}$&
$X_{d\bar{s}\bar{u}}^{-}\to   \Delta^{-}  \overline \Sigma_{\bar{c}}^{0} $ & $ -\sqrt{2} f_1 V^*_{cs}$\\\hline
$X_{d\bar{s}\bar{u}}^{-}\to   \Delta^{-}  \overline \Xi_{\bar{c}}^{\prime0} $ & $ -\frac{1}{2} \left(f_2+f_3\right) V^*_{cd}$&
$X_{d\bar{s}\bar{u}}^{-}\to   \Sigma^{\prime0}  \overline \Xi_{\bar{c}}^{\prime-} $ & $ -\frac{\left(2 f_1+f_2\right) V^*_{cs}}{\sqrt{6}}$\\\hline
$X_{d\bar{s}\bar{u}}^{-}\to   \Sigma^{\prime-}  \overline \Xi_{\bar{c}}^{\prime0} $ & $ -\frac{\left(4 f_1+f_2+f_3\right) V^*_{cs}}{2 \sqrt{3}}$&
$X_{d\bar{s}\bar{u}}^{-}\to   \Sigma^{\prime-}  \overline \Omega_{\bar{c}}^{0} $ & $ -\frac{\left(f_2+f_3\right) V^*_{cd}}{\sqrt{6}}$\\\hline
$X_{d\bar{s}\bar{u}}^{-}\to   \Xi^{\prime-}  \overline \Omega_{\bar{c}}^{0} $ & $ -\frac{\left(2 f_1+f_2+f_3\right) V^*_{cs}}{\sqrt{6}}$&
$X_{s\bar{u}\bar{d}}^{-}\to   \Sigma^{\prime+}  \overline \Sigma_{\bar{c}}^{--} $ & $ \frac{\left(2 f_1+f_2-f_3\right) V^*_{cd}}{\sqrt{6}}$\\\hline
$X_{s\bar{u}\bar{d}}^{-}\to   \Sigma^{\prime0}  \overline \Sigma_{\bar{c}}^- $ & $ \frac{\left(2 f_1+f_2\right) V^*_{cd}}{\sqrt{6}}$&
$X_{s\bar{u}\bar{d}}^{-}\to   \Sigma^{\prime-}  \overline \Sigma_{\bar{c}}^{0} $ & $ \frac{\left(2 f_1+f_2+f_3\right) V^*_{cd}}{\sqrt{6}}$\\\hline
$X_{s\bar{u}\bar{d}}^{-}\to   \Xi^{\prime0}  \overline \Sigma_{\bar{c}}^- $ & $ \frac{\left(f_2+f_3\right) V^*_{cs}}{2 \sqrt{3}}$&
$X_{s\bar{u}\bar{d}}^{-}\to   \Xi^{\prime0}  \overline \Xi_{\bar{c}}^{\prime-} $ & $ \frac{\left(4 f_1+f_2-f_3\right) V^*_{cd}}{2 \sqrt{3}}$\\\hline
$X_{s\bar{u}\bar{d}}^{-}\to   \Xi^{\prime-}  \overline \Sigma_{\bar{c}}^{0} $ & $ \frac{\left(f_2+f_3\right) V^*_{cs}}{\sqrt{6}}$&
$X_{s\bar{u}\bar{d}}^{-}\to   \Xi^{\prime-}  \overline \Xi_{\bar{c}}^{\prime0} $ & $ \frac{\left(4 f_1+f_2+f_3\right) V^*_{cd}}{2 \sqrt{3}}$\\\hline
$X_{s\bar{u}\bar{d}}^{-}\to   \Omega^-  \overline \Xi_{\bar{c}}^{\prime0} $ & $ \frac{1}{2} \left(f_2+f_3\right) V^*_{cs}$&
$X_{s\bar{u}\bar{d}}^{-}\to   \Omega^-  \overline \Omega_{\bar{c}}^{0} $ & $ \sqrt{2} f_1 V^*_{cd}$\\\hline
$Y_{(u\bar{u},s\bar{s})\bar{d}}^{0}\to   \Delta^{++}  \overline \Sigma_{\bar{c}}^{--} $ & $ \frac{1}{2} \left(2 f_1+f_2-f_3\right) V^*_{cd}$&
$Y_{(u\bar{u},s\bar{s})\bar{d}}^{0}\to   \Delta^{+}  \overline \Sigma_{\bar{c}}^- $ & $ \frac{\left(2 f_1+f_2\right) V^*_{cd}}{\sqrt{6}}$\\\hline
$Y_{(u\bar{u},s\bar{s})\bar{d}}^{0}\to   \Delta^{0}  \overline \Sigma_{\bar{c}}^{0} $ & $ \frac{\left(2 f_1+f_2+f_3\right) V^*_{cd}}{2 \sqrt{3}}$&
$Y_{(u\bar{u},s\bar{s})\bar{d}}^{0}\to   \Sigma^{\prime+}  \overline \Sigma_{\bar{c}}^- $ & $ \frac{f_2 V^*_{cs}}{\sqrt{6}}$\\\hline
$Y_{(u\bar{u},s\bar{s})\bar{d}}^{0}\to   \Sigma^{\prime+}  \overline \Xi_{\bar{c}}^{\prime-} $ & $ \sqrt{\frac{2}{3}} f_1 V^*_{cd}$&
$Y_{(u\bar{u},s\bar{s})\bar{d}}^{0}\to   \Sigma^{\prime0}  \overline \Sigma_{\bar{c}}^{0} $ & $ \frac{f_2 V^*_{cs}}{\sqrt{6}}$\\\hline
$Y_{(u\bar{u},s\bar{s})\bar{d}}^{0}\to   \Sigma^{\prime0}  \overline \Xi_{\bar{c}}^{\prime0} $ & $ \frac{\left(2 f_1+f_3\right) V^*_{cd}}{2 \sqrt{3}}$&
$Y_{(u\bar{u},s\bar{s})\bar{d}}^{0}\to   \Xi^{\prime0}  \overline \Xi_{\bar{c}}^{\prime0} $ & $ \frac{f_2 V^*_{cs}}{\sqrt{6}}$\\\hline
$Y_{(u\bar{u},s\bar{s})\bar{d}}^{0}\to   \Xi^{\prime0}  \overline \Omega_{\bar{c}}^{0} $ & $ \frac{\left(2 f_1-f_2+f_3\right) V^*_{cd}}{2 \sqrt{3}}$&
$Y_{(u\bar{u},d\bar{d})\bar{s}}^{0}\to   \Delta^{++}  \overline \Sigma_{\bar{c}}^{--} $ & $ -\frac{1}{2} \left(2 f_1+f_2-f_3\right) V^*_{cs}$\\\hline
$Y_{(u\bar{u},d\bar{d})\bar{s}}^{0}\to   \Delta^{+}  \overline \Sigma_{\bar{c}}^- $ & $ -\sqrt{\frac{2}{3}} f_1 V^*_{cs}$&
$Y_{(u\bar{u},d\bar{d})\bar{s}}^{0}\to   \Delta^{+}  \overline \Xi_{\bar{c}}^{\prime-} $ & $ -\frac{f_2 V^*_{cd}}{\sqrt{6}}$\\\hline
$Y_{(u\bar{u},d\bar{d})\bar{s}}^{0}\to   \Delta^{0}  \overline \Sigma_{\bar{c}}^{0} $ & $ -\frac{\left(2 f_1-f_2+f_3\right) V^*_{cs}}{2 \sqrt{3}}$&
$Y_{(u\bar{u},d\bar{d})\bar{s}}^{0}\to   \Delta^{0}  \overline \Xi_{\bar{c}}^{\prime0} $ & $ -\frac{f_2 V^*_{cd}}{\sqrt{6}}$\\\hline
$Y_{(u\bar{u},d\bar{d})\bar{s}}^{0}\to   \Sigma^{\prime+}  \overline \Xi_{\bar{c}}^{\prime-} $ & $ -\frac{\left(2 f_1+f_2\right) V^*_{cs}}{\sqrt{6}}$&
$Y_{(u\bar{u},d\bar{d})\bar{s}}^{0}\to   \Sigma^{\prime0}  \overline \Xi_{\bar{c}}^{\prime0} $ & $ -\frac{\left(2 f_1+f_3\right) V^*_{cs}}{2 \sqrt{3}}$\\\hline
$Y_{(u\bar{u},d\bar{d})\bar{s}}^{0}\to   \Sigma^{\prime0}  \overline \Omega_{\bar{c}}^{0} $ & $ -\frac{f_2 V^*_{cd}}{\sqrt{6}}$&
$Y_{(u\bar{u},d\bar{d})\bar{s}}^{0}\to   \Xi^{\prime0}  \overline \Omega_{\bar{c}}^{0} $ & $ -\frac{\left(2 f_1+f_2+f_3\right) V^*_{cs}}{2 \sqrt{3}}$\\\hline
$Y_{(d\bar{d},s\bar{s})\bar{u}}^{-}\to   \Delta^{+}  \overline \Sigma_{\bar{c}}^{--} $ & $ \frac{\left(2 f_1+f_2-f_3\right) V^*_{cd}}{2 \sqrt{3}}$&
$Y_{(d\bar{d},s\bar{s})\bar{u}}^{-}\to   \Delta^{0}  \overline \Sigma_{\bar{c}}^- $ & $ \frac{\left(2 f_1+f_2\right) V^*_{cd}}{\sqrt{6}}$\\\hline
$Y_{(d\bar{d},s\bar{s})\bar{u}}^{-}\to   \Delta^{-}  \overline \Sigma_{\bar{c}}^{0} $ & $ \frac{1}{2} \left(2 f_1+f_2+f_3\right) V^*_{cd}$&
$Y_{(d\bar{d},s\bar{s})\bar{u}}^{-}\to   \Sigma^{\prime+}  \overline \Sigma_{\bar{c}}^{--} $ & $ -\frac{\left(2 f_1+f_2-f_3\right) V^*_{cs}}{2 \sqrt{3}}$\\\hline
$Y_{(d\bar{d},s\bar{s})\bar{u}}^{-}\to   \Sigma^{\prime0}  \overline \Sigma_{\bar{c}}^- $ & $ \frac{\left(f_3-2 f_1\right) V^*_{cs}}{2 \sqrt{3}}$&
$Y_{(d\bar{d},s\bar{s})\bar{u}}^{-}\to   \Sigma^{\prime0}  \overline \Xi_{\bar{c}}^{\prime-} $ & $ \frac{\left(2 f_1-f_3\right) V^*_{cd}}{2 \sqrt{3}}$\\\hline
$Y_{(d\bar{d},s\bar{s})\bar{u}}^{-}\to   \Sigma^{\prime-}  \overline \Sigma_{\bar{c}}^{0} $ & $ \frac{\left(-2 f_1+f_2+f_3\right) V^*_{cs}}{2 \sqrt{3}}$&
$Y_{(d\bar{d},s\bar{s})\bar{u}}^{-}\to   \Sigma^{\prime-}  \overline \Xi_{\bar{c}}^{\prime0} $ & $ \sqrt{\frac{2}{3}} f_1 V^*_{cd}$\\\hline
$Y_{(d\bar{d},s\bar{s})\bar{u}}^{-}\to   \Xi^{\prime0}  \overline \Xi_{\bar{c}}^{\prime-} $ & $ -\frac{\left(2 f_1+f_2\right) V^*_{cs}}{\sqrt{6}}$&
$Y_{(d\bar{d},s\bar{s})\bar{u}}^{-}\to   \Xi^{\prime-}  \overline \Xi_{\bar{c}}^{\prime0} $ & $ -\sqrt{\frac{2}{3}} f_1 V^*_{cs}$\\\hline
$Y_{(d\bar{d},s\bar{s})\bar{u}}^{-}\to   \Xi^{\prime-}  \overline \Omega_{\bar{c}}^{0} $ & $ \frac{\left(2 f_1-f_2-f_3\right) V^*_{cd}}{2 \sqrt{3}}$&
$Y_{(d\bar{d},s\bar{s})\bar{u}}^{-}\to   \Omega^-  \overline \Omega_{\bar{c}}^{0} $ & $ -\frac{1}{2} \left(2 f_1+f_2+f_3\right) V^*_{cs}$\\\hline
\hline
\end{tabular}
\end{table}
\subsubsection{Three-body decays into mesons  }
For $X_{b6}$ decays into three meson modes, the hadron-level effective Hamiltonian can be constructed as
\begin{eqnarray}
  {\cal H}_{{eff}}&=& b_{1} (X_{b6})_{[jk]}^i (H_{\bar 3}'')^{[jk]} D_{i} M^l_m M^m_l + b_{2} (X_{b6})_{[jk]}^i (H_{\bar 3}'')^{[jl]} D_{i} M^k_m M^m_l  \nonumber\\
 &&+ b_{3} (X_{b6})_{[jk]}^i (H_{\bar 3}'')^{[lm]} D_{i} M^j_l M^k_m+ b_{4} (X_{b6})_{[jk]}^i (H_{\bar 3}'')^{[jk]} D_{m} M^l_i M^m_l   \nonumber\\
 &&+ b_{5} (X_{b6})_{[jk]}^i (H_{\bar 3}'')^{[jl]} D_{m} M^k_i M^m_l + b_{6} (X_{b6})_{[jk]}^i (H_{\bar 3}'')^{[jl]} D_{l} M^m_i M^k_m \nonumber\\
 &&+ b_{7} (X_{b6})_{[jk]}^i (H_{\bar 3}'')^{[jl]} D_{m} M^m_i M^k_l + b_{8} (X_{b6})_{[jk]}^i (H_{6}'')^{\{jl\}} D_{i} M^k_m M^m_l\nonumber\\
 &&+b_{9} (X_{b6})_{[jk]}^i (H_{6}'')^{\{jl\}} D_{m} M^k_i M^m_l + b_{10} (X_{b6})_{[jk]}^i (H_{6}'')^{\{jl\}} D_{l} M^m_i M^k_m \nonumber\\
 &&+ b_{11} (X_{b6})_{[jk]}^i (H_{6}'')^{\{jl\}} D_{m} M^m_i M^k_l .
\end{eqnarray}
The related amplitudes for different channels are presented in Tab.~\ref{tab:b6_antiD_2M_1} and Tab.~\ref{tab:b6_antiD_2M_2}, from which one obtain relations for decay widths
\begin{eqnarray*}
     \Gamma(X_{s\bar{u}\bar{d}}^{-}\to D^- \overline K^0 \overline K^0 )=2\Gamma(X_{s\bar{u}\bar{d}}^{-}\to \overline D^0 \overline K^0 K^- ),
     \Gamma(Y_{(u\bar{u},s\bar{s})\bar{d}}^{0}\to \overline D^0 \pi^0 \overline K^0 )= 2\Gamma(Y_{(u\bar{u},s\bar{s})\bar{d}}^{0}\to \overline D^0 \eta \overline K^0 ),\\
     \Gamma(Y_{(u\bar{u},s\bar{s})\bar{d}}^{0}\to D^- \pi^+ \pi^0 )= \Gamma(Y_{(d\bar{d},s\bar{s})\bar{u}}^{-}\to \overline D^0 \pi^0 \pi^- ),
     \Gamma(Y_{(u\bar{u},d\bar{d})\bar{s}}^{0}\to \overline D^0 \pi^0 K^0 )=
     { }\Gamma(Y_{(u\bar{u},d\bar{d})\bar{s}}^{0}\to \overline D^0 K^0 \eta ),\\
     \Gamma(X_{u\bar{d}\bar{s}}^{+}\to \overline D^0 \pi^+ \pi^0 )= \frac{1}{2}\Gamma(X_{u\bar{d}\bar{s}}^{+}\to  D^-_s \pi^+ K^+ ),
     \Gamma(X_{d\bar{s}\bar{u}}^{-}\to D^- \pi^0 \pi^0 )=\frac{1}{2}\Gamma(Y_{(u\bar{u},d\bar{d})\bar{s}}^{0}\to \overline D^0 \pi^+ \pi^- ),\\
     \Gamma(X_{s\bar{u}\bar{d}}^{-}\to  D^-_s \pi^0 \pi^0 )= \frac{1}{2}\Gamma(X_{s\bar{u}\bar{d}}^{-}\to  D^-_s \pi^+ \pi^- ),
     \Gamma(X_{s\bar{u}\bar{d}}^{-}\to  D^-_s \pi^0 \overline K^0 )= \frac{1}{2}\Gamma(X_{s\bar{u}\bar{d}}^{-}\to  D^-_s \pi^+ K^- ),\\
      \Gamma(Y_{(d\bar{d},s\bar{s})\bar{u}}^{-}\to  D^-_s \pi^0 \pi^0 )=\frac{1}{2}\Gamma(Y_{(d\bar{d},s\bar{s})\bar{u}}^{-}\to  D^-_s \pi^+ \pi^- ),
    \Gamma(X_{u\bar{d}\bar{s}}^{+}\to D^- \pi^+ K^+ )= \frac{1}{2}\Gamma(X_{u\bar{d}\bar{s}}^{+}\to  D^-_s K^+ K^+ ),\\
    \Gamma(X_{d\bar{s}\bar{u}}^{-}\to \overline D^0 \pi^- K^0 )= \frac{1}{2}\Gamma(X_{d\bar{s}\bar{u}}^{-}\to  D^-_s K^0 K^0 ),
     \Gamma(Y_{(u\bar{u},s\bar{s})\bar{d}}^{0}\to  D^-_s \pi^+ \eta )= { }\Gamma(Y_{(d\bar{d},s\bar{s})\bar{u}}^{-}\to  D^-_s \pi^0 \eta ),\\
      \Gamma(Y_{(u\bar{u},d\bar{d})\bar{s}}^{0}\to D^- \pi^+ \eta )= 2\Gamma(X_{d\bar{s}\bar{u}}^{-}\to D^- \pi^0 \eta ),
       \Gamma(Y_{(u\bar{u},d\bar{d})\bar{s}}^{0}\to  D^-_s \pi^+ K^0 )= { }\Gamma(X_{d\bar{s}\bar{u}}^{-}\to  D^-_s \pi^0 K^0 ),\\
       \Gamma(Y_{(u\bar{u},d\bar{d})\bar{s}}^{0}\to  D^-_s K^+ \eta )= { }\Gamma(X_{d\bar{s}\bar{u}}^{-}\to  D^-_s K^0 \eta ),
    \Gamma(Y_{(u\bar{u},d\bar{d})\bar{s}}^{0}\to \overline D^0 \eta \eta )= \frac{1}{2}\Gamma(X_{d\bar{s}\bar{u}}^{-}\to D^- \eta \eta ),\\
    \Gamma(Y_{(u\bar{u},d\bar{d})\bar{s}}^{0}\to D^- \pi^0 K^+ )= 3\Gamma(Y_{(u\bar{u},d\bar{d})\bar{s}}^{0}\to D^- K^+ \eta ).\\
    \Gamma(X_{d\bar{s}\bar{u}}^{-}\to \overline D^0 \pi^0 \pi^- )=
\frac{1}{4}\Gamma(X_{u\bar{d}\bar{s}}^{+}\to D^- \pi^+ \pi^+ )=
\Gamma(X_{u\bar{d}\bar{s}}^{+}\to \overline D^0 \pi^0 \pi^+ )=
\frac{1}{2}\Gamma(X_{u\bar{d}\bar{s}}^{+}\to  D^-_s K^+ \pi^+ )\\=
\frac{1}{2}\Gamma(Y_{(u\bar{u},d\bar{d})\bar{s}}^{0}\to D^- \pi^0 \pi^+ ),\\
\end{eqnarray*}
\begin{table}
  \footnotesize
    \newcommand{\tabincell}[2]{\begin{tabular}{@{}#1@{}}#2\end{tabular}}
\caption{Open bottom tetraquark $X_{b6}$ decays into an anti-charmed meson and two light meson.}\label{tab:b6_antiD_2M_1}\begin{tabular}{|cc|cc|}\hline\hline
channel & amplitude($/V_{ub}$) &channel & amplitude($/V_{ub}$) \\\hline
$X_{u\bar{d}\bar{s}}^{+}\to   \overline D^0  \pi^+   \pi^0  $ & $ \frac{(b_5+b_7-b_9-b_{11}) V^*_{cs}}{2}$&
$X_{u\bar{d}\bar{s}}^{+}\to   \overline D^0  \pi^+   K^0  $ & $ \frac{\sqrt{2} \left(- b_2 +2  b_3 - b_6 + b_8 + b_{10} \right)V^*_{cd}}{2}$\\\hline
$X_{u\bar{d}\bar{s}}^{+}\to   \overline D^0  \pi^+   \eta  $ & \tabincell{c}{$ \frac{\left(2 b_2-4 b_3+b_5+2 b_6+b_7\right) V^*_{cs}}{2 \sqrt{3}}$\\+$\frac{\left(-2 b_8-b_9-2 b_{10}-b_{11}\right) V^*_{cs}}{2 \sqrt{3}}$}&
$X_{u\bar{d}\bar{s}}^{+}\to   \overline D^0  \pi^0   K^+  $ & \tabincell{c}{$ \frac{\left(-b_2+2 b_3-b_5-b_6\right) V^*_{cd}}{2} $\\+$\frac{\left(-b_7+b_8+b_9+b_{10}+b_{11}\right) V^*_{cd}}{2} $}\\\hline
$X_{u\bar{d}\bar{s}}^{+}\to   \overline D^0  K^+   \overline K^0  $ & $ \frac{\sqrt{2} \left( b_2 -2  b_3 + b_6 - b_8 - b_{10}\right)V^*_{cs}}{2}$&
$X_{u\bar{d}\bar{s}}^{+}\to   \overline D^0  K^+   \eta  $ & \tabincell{c}{$ \frac{\sqrt{3}  \left(b_2 -2  b_3 - b_5 + b_6 - b_7\right)V^*_{cd}}{6}$\\+ $\frac{\sqrt{3}  \left(- b_8 + b_9 - b_{10} + b_{11} \right)V^*_{cd}}{6}$}\\\hline
$X_{u\bar{d}\bar{s}}^{+}\to   D^-  \pi^+   \pi^+  $ & $ \sqrt{2} \left(b_5+b_7-b_9-b_{11}\right) V^*_{cs}$&
$X_{u\bar{d}\bar{s}}^{+}\to   D^-  \pi^+   K^+  $ & $ \frac{\left(-b_5-b_7+b_9+b_{11}\right) V^*_{cd}}{\sqrt{2}}$\\\hline
$X_{u\bar{d}\bar{s}}^{+}\to    D^-_s  \pi^+   K^+  $ & $ \frac{\left(b_5+b_7-b_9-b_{11}\right) V^*_{cs}}{\sqrt{2}}$&
$X_{u\bar{d}\bar{s}}^{+}\to    D^-_s  K^+   K^+  $ & $ \sqrt{2}(- b_5 - b_7 + b_9 + b_{11} )V^*_{cd}$\\\hline
$X_{d\bar{s}\bar{u}}^{-}\to   \overline D^0  \pi^0   \pi^-  $ & $ \frac{\left(-b_5-b_7+b_9+b_{11}\right) V^*_{cs}}{2} $&
$X_{d\bar{s}\bar{u}}^{-}\to   \overline D^0  \pi^-   K^0  $ & $ -\frac{\left(b_5+b_7+b_9+b_{11}\right) V^*_{cd}}{\sqrt{2}}$\\\hline
$X_{d\bar{s}\bar{u}}^{-}\to   \overline D^0  \pi^-   \eta  $ & \tabincell{c}{$ \frac{\left(-4 b_4-b_5-2 b_6+b_7\right) V^*_{cs}}{2 \sqrt{3}}$\\+$\frac{\left(+b_9+2 b_{10}+3 b_{11}\right) V^*_{cs}}{2 \sqrt{3}}$}&
$X_{d\bar{s}\bar{u}}^{-}\to   \overline D^0  K^0   K^-  $ & $ -\frac{\left(2 b_4+b_5+b_6+b_9-b_{10}\right) V^*_{cs}}{\sqrt{2}}$\\\hline
$X_{d\bar{s}\bar{u}}^{-}\to   D^-  \pi^+   \pi^-  $ & $ -\frac{\left(4 b_1+b_2+2 b_4+b_5-b_8-b_9\right) V^*_{cs}}{\sqrt{2}}$&
$X_{d\bar{s}\bar{u}}^{-}\to   D^-  \pi^0   \pi^0  $ & $ \frac{\sqrt{2} \left(-4  b_1 - b_2 -2  b_4 + b_7 + b_8 - b_{11} \right)V^*_{cs}}{2}$\\\hline
$X_{d\bar{s}\bar{u}}^{-}\to   D^-  \pi^0   K^0  $ & \tabincell{c}{$ \frac{ \left(b_2-2 b_3+b_5+b_6+b_7\right) V^*_{cd}}{2}$\\+$\frac{ \left(b_8+b_9+b_{10}+b_{11}\right) V^*_{cd}}{2}$}&
$X_{d\bar{s}\bar{u}}^{-}\to   D^-  \pi^0   \eta  $ & $ -\frac{\left(b_2-2 b_3-2 b_4+b_7-b_8+b_{11}\right) V^*_{cs}}{\sqrt{6}}$\\\hline
$X_{d\bar{s}\bar{u}}^{-}\to   D^-  \pi^-   K^+  $ & $ \frac{\left(-b_2+2 b_3-b_6-b_8-b_{10}\right) V^*_{cd}}{\sqrt{2}}$&
$X_{d\bar{s}\bar{u}}^{-}\to   D^-  K^+   K^-  $ & $ \sqrt{2}(-2 b_1 - b_2 + b_3 )V^*_{cs}$\\\hline
$X_{d\bar{s}\bar{u}}^{-}\to   D^-  K^0   \overline K^0  $ & $ -\frac{\left(4 b_1+b_2+2 b_4+b_5+b_8+b_9\right) V^*_{cs}}{\sqrt{2}}$&
$X_{d\bar{s}\bar{u}}^{-}\to   D^-  K^0   \eta  $ & \tabincell{c}{$ \frac{\left(b_2-2 b_3-b_5+b_6-b_7\right) V^*_{cd}}{2 \sqrt{3}}$\\+$\frac{\left(b_8-b_9+b_{10}-b_{11}\right) V^*_{cd}}{2 \sqrt{3}}$}\\\hline
$X_{d\bar{s}\bar{u}}^{-}\to   D^-  \eta   \eta  $ &  \tabincell{c}{$ \frac{\sqrt{2} \left(-12 b_1 -5  b_2 +4 b_3\right)V^*_{cs}}{6}$\\ +$\frac{\sqrt{2} \left(-2 b_4 + b_7 -3 b_8 +3 b_{11} \right)V^*_{cs}}{6} $}&
$X_{d\bar{s}\bar{u}}^{-}\to    D^-_s  \pi^0   K^0  $ & $ \frac{\left(2 b_4+b_6-b_7+b_{10}+b_{11}\right) V^*_{cs}}{2} $\\\hline
$X_{d\bar{s}\bar{u}}^{-}\to    D^-_s  \pi^-   K^+  $ & $ -\frac{\left(2 b_4+b_5+b_6-b_9+b_{10}\right) V^*_{cs}}{\sqrt{2}}$&
$X_{d\bar{s}\bar{u}}^{-}\to    D^-_s  K^0   K^0  $ & $ -\sqrt{2} \left(b_5+b_7+b_9+b_{11}\right) V^*_{cd}$\\\hline
$X_{d\bar{s}\bar{u}}^{-}\to    D^-_s  K^0   \eta  $ & \tabincell{c}{$ \frac{\left(2 b_4+2 b_5+b_6+b_7\right) V^*_{cs}}{2 \sqrt{3}}$\\+$\frac{\left(2 b_9+b_{10}+3 b_{11}\right) V^*_{cs}}{2 \sqrt{3}}$}&
$X_{s\bar{u}\bar{d}}^{-}\to   \overline D^0  \pi^0   K^-  $ & $ \frac{ \left(2 b_4+b_5+b_6-b_9-b_{10}-2 b_{11}\right) V^*_{cd}}{2}$\\\hline
$X_{s\bar{u}\bar{d}}^{-}\to   \overline D^0  \pi^-   \overline K^0  $ & $ \frac{\left(2 b_4+b_5+b_6+b_9-b_{10}\right) V^*_{cd}}{\sqrt{2}}$&
$X_{s\bar{u}\bar{d}}^{-}\to   \overline D^0  \overline K^0   K^-  $ & $ \frac{\left(b_5+b_7+b_9+b_{11}\right) V^*_{cs}}{\sqrt{2}}$\\\hline
$X_{s\bar{u}\bar{d}}^{-}\to   \overline D^0  K^-   \eta  $ & $ \frac{\left(-2 b_4+b_5-b_6+2 b_7-b_9+b_{10}\right) V^*_{cd}}{2 \sqrt{3}}$&
$X_{s\bar{u}\bar{d}}^{-}\to   D^-  \pi^+   K^-  $ & $ \frac{\left(2 b_4+b_5+b_6-b_9+b_{10}\right) V^*_{cd}}{\sqrt{2}}$\\\hline
$X_{s\bar{u}\bar{d}}^{-}\to   D^-  \pi^0   \overline K^0  $ & $ -\frac{ \left(2 b_4+b_5+b_6+b_9+b_{10}+2 b_{11}\right) V^*_{cd}}{2}$&
$X_{s\bar{u}\bar{d}}^{-}\to   D^-  \overline K^0   \overline K^0  $ & $ \sqrt{2} \left(b_5+b_7+b_9+b_{11}\right) V^*_{cs}$\\\hline
$X_{s\bar{u}\bar{d}}^{-}\to   D^-  \overline K^0   \eta  $ & $ \frac{\sqrt{3} \left(-2  b_4 + b_5 - b_6 +2  b_7 + b_9 -b_{10} \right)V^*_{cd}}{6}$&
$X_{s\bar{u}\bar{d}}^{-}\to    D^-_s  \pi^+   \pi^-  $ & $ \sqrt{2} \left(2 b_1+b_2-b_3\right) V^*_{cd}$\\\hline
$X_{s\bar{u}\bar{d}}^{-}\to    D^-_s  \pi^+   K^-  $ & $ \frac{\left(b_2-2 b_3+b_6+b_8+b_{10}\right) V^*_{cs}}{\sqrt{2}}$&
$X_{s\bar{u}\bar{d}}^{-}\to    D^-_s  \pi^0   \pi^0  $ & $ \sqrt{2} \left(2 b_1+b_2-b_3\right) V^*_{cd}$\\\hline
$X_{s\bar{u}\bar{d}}^{-}\to    D^-_s  \pi^0   \overline K^0  $ & $ -\frac{ \left(b_2-2 b_3+b_6+b_8+b_{10}\right) V^*_{cs}}{2}$&
$X_{s\bar{u}\bar{d}}^{-}\to    D^-_s  \pi^0   \eta  $ & $ \sqrt{\frac{2}{3}} \left(b_{11}-b_8\right) V^*_{cd}$\\\hline
$X_{s\bar{u}\bar{d}}^{-}\to    D^-_s  K^+   K^-  $ & $ \frac{\left(4 b_1+b_2+2 b_4+b_5-b_8-b_9\right) V^*_{cd}}{\sqrt{2}}$&
$X_{s\bar{u}\bar{d}}^{-}\to    D^-_s  K^0   \overline K^0  $ & $ \frac{\left(4 b_1+b_2+2 b_4+b_5+b_8+b_9\right) V^*_{cd}}{\sqrt{2}}$\\\hline
$X_{s\bar{u}\bar{d}}^{-}\to    D^-_s  \overline K^0   \eta  $ & \tabincell{c}{$ \frac{\sqrt{3} \left(- b_2 +2 b_3 -2 b_5 - b_6 -2 b_7\right)V^*_{cs}}{6}$\\ +$\frac{\sqrt{3} \left(-b_8 -2  b_9 -b_{10} -2 b_{11} \right)V^*_{cs}}{6}$}&
$X_{s\bar{u}\bar{d}}^{-}\to    D^-_s  \eta   \eta  $ & $ \frac{ \sqrt{2} \left(6 b_1+b_2+b_3+4 b_4-2 b_7\right) V^*_{cd}}{3}$\\\hline
$Y_{(u\bar{u},s\bar{s})\bar{d}}^{0}\to   \overline D^0  \pi^+   \pi^-  $ & \tabincell{c}{$ \frac{ \left(4 b_1+2 b_2-2 b_3+2 b_4\right) V^*_{cd}}{2}$\\+$\frac{ \left(b_5+b_6+b_9-b_{10}\right) V^*_{cd}}{2}$}&
$Y_{(u\bar{u},s\bar{s})\bar{d}}^{0}\to   \overline D^0  \pi^+   K^-  $ &  \tabincell{c}{$ \frac{ \left(b_2-2 b_3+b_5+b_6+b_7\right) V^*_{cs}}{2}$\\+$\frac{ \left(b_8+b_9-b_{10}-b_{11}\right) V^*_{cs}}{2}$}\\\hline
$Y_{(u\bar{u},s\bar{s})\bar{d}}^{0}\to   \overline D^0  \pi^0   \pi^0  $ & \tabincell{c}{$ \frac{ \left(4 b_1+2 b_2-2 b_3+2 b_4+b_5\right) V^*_{cd}}{2}$\\+$\frac{ \left(b_6-b_9-b_{10}-2 b_{11}\right) V^*_{cd}}{2}$}&
$Y_{(u\bar{u},s\bar{s})\bar{d}}^{0}\to   \overline D^0  \pi^0   \overline K^0  $ & \tabincell{c}{$ \frac{\left(-b_2+2 b_3+b_5-b_6+b_7\right) V^*_{cs}}{2 \sqrt{2}}$\\-$\frac{\left(b_8-b_9+b_{10}+b_{11}\right) V^*_{cs}}{2 \sqrt{2}}$}\\\hline
$Y_{(u\bar{u},s\bar{s})\bar{d}}^{0}\to   \overline D^0  \pi^0   \eta  $ & \tabincell{c}{$ \frac{\sqrt{3} \left(2  b_4 +2  b_5 + b_6 + b_7 \right)V^*_{cd}}{6}$\\ -$\frac{\sqrt{3} \left(2  b_8 -2  b_9 -b_{10} - b_{11} \right)V^*_{cd}}{6}$}&
$Y_{(u\bar{u},s\bar{s})\bar{d}}^{0}\to   \overline D^0  K^+   K^-  $ & $ \frac{ \left(4 b_1+b_2+2 b_4-b_7-b_8+b_{11}\right) V^*_{cd}}{2}$\\\hline
$Y_{(u\bar{u},s\bar{s})\bar{d}}^{0}\to   \overline D^0  K^0   \overline K^0  $ & $ \frac{1}{2} \left(4 b_1+b_2-b_6+b_8+b_{10}\right) V^*_{cd}$&
$Y_{(u\bar{u},s\bar{s})\bar{d}}^{0}\to   \overline D^0  \overline K^0   \eta  $ & \tabincell{c}{$ \frac{\sqrt{6} \left(- b_2 +2  b_3 + b_5 - b_6 + b_7\right)V^*_{cs}}{12}$\\ +$ \frac{\sqrt{6} \left(- b_8 - b_9 +b_{10} + b_{11} \right)V^*_{cs}}{12}$}\\\hline
$Y_{(u\bar{u},s\bar{s})\bar{d}}^{0}\to   \overline D^0  \eta   \eta  $ & \tabincell{c}{$ \frac{ \left(12 b_1+2 b_2+2 b_3+2 b_4\right) V^*_{cd}}{6}$\\+$\frac{ \left(3 b_5-3 b_6+2 b_7-3 b_9+3 b_{10}\right) V^*_{cd}}{6}$}&
$Y_{(u\bar{u},s\bar{s})\bar{d}}^{0}\to   D^-  \pi^+   \pi^0  $ & $ -\frac{\left(b_9+b_{11}\right) V^*_{cd}}{\sqrt{2}}$\\\hline
$Y_{(u\bar{u},s\bar{s})\bar{d}}^{0}\to   D^-  \pi^+   \overline K^0  $ & $ \left(b_5+b_7\right) V^*_{cs}$&
$Y_{(u\bar{u},s\bar{s})\bar{d}}^{0}\to   D^-  \pi^+   \eta  $ & $ \frac{\left(2 b_4+2 b_5+b_6+b_7-b_9+b_{10}\right) V^*_{cd}}{\sqrt{6}}$\\\hline
$Y_{(u\bar{u},s\bar{s})\bar{d}}^{0}\to   D^-  K^+   \overline K^0  $ & $ \frac{ \left(2 b_4 +b_6 -b_7 +b_{10} +b_{11} \right)V^*_{cd}}{2}$&
$Y_{(u\bar{u},s\bar{s})\bar{d}}^{0}\to    D^-_s  \pi^+   K^0  $ & $ \frac{ \left(-b_2+2 b_3+2 b_4+b_5+b_8+b_9\right) V^*_{cd}}{2}$\\\hline
\end{tabular}
\end{table}
\begin{table}
  \footnotesize
  \newcommand{\tabincell}[2]{\begin{tabular}{@{}#1@{}}#2\end{tabular}}
\caption{Open bottom tetraquark $X_{b6}$ decays into an anti-charmed meson and two light mesons.}\label{tab:b6_antiD_2M_2}\begin{tabular}{|cc|cc|}\hline\hline
channel & amplitude($/V_{ub}$) &channel & amplitude($/V_{ub}$) \\\hline
$Y_{(u\bar{u},s\bar{s})\bar{d}}^{0}\to    D^-_s  \pi^+   \eta  $ & \tabincell{c}{$ \frac{\sqrt{6} \left( b_2 -2 b_3 - b_5 +b_6 - b_7 \right)V^*_{cs}}{6}$\\+$\frac{\sqrt{6} \left(- b_8 - b_9 + b_{10} + b_{11} \right)V^*_{cs}}{6}$}&
$Y_{(u\bar{u},s\bar{s})\bar{d}}^{0}\to    D^-_s  \pi^0   K^+  $ & \tabincell{c}{$ \frac{\left(-b_2+2 b_3+2 b_4+b_5\right) V^*_{cd}}{2 \sqrt{2}}$\\+$ \frac{\left(b_8-b_9-2 b_{11}\right) V^*_{cd}}{2 \sqrt{2}}$}\\\hline
$Y_{(u\bar{u},s\bar{s})\bar{d}}^{0}\to    D^-_s  K^+   \overline K^0  $ & $ \frac{ \left(b_2-2 b_3+b_5+b_6+b_7-b_8-b_9+b_{10}+b_{11}\right) V^*_{cs}}{2}$&
$Y_{(u\bar{u},s\bar{s})\bar{d}}^{0}\to    D^-_s  K^+   \eta  $ & \tabincell{c}{$ \frac{\sqrt{6} \left( b_2 -2  b_3 -2  b_4 +3  b_5\right)V^*_{cd}}{12}$\\+$\frac{\sqrt{6} \left(4 b_7 - b_8 -3  b_9 -2  b_{11} \right)V^*_{cd}}{12}$}\\\hline
$Y_{(u\bar{u},d\bar{d})\bar{s}}^{0}\to   \overline D^0  \pi^+   \pi^-  $ & $ -\frac{ \left(4 b_1+b_2+2 b_4-b_7-b_8+b_{11}\right) V^*_{cs}}{2}$&
$Y_{(u\bar{u},d\bar{d})\bar{s}}^{0}\to   \overline D^0  \pi^0   \pi^0  $ & \tabincell{c}{$ -\frac{ \left(4 b_1+b_2+2 b_4+2 b_5\right) V^*_{cs}}{2}$\\-$\frac{ \left(b_7-b_8-2 b_9-b_{11}\right) V^*_{cs}}{2}$}\\\hline
$Y_{(u\bar{u},d\bar{d})\bar{s}}^{0}\to   \overline D^0  \pi^0   K^0  $ & \tabincell{c}{$ \frac{\left(b_2-2 b_3-b_5+b_6-b_7\right) V^*_{cd}}{2 \sqrt{2}}$\\+$ \frac{\left(b_8+b_9-b_{10}-b_{11}\right) V^*_{cd}}{2 \sqrt{2}}$}&
$Y_{(u\bar{u},d\bar{d})\bar{s}}^{0}\to   \overline D^0  \pi^0   \eta  $ & \tabincell{c}{$ \frac{\sqrt{3} \left(-  b_2 +2  b_3 -2 b_4 - b_5 -2 b_6\right)V^*_{cs}}{6}$\\ +  $\frac{\sqrt{3} \left(b_8 +  b_9 +2 b_{10} +2 b_{11} \right)V^*_{cs}}{6}$}\\\hline
$Y_{(u\bar{u},d\bar{d})\bar{s}}^{0}\to   \overline D^0  \pi^-   K^+  $ & \tabincell{c}{$ -\frac{ \left(b_2-2 b_3+b_5+b_6+b_7\right) V^*_{cd}}{2}$\\-$\frac{ \left(b_8+b_9-b_{10}-b_{11}\right) V^*_{cd}}{2}$}&
$Y_{(u\bar{u},d\bar{d})\bar{s}}^{0}\to   \overline D^0  K^+   K^-  $ & \tabincell{c}{$ -\frac{ \left(4 b_1+2 b_2-2 b_3+2 b_4\right) V^*_{cs}}{2}$\\-$\frac{ \left(b_5+b_6+b_9-b_{10}\right) V^*_{cs}}{2}$}\\\hline
$Y_{(u\bar{u},d\bar{d})\bar{s}}^{0}\to   \overline D^0  K^0   \overline K^0  $ & $ -\frac{ \left(4 b_1+b_2-b_6+b_8+b_{10}\right) V^*_{cs}}{2}$&
$Y_{(u\bar{u},d\bar{d})\bar{s}}^{0}\to   \overline D^0  K^0   \eta  $ & \tabincell{c}{$ \frac{\sqrt{6} \left( b_2 -2  b_3 - b_5 + b_6- b_7\right)V^*_{cd}}{12}$\\ +$\frac{\sqrt{6} \left( b_8+ b_9 - b_{10} - b_{11} \right)V^*_{cd}}{12}$}\\\hline
$Y_{(u\bar{u},d\bar{d})\bar{s}}^{0}\to   \overline D^0  \eta   \eta  $ & \tabincell{c}{$ \frac{ \left(-12 b_1-5 b_2+4 b_3\right) V^*_{cs}}{6}$\\+$\frac{ \left(-2 b_4+b_7-3 b_8+3 b_{11}\right) V^*_{cs}}{6}$}&
$Y_{(u\bar{u},d\bar{d})\bar{s}}^{0}\to   D^-  \pi^+   \pi^0  $ & $ \frac{\left(-b_5-b_7+b_9+b_{11}\right) V^*_{cs}}{\sqrt{2}}$\\\hline
$Y_{(u\bar{u},d\bar{d})\bar{s}}^{0}\to   D^-  \pi^+   K^0  $ & $ -\frac{ \left(b_2-2 b_3+b_5+b_6+b_7-b_8-b_9+b_{10}+b_{11}\right) V^*_{cd}}{2}$&
$Y_{(u\bar{u},d\bar{d})\bar{s}}^{0}\to   D^-  \pi^+   \eta  $ & \tabincell{c}{$ \frac{\sqrt{6} \left( b_2 -2  b_3 -2  b_4 \right)V^*_{cs}}{6}$\\+$ \frac{\sqrt{6} \left(b_7 - b_8 + b_{11} \right)V^*_{cs}}{6}$}\\\hline
$Y_{(u\bar{u},d\bar{d})\bar{s}}^{0}\to   D^-  \pi^0   K^+  $ & \tabincell{c}{$ \frac{\left(-b_2+2 b_3+b_5-b_6+b_7\right) V^*_{cd}}{2 \sqrt{2}}$\\+$\frac{\left(b_8+b_9-b_{10}-b_{11}\right) V^*_{cd}}{2 \sqrt{2}}$}&
$Y_{(u\bar{u},d\bar{d})\bar{s}}^{0}\to   D^-  K^+   \overline K^0  $ & $ \frac{ \left(b_2-2 b_3-2 b_4-b_5-b_8-b_9\right) V^*_{cs}}{2}$\\\hline
$Y_{(u\bar{u},d\bar{d})\bar{s}}^{0}\to   D^-  K^+   \eta  $ & \tabincell{c}{$ \frac{\left(b_2-2 b_3-b_5+b_6-b_7\right) V^*_{cd}}{2 \sqrt{6}}$\\-$\frac{\left(b_8-b_9+b_{10}+b_{11}\right) V^*_{cd}}{2 \sqrt{6}}$}&
$Y_{(u\bar{u},d\bar{d})\bar{s}}^{0}\to    D^-_s  \pi^+   K^0  $ & $ -\frac{ \left(2 b_4+b_6-b_7+b_{10}+b_{11}\right) V^*_{cs}}{2}$\\\hline
$Y_{(u\bar{u},d\bar{d})\bar{s}}^{0}\to    D^-_s  \pi^0   K^+  $ & \tabincell{c}{$ -\frac{\left(2 b_4+2 b_5+b_6+b_7\right) V^*_{cs}}{2 \sqrt{2}}$\\ -$\frac{\left(-2 b_9+b_{10}-b_{11}\right) V^*_{cs}}{2 \sqrt{2}}$}&
$Y_{(u\bar{u},d\bar{d})\bar{s}}^{0}\to    D^-_s  K^+   K^0  $ & $ -\left(b_5+b_7\right) V^*_{cd}$\\\hline
$Y_{(u\bar{u},d\bar{d})\bar{s}}^{0}\to    D^-_s  K^+   \eta  $ & \tabincell{c}{$ \frac{\sqrt{6}  \left(2 b_4 +2  b_5 + b_6 + b_7\right)V^*_{cs}}{12}$\\ +$ \frac{\sqrt{6}  \left(2 b_9 + b_{10} +3  b_{11} \right)V^*_{cs}}{12}$}&
$Y_{(d\bar{d},s\bar{s})\bar{u}}^{-}\to   \overline D^0  \pi^0   \pi^-  $ & $ -\frac{\left(b_9+b_{11}\right) V^*_{cd}}{\sqrt{2}}$\\\hline
$Y_{(d\bar{d},s\bar{s})\bar{u}}^{-}\to   \overline D^0  \pi^0   K^-  $ & $ -\frac{\left(2 b_4+2 b_5+b_6+b_7-b_{10}-b_{11}\right) V^*_{cs}}{2 \sqrt{2}}$&
$Y_{(d\bar{d},s\bar{s})\bar{u}}^{-}\to   \overline D^0  \pi^-   \overline K^0  $ & $\frac{ \left(-2 b_4-b_6+b_7+b_{10}+b_{11}\right) V^*_{cs}}{2}$\\\hline
$Y_{(d\bar{d},s\bar{s})\bar{u}}^{-}\to   \overline D^0  \pi^-   \eta  $ & $ \frac{\sqrt{6} \left(2  b_4 +2  b_5 + b_6 + b_7 + b_9 - b_{10} \right)V^*_{cd}}{6}$&
$Y_{(d\bar{d},s\bar{s})\bar{u}}^{-}\to   \overline D^0  K^0   K^-  $ & $ \frac{ \left(2 b_4+b_6-b_7-b_{10}-b_{11}\right) V^*_{cd}}{2}$\\\hline
$Y_{(d\bar{d},s\bar{s})\bar{u}}^{-}\to   \overline D^0  K^-   \eta  $ & \tabincell{c}{$ \frac{\sqrt{6} \left(2  b_4 +2  b_5 +b_6 + b_7\right)V^*_{cs}}{12}$\\ +$ \frac{\sqrt{6} \left(4 b_9-b_{10} +3  b_{11} \right)V^*_{cs}}{12}$}&
$Y_{(d\bar{d},s\bar{s})\bar{u}}^{-}\to   D^-  \pi^+   \pi^-  $ & \tabincell{c}{$ \frac{ \left(4 b_1+2 b_2-2 b_3+2 b_4\right) V^*_{cd}}{2}$\\+$ \frac{ \left(b_5+b_6-b_9+b_{10}\right) V^*_{cd}}{2}$}\\\hline
$Y_{(d\bar{d},s\bar{s})\bar{u}}^{-}\to   D^-  \pi^+   K^-  $ & $ \frac{ \left(b_2-2 b_3-2 b_4-b_5+b_8+b_9\right) V^*_{cs}}{2}$&
$Y_{(d\bar{d},s\bar{s})\bar{u}}^{-}\to   D^-  \pi^0   \pi^0  $ & \tabincell{c}{$ \frac{ \left(4 b_1+2 b_2-2 b_3+2 b_4+b_5\right) V^*_{cd}}{2}$\\+$ \frac{ \left(b_6+b_9+b_{10}+2 b_{11}\right) V^*_{cd}}{2}$}\\\hline
$Y_{(d\bar{d},s\bar{s})\bar{u}}^{-}\to   D^-  \pi^0   \overline K^0  $ & $ \frac{\sqrt{2} \left(- b_2 +2  b_3 +2  b_4 - b_5 -2 b_7- b_8 - b_9 \right)V^*_{cs}}{4}$&
$Y_{(d\bar{d},s\bar{s})\bar{u}}^{-}\to   D^-  \pi^0   \eta  $ & \tabincell{c}{$ -\frac{\left(2 b_4+2 b_5+b_6+b_7\right) V^*_{cd}}{2 \sqrt{3}}$\\ -$\frac{\left(2 b_8+2 b_9+b_{10}+b_{11}\right) V^*_{cd}}{2 \sqrt{3}}$}\\\hline
$Y_{(d\bar{d},s\bar{s})\bar{u}}^{-}\to   D^-  K^+   K^-  $ & $ \frac{ \left(4 b_1+b_2-b_6-b_8-b_{10}\right) V^*_{cd}}{2}$&
$Y_{(d\bar{d},s\bar{s})\bar{u}}^{-}\to   D^-  K^0   \overline K^0  $ & $ \frac{ \left(4 b_1+b_2+2 b_4-b_7+b_8-b_{11}\right) V^*_{cd}}{2}$\\\hline
$Y_{(d\bar{d},s\bar{s})\bar{u}}^{-}\to   D^-  \overline K^0   \eta  $ & \tabincell{c}{$ \frac{\left(-b_2+2 b_3+2 b_4+3 b_5\right) V^*_{cs}}{2 \sqrt{6}}$\\+$\frac{\left(2 b_7-b_8+3 b_9+4 b_{11}\right) V^*_{cs}}{2 \sqrt{6}}$}&
$Y_{(d\bar{d},s\bar{s})\bar{u}}^{-}\to   D^-  \eta   \eta  $ & \tabincell{c}{$ \frac{ \left(12 b_1+2 b_2+2 b_3+2 b_4+3 b_5\right) V^*_{cd}}{6}$\\+$\frac{ \left(-3 b_6+2 b_7+3 b_9-3 b_{10}\right) V^*_{cd}}{6}$}\\\hline
$Y_{(d\bar{d},s\bar{s})\bar{u}}^{-}\to    D^-_s  \pi^+   \pi^-  $ & $ \frac{ \left(-4 b_1-b_2+b_6+b_8+b_{10}\right) V^*_{cs}}{2}$&
$Y_{(d\bar{d},s\bar{s})\bar{u}}^{-}\to    D^-_s  \pi^0   \pi^0  $ & $ \frac{ \left(-4 b_1-b_2+b_6+b_8+b_{10}\right) V^*_{cs}}{2}$\\\hline
$Y_{(d\bar{d},s\bar{s})\bar{u}}^{-}\to    D^-_s  \pi^0   K^0  $ & $ \frac{\left(b_2-2 b_3-2 b_4-b_5+b_8-b_9-2 b_{11}\right) V^*_{cd}}{2 \sqrt{2}}$&
$Y_{(d\bar{d},s\bar{s})\bar{u}}^{-}\to    D^-_s  \pi^0   \eta  $ & \tabincell{c}{$ \frac{\sqrt{3} \left(-  b_2 +2  b_3 +  b_5 - b_6 + b_7\right)V^*_{cs}}{6}$\\ + $ \frac{\sqrt{3} \left( b_8 +  b_9 -  b_{10} -  b_{11} \right)V^*_{cs}}{6}$}\\\hline
$Y_{(d\bar{d},s\bar{s})\bar{u}}^{-}\to    D^-_s  \pi^-   K^+  $ & $ -\frac{ \left(b_2-2 b_3-2 b_4-b_5+b_8+b_9\right) V^*_{cd}}{2}$&
$Y_{(d\bar{d},s\bar{s})\bar{u}}^{-}\to    D^-_s  K^+   K^-  $ & \tabincell{c}{$ -\frac{ \left(4 b_1+2 b_2-2 b_3+2 b_4\right) V^*_{cs}}{2}$\\-$\frac{ \left(b_5+b_6-b_9+b_{10}\right) V^*_{cs}}{2}$}\\\hline
$Y_{(d\bar{d},s\bar{s})\bar{u}}^{-}\to    D^-_s  K^0   \overline K^0  $ & $ \frac{ \left(-4 b_1 -b_2 -2 b_4 +b_7 -b_8 +b_{11} \right)V^*_{cs}}{2}$&
$Y_{(d\bar{d},s\bar{s})\bar{u}}^{-}\to    D^-_s  K^0   \eta  $ & \tabincell{c}{$ \frac{\left(b_2-2 b_3-2 b_4+3 b_5\right) V^*_{cd}}{2 \sqrt{6}}$\\+$ \frac{\left(4 b_7+b_8+3 b_9+2 b_{11}\right) V^*_{cd}}{2 \sqrt{6}}$}\\\hline
$Y_{(d\bar{d},s\bar{s})\bar{u}}^{-}\to    D^-_s  \eta   \eta  $ &  \tabincell{c}{$-\frac{ \left(12 b_1+5 b_2-4 b_3+8 b_4\right) V^*_{cs}}{6}$\\-$\frac{ \left(6 b_5+3 b_6+2 b_7+3 b_8+6 b_9+3 b_{10}+6 b_{11}\right) V^*_{cs}}{6}$}& &\\
\hline
\end{tabular}
\end{table}

\subsection{Charmless $b\to q_1 \bar q_2 q_3$ transition}

\subsubsection{Two-body decays into mesons}
The charmless tree operator of $(\bar q_1 b)(\bar q_2 q_3)$ can lead to a triple $H_{\bf 3}$, a sextet $H_{\bf\overline6}$ with antisymmetric  upper indices, and a traceless  $H_{\bf{15}}$ with symmetric upper indices. The penguin  operators  behave as the triplet $H_{\bf 3}$.
 The nonzero components of the irreducible tensor for the $\Delta S=0 (b\to d)$ transition are
\begin{eqnarray}
 (H_3)^2=(H_3)_{31}=-(H_3)_{13}=1,\;\;\;(H_{\overline6})^{12}_1=-(H_{\overline6})^{21}_1=(H_{\overline6})^{23}_3=-(H_{\overline6})^{32}_3=1,\nonumber\\
 2(H_{15})^{12}_1= 2(H_{15})^{21}_1=-3(H_{15})^{22}_2=
 -6(H_{15})^{23}_3=-6(H_{15})^{32}_3=6.\label{eq:H3615_bb}
\end{eqnarray}
 For the $\Delta S=1(b\to s)$ transition, the nonzero entries can be obtained from Eq.~\eqref{eq:H3615_bb}
by the exchange  $2\leftrightarrow 3$.
The effective Hamiltonian for $X_{b6}$ decays into two mesons in the charmless transition is
  \begin{eqnarray}
  \mathcal{H}_{eff}&=&b_{1} (X_{b6})^{\{ij\}} (H_3)_{[il]} M^k_j  M^l_k
  -b_{2} (X_{b6})^{\{ij\}} (H_{\overline 6})_{\{ij\}} M^k_l  M^l_k-b_{3} (X_{b6})^{\{ij\}} (H_{\overline 6})_{\{il\}} M^k_j  M^l_k\nonumber\\
  &&-b_{4} (X_{b6})^{\{ij\}} (H_{\overline 6})_{\{kl\}} M^k_i  M^l_j+b_{5} (X_{b6})_{[jk]}^i (H_{15})^{\{jl\}}_i M^k_m  M^m_l+b_{6} (X_{b6})_{[jk]}^i (H_{15})^{\{jl\}}_m M^k_i  M^m_l\nonumber\\
  &&+b_{7} (X_{b6})_{[jk]}^i (H_{15})^{\{jl\}}_m M^m_i  M^k_l.
  \end{eqnarray}
The related decay amplitudes are given in Tab.~\ref{tab:b6_2Md} for the  $b\to d$ transition and Tab.~\ref{tab:b6_2Ms} for the  $b\to s$ transition.  The relations of decay widths become
\begin{eqnarray*}
    \Gamma(Y_{(u\bar{u},d\bar{d})\bar{s}}^{0}\to \pi^0 K^0 )= 3\Gamma(Y_{(u\bar{u},d\bar{d})\bar{s}}^{0}\to \eta K^0 ),
\end{eqnarray*}
and
\begin{eqnarray*}
    \Gamma(X_{u\bar{d}\bar{s}}^{+}\to \pi^+ \pi^0 )= { }\Gamma(X_{d\bar{s}\bar{u}}^{-}\to \pi^- \pi^0 ), \Gamma(Y_{(u\bar{u},s\bar{s})\bar{d}}^{0}\to \pi^0 \overline K^0 )= 3\Gamma(Y_{(u\bar{u},s\bar{s})\bar{d}}^{0}\to \eta \overline K^0 ).
\end{eqnarray*}
\begin{table}
\caption{Open bottom tetraquark $X_{b6}$ decays into two light mesons induced by the charmless $b\to d$ transition.}\label{tab:b6_2Md}\begin{tabular}{|cc|cc|}\hline\hline
channel & amplitude &channel & amplitude \\\hline
$X_{u\bar{d}\bar{s}}^{+}\to   \pi^+   K^0  $ & $ \frac{b_1+b_3+3 b_5+b_6-2 b_7}{\sqrt{2}}$&
$X_{u\bar{d}\bar{s}}^{+}\to   \pi^0   K^+  $ & $ \frac{b_1+b_3+2 b_4+3 b_5+5 b_6+2 b_7}{2}$\\\hline
$X_{u\bar{d}\bar{s}}^{+}\to   K^+   \eta  $ & $ -\frac{b_1+b_3-2 b_4+3 b_5-3 b_6-6 b_7}{2 \sqrt{3}}$&
$X_{d\bar{s}\bar{u}}^{-}\to   \pi^-   K^0  $ & $ \sqrt{2} \left(b_4-2 \left(b_6+b_7\right)\right)$\\\hline
$X_{s\bar{u}\bar{d}}^{-}\to   \pi^0   K^-  $ & $ \frac{-b_1+b_3+b_5-5 b_6-6 b_7}{2}$&
$X_{s\bar{u}\bar{d}}^{-}\to   \pi^-   \overline K^0  $ & $ \frac{-b_1+b_3+b_5+3 b_6+2 b_7}{\sqrt{2}}$\\\hline
$X_{s\bar{u}\bar{d}}^{-}\to   K^-   \eta  $ & $ \frac{b_1-b_3-4 b_4-b_5-3 b_6-2 b_7}{2 \sqrt{3}}$&
$Y_{(u\bar{u},s\bar{s})\bar{d}}^{0}\to   \pi^+   \pi^-  $ & $ \frac{-b_1+4 b_2+b_3+b_5+3 b_6+2 b_7}{2}$\\\hline
$Y_{(u\bar{u},s\bar{s})\bar{d}}^{0}\to   \pi^0   \pi^0  $ & $ \frac{-b_1+4 b_2+b_3+b_5-5 b_6-6 b_7}{2}$&
$Y_{(u\bar{u},s\bar{s})\bar{d}}^{0}\to   \pi^0   \eta  $ & $ -\frac{b_1-b_3+2 b_4+7 b_5+9 b_6+2 b_7}{2 \sqrt{3}}$\\\hline
$Y_{(u\bar{u},s\bar{s})\bar{d}}^{0}\to   K^+   K^-  $ & $ 2 b_2+b_3+b_4-2 b_5+2 b_7$&
$Y_{(u\bar{u},s\bar{s})\bar{d}}^{0}\to   K^0   \overline K^0  $ & $ \frac{b_1+4 b_2+b_3+3 b_5+b_6-2 b_7}{2}$\\\hline
$Y_{(u\bar{u},s\bar{s})\bar{d}}^{0}\to   \eta   \eta  $ & $ \frac{3 b_1+12 b_2+5 b_3-4 b_4-3 b_5-9 b_6-6 b_7}{6}$&
$Y_{(u\bar{u},d\bar{d})\bar{s}}^{0}\to   \pi^0   K^0  $ & $ -\frac{b_1+b_3-2 b_4-5 b_5-3 b_6+2 b_7}{2 \sqrt{2}}$\\\hline
$Y_{(u\bar{u},d\bar{d})\bar{s}}^{0}\to   \pi^-   K^+  $ & $ \frac{b_1+b_3+2 b_4-5 b_5-3 b_6+2 b_7}{2} $&
$Y_{(u\bar{u},d\bar{d})\bar{s}}^{0}\to   K^0   \eta  $ & $ -\frac{b_1+b_3-2 b_4-5 b_5-3 b_6+2 b_7}{2 \sqrt{6}}$\\\hline
$Y_{(d\bar{d},s\bar{s})\bar{u}}^{-}\to   \pi^0   \pi^-  $ & $ -2 \sqrt{2} \left(b_6+b_7\right)$&
$Y_{(d\bar{d},s\bar{s})\bar{u}}^{-}\to   \pi^-   \eta  $ & $ \frac{-b_1+b_3-2 b_4+b_5+3 b_6+2 b_7}{\sqrt{6}}$\\\hline
$Y_{(d\bar{d},s\bar{s})\bar{u}}^{-}\to   K^0   K^-  $ & $ \frac{-b_1+b_3+2 b_4+b_5-b_6-2 b_7}{2} $& &\\
\hline
\end{tabular}
\end{table}
\begin{table}
\caption{Open bottom tetraquark $X_{b6}$ decays into two light mesons induced by the charmless $b\to s$ transition.}\label{tab:b6_2Ms}\begin{tabular}{|cc|cc|}\hline\hline
channel & amplitude &channel & amplitude\\\hline
$X_{u\bar{d}\bar{s}}^{+}\to   \pi^+   \pi^0  $ & $ -b_4-2 \left(b_6+b_7\right)$&
$X_{u\bar{d}\bar{s}}^{+}\to   \pi^+   \eta  $ & $ -\frac{b_1+b_3+b_4+3 b_5+3 b_6}{\sqrt{3}}$\\\hline
$X_{u\bar{d}\bar{s}}^{+}\to   K^+   \overline K^0  $ & $ -\frac{b_1+b_3+3 b_5+b_6-2 b_7}{\sqrt{2}}$&
$X_{d\bar{s}\bar{u}}^{-}\to   \pi^0   \pi^-  $ & $ b_4+2 \left(b_6+b_7\right)$\\\hline
$X_{d\bar{s}\bar{u}}^{-}\to   \pi^-   \eta  $ & $ \frac{b_1-b_3-b_4-b_5+3 b_6+4 b_7}{\sqrt{3}}$&
$X_{d\bar{s}\bar{u}}^{-}\to   K^0   K^-  $ & $ \frac{b_1-b_3-b_5-3 b_6-2 b_7}{\sqrt{2}}$\\\hline
$X_{s\bar{u}\bar{d}}^{-}\to   \overline K^0   K^-  $ & $ \sqrt{2} \left(2 \left(b_6+b_7\right)-b_4\right)$&
$Y_{(u\bar{u},s\bar{s})\bar{d}}^{0}\to   \pi^+   K^-  $ & $ \frac{-b_1-b_3-2 b_4+5 b_5+3 b_6-2 b_7}{2}$\\\hline
$Y_{(u\bar{u},s\bar{s})\bar{d}}^{0}\to   \pi^0   \overline K^0  $ & $ \frac{b_1+b_3-2 b_4-5 b_5-3 b_6+2 b_7}{2 \sqrt{2}}$&
$Y_{(u\bar{u},s\bar{s})\bar{d}}^{0}\to   \overline K^0   \eta  $ & $ \frac{b_1+b_3-2 b_4-5 b_5-3 b_6+2 b_7}{2 \sqrt{6}}$\\\hline
$Y_{(u\bar{u},d\bar{d})\bar{s}}^{0}\to   \pi^+   \pi^-  $ & $ -2 b_2-b_3-b_4+2 b_5-2 b_7$&
$Y_{(u\bar{u},d\bar{d})\bar{s}}^{0}\to   \pi^0   \pi^0  $ & $ -2b_2-\frac{b_3}+\frac{b_4}+2b_5+4 b_6+2b_7$\\\hline
$Y_{(u\bar{u},d\bar{d})\bar{s}}^{0}\to   \pi^0   \eta  $ & $ \frac{b_1+b_5+3 b_6+2 b_7}{\sqrt{3}}$&
$Y_{(u\bar{u},d\bar{d})\bar{s}}^{0}\to   K^+   K^-  $ & $ \frac{b_1-4 b_2-b_3-b_5-3 b_6-2 b_7}{2}$\\\hline
$Y_{(u\bar{u},d\bar{d})\bar{s}}^{0}\to   K^0   \overline K^0  $ & $ \frac{-b_1-4 b_2-b_3-3 b_5-b_6+2 b_7}{2}$&
$Y_{(u\bar{u},d\bar{d})\bar{s}}^{0}\to   \eta   \eta  $ & $ 2 \left(-b_2-\frac{b_3}{6}-\frac{b_4}{6}-b_5+b_7\right)$\\\hline
$Y_{(d\bar{d},s\bar{s})\bar{u}}^{-}\to   \pi^0   K^-  $ & $ \frac{b_1-b_3+2 b_4-b_5+b_6+2 b_7}{2 \sqrt{2}}$&
$Y_{(d\bar{d},s\bar{s})\bar{u}}^{-}\to   \pi^-   \overline K^0  $ & $ \frac{b_1-b_3-2 b_4-b_5+b_6+2 b_7}{2}$\\\hline
$Y_{(d\bar{d},s\bar{s})\bar{u}}^{-}\to   K^-   \eta  $ & $ \frac{-b_1+b_3-2 b_4+b_5+15 b_6+14 b_7}{2 \sqrt{6}}$& &\\
\hline
\end{tabular}
\end{table}
\subsubsection{Two-body decays into a baryon and an anti-baryon}

 To construct the  hadron-level effective Hamiltonian, one need contain all possible combinations for final states.
\begin{eqnarray}
  \mathcal{H}_{eff}&=&b_1 (X_{b6})_{[jk]}^i (H_3)^j (T_8)^k_l (\overline T_8)^l_i +b_2 (X_{b6})_{[jk]}^i (H_3)^j (T_8)^l_i (\overline T_8)^k_l +b_3 (X_{b6})_{[jk]}^i (H_3)^l (T_8)^j_l (\overline T_8)^k_i \nonumber\\
  &&+b_4 (X_{b6})_{[jk]}^i (H_3)^l (T_8)^j_i (\overline T_8)^k_l
  -b_5 (X_{b6})^{\{ij\}} (H_{\overline 6})_{\{ij\}} (T_8)^k_l (\overline T_8)^l_k -b_6 (X_{b6})^{\{ij\}} (H_{\overline 6})_{\{il\}} (T_8)^k_j (\overline T_8)^l_k \nonumber\\
  &&-b_7 (X_{b6})^{\{ij\}} (H_{\overline 6})_{\{ik\}} (T_8)^k_l (\overline T_8)^l_j -b_8 (X_{b6})^{\{ij\}} (H_{\overline 6})_{\{kl\}} (T_8)^k_i (\overline T_8)^l_j\nonumber\\
  &&+b_{9} (X_{b6})_{[jk]}^i (H_{15})^{\{jl\}}_i (T_8)^k_m (\overline T_8)^m_l+b_{10} (X_{b6})_{[jk]}^i (H_{15})^{\{jl\}}_i (T_8)^m_l (\overline T_8)^k_m\nonumber\\
  &&+b_{11} (X_{b6})_{[jk]}^i (H_{15})^{\{lm\}}_i (T_8)^j_l (\overline T_8)^k_m+b_{12} (X_{b6})_{[jk]}^i (H_{15})^{\{jm\}}_l (T_8)^k_i (\overline T_8)^l_m\nonumber\\
  &&+b_{13} (X_{b6})_{[jk]}^i (H_{15})^{\{jm\}}_l (T_8)^l_i (\overline T_8)^k_m+b_{14} (X_{b6})_{[jk]}^i (H_{15})^{\{jm\}}_l (T_8)^k_m (\overline T_8)^l_i\nonumber\\
  &&+b_{15} (X_{b6})_{[jk]}^i (H_{15})^{\{jm\}}_l (T_8)^l_m (\overline T_8)^k_i-c_1 (X_{b6})^{\{ij\}} (H_3)^k (T_{10})_{\{ijl\}} (\overline T_{8})^l_k \nonumber\\
  &&-c_2 (X_{b6})^{\{ij\}} (H_3)^k (T_{10})_{\{ikl\}} (\overline T_{8})^l_j-c_3 (X_{b6})^{\{ij\}} (H_{\overline 6})^{[kl]}_i (T_{10})_{\{mjk\}} (\overline T_8)^m_l\nonumber\\
  &&-c_4 (X_{b6})^{\{ij\}} (H_{\overline 6})^{[kl]}_m (T_{10})_{\{ijk\}} (\overline T_8)^m_l
  -c_5 (X_{b6})^{\{ij\}} (H_{15})^{\{kl\}}_i (T_{10})_{\{mjk\}} (\overline T_8)^m_l\nonumber\\
  &&-c_6 (X_{b6})^{\{ij\}} (H_{15})^{\{kl\}}_i (T_{10})_{\{mkl\}} (\overline T_8)^m_j-c_7 (X_{b6})^{\{ij\}} (H_{15})^{\{kl\}}_m (T_{10})_{\{ijk\}} (\overline T_8)^m_l\nonumber\\
  &&-c_{8} (X_{b6})^{\{ij\}} (H_{15})^{\{kl\}}_m (T_{10})_{\{ikl\}} (\overline T_8)^m_j
  +d_1 (X_{b6})_{[ij],[kl]} (H_{\overline 6})^{[ij]}_m (T_{8})^k_o (\overline T_{\overline {10}})^{\{lmo\}}\nonumber\\
  &&+d_2 (X_{b6})_{[ij],[kl]} (H_{\overline 6})^{[io]}_m (T_{8})^k_o (\overline T_{\overline {10}})^{\{jlm\}}
  +d_3 (X_{b6})_{[ij],[kl]} (H_{15})^{\{ik\}}_m (T_{8})^j_o (\overline T_{\overline {10}})^{\{lmo\}}\nonumber\\
  &&+d_{4} (X_{b6})_{[ij],[kl]} (H_{15})^{\{ik\}}_m (T_{8})^m_o (\overline T_{\overline {10}})^{\{jlo\}}
  +d_{5} (X_{b6})_{[ij],[kl]} (H_{15})^{\{io\}}_m (T_{8})^k_o (\overline T_{\overline {10}})^{\{jlm\}}\nonumber\\
  &&+f_1 (X_{b6})^{i}_{[jk]} (H_3)^j (T_{10})_{\{ilm\}} (\overline T_{\overline {10}})^{\{klm\}}-f_2 (X_{b6})^{\{ij\}} (H_{\overline 6})_{\{ij\}} (T_{10})_{\{klm\}} (\overline T_{\overline {10}})^{\{klm\}}\nonumber\\
  &&-f_3 (X_{b6})^{\{ij\}} (H_{\overline 6})_{\{ik\}} (T_{10})_{\{jlm\}} (\overline T_{\overline {10}})^{\{klm\}}-f_4 (X_{b6})^{\{ij\}} (H_{\overline 6})_{\{kl\}} (T_{10})_{\{ijm\}} (\overline T_{\overline {10}})^{\{klm\}}\nonumber\\
  &&+f_5 (X_{b6})^i_{[jk]} (H_{15})^{\{jl\}}_i (T_{10})_{\{lmo\}} (\overline T_{\overline {10}})^{\{kmo\}}+f_6 (X_{b6})^i_{[jk]} (H_{15})^{\{jm\}}_l (T_{10})_{\{imo\}} (\overline T_{\overline {10}})^{\{klo\}}.
\end{eqnarray}
In this case, the decay amplitudes are given in Tab.~\ref{tab:b6_T8_Tbar8_bd}, Tab.~\ref{tab:b6_T10_Tbar8_bd} and Tab.~\ref{tab:b6_T8_Tbar10bar_bd} for the $b\to d$ transition, Tab.~\ref{tab:b6_T8_Tbar8_bs}, Tab.~\ref{tab:b6_T10_Tbar8_bs} and Tab.~\ref{tab:b6_T8_Tbar10bar_bs} for the $b\to s$ transition.
For $b\to d$  transition, the relations of decay widths for Class $\uppercase\expandafter{\romannumeral2}$ are
\begin{eqnarray*}
    \Gamma(X_{s\bar{u}\bar{d}}^{-}\to \Sigma^{\prime0}\overline p)=
     \frac{1}{2}\Gamma(X_{s\bar{u}\bar{d}}^{-}\to \Sigma^{\prime-}\overline n)=
 \frac{1}{3}\Gamma(Y_{(d\bar{d},s\bar{s})\bar{u}}^{-}\to \Delta^{-}\overline n)=\Gamma(Y_{(d\bar{d},s\bar{s})\bar{u}}^{-}\to \Delta^{0}\overline p),\\
      \Gamma(Y_{(u\bar{u},s\bar{s})\bar{d}}^{0}\to \Sigma^{\prime-}\overline \Sigma^+)= { }\Gamma(Y_{(u\bar{u},s\bar{s})\bar{d}}^{0}\to \Xi^{\prime-}\overline \Xi^+)=
\Gamma(Y_{(u\bar{u},d\bar{d})\bar{s}}^{0}\to \Sigma^{\prime-}\overline \Xi^+)=
\frac{1}{3}\Gamma(Y_{(u\bar{u},d\bar{d})\bar{s}}^{0}\to \Delta^{-}\overline \Sigma^+),\\
     \Gamma(Y_{(u\bar{u},s\bar{s})\bar{d}}^{0}\to \Delta^{+}\overline p)= { }\Gamma(Y_{(u\bar{u},s\bar{s})\bar{d}}^{0}\to \Delta^{0}\overline n), \Gamma(X_{u\bar{d}\bar{s}}^{+}\to \Delta^{0}\overline \Sigma^+)= 2\Gamma(X_{u\bar{d}\bar{s}}^{+}\to \Sigma^{\prime0}\overline \Xi^+).
\end{eqnarray*}
The relations of decay widths for Class $\uppercase\expandafter{\romannumeral3}$   are
\begin{eqnarray*}
    \Gamma(X_{u\bar{d}\bar{s}}^{+}\to \Lambda^0\overline \Xi^{\prime+})= \frac{1}{2}\Gamma(X_{u\bar{d}\bar{s}}^{+}\to \Xi^0\overline\Omega^+),
3\Gamma(X_{u\bar{d}\bar{s}}^{+}\to \Sigma^0\overline\Xi^{\prime+})=\frac{3}{2}\Gamma(X_{u\bar{d}\bar{s}}^{+}\to {n}\overline \Sigma^{\prime+}),\\
        \Gamma(Y_{(u\bar{u},s\bar{s})\bar{d}}^{0}\to \Sigma^-\overline \Sigma^{\prime+})= { }\Gamma(Y_{(u\bar{u},s\bar{s})\bar{d}}^{0}\to \Xi^-\overline\Xi^{\prime+}),
    \Gamma(X_{s\bar{u}\bar{d}}^{-}\to \Sigma^+\overline\Delta^{--})= 2\Gamma(Y_{(d\bar{d},s\bar{s})\bar{u}}^{-}\to {p}\overline\Delta^{--}),\\
      \Gamma(X_{u\bar{d}\bar{s}}^{+}\to {p}\overline \Sigma^{\prime0})= \frac{1}{2}\Gamma(X_{u\bar{d}\bar{s}}^{+}\to \Sigma^+\overline \Xi^{\prime0}),
      \Gamma(X_{d\bar{s}\bar{u}}^{-}\to {n}\overline\Sigma^{\prime-})= { }\Gamma(X_{d\bar{s}\bar{u}}^{-}\to \Sigma^-\overline \Xi^{\prime0}),\\
      \Gamma(X_{s\bar{u}\bar{d}}^{-}\to \Xi^-\overline \Sigma^{\prime0})= \Gamma(Y_{(d\bar{d},s\bar{s})\bar{u}}^{-}\to \Xi^-\overline \Xi^{\prime0})=\Gamma(Y_{(u\bar{u},d\bar{d})\bar{s}}^{0}\to \Sigma^-\overline\Xi^{\prime+})=
 \frac{1}{3}\Gamma(Y_{(u\bar{u},d\bar{d})\bar{s}}^{0}\to \Xi^-\overline\Omega^+)\\=
\Gamma(Y_{(u\bar{u},s\bar{s})\bar{d}}^{0}\to \Xi^-\overline\Xi^{\prime+}).
 \end{eqnarray*}
The relations of decay widths for Class $\uppercase\expandafter{\romannumeral4}$  in $b\to d$ transition are given as:
\begin{eqnarray*}
    \Gamma(X_{d\bar{s}\bar{u}}^{-}\to \Delta^{0}\overline\Sigma^{\prime-})= \frac{2}{3}\Gamma(X_{d\bar{s}\bar{u}}^{-}\to \Delta^{-}\overline \Sigma^{\prime0})=\Gamma(X_{d\bar{s}\bar{u}}^{-}\to \Sigma^{\prime-}\overline \Xi^{\prime0}),\\
    \Gamma(Y_{(u\bar{u},d\bar{d})\bar{s}}^{0}\to \Delta^{-}\overline \Sigma^{\prime+})= \frac{3}{4}\Gamma(Y_{(u\bar{u},d\bar{d})\bar{s}}^{0}\to \Sigma^{\prime-}\overline\Xi^{\prime+})=\Gamma(Y_{(u\bar{u},d\bar{d})\bar{s}}^{0}\to \Xi^{\prime-}\overline\Omega^+),\\
    \Gamma(X_{u\bar{d}\bar{s}}^{+}\to \Delta^{0}\overline \Sigma^{\prime+})= \frac{1}{2}\Gamma(X_{u\bar{d}\bar{s}}^{+}\to \Sigma^{\prime0}\overline\Xi^{\prime+})=\frac{1}{3}\Gamma(X_{u\bar{d}\bar{s}}^{+}\to \Xi^{\prime0}\overline\Omega^+),\\
      \Gamma(Y_{(u\bar{u},d\bar{d})\bar{s}}^{0}\to \Delta^{0}\overline \Sigma^{\prime0})= { }\Gamma(Y_{(u\bar{u},d\bar{d})\bar{s}}^{0}\to \Sigma^{\prime0}\overline \Xi^{\prime0}),
        \Gamma(Y_{(d\bar{d},s\bar{s})\bar{u}}^{-}\to \Sigma^{\prime-}\overline \Sigma^{\prime0})= \frac{1}{3}\Gamma(X_{s\bar{u}\bar{d}}^{-}\to \Omega^-\overline \Xi^{\prime0}),\\
    \Gamma(Y_{(d\bar{d},s\bar{s})\bar{u}}^{-}\to \Delta^{+}\overline\Delta^{--})= \frac{1}{2}\Gamma(X_{s\bar{u}\bar{d}}^{-}\to \Sigma^{\prime+}\overline\Delta^{--}),
     \Gamma(Y_{(d\bar{d},s\bar{s})\bar{u}}^{-}\to \Delta^{0}\overline\Delta^{-})= { }\Gamma(X_{s\bar{u}\bar{d}}^{-}\to \Sigma^{\prime0}\overline\Delta^{-}),\\
      \Gamma(Y_{(d\bar{d},s\bar{s})\bar{u}}^{-}\to \Delta^{-}\overline \Delta^{0})= \frac{3}{2}\Gamma(X_{s\bar{u}\bar{d}}^{-}\to \Sigma^{\prime-}\overline \Delta^{0}),
      \Gamma(X_{u\bar{d}\bar{s}}^{+}\to \Delta^{+}\overline \Sigma^{\prime0})= \frac{1}{2}\Gamma(X_{u\bar{d}\bar{s}}^{+}\to \Sigma^{\prime+}\overline \Xi^{\prime0}).
\end{eqnarray*}

For $b\to s$ transition,  the relations of decay widths for Class $\uppercase\expandafter{\romannumeral2}$ are
\begin{eqnarray*}
     \Gamma(X_{d\bar{s}\bar{u}}^{-}\to \Sigma^{\prime0}\overline \Sigma^-)=
     \frac{1}{2}\Gamma(X_{d\bar{s}\bar{u}}^{-}\to \Xi^{\prime-}\overline \Xi^0)=
 \frac{1}{3}\Gamma(Y_{(d\bar{d},s\bar{s})\bar{u}}^{-}\to \Omega^-\overline \Xi^0)=
 \Gamma(Y_{(d\bar{d},s\bar{s})\bar{u}}^{-}\to \Xi^{\prime0}\overline \Sigma^-),\\
     \Gamma(Y_{(u\bar{u},s\bar{s})\bar{d}}^{0}\to \Xi^{\prime-}\overline \Sigma^+)= \frac{1}{3}\Gamma(Y_{(u\bar{u},s\bar{s})\bar{d}}^{0}\to \Omega^-\overline \Xi^+)=
\Gamma(Y_{(u\bar{u},d\bar{d})\bar{s}}^{0}\to \Xi^{\prime-}\overline \Xi^+)=\Gamma(Y_{(u\bar{u},d\bar{d})\bar{s}}^{0}\to \Sigma^{\prime-}\overline \Sigma^+),\\
      \Gamma(Y_{(u\bar{u},d\bar{d})\bar{s}}^{0}\to \Sigma^{\prime+}\overline \Sigma^-)= { }\Gamma(Y_{(u\bar{u},d\bar{d})\bar{s}}^{0}\to \Xi^{\prime0}\overline \Xi^0), \Gamma(X_{u\bar{d}\bar{s}}^{+}\to \Sigma^{\prime0}\overline \Sigma^+)= \frac{1}{2}\Gamma(X_{u\bar{d}\bar{s}}^{+}\to \Xi^{\prime0}\overline \Xi^+).
\end{eqnarray*}
The relations of decay widths for Class $\uppercase\expandafter{\romannumeral3}$ are given as
\begin{eqnarray*}
    \Gamma(X_{u\bar{d}\bar{s}}^{+}\to \Lambda^0\overline \Sigma^{\prime+})=
     \frac{3}{2}\Gamma(X_{u\bar{d}\bar{s}}^{+}\to \Xi^0\overline\Xi^{\prime+})=
3\Gamma(X_{u\bar{d}\bar{s}}^{+}\to \Sigma^0\overline \Sigma^{\prime+})=\frac{1}{2}\Gamma(X_{u\bar{d}\bar{s}}^{+}\to {n}\overline\Delta^{+}),\\
     \Gamma(Y_{(u\bar{u},s\bar{s})\bar{d}}^{0}\to \Sigma^-\overline\Delta^{+})=
      3\Gamma(Y_{(u\bar{u},s\bar{s})\bar{d}}^{0}\to \Xi^-\overline \Sigma^{\prime+})=
3\Gamma(Y_{(u\bar{u},d\bar{d})\bar{s}}^{0}\to \Sigma^-\overline \Sigma^{\prime+})=
 3\Gamma(Y_{(u\bar{u},d\bar{d})\bar{s}}^{0}\to \Xi^-\overline\Xi^{\prime+})\\=
3\Gamma(Y_{(u\bar{u},d\bar{d})\bar{s}}^{0}\to \Xi^-\overline\Xi^{\prime+}),\ \
    \Gamma(Y_{(d\bar{d},s\bar{s})\bar{u}}^{-}\to \Sigma^+\overline\Delta^{--})= \frac{1}{2}\Gamma(X_{d\bar{s}\bar{u}}^{-}\to {p}\overline\Delta^{--}),\\
        \Gamma(X_{u\bar{d}\bar{s}}^{+}\to {p}\overline \Delta^{0})= 2\Gamma(X_{u\bar{d}\bar{s}}^{+}\to \Sigma^+\overline \Sigma^{\prime0}),
       \Gamma(X_{s\bar{u}\bar{d}}^{-}\to \Xi^0\overline\Delta^{-})= { }\Gamma(X_{s\bar{u}\bar{d}}^{-}\to \Xi^-\overline \Delta^{0}),\\
        \Gamma(Y_{(d\bar{d},s\bar{s})\bar{u}}^{-}\to \Lambda^0\overline\Delta^{-})= { }\Gamma(Y_{(u\bar{u},s\bar{s})\bar{d}}^{0}\to \Lambda^0\overline \Delta^{0}),
         \Gamma(Y_{(d\bar{d},s\bar{s})\bar{u}}^{-}\to \Sigma^-\overline \Delta^{0})= { }\Gamma(X_{d\bar{s}\bar{u}}^{-}\to \Sigma^-\overline \Sigma^{\prime0}).
\end{eqnarray*}
The relations of decay widths for Class $\uppercase\expandafter{\romannumeral4}$ become
\begin{eqnarray*}
     \Gamma(X_{u\bar{d}\bar{s}}^{+}\to \Delta^{0}\overline \Delta^{+})= \frac{3}{2}\Gamma(X_{u\bar{d}\bar{s}}^{+}\to \Sigma^{\prime0}\overline \Sigma^{\prime+})=3\Gamma(X_{u\bar{d}\bar{s}}^{+}\to \Xi^{\prime0}\overline \Xi^{\prime+}),\\
      \Gamma(X_{s\bar{u}\bar{d}}^{-}\to \Xi^{\prime0}\overline \Delta^{-})= { }\Gamma(X_{s\bar{u}\bar{d}}^{-}\to \Xi^{\prime-}\overline \Delta^{0})=\frac{2}{3}\Gamma(X_{s\bar{u}\bar{d}}^{-}\to \Omega^-\overline \Sigma^{\prime0}),\\
       \Gamma(Y_{(u\bar{u},s\bar{s})\bar{d}}^{0}\to \Sigma^{\prime-}\overline \Delta^{+})= \frac{3}{4}\Gamma(Y_{(u\bar{u},s\bar{s})\bar{d}}^{0}\to \Xi^{\prime-}\overline \Sigma^{\prime+})= \Gamma(Y_{(u\bar{u},s\bar{s})\bar{d}}^{0}\to \Omega^-\overline \Xi^{\prime+}),\\
   \Gamma(X_{u\bar{d}\bar{s}}^{+}\to \Delta^{+}\overline \Delta^{0})= 2\Gamma(X_{u\bar{d}\bar{s}}^{+}\to \Sigma^{\prime+}\overline \Sigma^{\prime0}),
    \Gamma(X_{d\bar{s}\bar{u}}^{-}\to \Delta^{+}\overline \Delta^{--})= 2\Gamma(Y_{(d\bar{d},s\bar{s})\bar{u}}^{-}\to \Sigma^{\prime+}\overline \Delta^{--}),\\
    \Gamma(X_{d\bar{s}\bar{u}}^{-}\to \Sigma^{\prime0}\overline \Sigma^{\prime-})= { }\Gamma(Y_{(d\bar{d},s\bar{s})\bar{u}}^{-}\to \Xi^{\prime0}\overline \Sigma^{\prime-}),
     \Gamma(X_{d\bar{s}\bar{u}}^{-}\to \Xi^{\prime-}\overline \Xi^{\prime0})= \frac{2}{3}\Gamma(Y_{(d\bar{d},s\bar{s})\bar{u}}^{-}\to \Omega^-\overline \Xi^{\prime0}),\\
      \Gamma(X_{d\bar{s}\bar{u}}^{-}\to \Delta^{-}\overline \Delta^{0})= 3\Gamma(Y_{(d\bar{d},s\bar{s})\bar{u}}^{-}\to \Xi^{\prime-}\overline \Sigma^{\prime0}),
      \Gamma(Y_{(u\bar{u},s\bar{s})\bar{d}}^{0}\to \Sigma^{\prime0}\overline \Delta^{0})= { }\Gamma(Y_{(u\bar{u},s\bar{s})\bar{d}}^{0}\to \Xi^{\prime0}\overline \Sigma^{\prime0}),\\
\end{eqnarray*}
\begin{table}
\caption{Open bottom tetraquark $X_{b6}$ decays into a light anti-baryon octet and a light baryon octet induced by the charmless $b\to d$ transition. }\label{tab:b6_T8_Tbar8_bd}
\begin{tabular}{|c|c|c|c|c|c|c|c}\hline\hline
\multicolumn{2}{|c|} {Class \uppercase\expandafter{\romannumeral1}} \\\hline
channel & amplitude \\\hline
$X_{u\bar{d}\bar{s}}^{+}\to   \Lambda^0  \overline \Xi^+ $ & $ \frac{-2 b_1+b_2+b_3+b_6-2 b_7+b_8-6 b_9+3 b_{10}+3 b_{11}+3 b_{13}+3 b_{14}+3 b_{15}}{2 \sqrt{3}}$\\\hline
$X_{u\bar{d}\bar{s}}^{+}\to   \Sigma^+  \overline \Xi^0 $ & $ \frac{b_2+b_4+b_6+3 b_{10}+3 b_{11}+b_{12}-2 b_{13}}{\sqrt{2}}$\\\hline
$X_{u\bar{d}\bar{s}}^{+}\to   \Sigma^0  \overline \Xi^+ $ & $ \frac{ b_2-b_3+b_6+b_8+3 b_{10}-3 b_{11}+3 b_{13}-b_{14}+5 b_{15}}{2}$\\\hline
$X_{u\bar{d}\bar{s}}^{+}\to   {p}  \overline \Lambda^0 $ & $ \frac{b_1-2 b_2-b_4-2 b_6+b_7+b_8+3 b_9-6 b_{10}-3 b_{11}+3 b_{12}+3 b_{13}+3 b_{14}}{2 \sqrt{3}}$\\\hline
$X_{u\bar{d}\bar{s}}^{+}\to   {p}  \overline \Sigma^0 $ & $ \frac{b_1+b_4+b_7+b_8+3 b_9+3 b_{11}+5 b_{12}-b_{13}+3 b_{14}}{2}$\\\hline
$X_{u\bar{d}\bar{s}}^{+}\to   {n}  \overline \Sigma^+ $ & $ \frac{b_1-b_3+b_7+3 b_9-3 b_{11}-2 b_{14}+b_{15}}{\sqrt{2}}$\\
\hline
\hline
\end{tabular}
\end{table}
\begin{table}
\caption{Open bottom tetraquark $X_{b6}$ decays into a light anti-baryon octet and a light baryon decuplet induced by the charmless $b\to d$ transition. }\label{tab:b6_T10_Tbar8_bd}\begin{tabular}{|cc|cc|}\hline\hline
\multicolumn{4}{|c|} {Class \uppercase\expandafter{\romannumeral2}} \\\hline
channel & amplitude&channel & amplitude \\\hline
$X_{u\bar{d}\bar{s}}^{+}\to   \Delta^{++}  \overline \Sigma^- $ & $ \frac{c_1+c_3+c_4+3 c_5+3 c_7}{\sqrt{2}}$&
$X_{u\bar{d}\bar{s}}^{+}\to   \Delta^{+}  \overline \Lambda^0 $ & $ \frac{c_1+c_2-3 c_4+6 c_5+6 c_6+3 c_7}{6} $\\\hline
$X_{u\bar{d}\bar{s}}^{+}\to   \Delta^{+}  \overline \Sigma^0 $ & $ -\frac{c_1-c_2+2 c_3+c_4-6 c_6-5 c_7}{2 \sqrt{3}}$&
$X_{u\bar{d}\bar{s}}^{+}\to   \Delta^{0}  \overline \Sigma^+ $ & $ \frac{c_2-c_3+3 c_5+6 c_6}{\sqrt{6}}$\\\hline
$X_{u\bar{d}\bar{s}}^{+}\to   \Sigma^{\prime+}  \overline \Xi^0 $ & $ \frac{c_1+c_3-c_4+3 c_5-c_7}{\sqrt{6}}$&
$X_{u\bar{d}\bar{s}}^{+}\to   \Sigma^{\prime0}  \overline \Xi^+ $ & $ \frac{c_2-c_3+3 c_5+6 c_6}{2 \sqrt{3}}$\\\hline
$X_{d\bar{s}\bar{u}}^{-}\to   \Delta^{0}  \overline \Sigma^- $ & $ \frac{c_1+c_2+c_4-2 c_5-2 c_6+3 c_7}{\sqrt{6}}$&
$X_{d\bar{s}\bar{u}}^{-}\to   \Delta^{-}  \overline \Lambda^0 $ & $ \frac{c_1+c_2-3 c_4-2 c_5-2 c_6+3 c_7}{2 \sqrt{3}}$\\\hline
$X_{d\bar{s}\bar{u}}^{-}\to   \Delta^{-}  \overline \Sigma^0 $ & $ \frac{-c_1-c_2-c_4+2 c_5+2 c_6+5 c_7}{2}$&
$X_{d\bar{s}\bar{u}}^{-}\to   \Sigma^{\prime-}  \overline \Xi^0 $ & $ \frac{c_1+c_2-c_4-2 c_5-2 c_6-c_7}{\sqrt{6}}$\\\hline
$X_{s\bar{u}\bar{d}}^{-}\to   \Sigma^{\prime0}  \overline p $ & $ \frac{c_2+c_3-c_5-2 c_6}{2 \sqrt{3}}$&
$X_{s\bar{u}\bar{d}}^{-}\to   \Sigma^{\prime-}  \overline n $ & $ \frac{c_2+c_3-c_5-2 c_6}{\sqrt{6}}$\\
\hline
\hline
\end{tabular}
\end{table}
\begin{table}
\caption{Open bottom tetraquark $X_{b6}$ decays into a light anti-baryon anti-decuplet plus a light baryon octet for class III and a light anti-baryon  anti-decuplet plus a light baryon decuplet for class IV induced by the charmless $b\to d$ transition. }\label{tab:b6_T8_Tbar10bar_bd}\begin{tabular}{|cc|cc|}\hline\hline
\multicolumn{4}{|c|} {Class \uppercase\expandafter{\romannumeral3}} \\\hline
channel & amplitude&channel & amplitude \\\hline
$X_{u\bar{d}\bar{s}}^{+}\to   \Lambda^0  \overline \Xi^{\prime+} $ & $ \frac{2 d_1+d_2+d_3-2 d_4-d_5}{2}$&
$X_{u\bar{d}\bar{s}}^{+}\to   \Sigma^+  \overline \Xi^{\prime0} $ & $ \frac{2 d_1-d_2+d_3-2 d_4+3 d_5}{\sqrt{6}}$\\\hline
$X_{u\bar{d}\bar{s}}^{+}\to   \Sigma^0  \overline \Xi^{\prime+} $ & $ -\frac{2 d_1+d_2+d_3-2 d_4-d_5}{2 \sqrt{3}}$&
$X_{u\bar{d}\bar{s}}^{+}\to   {p}  \overline \Sigma^{\prime0} $ & $ -\frac{2 d_1-d_2+d_3-2 d_4+3 d_5}{2 \sqrt{3}}$\\\hline
$X_{u\bar{d}\bar{s}}^{+}\to   {n}  \overline \Sigma^{\prime+} $ & $ -\frac{2 d_1+d_2+d_3-2 d_4-d_5}{\sqrt{6}}$&
$X_{u\bar{d}\bar{s}}^{+}\to   \Xi^0  \overline \Omega^+ $ & $ \frac{2 d_1+d_2+d_3-2 d_4-d_5}{\sqrt{2}}$\\\hline
$X_{d\bar{s}\bar{u}}^{-}\to   \Sigma^-  \overline \Xi^{\prime0} $ & $ \sqrt{\frac{2}{3}} \left(d_2+2 d_5\right)$&
$X_{d\bar{s}\bar{u}}^{-}\to   {n}  \overline \Sigma^{\prime-} $ & $ -\sqrt{\frac{2}{3}} \left(d_2+2 d_5\right)$\\\hline
$X_{s\bar{u}\bar{d}}^{-}\to   \Lambda^0  \overline \Delta^{-} $ & $ -\frac{4\left(d_3+d_4+d_5\right)}{3} $&
$X_{s\bar{u}\bar{d}}^{-}\to   \Sigma^+  \overline \Delta^{--} $ & $ -\frac{2 d_1+d_2+3 d_3+2 d_4-3 d_5}{\sqrt{2}}$\\\hline
$X_{s\bar{u}\bar{d}}^{-}\to   \Sigma^0  \overline \Delta^{-} $ & $ \frac{2 d_1+d_2-d_3-2 d_4+d_5}{\sqrt{3}}$&
$X_{s\bar{u}\bar{d}}^{-}\to   \Sigma^-  \overline \Delta^{0} $ & $ \frac{2 d_1+d_2-5 d_3-6 d_4+5 d_5}{\sqrt{6}}$\\
\end{tabular}
\multicolumn{4}{|c|} {Class \uppercase\expandafter{\romannumeral4}} \\\hline
$X_{u\bar{d}\bar{s}}^{+}\to   \Delta^{++}  \overline\Sigma^{\prime-} $ & $ \frac{f_1+f_3+2 f_4+3 f_5+3 f_6}{\sqrt{6}}$&
$X_{u\bar{d}\bar{s}}^{+}\to   \Delta^{+}  \overline \Sigma^{\prime0} $ & $ \frac{f_1+f_3+f_4+3 f_5+f_6}{3}$\\\hline
$X_{u\bar{d}\bar{s}}^{+}\to   \Delta^{0}  \overline \Sigma^{\prime+} $ & $ \frac{f_1+f_3+3 f_5-f_6}{3 \sqrt{2}}$&
$X_{u\bar{d}\bar{s}}^{+}\to   \Sigma^{\prime+}  \overline \Xi^{\prime0} $ & $ \frac{\sqrt{2} \left(f_1+f_3+f_4+3 f_5+f_6\right)}{3}$\\\hline
$X_{u\bar{d}\bar{s}}^{+}\to   \Sigma^{\prime0}  \overline\Xi^{\prime+} $ & $ \frac{f_1+f_3+3 f_5-f_6}{3} $&
$X_{u\bar{d}\bar{s}}^{+}\to   \Xi^{\prime0}  \overline\Omega^+ $ & $ \frac{f_1+f_3+3 f_5-f_6}{\sqrt{6}}$\\\hline
$X_{d\bar{s}\bar{u}}^{-}\to   \Delta^{0}  \overline\Sigma^{\prime-} $ & $ \frac{\sqrt{2} \left(f_4-2 f_6\right)}{3} $&
$X_{d\bar{s}\bar{u}}^{-}\to   \Delta^{-}  \overline \Sigma^{\prime0} $ & $ \frac{f_4-2 f_6}{\sqrt{3}}$\\\hline
$X_{d\bar{s}\bar{u}}^{-}\to   \Sigma^{\prime-}  \overline \Xi^{\prime0} $ & $ \frac{ \sqrt{2} \left(f_4-2 f_6\right)}{3} $&
$X_{s\bar{u}\bar{d}}^{-}\to   \Sigma^{\prime+}  \overline\Delta^{--} $ & $ \frac{-f_1+f_3+f_5-3 f_6}{\sqrt{6}}$\\\hline
$X_{s\bar{u}\bar{d}}^{-}\to   \Sigma^{\prime0}  \overline\Delta^{-} $ & $ \frac{ \left(-f_1+f_3+f_5+f_6\right)}{3} $&
$X_{s\bar{u}\bar{d}}^{-}\to   \Sigma^{\prime-}  \overline \Delta^{0} $ & $ \frac{-f_1+f_3+f_5+5 f_6}{3 \sqrt{2}}$\\\hline
$X_{s\bar{u}\bar{d}}^{-}\to   \Xi^{\prime0}  \overline\Sigma^{\prime-} $ & $ \frac{ \sqrt{2} \left(-f_1+f_3+f_4+f_5-f_6\right)}{3} $&
$X_{s\bar{u}\bar{d}}^{-}\to   \Xi^{\prime-}  \overline \Sigma^{\prime0} $ & $ \frac{ \left(-f_1+f_3+f_4+f_5+3 f_6\right)}{3} $\\
\hline
\end{tabular}
\end{table}


\begin{table}
\caption{Open bottom tetraquark $X_{b6}$ decays into a light anti-baryon octet and a light baryon octet induced by the charmless $b\to s$ transition. }\label{tab:b6_T8_Tbar8_bs}\begin{tabular}{|c|c|c|c|c|c|c|c}\hline\hline
\multicolumn{2}{|c|} {Class \uppercase\expandafter{\romannumeral1}} \\\hline
channel & amplitude \\\hline
$X_{u\bar{d}\bar{s}}^{+}\to   \Lambda^0  \overline \Sigma^+ $ & $ \frac{ \sqrt{3}\left(- b_1- b_2+2  b_3- b_6- b_7- b_8-3 b_9-3 b_{10}+6 b_{11}-3 b_{13}+3  b_{14}-6  b_{15}\right)}{6}$\\\hline
$X_{u\bar{d}\bar{s}}^{+}\to   \Sigma^+  \overline \Lambda^0 $ & $ -\frac{b_1+b_2+2 b_4+b_6+b_7+b_8+3 b_9+3 b_{10}+6 b_{11}+6 b_{12}-3 b_{13}+3 b_{14}}{2 \sqrt{3}}$\\\hline
$X_{u\bar{d}\bar{s}}^{+}\to   \Sigma^+  \overline \Sigma^0 $ & $ \frac{-b_1+b_2+b_6-b_7-b_8-3 b_9+3 b_{10}-4 b_{12}-b_{13}-3 b_{14}}{2}$\\\hline
$X_{u\bar{d}\bar{s}}^{+}\to   \Sigma^0  \overline \Sigma^+ $ & $ \frac{b_1-b_2-b_6+b_7-b_8+3 b_9-3 b_{10}-3 b_{13}-b_{14}-4 b_{15}}{2}$\\\hline
$X_{u\bar{d}\bar{s}}^{+}\to   {p}  \overline n $ & $ -\frac{b_2+b_4+b_6+3 b_{10}+3 b_{11}+b_{12}-2 b_{13}}{\sqrt{2}}$\\\hline
$X_{u\bar{d}\bar{s}}^{+}\to   \Xi^0  \overline \Xi^+ $ & $ -\frac{b_1-b_3+b_7+3 b_9-3 b_{11}-2 b_{14}+b_{15}}{\sqrt{2}}$\\
\hline
\hline
\end{tabular}
\end{table}
\begin{table}
\caption{Open bottom tetraquark $X_{b6}$ decays into a light anti-baryon octet and a light baryon decuplet induced by the charmless $b\to s$ transition.}\label{tab:b6_T10_Tbar8_bs}\begin{tabular}{|cc|cc|}\hline\hline
\multicolumn{4}{|c|} {Class \uppercase\expandafter{\romannumeral2}} \\\hline
channel & amplitude&channel & amplitude \\\hline
$X_{u\bar{d}\bar{s}}^{+}\to   \Delta^{++}  \overline p $ & $ \frac{c_1+c_3+c_4+3 c_5+3 c_7}{\sqrt{2}}$&
$X_{u\bar{d}\bar{s}}^{+}\to   \Delta^{+}  \overline n $ & $ \frac{c_1+c_3-c_4+3 c_5-c_7}{\sqrt{6}}$\\\hline
$X_{u\bar{d}\bar{s}}^{+}\to   \Sigma^{\prime+}  \overline \Lambda^0 $ & $ \frac{-2 c_1+c_2-3 c_3-3 c_5+6 c_6+6 c_7}{6}$&
$X_{u\bar{d}\bar{s}}^{+}\to   \Sigma^{\prime+}  \overline \Sigma^0 $ & $ \frac{c_2-c_3-2 c_4+3 c_5+6 c_6+4 c_7}{2 \sqrt{3}}$\\\hline
$X_{u\bar{d}\bar{s}}^{+}\to   \Sigma^{\prime0}  \overline \Sigma^+ $ & $ \frac{c_2-c_3+3 c_5+6 c_6}{2 \sqrt{3}}$&
$X_{u\bar{d}\bar{s}}^{+}\to   \Xi^{\prime0}  \overline \Xi^+ $ & $ \frac{c_2-c_3+3 c_5+6 c_6}{\sqrt{6}}$\\\hline
$X_{d\bar{s}\bar{u}}^{-}\to   \Delta^{0}  \overline p $ & $ \frac{c_1-c_3+c_4-c_5+3 c_7}{\sqrt{6}}$&
$X_{d\bar{s}\bar{u}}^{-}\to   \Delta^{-}  \overline n $ & $ \frac{c_1-c_3-c_4-c_5-c_7}{\sqrt{2}}$\\\hline
$X_{d\bar{s}\bar{u}}^{-}\to   \Sigma^{\prime0}  \overline \Sigma^- $ & $ \frac{c_2+c_3-c_5-2 c_6}{2 \sqrt{3}}$&
$X_{d\bar{s}\bar{u}}^{-}\to   \Sigma^{\prime-}  \overline \Lambda^0 $ & $ \frac{-2 c_1+c_2+3 c_3+c_5-2 c_6+6 c_7}{6}$\\\hline
$X_{d\bar{s}\bar{u}}^{-}\to   \Sigma^{\prime-}  \overline \Sigma^0 $ & $ -\frac{c_2+c_3+2 c_4-c_5-2 c_6-4 c_7}{2 \sqrt{3}}$&
$X_{d\bar{s}\bar{u}}^{-}\to   \Xi^{\prime-}  \overline \Xi^0 $ & $ \frac{c_2+c_3-c_5-2 c_6}{\sqrt{6}}$\\
\hline
\hline
\end{tabular}
\end{table}
\begin{table}
\caption{Open bottom tetraquark $X_{b6}$ decays into a light anti-baryon anti-decuplet plus a light baryon octet for class III and  a light anti-baryon  anti-decuplet and a light baryon decuplet for class IV induced by the charmless $b\to s$ transition. }\label{tab:b6_T8_Tbar10bar_bs}\begin{tabular}{|cc|cc|}\hline\hline
\multicolumn{4}{|c|} {Class \uppercase\expandafter{\romannumeral3}} \\\hline
channel & amplitude&channel & amplitude \\\hline
$X_{u\bar{d}\bar{s}}^{+}\to   \Lambda^0  \overline \Sigma^{\prime+} $ & $ \frac{-2 d_1-d_2-d_3+2 d_4+d_5}{2}$&
$X_{u\bar{d}\bar{s}}^{+}\to   \Sigma^+  \overline \Sigma^{\prime0} $ & $ -\frac{2 d_1-d_2+d_3-2 d_4+3 d_5}{2 \sqrt{3}}$\\\hline
$X_{u\bar{d}\bar{s}}^{+}\to   \Sigma^0  \overline \Sigma^{\prime+} $ & $ \frac{2 d_1+d_2+d_3-2 d_4-d_5}{2 \sqrt{3}}$&
$X_{u\bar{d}\bar{s}}^{+}\to   {p}  \overline \Delta^{0} $ & $ \frac{2 d_1-d_2+d_3-2 d_4+3 d_5}{\sqrt{6}}$\\\hline
$X_{u\bar{d}\bar{s}}^{+}\to   {n}  \overline\Delta^{+} $ & $ \frac{2 d_1+d_2+d_3-2 d_4-d_5}{\sqrt{2}}$&
$X_{u\bar{d}\bar{s}}^{+}\to   \Xi^0  \overline\Xi^{\prime+} $ & $ -\frac{2 d_1+d_2+d_3-2 d_4-d_5}{\sqrt{6}}$\\\hline
$X_{d\bar{s}\bar{u}}^{-}\to   \Lambda^0  \overline\Sigma^{\prime-} $ & $ \frac{6 d_1+3 d_2+d_3-2 d_4+7 d_5}{6}$&
$X_{d\bar{s}\bar{u}}^{-}\to   \Sigma^0  \overline\Sigma^{\prime-} $ & $ \frac{2 d_1+d_2-5 d_3-6 d_4-3 d_5}{2 \sqrt{3}}$\\\hline
$X_{d\bar{s}\bar{u}}^{-}\to   \Sigma^-  \overline \Sigma^{\prime0} $ & $ \frac{2 d_1-d_2-5 d_3-6 d_4+d_5}{2 \sqrt{3}}$&
$X_{d\bar{s}\bar{u}}^{-}\to   {p}  \overline\Delta^{--} $ & $ -\frac{2 d_1+d_2+3 d_3+2 d_4-3 d_5}{\sqrt{2}}$\\\hline
$X_{d\bar{s}\bar{u}}^{-}\to   {n}  \overline\Delta^{-} $ & $ -\frac{2 d_1-d_2+3 d_3+2 d_4+d_5}{\sqrt{6}}$&
$X_{d\bar{s}\bar{u}}^{-}\to   \Xi^-  \overline \Xi^{\prime0} $ & $ \frac{2 d_1+d_2-5 d_3-6 d_4+5 d_5}{\sqrt{6}}$\\
\end{tabular}
\multicolumn{4}{|c|} {Class \uppercase\expandafter{\romannumeral4}} \\\hline
$X_{u\bar{d}\bar{s}}^{+}\to   \Delta^{++}  \overline\Delta^{-} $ & $ -\frac{f_1+f_3+2 f_4+3 f_5+3 f_6}{\sqrt{6}}$&
$X_{u\bar{d}\bar{s}}^{+}\to   \Delta^{+}  \overline \Delta^{0} $ & $ -\frac{ \sqrt{2} \left(f_1+f_3+f_4+3 f_5+f_6\right)}{3}$\\\hline
$X_{u\bar{d}\bar{s}}^{+}\to   \Delta^{0}  \overline\Delta^{+} $ & $ -\frac{f_1+f_3+3 f_5-f_6}{\sqrt{6}}$&
$X_{u\bar{d}\bar{s}}^{+}\to   \Sigma^{\prime+}  \overline \Sigma^{\prime0} $ & $ \frac{-f_1-f_3-f_4-3 f_5-f_6}{3} $\\\hline
$X_{u\bar{d}\bar{s}}^{+}\to   \Sigma^{\prime0}  \overline \Sigma^{\prime+} $ & $ \frac{-f_1-f_3-3 f_5+f_6}{3} $&
$X_{u\bar{d}\bar{s}}^{+}\to   \Xi^{\prime0}  \overline\Xi^{\prime+} $ & $ -\frac{f_1+f_3+3 f_5-f_6}{3 \sqrt{2}}$\\\hline
$X_{d\bar{s}\bar{u}}^{-}\to   \Delta^{+}  \overline\Delta^{--} $ & $ \frac{f_1-f_3-f_5+3 f_6}{\sqrt{6}}$&
$X_{d\bar{s}\bar{u}}^{-}\to   \Delta^{0}  \overline\Delta^{-} $ & $ \frac{ \sqrt{2} \left(f_1-f_3-f_4-f_5+f_6\right)}{3} $\\\hline
$X_{d\bar{s}\bar{u}}^{-}\to   \Delta^{-}  \overline \Delta^{0} $ & $ \frac{f_1-f_3-2 f_4-f_5-f_6}{\sqrt{6}}$&
$X_{d\bar{s}\bar{u}}^{-}\to   \Sigma^{\prime0}  \overline\Sigma^{\prime-} $ & $ \frac{f_1-f_3-f_5-f_6}{3} $\\\hline
$X_{d\bar{s}\bar{u}}^{-}\to   \Sigma^{\prime-}  \overline \Sigma^{\prime0} $ & $ \frac{f_1-f_3-f_4-f_5-3 f_6}{3} $&
$X_{d\bar{s}\bar{u}}^{-}\to   \Xi^{\prime-}  \overline \Xi^{\prime0} $ & $ \frac{f_1-f_3-f_5-5 f_6}{3 \sqrt{2}}$\\\hline
$X_{s\bar{u}\bar{d}}^{-}\to   \Xi^{\prime0}  \overline\Delta^{-} $ & $ -\frac{ \sqrt{2} \left(f_4-2 f_6\right)}{3} $&
$X_{s\bar{u}\bar{d}}^{-}\to   \Xi^{\prime-}  \overline \Delta^{0} $ & $ -\frac{\sqrt{2} \left(f_4-2 f_6\right)}{3} $\\
\hline
\end{tabular}
\end{table}

\begin{table}
\caption{Open bottom tetraquark $X_{b6}$ decays into three light mesons induced by the charmless $b\to d$ transition (part I).}\label{tab:b6_3Md_1}\begin{tabular}{|c|c|c|c|}\hline\hline
channel & amplitude \\\hline
$X_{u\bar{d}\bar{s}}^{+}\to   \pi^+   \pi^0   K^0  $ & $ \frac{1}{2} \left(-c_3-c_4+c_8-c_{12}-c_{13}-c_{14}-c_{15}\right)$\\\hline
$X_{u\bar{d}\bar{s}}^{+}\to   \pi^+   \pi^-   K^+  $ & $ \frac{2 c_1+c_2+2 c_6+c_7+c_8-2 c_9-c_{10}-c_{13}+c_{14}}{\sqrt{2}}$\\\hline
$X_{u\bar{d}\bar{s}}^{+}\to   \pi^+   K^0   \eta  $ & $ -\frac{c_3+c_4-c_8+c_{12}+c_{13}+c_{14}+c_{15}}{2 \sqrt{3}}$\\\hline
$X_{u\bar{d}\bar{s}}^{+}\to   \pi^0   \pi^0   K^+  $ & $ \frac{2 c_1+c_2-c_3-c_4+2 c_6+c_7+2 c_8-2 c_9-c_{10}-c_{12}-2 c_{13}-c_{15}}{\sqrt{2}}$\\\hline
$X_{u\bar{d}\bar{s}}^{+}\to   \pi^0   K^+   \eta  $ & $ \frac{\sqrt{6} \left(c_3+ c_4+c_8- c_{12}- c_{13}- c_{14}- c_{15}\right)}{6} $\\\hline
$X_{u\bar{d}\bar{s}}^{+}\to   K^+   K^+   K^-  $ & $ \sqrt{2} \left(2 c_1+c_2+2 c_6+c_7+c_8-2 c_9-c_{10}-c_{13}+c_{14}\right)$\\\hline
$X_{u\bar{d}\bar{s}}^{+}\to   K^+   K^0   \overline K^0  $ & $ \frac{\sqrt{2} \left(2  c_1+ c_2- c_3-c_4+2  c_6+ c_7-2 c_9- c_{10}+ c_{12}+2 c_{14}+ c_{15}\right)}{2}$\\
\hline
\hline
\end{tabular}
\end{table}
\begin{table}
\caption{Open bottom tetraquark $X_{b6}$ decays into three light mesons induced by the charmless $b\to d$ transition (part II).}\label{tab:b6_3Md_2}\begin{tabular}{|c|c|c|c|}\hline\hline
channel & amplitude \\\hline
$Y_{(u\bar{u},d\bar{d})\bar{s}}^{0}\to   K^+   K^0   K^-  $ & $ \frac{2 c_1+c_2+c_3+c_4+2 c_6+c_7+2 c_8-2 c_9-c_{10}-c_{12}-2 c_{13}-c_{15}}{2}$\\\hline
$Y_{(u\bar{u},d\bar{d})\bar{s}}^{0}\to   K^0   K^0   \overline K^0  $ & $ 2 c_1+c_2-c_3-c_4+2 c_6+c_7-2 c_9-c_{10}+c_{12}+2 c_{14}+c_{15}$\\\hline
$Y_{(u\bar{u},d\bar{d})\bar{s}}^{0}\to   K^0   \eta   \eta  $ & $ \frac{6 c_1+3 c_2-c_3-c_4+6 c_6+3 c_7-6 c_9-3 c_{10}+3 c_{12}+6 c_{14}+3 c_{15}}{6}$\\\hline
$Y_{(d\bar{d},s\bar{s})\bar{u}}^{-}\to   \pi^+   \pi^-   \pi^-  $ & $ -2 c_1-c_2+c_3+c_4+2 c_6+c_7-2 c_9-c_{10}+c_{12}+2 c_{14}+c_{15}$\\\hline
$Y_{(d\bar{d},s\bar{s})\bar{u}}^{-}\to   \pi^0   \pi^0   \pi^-  $ & $ \frac{-2 c_1-c_2+c_3+c_4+2 c_6+c_7-2 c_9-c_{10}+c_{12}+2 c_{14}+c_{15}}{2}$\\\hline
$Y_{(d\bar{d},s\bar{s})\bar{u}}^{-}\to   \pi^0   K^0   K^-  $ & $ \frac{c_3+c_4}{\sqrt{2}}$\\\hline
$Y_{(d\bar{d},s\bar{s})\bar{u}}^{-}\to   \pi^-   K^+   K^-  $ & $ \frac{ -2 c_1-c_2-c_3-c_4+2 c_6+c_7+2 c_8-2 c_9-c_{10}-c_{12}-2 c_{13}-c_{15}}{2}$\\\hline
$Y_{(d\bar{d},s\bar{s})\bar{u}}^{-}\to   \pi^-   K^0   \overline K^0  $ & $ \frac{-2 c_1-c_2+c_3+c_4+2 c_6+c_7+2 c_8-2 c_9-c_{10}-c_{12}-2 c_{13}-c_{15}}{2}$\\\hline
$Y_{(d\bar{d},s\bar{s})\bar{u}}^{-}\to   \pi^-   \eta   \eta  $ & $ \frac{-6 c_1-3 c_2-5 c_3-5 c_4+6 c_6+3 c_7-6 c_9-3 c_{10}+3 c_{12}+6 c_{14}+3 c_{15}}{6}$\\\hline
$Y_{(d\bar{d},s\bar{s})\bar{u}}^{-}\to   K^0   K^-   \eta  $ & $ -\frac{c_3+c_4+2 c_8-2 c_{12}-2 c_{13}-2 c_{14}-2 c_{15}}{\sqrt{6}}$\\\hline
\hline
\end{tabular}
\end{table}
\begin{table}
\caption{Open bottom tetraquark $X_{b6}$ decays into three light mesons induced by the charmless $b\to s$ transition (part I).}\label{tab:b6_3Ms_1}\begin{tabular}{|c|c|c|c|}\hline\hline
channel & amplitude  \\ \hline
$X_{u\bar{d}\bar{s}}^{+}\to   \pi^+   \pi^+   \pi^-  $ & $ \sqrt{2}(-2  c_1- c_2-2  c_6- c_7-c_8+2 c_9+ c_{10}+ c_{13}- c_{14})$\\\hline
$X_{u\bar{d}\bar{s}}^{+}\to   \pi^+   \pi^0   \pi^0  $ & $ \frac{-2 c_1-c_2-2  c_6- c_7- c_8+2 c_9+ c_{10}+c_{13}- c_{14}}{\sqrt{2}} $\\\hline
$X_{u\bar{d}\bar{s}}^{+}\to   \pi^+   \pi^0   \eta  $ & $ \sqrt{\frac{2}{3}} \left(-c_8+c_{12}+c_{13}+c_{14}+c_{15}\right)$\\\hline
$X_{u\bar{d}\bar{s}}^{+}\to   \pi^+   K^+   K^-  $ & $ \frac{-2 c_1- c_2-2  c_6- c_7- c_8+2  c_9+ c_{10}+c_{13}- c_{14}}{\sqrt{2}} $\\\hline
$X_{u\bar{d}\bar{s}}^{+}\to   \pi^+   K^0   \overline K^0  $ & $ \frac{-2  c_1-  c_2+ c_3+ c_4-2 c_6-  c_7+2  c_9+ c_{10}- c_{12}-2 c_{14}- c_{15}}{\sqrt{2}}$\\\hline
$X_{u\bar{d}\bar{s}}^{+}\to   \pi^+   \eta   \eta  $ & $ \frac{-6 c_1-3 c_2+4 c_3+4 c_4-6 c_6-3 c_7-3 c_8+6 c_9+3 c_{10}+3 c_{13}-3 c_{14}}{3 \sqrt{2}}$\\\hline
$X_{u\bar{d}\bar{s}}^{+}\to   \pi^0   K^+   \overline K^0  $ & $ \frac{c_3+c_4-c_8+c_{12}+c_{13}+c_{14}+c_{15}}{2}$\\
\hline
\hline
\end{tabular}
\end{table}
\begin{table}
\caption{Open bottom tetraquark $X_{b6}$ decays into three light mesons induced by the charmless $b\to s$ transition (part II).}\label{tab:b6_3Ms_2}\begin{tabular}{|c|c|c|c|}\hline\hline
channel & amplitude  \\ \hline
$Y_{(u\bar{u},s\bar{s})\bar{d}}^{0}\to   \pi^+   \pi^-   \overline K^0  $ & $ \frac{-2 c_1-c_2-c_3-c_4-2 c_6-c_7-2 c_8+2 c_9+c_{10}+c_{12}+2 c_{13}+c_{15}}{2}$\\\hline
$Y_{(u\bar{u},s\bar{s})\bar{d}}^{0}\to   \pi^+   K^-   \eta  $ & $ \frac{2 c_3+2 c_4-c_8+c_{12}+c_{13}+c_{14}+c_{15}}{\sqrt{6}}$\\\hline
$Y_{(u\bar{u},s\bar{s})\bar{d}}^{0}\to   \pi^0   \pi^0   \overline K^0  $ & $ \frac{-2 c_1-c_2-c_3-c_4-2 c_6-c_7+2 c_9+c_{10}-c_{12}-2 c_{14}-c_{15}}{2}$\\\hline
$Y_{(u\bar{u},s\bar{s})\bar{d}}^{0}\to   \pi^0   \overline K^0   \eta  $ & $ -\frac{c_3+c_4}{\sqrt{3}}$\\\hline
$Y_{(u\bar{u},s\bar{s})\bar{d}}^{0}\to   K^+   \overline K^0   K^-  $ & $ \frac{-2 c_1-c_2+c_3+c_4-2 c_6-c_7-2 c_8+2 c_9+c_{10}+c_{12}+2 c_{13}+c_{15}}{2} $\\\hline
$Y_{(u\bar{u},s\bar{s})\bar{d}}^{0}\to   K^0   \overline K^0   \overline K^0  $ & $ -2 c_1-c_2+c_3+c_4-2 c_6-c_7+2 c_9+c_{10}-c_{12}-2 c_{14}-c_{15}$\\\hline
$Y_{(u\bar{u},s\bar{s})\bar{d}}^{0}\to   \overline K^0   \eta   \eta  $ & $ \frac{-6 c_1-3 c_2+c_3+c_4-6 c_6-3 c_7+6 c_9+3 c_{10}-3 c_{12}-6 c_{14}-3 c_{15}}{6} $\\
\hline
\hline
\end{tabular}
\end{table}

\subsubsection{Three-body decays into mesons}
The effective hadron-level Hamiltonian for $X_{b6}$ decays into three light mesons is written as
\begin{eqnarray}
 {\cal H}_{eff}&=&c_1(X_{b6})_{[jk]}^i (H_{3})^j  M^k_i M^l_m M^m_l +c_2 (X_{b6})_{[jk]}^i (H_{3})^j  M^l_i M^k_m M^m_l  \nonumber\\
 && + c_3 (X_{b6})_{[jk]}^i (H_{3})^l  M^j_i M^k_m M^m_l +c_4 (X_{b6})_{[jk]}^i (H_{3})^l  M^m_i M^j_l M^k_m \nonumber\\
 &&-  c_5 (X_{b6})^{\{ij\}} (H_{\overline6})_{\{ij\}}  M^k_l M^l_m M^m_k  -c_6 (X_{b6})^{\{ij\}} (H_{\overline6})_{\{ik\}}  M^k_j M^l_m M^m_l\nonumber\\
 &&-c_7 (X_{b6})^{\{ij\}} (H_{\overline6})_{\{il\}}  M^k_j M^l_m M^m_k  -c_8 (X_{b6})^{\{ij\}} (H_{\overline6})_{\{lm\}}  M^k_i M^l_j M^m_k \nonumber\\
 &&+ c_{9} (X_{b6})_{[jk]}^i (H_{15})^{\{jl\}}_{i}  M^k_l M^m_o M^o_m+ c_{10} (X_{b6})_{[jk]}^i (H_{15})^{\{jl\}}_{i}  M^k_o M^m_l M^o_m \nonumber\\
 &&+c_{11} (X_{b6})_{[jk]}^i (H_{15})^{\{lm\}}_{i}  M^j_o M^k_l M^o_m+ c_{12} (X_{b6})_{[jk]}^i (H_{15})^{\{jl\}}_{o}  M^k_i M^m_l M^o_m \nonumber\\
 &&+c_{13} (X_{b6})_{[jk]}^i (H_{15})^{\{jl\}}_{o}  M^m_i M^k_l M^o_m+c_{14} (X_{b6})_{[jk]}^i (H_{15})^{\{jl\}}_{m}  M^m_i M^k_o M^o_l \nonumber\\
 &&+c_{15} (X_{b6})_{[jk]}^i (H_{15})^{\{lm\}}_{o}  M^j_i M^k_l M^o_m .
\end{eqnarray}
The decay amplitudes are given in Tab.~\ref{tab:b6_3Md_1} and Tab.~\ref{tab:b6_3Md_2} for the $b\to d$  transition, and Tab.~\ref{tab:b6_3Ms_1} and Tab.~\ref{tab:b6_3Ms_2} for the  $b\to s$ transition. The corresponding relations of decay widths are
\begin{eqnarray*}
2\Gamma(X_{u\bar{d}\bar{s}}^{+}\to   \pi^+   \pi^0   K^0 )=6\Gamma(X_{u\bar{d}\bar{s}}^{+}\to   \pi^+   K^0   \eta )=3\Gamma(X_{u\bar{d}\bar{s}}^{+}\to   \pi^0   K^+   \eta )=2\Gamma(X_{d\bar{s}\bar{u}}^{-}\to   \pi^0   \pi^-   K^0 )\\
=\frac{1}{2}\Gamma(X_{d\bar{s}\bar{u}}^{-}\to   \pi^-   \pi^-   K^+ )=\Gamma(Y_{(u\bar{u},d\bar{d})\bar{s}}^{0}\to   \pi^0   \pi^-   K^+ )=\frac{3}{2}\Gamma(Y_{(u\bar{u},d\bar{d})\bar{s}}^{0}\to   \pi^-   K^+   \eta  ),\\
\frac{1}{3}\Gamma(X_{d\bar{s}\bar{u}}^{-}\to   K^0   K^0   K^-)=2\Gamma(X_{s\bar{u}\bar{d}}^{-}\to   \pi^0   K^-   \eta)=\Gamma(X_{s\bar{u}\bar{d}}^{-}\to   \pi^-   \overline K^0   \eta),\\
\Gamma(X_{s\bar{u}\bar{d}}^{-}\to   \pi^+   \pi^-   K^-)=2\Gamma(X_{s\bar{u}\bar{d}}^{-}\to   \pi^0   \pi^0   K^-)=4\Gamma(Y_{(u\bar{u},s\bar{s})\bar{d}}^{0}\to   \pi^+   \pi^0   \pi^-)=8\Gamma(Y_{(u\bar{u},s\bar{s})\bar{d}}^{0}\to   \pi^0   \pi^0   \pi^0)\\
=\Gamma(Y_{(d\bar{d},s\bar{s})\bar{u}}^{-}\to   \pi^+   \pi^-   \pi^-)=4\Gamma(Y_{(d\bar{d},s\bar{s})\bar{u}}^{-}\to   \pi^0   \pi^0   \pi^-),\\
\Gamma(X_{u\bar{d}\bar{s}}^{+}\to   \pi^0   \pi^0   K^+)=\Gamma(Y_{(u\bar{u},d\bar{d})\bar{s}}^{0}\to   \pi^+   \pi^-   K^0),
\Gamma(X_{u\bar{d}\bar{s}}^{+}\to   K^+   K^0   \overline K^0)=\Gamma(Y_{(u\bar{u},d\bar{d})\bar{s}}^{0}\to   K^0   K^0   \overline K^0),\\
\Gamma(X_{u\bar{d}\bar{s}}^{+}\to   K^+   \eta   \eta)=2\Gamma(Y_{(u\bar{u},d\bar{d})\bar{s}}^{0}\to   K^0   \eta   \eta),
\Gamma(Y_{(u\bar{u},s\bar{s})\bar{d}}^{0}\to   \pi^+   \pi^-   \eta)=2\Gamma(Y_{(u\bar{u},s\bar{s})\bar{d}}^{0}\to   \pi^0   \pi^0   \eta),\\
\Gamma(Y_{(u\bar{u},s\bar{s})\bar{d}}^{0}\to   \pi^0   \eta   \eta)=\frac{1}{2}\Gamma(Y_{(d\bar{d},s\bar{s})\bar{u}}^{-}\to   \pi^-   \eta   \eta),
\frac{3}{2}\Gamma(Y_{(u\bar{u},d\bar{d})\bar{s}}^{0}\to   \pi^0   K^0   \eta)=\Gamma(Y_{(d\bar{d},s\bar{s})\bar{u}}^{-}\to   \pi^0   K^0   K^-),\\
\Gamma(X_{u\bar{d}\bar{s}}^{+}\to   \pi^+   \pi^-   K^+)=\frac{1}{2}\Gamma(X_{u\bar{d}\bar{s}}^{+}\to   K^+   K^+   K^-),
\frac{1}{2}\Gamma(X_{s\bar{u}\bar{d}}^{-}\to   K^+   K^-   K^-)=\Gamma(X_{s\bar{u}\bar{d}}^{-}\to   K^0   \overline K^0   K^-).
\end{eqnarray*}
and
\begin{eqnarray*}
\Gamma(X_{u\bar{d}\bar{s}}^{+}\to   \pi^+   \pi^+   \pi^- )=4\Gamma(X_{u\bar{d}\bar{s}}^{+}\to   \pi^+   \pi^0   \pi^0 )=2\Gamma(X_{u\bar{d}\bar{s}}^{+}\to   \pi^+   K^+   K^- ),\\
\frac{3}{4}\Gamma(X_{u\bar{d}\bar{s}}^{+}\to   \pi^+   \pi^0   \eta )=\frac{3}{4}\Gamma(X_{d\bar{s}\bar{u}}^{-}\to   \pi^0   \pi^-   \eta  )=\frac{3}{2}\Gamma(X_{s\bar{u}\bar{d}}^{-}\to   \overline K^0   K^-   \eta  )=\Gamma(Y_{(u\bar{u},s\bar{s})\bar{d}}^{0}\to   \pi^+   \pi^0   K^-  )\\
=\Gamma(Y_{(d\bar{d},s\bar{s})\bar{u}}^{-}\to   \pi^0   \pi^-   \overline K^0 ),\  \
\Gamma(X_{u\bar{d}\bar{s}}^{+}\to   \pi^0   K^+   \overline K^0  )=3\Gamma(X_{u\bar{d}\bar{s}}^{+}\to   K^+   \overline K^0   \eta  )=\frac{1}{4}\Gamma(X_{s\bar{u}\bar{d}}^{-}\to   \pi^+   K^-   K^-  ),\\
\Gamma(X_{d\bar{s}\bar{u}}^{-}\to   \pi^0   K^0   K^-  )=3\Gamma(X_{d\bar{s}\bar{u}}^{-}\to   K^0   K^-   \eta  )=\frac{1}{4}\Gamma(X_{s\bar{u}\bar{d}}^{-}\to   \pi^-   \overline K^0   \overline K^0  ),\\
\frac{1}{3}\Gamma(X_{s\bar{u}\bar{d}}^{-}\to   \pi^0   \overline K^0   K^- )=\Gamma(Y_{(u\bar{u},s\bar{s})\bar{d}}^{0}\to   \pi^0   \overline K^0   \eta  )=\Gamma(Y_{(d\bar{d},s\bar{s})\bar{u}}^{-}\to   \pi^0   K^-   \eta  ),\\
\Gamma(X_{u\bar{d}\bar{s}}^{+}\to   \pi^+   K^0   \overline K^0 )=\Gamma(Y_{(u\bar{u},s\bar{s})\bar{d}}^{0}\to   K^0   \overline K^0   \overline K^0 ),
\Gamma(X_{d\bar{s}\bar{u}}^{-}\to   \pi^+   \pi^-   \pi^- )=4\Gamma(X_{d\bar{s}\bar{u}}^{-}\to   \pi^0   \pi^0   \pi^- ),\\
\Gamma(Y_{(u\bar{u},d\bar{d})\bar{s}}^{0}\to   \pi^+   \pi^0   \pi^- )=\frac{2}{3}\Gamma(Y_{(u\bar{u},d\bar{d})\bar{s}}^{0}\to   \pi^0   \pi^0   \pi^0 ),
\Gamma(X_{d\bar{s}\bar{u}}^{-}\to   \pi^-   K^+   K^- )=6\Gamma(Y_{(d\bar{d},s\bar{s})\bar{u}}^{-}\to   K^+   K^-   K^-  ).
\end{eqnarray*}


\section{Golden  $X_{b6}$ Decay Channels}
\label{sec:golden_channels}
The golden decay channels will have a large branching ratios and also have relatively clear final products, which are very helpful to hunting for the stable $X_{b6}$ states~\cite{Yu:2017zst}.  In this section, we will list the golden channels of $X_{b6}$ decays. Under the SU(3) flavor analysis, the conclusions will not be changed when one hadron is replaced by the excited state with the same quark structure.  For instance one may  replace  the pseudoscalar $\overline K^0$ by the vector $\overline K^{*0}$ which decays into $K^-\pi^+$. Considering some neutral mesons and baryons are very difficult to reconstruct at large hadron colliders, the channels including the $\pi^0, \eta, \rho^0, \omega$ and $n,\bar n$ are not employed.

\begin{table}
\footnotesize
 \caption{Cabibbo allowed $X_{b6}$ decays, in which $\bar{K}^0$ can be replaced by the vector meson $\bar{K}^{*0}$.  }\label{tab:Xb6_golden_meson}\begin{tabular}{|c  c   c  c |}\hline\hline
 \multicolumn{4}{|l|}{\qquad \textbf{Two body decays}}\\\hline
$X_{u\bar d\bar s}^+ \to \pi^+ J/\psi$ &
$X_{u\bar d\bar s}^+ \to D^+ \overline D^0 $&
$X_{u\bar d\bar s}^+ \to D^0 K^+$&
$ X_{u\bar d\bar s}^+ \to D^+ K^0$ \\
$X_{u\bar d\bar s}^+ \to \overline D^0 \pi^+ $ &&&\\

\hline
$X_{d\bar s\bar u}^- \to \pi^- J/\psi$ &
$X_{d\bar s\bar u}^- \to D^0 D^-$ &
$X_{d\bar s\bar u}^- \to \overline D^0 \pi^-$&
$X_{d\bar s\bar u}^- \to D_s^- K^0$\\

\hline
$X_{s\bar u\bar d}^- \to D^0 K^- $&
$X_{s\bar u\bar d}^- \to D_s^- \overline K^0$&&\\

\hline
$Y_{\{u\bar u,s\bar s\}\bar d}^0 \to \overline K^0  J/\psi $&
$ Y_{\{u\bar u,s\bar s\}\bar d}^0 \to D^+  D_s^-$&
$ Y_{\{u\bar u,s\bar s\}\bar d}^0 \to D^+ \pi^- $&
$ Y_{\{u\bar u,s\bar s\}\bar d}^0 \to D_s^+ K^-$ \\
$Y_{\{u\bar u,s\bar s\}\bar d}^0 \to \overline D^0 \overline K^0$&
$Y_{\{u\bar u,s\bar s\}\bar d}^0 \to D_s^- \pi^+$&&\\

\hline

$Y_{\{u\bar u ,d\bar d\}\bar s}^0 \to D^+ D^- $&
$Y_{\{u\bar u ,d\bar d\}\bar s}^0 \to D^0 \overline D^0 $&
$Y_{\{u\bar u ,d\bar d\}\bar s}^0 \to D^0 K^0 $&
$ Y_{\{u\bar u ,d\bar d\}\bar s}^0 \to D_s^+ \pi^-$\\
$Y_{\{u\bar u ,d\bar d\}\bar s}^0 \to D^- \pi^+$&
$Y_{\{u\bar u ,d\bar d\}\bar s}^0 \to D_s^- K^+$&&\\

\hline
$Y_{\{d\bar d ,s\bar s\}\bar u}^- \to K^- J/\psi $&
$Y_{\{d\bar d ,s\bar s\}\bar u}^- \to D^0 D_s^-$&
$Y_{\{d\bar d ,s\bar s\}\bar u}^- \to \overline D^0 K^- $&
$Y_{\{d\bar d ,s\bar s\}\bar u}^- \to  D^- \overline K^0$  \\

\hline\hline

 \multicolumn{4}{|l|}{\qquad \textbf{Three body decays}}\\\hline

$X_{u\bar d\bar s}^+ \to  D^0 \pi^+ \overline D^0 $&
$X_{u\bar d\bar s}^+ \to D^+ \pi^+ D^-$&
$X_{u\bar d\bar s}^+ \to D_s^+ \pi^+ D_s^-$&
$X_{u\bar d\bar s}^+ \to  D^+ K^+ D_s^-$\\
$X_{u\bar d\bar s}^+ \to  D_s^+ \overline K^0 \overline D^0$&
$X_{u\bar d\bar s}^+ \to D^0 \pi^+ K^0$&
$X_{u\bar d\bar s}^+ \to D^+ \pi^- K^0$&
$X_{u\bar d\bar s}^+ \to D_s^+ \pi^+\pi^- $\\
$X_{u\bar d\bar s}^+ \to D_s^+ K^+K^- $&
$X_{u\bar d\bar s}^+ \to D_s^+ K^0 \overline K^0$&
$X_{u\bar d\bar s}^+ \to \overline D^0 K^+ \overline K^0$&
$X_{u\bar d\bar s}^+ \to  D^- \pi^+\pi^+$\\
$X_{u\bar d\bar s}^+ \to D_s^- \pi^+ K^+$&
$X_{u\bar d\bar s}^+ \to K^+ \overline K^0 J/\psi$ &$X_{u\bar{d}\bar{s}}^{+}\to \pi^+    D^+_s l^-\bar\nu $&
$X_{u\bar{d}\bar{s}}^{+}\to K^+    D^+ l^-\bar\nu $ \\
\hline
$X_{d\bar s\bar u}^- \to D_s^+ \pi^- D_s^- $&
$X_{d\bar s\bar u}^- \to D^0 \pi^- \overline D^0 $ &
$X_{d\bar s\bar u}^- \to D^0 K^0 D_s^- $&
$X_{d\bar s\bar u}^- \to D^+ \pi^- D^- $\\
$X_{d\bar s\bar u}^- \to D_s^+ K^- D^- $&
$X_{d\bar s\bar u}^- \to D^0 \pi^- K^0$&
$X_{d\bar s\bar u}^- \to D_s^+ \pi^-\pi^- $&
$X_{d\bar s\bar u}^- \to \overline D^0 K^0 K^-$\\
$X_{d\bar s\bar u}^- \to D^- \pi^+\pi^-$&
$X_{d\bar s\bar u}^- \to D^- K^+ K^- $&
$X_{d\bar s\bar u}^- \to D_s^- \pi^- K^+$&
$X_{d\bar s\bar u}^- \to D^- K^0 \overline K^0$\\
$X_{d\bar s\bar u}^- \to K^0 K^- J/\psi $&$X_{d\bar{s}\bar{u}}^{-}\to \pi^-    D^+_s l^-\bar\nu $&
$X_{d\bar{s}\bar{u}}^{-}\to K^0    D^0 l^-\bar\nu $ &\\
\hline
$X_{s\bar u\bar d}^- \to D^0 \overline K^0 D_s^-$&
$X_{s\bar u\bar d}^- \to D^+ K^- D_s^-$&
$X_{s\bar u\bar d}^- \to D^0 \pi^- \overline K^0$&
$X_{s\bar u\bar d}^- \to D^+ \pi^- K^-$\\
$X_{s\bar u\bar d}^- \to \overline D^0 \overline K^0 K^-$&
$X_{s\bar u\bar d}^- \to D^- \overline K^0 \overline K^0$&
$X_{s\bar u\bar d}^- \to D_s^- \pi^+ K^-$&$X_{s\bar{u}\bar{d}}^{-}\to K^- \overline K^0    J/\psi $ \\
$X_{s\bar{u}\bar{d}}^{-}\to K^-    D^+ l^-\bar\nu $&$X_{s\bar{u}\bar{d}}^{-}\to \overline K^0    D^0 l^-\bar\nu $&&\\

\hline
$Y_{\{u\bar u,s\bar s\}\bar d}^0 \to D^0 \overline K^0 \overline D^0 $&
$Y_{\{u\bar u,s\bar s\}\bar d}^0 \to D_s^+ \overline K^0  D_s^- $&
$Y_{\{u\bar u,s\bar s\}\bar d}^0 \to D^+ \overline K^0 D^- $ &
$Y_{\{u\bar u,s\bar s\}\bar d}^0 \to D^+ K^- \overline D^0 $\\
$Y_{\{u\bar u,s\bar s\}\bar d}^0 \to D^0 \pi^+ D_s^-$&
$ Y_{\{u\bar u,s\bar s\}\bar d}^0 \to D^0 \pi^+\pi^- $&
$Y_{\{u\bar u,s\bar s\}\bar d}^0 \to D^0 K^+K^- $&
$Y_{\{u\bar u,s\bar s\}\bar d}^0 \to D^0 K^0 \overline K^0$\\
$Y_{\{u\bar u,s\bar s\}\bar d}^0 \to D^+ K^0 K^- $&
$ Y_{\{u\bar u,s\bar s\}\bar d}^0 \to D_s^+ \pi^- \overline K^0 $&
$Y_{\{u\bar u,s\bar s\}\bar d}^0 \to  \overline D^0 \pi^+ K^-$&
$Y_{\{u\bar u,s\bar s\}\bar d}^0 \to D^-\pi^+ \overline K^0$\\
$Y_{\{u\bar u,s\bar s\}\bar d}^0 \to D_s^- K^+ \overline K^0 $ &
$Y_{\{u\bar u,s\bar s\}\bar d}^0 \to  K^- \pi^+ J/\psi $ &
$Y_{(u\bar{u},s\bar{s})\bar{d}}^{0}\to \pi^+    D^0 l^-\bar\nu $ &
$Y_{(u\bar{u},s\bar{s})\bar{d}}^{0}\to \pi^0    D^+ l^-\bar\nu $ \\
$Y_{(u\bar{u},s\bar{s})\bar{d}}^{0}\to \overline K^0    D^+_s l^-\bar\nu $&
$Y_{(u\bar{u},s\bar{s})\bar{d}}^{0}\to \eta    D^+ l^-\bar\nu $ &&\\
\hline

$Y_{\{u\bar u ,d\bar d\}\bar s}^0 \to D^0 \pi^+ D^-$&
$Y_{\{u\bar u,d\bar d\}\bar s}^0 \to  D^0 K^+ D_s^-$&
$Y_{\{u\bar u,d\bar d\}\bar s}^0 \to  D^+ \pi^- \overline D^0$ &
$Y_{\{u\bar u,d\bar d\}\bar s}^0 \to  D^+ K^0 D_s^-$\\
$Y_{\{u\bar u,d\bar d\}\bar s}^0 \to  D_s^+ \overline K^0 D^-$&
$Y_{\{u\bar u,d\bar d\}\bar s}^0 \to  D_s^+ K^- \overline D^0$&
$Y_{\{u\bar u ,d\bar d\}\bar s}^0 \to D^0 \pi^- K^+$&
$Y_{\{u\bar u ,d\bar d\}\bar s}^0 \to D^+ \pi^- K^0$\\
$Y_{\{u\bar u ,d\bar d\}\bar s}^0 \to D_s^+ K^0 K^-$&
$Y_{\{u\bar u,d\bar d\}\bar s}^0 \to  \overline D^0 \pi^+ \pi^-$&
$Y_{\{u\bar u,d\bar d\}\bar s}^0 \to \overline D^0 K^+ K^-$&
$Y_{\{u\bar u,d\bar d\}\bar s}^0 \to \overline D^0 K^0 \overline K^0 $\\
$Y_{\{u\bar u,d\bar d\}\bar s}^0 \to D^- K^+ \overline K^0$&
$Y_{\{u\bar u,d\bar d\}\bar s}^0 \to D_s^- \pi^+ K^0 $&
$Y_{\{u\bar u,d\bar d\}\bar s}^0 \to K^+ K^- J/\psi$&
$Y_{\{u\bar u,d\bar d\}\bar s}^0 \to  K^0 \overline K^0 J/\psi$\\
$Y_{(u\bar{u},d\bar{d})\bar{s}}^{0}\to \pi^0    D^+_s l^-\bar\nu $ &
$Y_{(u\bar{u},d\bar{d})\bar{s}}^{0}\to K^+    D^0 l^-\bar\nu $&
$Y_{(u\bar{u},d\bar{d})\bar{s}}^{0}\to K^0    D^+ l^-\bar\nu $ &\\
\hline
$Y_{\{d\bar d ,s\bar s\}\bar u}^- \to D^+ \pi^- D_s^- $&
$Y_{\{d\bar d ,s\bar s\}\bar u}^- \to  D^0 \overline K^0 D^-$&
$Y_{\{d\bar d ,s\bar s\}\bar u}^- \to  D^0 K^- \overline D^0$&
$Y_{\{d\bar d ,s\bar s\}\bar u}^- \to  D^+ K^- D^-$\\
$Y_{\{d\bar d ,s\bar s\}\bar u}^- \to  D_s^+ K^- D_s^-$&
$Y_{\{d\bar d ,s\bar s\}\bar u}^- \to D^0 K^0 K^-$&
$Y_{\{d\bar d ,s\bar s\}\bar u}^- \to D^+ \pi^- \pi^- $&
$Y_{\{d\bar d ,s\bar s\}\bar u}^- \to D_s+\pi^- K^-$\\
$Y_{\{d\bar d ,s\bar s\}\bar u}^- \to \overline D^0 \pi^- \overline K^0$&
$Y_{\{d\bar d ,s\bar s\}\bar u}^- \to D^- \pi^+ K^- $&
$Y_{\{d\bar d ,s\bar s\}\bar u}^- \to D_s^- \pi^+ \pi^- $&
$Y_{\{d\bar d ,s\bar s\}\bar u}^- \to D_s^- K^+ K^-$\\
$Y_{\{d\bar d ,s\bar s\}\bar u}^- \to D_s^- K^0 \overline K^0$&
$Y_{\{d\bar d ,s\bar s\}\bar u}^- \to  K^0 K^- J/\psi$ &$Y_{(d\bar{d},s\bar{s})\bar{u}}^{-}\to \pi^0    D^0 l^-\bar\nu $ &
$Y_{(d\bar{d},s\bar{s})\bar{u}}^{-}\to \pi^-    D^+ l^-\bar\nu $ \\
$Y_{(d\bar{d},s\bar{s})\bar{u}}^{-}\to K^-    D^+_s l^-\bar\nu $ &
$Y_{(d\bar{d},s\bar{s})\bar{u}}^{-}\to \eta    D^0 l^-\bar\nu $ &&\\
\hline
\end{tabular}
\end{table}

\begin{table}
 \caption{Cabibbo allowed $X_{b6}$ decays in which $\bar{K}^0$ can be replaced by the vector meson $\bar{K}^{*0}$.  }\label{tab:Xb6_golden_baryon}\begin{tabular}{|c  c   c  c  c|}\hline\hline
 \multicolumn{5}{|l|}{\qquad \textbf{Two body decays}}\\\hline
$X_{u\bar{d}\bar{s}}^{+}\to   \Xi_c^+  \overline \Xi_{\bar{c}}^0 $&
$X_{u\bar{d}\bar{s}}^{+}\to   \Sigma_{c}^{++}  \overline \Lambda_{\bar{c}}^- $&
$X_{u\bar{d}\bar{s}}^{+}\to   \Xi_{c}^{\prime+}  \overline \Xi_{\bar{c}}^0 $&
$X_{u\bar{d}\bar{s}}^{+}\to   \Lambda_c^+  \overline \Sigma_{\bar{c}}^{0} $&
$X_{u\bar{d}\bar{s}}^{+}\to   \Xi_c^+  \overline \Xi_{\bar{c}}^{\prime0} $\\
$X_{u\bar{d}\bar{s}}^{+}\to   \Sigma_{c}^{++}  \overline \Sigma_{\bar{c}}^- $&
$X_{u\bar{d}\bar{s}}^{+}\to   \Sigma_{c}^{+}  \overline \Sigma_{\bar{c}}^{0} $&
$X_{u\bar{d}\bar{s}}^{+}\to   \Xi_{c}^{\prime+}  \overline \Xi_{\bar{c}}^{\prime0} $&
$X_{u\bar{d}\bar{s}}^{+}\to   \Lambda_c^+  \overline \Lambda^0 $&
$X_{u\bar{d}\bar{s}}^{+}\to   \Lambda_c^+  \overline \Sigma^0 $\\
$X_{u\bar{d}\bar{s}}^{+}\to   \Xi_c^+  \overline \Xi^0 $&
$X_{u\bar{d}\bar{s}}^{+}\to   \Xi_c^0  \overline \Xi^+ $&
$X_{u\bar{d}\bar{s}}^{+}\to   \Sigma_{c}^{++}  \overline \Sigma^- $&
$X_{u\bar{d}\bar{s}}^{+}\to   \Sigma_{c}^{+}  \overline \Lambda^0 $&
$X_{u\bar{d}\bar{s}}^{+}\to   \Sigma_{c}^{+}  \overline \Sigma^0 $\\
$X_{u\bar{d}\bar{s}}^{+}\to   \Sigma_{c}^{0}  \overline \Sigma^+ $&
$X_{u\bar{d}\bar{s}}^{+}\to   \Xi_{c}^{\prime+}  \overline \Xi^0 $&
$X_{u\bar{d}\bar{s}}^{+}\to   \Xi_{c}^{\prime0}  \overline \Xi^+ $&
$X_{u\bar{d}\bar{s}}^{+}\to   \Lambda_c^+  \overline \Sigma^{\prime0} $&
$X_{u\bar{d}\bar{s}}^{+}\to   \Xi_c^+  \overline \Xi^{\prime0} $\\
$X_{u\bar{d}\bar{s}}^{+}\to   \Sigma_{c}^{++}  \overline \Sigma^{\prime-} $&
$X_{u\bar{d}\bar{s}}^{+}\to   \Sigma_{c}^{+}  \overline \Sigma^{\prime0} $&
$X_{u\bar{d}\bar{s}}^{+}\to   \Sigma_{c}^{0}  \overline \Sigma^{\prime+} $&
$X_{u\bar{d}\bar{s}}^{+}\to   \Xi_{c}^{\prime+}  \overline \Xi^{\prime0} $&
$X_{u\bar{d}\bar{s}}^{+}\to   \Xi_{c}^{\prime0}  \overline \Xi^{\prime+} $\\
$X_{u\bar{d}\bar{s}}^{+}\to   \Omega_{c}^{0}  \overline \Omega^+ $&
$X_{u\bar{d}\bar{s}}^{+}\to   \Sigma^+  \overline \Xi_{\bar{c}}^0 $&
$X_{u\bar{d}\bar{s}}^{+}\to   \Sigma^+  \overline \Xi_{\bar{c}}^{\prime0} $&
$X_{u\bar{d}\bar{s}}^{+}\to   {p}  \overline \Sigma_{\bar{c}}^{0} $&
$X_{u\bar{d}\bar{s}}^{+}\to   \Delta^{++}  \overline \Lambda_{\bar{c}}^- $\\
$X_{u\bar{d}\bar{s}}^{+}\to   \Sigma^{\prime+}  \overline \Xi_{\bar{c}}^0 $&
$X_{u\bar{d}\bar{s}}^{+}\to   \Delta^{++}  \overline \Sigma_{\bar{c}}^- $&
$X_{u\bar{d}\bar{s}}^{+}\to   \Delta^{+}  \overline \Sigma_{\bar{c}}^{0} $&
$X_{u\bar{d}\bar{s}}^{+}\to   \Sigma^{\prime+}  \overline \Xi_{\bar{c}}^{\prime0} $&
 \\
\hline

$X_{d\bar{s}\bar{u}}^{-}\to   \Xi_c^0  \overline \Xi_{\bar{c}}^- $&
$X_{d\bar{s}\bar{u}}^{-}\to   \Sigma_{c}^{0}  \overline \Lambda_{\bar{c}}^- $&
$X_{d\bar{s}\bar{u}}^{-}\to   \Xi_{c}^{\prime0}  \overline \Xi_{\bar{c}}^- $&
$X_{d\bar{s}\bar{u}}^{-}\to   \Lambda_c^+  \overline \Sigma_{\bar{c}}^{--} $&
$X_{d\bar{s}\bar{u}}^{-}\to   \Xi_c^0  \overline \Xi_{\bar{c}}^{\prime-} $\\
$X_{d\bar{s}\bar{u}}^{-}\to   \Sigma_{c}^{+}  \overline \Sigma_{\bar{c}}^{--} $&
$X_{d\bar{s}\bar{u}}^{-}\to   \Sigma_{c}^{0}  \overline \Sigma_{\bar{c}}^- $&
$X_{d\bar{s}\bar{u}}^{-}\to   \Xi_{c}^{\prime0}  \overline \Xi_{\bar{c}}^{\prime-} $&
$X_{d\bar{s}\bar{u}}^{-}\to   \Sigma_{c}^{0}  \overline \Sigma^- $&
$X_{d\bar{s}\bar{u}}^{-}\to   \Lambda^0  \overline \Xi_{\bar{c}}^- $\\
$X_{d\bar{s}\bar{u}}^{-}\to   \Sigma^0  \overline \Xi_{\bar{c}}^- $&
$X_{d\bar{s}\bar{u}}^{-}\to   \Sigma^-  \overline \Xi_{\bar{c}}^0 $&
$X_{d\bar{s}\bar{u}}^{-}\to   \Lambda^0  \overline \Xi_{\bar{c}}^{\prime-} $&
$X_{d\bar{s}\bar{u}}^{-}\to   \Sigma^0  \overline \Xi_{\bar{c}}^{\prime-} $&
$X_{d\bar{s}\bar{u}}^{-}\to   \Sigma^-  \overline \Xi_{\bar{c}}^{\prime0} $\\
$X_{d\bar{s}\bar{u}}^{-}\to   {p}  \overline \Sigma_{\bar{c}}^{--} $&
$X_{d\bar{s}\bar{u}}^{-}\to   \Xi^-  \overline \Omega_{\bar{c}}^{0} $&
$X_{d\bar{s}\bar{u}}^{-}\to   \Delta^{0}  \overline \Lambda_{\bar{c}}^- $&
$X_{d\bar{s}\bar{u}}^{-}\to   \Sigma^{\prime0}  \overline \Xi_{\bar{c}}^- $&
$X_{d\bar{s}\bar{u}}^{-}\to   \Sigma^{\prime-}  \overline \Xi_{\bar{c}}^0 $\\
$X_{d\bar{s}\bar{u}}^{-}\to   \Delta^{+}  \overline \Sigma_{\bar{c}}^{--} $&
$X_{d\bar{s}\bar{u}}^{-}\to   \Delta^{0}  \overline \Sigma_{\bar{c}}^- $&
$X_{d\bar{s}\bar{u}}^{-}\to   \Delta^{-}  \overline \Sigma_{\bar{c}}^{0} $&
$X_{d\bar{s}\bar{u}}^{-}\to   \Sigma^{\prime0}  \overline \Xi_{\bar{c}}^{\prime-} $&
$X_{d\bar{s}\bar{u}}^{-}\to   \Sigma^{\prime-}  \overline \Xi_{\bar{c}}^{\prime0} $\\
$X_{d\bar{s}\bar{u}}^{-}\to   \Xi^{\prime-}  \overline \Omega_{\bar{c}}^{0} $&
& & &\\
\hline

$X_{s\bar{u}\bar{d}}^{-}\to   \Omega_{c}^{0}  \overline \Lambda_{\bar{c}}^- $&
$X_{s\bar{u}\bar{d}}^{-}\to   \Xi_c^0  \overline p $&
$X_{s\bar{u}\bar{d}}^{-}\to   \Xi_{c}^{\prime0}  \overline p $&
$X_{s\bar{u}\bar{d}}^{-}\to   \Omega_{c}^{0}  \overline \Sigma^- $&
$X_{s\bar{u}\bar{d}}^{-}\to   \Xi_c^+  \overline \Delta^{--} $\\
$X_{s\bar{u}\bar{d}}^{-}\to   \Xi_c^0  \overline \Delta^{-} $&
$X_{s\bar{u}\bar{d}}^{-}\to   \Xi_{c}^{\prime+}  \overline \Delta^{--} $&
$X_{s\bar{u}\bar{d}}^{-}\to   \Omega_{c}^{0}  \overline \Sigma^{\prime-} $&
$X_{s\bar{u}\bar{d}}^{-}\to   \Xi^0  \overline \Lambda_{\bar{c}}^- $&
$X_{s\bar{u}\bar{d}}^{-}\to   \Xi^-  \overline \Sigma_{\bar{c}}^{0} $\\
$X_{s\bar{u}\bar{d}}^{-}\to   \Xi^0  \overline \Sigma_{\bar{c}}^- $&
$X_{s\bar{u}\bar{d}}^{-}\to   \Xi^{\prime0}  \overline \Lambda_{\bar{c}}^- $&
$X_{s\bar{u}\bar{d}}^{-}\to   \Omega^-  \overline \Xi_{\bar{c}}^0 $&
$X_{s\bar{u}\bar{d}}^{-}\to   \Xi^{\prime0}  \overline \Sigma_{\bar{c}}^- $&
$X_{s\bar{u}\bar{d}}^{-}\to   \Xi^{\prime-}  \overline \Sigma_{\bar{c}}^{0} $\\
$X_{s\bar{u}\bar{d}}^{-}\to   \Omega^-  \overline \Xi_{\bar{c}}^{\prime0} $ & &&&\\
\hline

$Y_{(u\bar{u},s\bar{s})\bar{d}}^{0}\to   \Xi_c^+  \overline \Lambda_{\bar{c}}^- $&
$Y_{(u\bar{u},s\bar{s})\bar{d}}^{0}\to   \Xi_{c}^{\prime+}  \overline \Lambda_{\bar{c}}^- $&
$Y_{(u\bar{u},s\bar{s})\bar{d}}^{0}\to   \Omega_{c}^{0}  \overline \Xi_{\bar{c}}^0 $&
$Y_{(u\bar{u},s\bar{s})\bar{d}}^{0}\to   \Xi_c^+  \overline \Sigma_{\bar{c}}^- $&
$Y_{(u\bar{u},s\bar{s})\bar{d}}^{0}\to   \Xi_c^0  \overline \Sigma_{\bar{c}}^{0} $\\
$Y_{(u\bar{u},s\bar{s})\bar{d}}^{0}\to   \Xi_{c}^{\prime+}  \overline \Sigma_{\bar{c}}^- $&
$Y_{(u\bar{u},s\bar{s})\bar{d}}^{0}\to   \Xi_{c}^{\prime0}  \overline \Sigma_{\bar{c}}^{0} $&
$Y_{(u\bar{u},s\bar{s})\bar{d}}^{0}\to   \Omega_{c}^{0}  \overline \Xi_{\bar{c}}^{\prime0} $&
$Y_{(u\bar{u},s\bar{s})\bar{d}}^{0}\to   \Lambda_c^+  \overline p $&
$Y_{(u\bar{u},s\bar{s})\bar{d}}^{0}\to   \Xi_c^+  \overline \Sigma^- $\\
$Y_{(u\bar{u},s\bar{s})\bar{d}}^{0}\to   \Xi_c^0  \overline \Sigma^0 $&
$Y_{(u\bar{u},s\bar{s})\bar{d}}^{0}\to   \Sigma_{c}^{+}  \overline p $&
$Y_{(u\bar{u},s\bar{s})\bar{d}}^{0}\to   \Xi_{c}^{\prime+}  \overline \Sigma^- $&
$Y_{(u\bar{u},s\bar{s})\bar{d}}^{0}\to   \Xi_{c}^{\prime0}  \overline \Lambda^0 $&
$Y_{(u\bar{u},s\bar{s})\bar{d}}^{0}\to   \Xi_{c}^{\prime0}  \overline \Sigma^0 $\\
$Y_{(u\bar{u},s\bar{s})\bar{d}}^{0}\to   \Omega_{c}^{0}  \overline \Xi^0 $&
$Y_{(u\bar{u},s\bar{s})\bar{d}}^{0}\to   \Lambda_c^+  \overline \Delta^{-} $&
$Y_{(u\bar{u},s\bar{s})\bar{d}}^{0}\to   \Xi_c^+  \overline \Sigma^{\prime-} $&
$Y_{(u\bar{u},s\bar{s})\bar{d}}^{0}\to   \Xi_c^0  \overline \Sigma^{\prime0} $&
$Y_{(u\bar{u},s\bar{s})\bar{d}}^{0}\to   \Sigma_{c}^{++}  \overline \Delta^{--} $\\
$Y_{(u\bar{u},s\bar{s})\bar{d}}^{0}\to   \Sigma_{c}^{+}  \overline \Delta^{-} $&
$Y_{(u\bar{u},s\bar{s})\bar{d}}^{0}\to   \Sigma_{c}^{0}  \overline \Delta^{0} $&
$Y_{(u\bar{u},s\bar{s})\bar{d}}^{0}\to   \Xi_{c}^{\prime+}  \overline \Sigma^{\prime-} $&
$Y_{(u\bar{u},s\bar{s})\bar{d}}^{0}\to   \Xi_{c}^{\prime0}  \overline \Sigma^{\prime0} $&
$Y_{(u\bar{u},s\bar{s})\bar{d}}^{0}\to   \Omega_{c}^{0}  \overline \Xi^{\prime0} $\\
$Y_{(u\bar{u},s\bar{s})\bar{d}}^{0}\to   \Sigma^+  \overline \Lambda_{\bar{c}}^- $&
$Y_{(u\bar{u},s\bar{s})\bar{d}}^{0}\to   \Lambda^0  \overline \Sigma_{\bar{c}}^{0} $&
$Y_{(u\bar{u},s\bar{s})\bar{d}}^{0}\to   \Sigma^+  \overline \Sigma_{\bar{c}}^- $&
$Y_{(u\bar{u},s\bar{s})\bar{d}}^{0}\to   \Sigma^0  \overline \Sigma_{\bar{c}}^{0} $&
$Y_{(u\bar{u},s\bar{s})\bar{d}}^{0}\to   \Xi^0  \overline \Xi_{\bar{c}}^{\prime0} $\\
$Y_{(u\bar{u},s\bar{s})\bar{d}}^{0}\to   \Sigma^{\prime+}  \overline \Lambda_{\bar{c}}^- $&
$Y_{(u\bar{u},s\bar{s})\bar{d}}^{0}\to   \Xi^{\prime0}  \overline \Xi_{\bar{c}}^0 $&
$Y_{(u\bar{u},s\bar{s})\bar{d}}^{0}\to   \Sigma^{\prime+}  \overline \Sigma_{\bar{c}}^- $&
$Y_{(u\bar{u},s\bar{s})\bar{d}}^{0}\to   \Sigma^{\prime0}  \overline \Sigma_{\bar{c}}^{0} $&
$Y_{(u\bar{u},s\bar{s})\bar{d}}^{0}\to   \Xi^{\prime0}  \overline \Xi_{\bar{c}}^{\prime0} $ \\
\hline

$Y_{(u\bar{u},d\bar{d})\bar{s}}^{0}\to   \Xi_c^+  \overline \Xi_{\bar{c}}^- $&
$Y_{(u\bar{u},d\bar{d})\bar{s}}^{0}\to   \Xi_c^0  \overline \Xi_{\bar{c}}^0 $&
$Y_{(u\bar{u},d\bar{d})\bar{s}}^{0}\to   \Sigma_{c}^{+}  \overline \Lambda_{\bar{c}}^- $&
$Y_{(u\bar{u},d\bar{d})\bar{s}}^{0}\to   \Xi_{c}^{\prime+}  \overline \Xi_{\bar{c}}^- $&
$Y_{(u\bar{u},d\bar{d})\bar{s}}^{0}\to   \Xi_{c}^{\prime0}  \overline \Xi_{\bar{c}}^0 $\\
$Y_{(u\bar{u},d\bar{d})\bar{s}}^{0}\to   \Lambda_c^+  \overline \Sigma_{\bar{c}}^- $&
$Y_{(u\bar{u},d\bar{d})\bar{s}}^{0}\to   \Xi_c^+  \overline \Xi_{\bar{c}}^{\prime-} $&
$Y_{(u\bar{u},d\bar{d})\bar{s}}^{0}\to   \Xi_c^0  \overline \Xi_{\bar{c}}^{\prime0} $&
$Y_{(u\bar{u},d\bar{d})\bar{s}}^{0}\to   \Sigma_{c}^{++}  \overline \Sigma_{\bar{c}}^{--} $&
$Y_{(u\bar{u},d\bar{d})\bar{s}}^{0}\to   \Sigma_{c}^{0}  \overline \Sigma_{\bar{c}}^{0} $\\
$Y_{(u\bar{u},d\bar{d})\bar{s}}^{0}\to   \Xi_{c}^{\prime+}  \overline \Xi_{\bar{c}}^{\prime-} $&
$Y_{(u\bar{u},d\bar{d})\bar{s}}^{0}\to   \Xi_{c}^{\prime0}  \overline \Xi_{\bar{c}}^{\prime0} $&
$Y_{(u\bar{u},d\bar{d})\bar{s}}^{0}\to   \Lambda_c^+  \overline \Sigma^- $&
$Y_{(u\bar{u},d\bar{d})\bar{s}}^{0}\to   \Xi_c^0  \overline \Xi^0 $&
$Y_{(u\bar{u},d\bar{d})\bar{s}}^{0}\to   \Sigma_{c}^{+}  \overline \Sigma^- $\\
$Y_{(u\bar{u},d\bar{d})\bar{s}}^{0}\to   \Sigma_{c}^{0}  \overline \Lambda^0 $&
$Y_{(u\bar{u},d\bar{d})\bar{s}}^{0}\to   \Sigma_{c}^{0}  \overline \Sigma^0 $&
$Y_{(u\bar{u},d\bar{d})\bar{s}}^{0}\to   \Xi_{c}^{\prime0}  \overline \Xi^0 $&
$Y_{(u\bar{u},d\bar{d})\bar{s}}^{0}\to   \Lambda_c^+  \overline \Sigma^{\prime-} $&
$Y_{(u\bar{u},d\bar{d})\bar{s}}^{0}\to   \Xi_c^0  \overline \Xi^{\prime0} $\\
$Y_{(u\bar{u},d\bar{d})\bar{s}}^{0}\to   \Sigma_{c}^{+}  \overline \Sigma^{\prime-} $&
$Y_{(u\bar{u},d\bar{d})\bar{s}}^{0}\to   \Sigma_{c}^{0}  \overline \Sigma^{\prime0} $&
$Y_{(u\bar{u},d\bar{d})\bar{s}}^{0}\to   \Xi_{c}^{\prime0}  \overline \Xi^{\prime0} $&
$Y_{(u\bar{u},d\bar{d})\bar{s}}^{0}\to   \Lambda^0  \overline \Xi_{\bar{c}}^0 $&
$Y_{(u\bar{u},d\bar{d})\bar{s}}^{0}\to   \Sigma^+  \overline \Xi_{\bar{c}}^- $\\
$Y_{(u\bar{u},d\bar{d})\bar{s}}^{0}\to   \Sigma^0  \overline \Xi_{\bar{c}}^0 $&
$Y_{(u\bar{u},d\bar{d})\bar{s}}^{0}\to   {p}  \overline \Lambda_{\bar{c}}^- $&
$Y_{(u\bar{u},d\bar{d})\bar{s}}^{0}\to   \Lambda^0  \overline \Xi_{\bar{c}}^{\prime0} $&
$Y_{(u\bar{u},d\bar{d})\bar{s}}^{0}\to   \Sigma^+  \overline \Xi_{\bar{c}}^{\prime-} $&
$Y_{(u\bar{u},d\bar{d})\bar{s}}^{0}\to   \Sigma^0  \overline \Xi_{\bar{c}}^{\prime0} $\\
$Y_{(u\bar{u},d\bar{d})\bar{s}}^{0}\to   {p}  \overline \Sigma_{\bar{c}}^- $&
$Y_{(u\bar{u},d\bar{d})\bar{s}}^{0}\to   \Xi^0  \overline \Omega_{\bar{c}}^{0} $&
$Y_{(u\bar{u},d\bar{d})\bar{s}}^{0}\to   \Delta^{+}  \overline \Lambda_{\bar{c}}^- $&
$Y_{(u\bar{u},d\bar{d})\bar{s}}^{0}\to   \Sigma^{\prime+}  \overline \Xi_{\bar{c}}^- $&
$Y_{(u\bar{u},d\bar{d})\bar{s}}^{0}\to   \Sigma^{\prime0}  \overline \Xi_{\bar{c}}^0 $\\
$Y_{(u\bar{u},d\bar{d})\bar{s}}^{0}\to   \Delta^{++}  \overline \Sigma_{\bar{c}}^{--} $&
$Y_{(u\bar{u},d\bar{d})\bar{s}}^{0}\to   \Delta^{+}  \overline \Sigma_{\bar{c}}^- $&
$Y_{(u\bar{u},d\bar{d})\bar{s}}^{0}\to   \Delta^{0}  \overline \Sigma_{\bar{c}}^{0} $&
$Y_{(u\bar{u},d\bar{d})\bar{s}}^{0}\to   \Sigma^{\prime+}  \overline \Xi_{\bar{c}}^{\prime-} $&
$Y_{(u\bar{u},d\bar{d})\bar{s}}^{0}\to   \Sigma^{\prime0}  \overline \Xi_{\bar{c}}^{\prime0} $\\
$Y_{(u\bar{u},d\bar{d})\bar{s}}^{0}\to   \Xi^{\prime0}  \overline \Omega_{\bar{c}}^{0} $& & &
&\\
\hline

$Y_{(d\bar{d},s\bar{s})\bar{u}}^{-}\to   \Xi_c^0  \overline \Lambda_{\bar{c}}^- $&
$Y_{(d\bar{d},s\bar{s})\bar{u}}^{-}\to   \Xi_{c}^{\prime0}  \overline \Lambda_{\bar{c}}^- $&
$Y_{(d\bar{d},s\bar{s})\bar{u}}^{-}\to   \Omega_{c}^{0}  \overline \Xi_{\bar{c}}^- $&
$Y_{(d\bar{d},s\bar{s})\bar{u}}^{-}\to   \Xi_c^+  \overline \Sigma_{\bar{c}}^{--} $&
$Y_{(d\bar{d},s\bar{s})\bar{u}}^{-}\to   \Xi_c^0  \overline \Sigma_{\bar{c}}^- $\\
$Y_{(d\bar{d},s\bar{s})\bar{u}}^{-}\to   \Xi_{c}^{\prime+}  \overline \Sigma_{\bar{c}}^{--} $&
$Y_{(d\bar{d},s\bar{s})\bar{u}}^{-}\to   \Xi_{c}^{\prime0}  \overline \Sigma_{\bar{c}}^- $&
$Y_{(d\bar{d},s\bar{s})\bar{u}}^{-}\to   \Omega_{c}^{0}  \overline \Xi_{\bar{c}}^{\prime-} $&
$Y_{(d\bar{d},s\bar{s})\bar{u}}^{-}\to   \Xi_c^0  \overline \Sigma^- $&
$Y_{(d\bar{d},s\bar{s})\bar{u}}^{-}\to   \Sigma_{c}^{0}  \overline p $\\
$Y_{(d\bar{d},s\bar{s})\bar{u}}^{-}\to   \Xi_{c}^{\prime0}  \overline \Sigma^- $&
$Y_{(d\bar{d},s\bar{s})\bar{u}}^{-}\to   \Lambda_c^+  \overline \Delta^{--} $&
$Y_{(d\bar{d},s\bar{s})\bar{u}}^{-}\to   \Xi_c^0  \overline \Sigma^{\prime-} $&
$Y_{(d\bar{d},s\bar{s})\bar{u}}^{-}\to   \Sigma_{c}^{+}  \overline \Delta^{--} $&
$Y_{(d\bar{d},s\bar{s})\bar{u}}^{-}\to   \Sigma_{c}^{0}  \overline \Delta^{-} $\\
$Y_{(d\bar{d},s\bar{s})\bar{u}}^{-}\to   \Xi_{c}^{\prime0}  \overline \Sigma^{\prime-} $&
$Y_{(d\bar{d},s\bar{s})\bar{u}}^{-}\to   \Lambda^0  \overline \Lambda_{\bar{c}}^- $&
$Y_{(d\bar{d},s\bar{s})\bar{u}}^{-}\to   \Sigma^0  \overline \Lambda_{\bar{c}}^- $&
$Y_{(d\bar{d},s\bar{s})\bar{u}}^{-}\to   \Xi^0  \overline \Xi_{\bar{c}}^- $&
$Y_{(d\bar{d},s\bar{s})\bar{u}}^{-}\to   \Lambda^0  \overline \Sigma_{\bar{c}}^- $\\
$Y_{(d\bar{d},s\bar{s})\bar{u}}^{-}\to   \Sigma^+  \overline \Sigma_{\bar{c}}^{--} $&
$Y_{(d\bar{d},s\bar{s})\bar{u}}^{-}\to   \Sigma^0  \overline \Sigma_{\bar{c}}^- $&
$Y_{(d\bar{d},s\bar{s})\bar{u}}^{-}\to   \Sigma^-  \overline \Sigma_{\bar{c}}^{0} $&
$Y_{(d\bar{d},s\bar{s})\bar{u}}^{-}\to   \Xi^-  \overline \Xi_{\bar{c}}^{\prime0} $&
$Y_{(d\bar{d},s\bar{s})\bar{u}}^{-}\to   \Xi^0  \overline \Xi_{\bar{c}}^{\prime-} $\\
$Y_{(d\bar{d},s\bar{s})\bar{u}}^{-}\to   \Sigma^{\prime0}  \overline \Lambda_{\bar{c}}^- $&
$Y_{(d\bar{d},s\bar{s})\bar{u}}^{-}\to   \Xi^{\prime0}  \overline \Xi_{\bar{c}}^- $&
$Y_{(d\bar{d},s\bar{s})\bar{u}}^{-}\to   \Xi^{\prime-}  \overline \Xi_{\bar{c}}^0 $&
$Y_{(d\bar{d},s\bar{s})\bar{u}}^{-}\to   \Sigma^{\prime+}  \overline \Sigma_{\bar{c}}^{--} $&
$Y_{(d\bar{d},s\bar{s})\bar{u}}^{-}\to   \Sigma^{\prime0}  \overline \Sigma_{\bar{c}}^- $\\
$Y_{(d\bar{d},s\bar{s})\bar{u}}^{-}\to   \Sigma^{\prime-}  \overline \Sigma_{\bar{c}}^{0} $&
$Y_{(d\bar{d},s\bar{s})\bar{u}}^{-}\to   \Xi^{\prime0}  \overline \Xi_{\bar{c}}^{\prime-} $&
$Y_{(d\bar{d},s\bar{s})\bar{u}}^{-}\to   \Xi^{\prime-}  \overline \Xi_{\bar{c}}^{\prime0} $&
$Y_{(d\bar{d},s\bar{s})\bar{u}}^{-}\to   \Omega^-  \overline \Omega_{\bar{c}}^{0} $&  \\
\hline

\end{tabular}
\end{table}
To reconstruct the $X_{b6}$ via the decays into mesons, we collect the golden channels  in Tab.~\ref{tab:Xb6_golden_meson}, while we collect the golden channels for the $X_{b6}$ via the decays into baryons in Tab.~\ref{tab:Xb6_golden_baryon}. The typical branching fraction of stable $X_{b6}$ decays is the order of $10^{-3}$. To construct the final states such as $J/\psi$, $D$ and charmed baryons, another factor with the order of $10^{-2}$ is needed. Therefore, the largest branching rates of stable $X_{b6}$ tetraquark decays to the final detected particles may be around $10^{-5}$.


\section{Conclusions}
\label{sec:conclusions}
We studied the weak decay properties of open-bottom tetraquark $X_{b6}$ states under the SU(3) flavor symmetry. The states with the quark component $Qq_i \bar {q_j}\bar {q_k}$ can form $\bar 3$, $6$ and $\overline {15}$ representation by the decomposition of $3\bigotimes \bar 3\bigotimes\bar 3=\bar 3\bigoplus\bar 3\bigoplus 6\bigoplus \overline {15}$.  But only the sextet $X_{b6}$ states may be stable.
We focused on semi-leptonic and non-leptonic weak decays of the ground states of sextet representations whose masses are below the thresholds of strong and electromagnetic  decays. Their decay amplitudes were discussed  by constructing the relevant Hamiltonian at the hadronic level and parameterizing the interactions into some constants ($a_{i},b_{j},...$). It is easily to obtain  the relations of different channels when we ignore the small effect of phase space. We have given the Cabibbo allowed two-body and three-body decay channels, which shall play an important role to hunting for the stable open-bottom tetraquark $X_{b6}$ states.

\section*{Acknowledgments}
We thank Prof. Wei Wang for useful discussions  and the collaboration at the early stage of this work.
 This work was
supported in part by the National Natural Science Foundation of
China under Grant No.~11575110, 11655002, 11705092 and U1732101, by Natural
Science Foundation of Shanghai under Grant No.~15DZ2272100 and
No.~15ZR1423100,  by Natural Science Foundation of Jiangsu under
Grant No.~BK20171471, by Natural Science Foundation of Gansu under Grant No.~18JR3RA265, by the Young Thousand Talents Plan,   by Key
Laboratory for Particle Physics, Astrophysics and Cosmology,
Ministry of Education.

\end{document}